%% file: main.tex
\documentclass[final,5p,times,twocolumn,authoryear]{elsarticle}

\usepackage{amssymb}
\usepackage{amsmath}
\usepackage{lipsum}
\usepackage{comment}
\usepackage{soul}
\usepackage{aas_macros}

\usepackage{xcolor}
\usepackage{hyperref}

\journal{New Astronomy Reviews}

\begin{document}

\begin{frontmatter}



\title{What drives the growth of black holes: a decade of progress}

\author[CEA]{D.~M.~Alexander}
\author[Dartmouth]{R.~C.~Hickox}
\author[IfA]{J.~Aird}
\author[Obs_Paris]{F.~Combes}
\author[Newcastle]{T.~Costa}
\author[Geneva,MPE]{M.~Habouzit}
\author[Newcastle]{C.~M.~Harrison}
\author[Edinburgh]{R.~I.~Leng}
\author[CEA,ICC]{L.~K.~Morabito}
\author[Dur_Phil]{S.~L.~Uckelman}
\author[Dur_Phil]{P.~Vickers}

\affiliation[CEA]{organization={Centre for Extragalactic Astronomy, Department of Physics. Durham University},
             addressline={South Road},
             city={Durham},
             postcode={DH1 3LE},
             country={UK}} 
\affiliation[Dartmouth]{organization={Department of Physics and Astronomy, Dartmouth College},
             addressline={6127 Wilder Laboratory},
             city={Hanover},
             postcode={NH 03755},
             country={USA}}
\affiliation[IfA]{organization={Institute for Astronomy, University of Edinburgh},
             addressline={Royal Observatory},
             city={Edinburgh},
             postcode={EH9 3HJ},
             country={UK}}
\affiliation[Obs_Paris]{organization={Observatoire de Paris, LERMA, Collège de France, CNRS, PSL University, Sorbonne University},
             city={Paris},
             postcode={F-75014},
             country={France}}
\affiliation[Newcastle]{organization={School of Mathematics, Statistics and Physics, Newcastle University},
             addressline={},
             city={Newcastle upon Tyne},
             postcode={NE1 7RU},
             country={UK}} 
\affiliation[Geneva]{organization={Department of Astronomy, University of Geneva},
             addressline={Ch. d'Ecogia 16},
             city={Versoix},
             postcode={CH-1290},
             country={Switzerland}}
\affiliation[MPE]{organization={Max-Planck-Institut für Astronomie},
             addressline={Königstuhl 17},
             city={Heidelberg},
             postcode={D-69117},
             country={Germany}}
\affiliation[Edinburgh]{organization={School of Social Science and Political Science, University of Edinburgh},
             addressline={15a George Square},
             city={Edinburgh},
             postcode={EH8 9LD},
             country={UK}}
\affiliation[ICC]{organization={Institute of Computational Cosmology, Department of Physics, Durham University},
             addressline={South Road},
             city={Durham},
             postcode={DH1 3LE},
             country={UK}}
\affiliation[Dur_Phil]{organization={Department of Philosophy, Durham University},
             addressline={50 Old Elvet},
             city={Durham},
             postcode={DH1 3HN},
             country={UK}}
             


\begin{abstract}

\noindent The last decade has witnessed significant progress in our understanding of the growth of super-massive black holes (SMBHs). It is now clear that an Active Galactic Nucleus (AGN: the observed manifestation of a growing SMBH) is an ``event" within the broader lifecycle of a galaxy, which can significantly influence the shape and evolution of the galaxy itself. Our view of the obscuring medium that affects the observed properties of an AGN has also undergone a revolution, and we now have a more physical understanding of the connection between the fuelling of (and feedback from) the SMBH and the broader host-galaxy and larger-scale environment. We have a greater understanding of the physics of SMBH accretion, can identify AGNs out to $z=$~8--10 witnessing the very earliest phases of SMBH growth, and have a more complete census of AGN activity than ever before. This great progress has been enabled by new innovative facilities, an ever-increasing quantity of multi-wavelength data, the exploitation and development of new techniques, and greater community-wide engagement. In this article we review our understanding of AGNs and the growth of SMBHs, providing an update of the earlier \cite{AH12} review. Using citation-network analyses we also show where this review fits within the broader black-hole research literature and, adopting the previous article as a snapshot of the field over a decade ago, identify the drivers that have enabled the greatest scientific progress.

\end{abstract}




\end{frontmatter}




\section{Introduction}
\label{introduction}


Once considered exceptional events occuring within the centres of galaxies, Active Galactic Nuclei (AGNs) are now understood to be an integral component in the formation and evolution of galaxies. AGN activity is driven by the accretion of gas onto a Super-Massive Black Hole \citep[SMBH;][]{Shakura1973,Soltan1982,Rees1984}, which gives rise to a broad multi-wavelength spectral energy distribution (SED) extending from radio to X-ray 
\citep[and gamma rays in some cases; e.g.,][]{Elvis1994,padovani_active_2017}. During this mass-accretion phase, a substantial fraction of the mass and energy can also be injected into the broader host-galaxy environment (and even beyond) in the form of radiation, accretion-disk driven winds, and energetic jets, allowing the AGN to influence the growth of the galaxy \citep[e.g.,\ ``AGN feedback";][]{Fabian:12,King:15,Harrison:17}.

\begin{figure*}[t]
	\centering 
    \includegraphics[width=0.99\textwidth, angle=0]{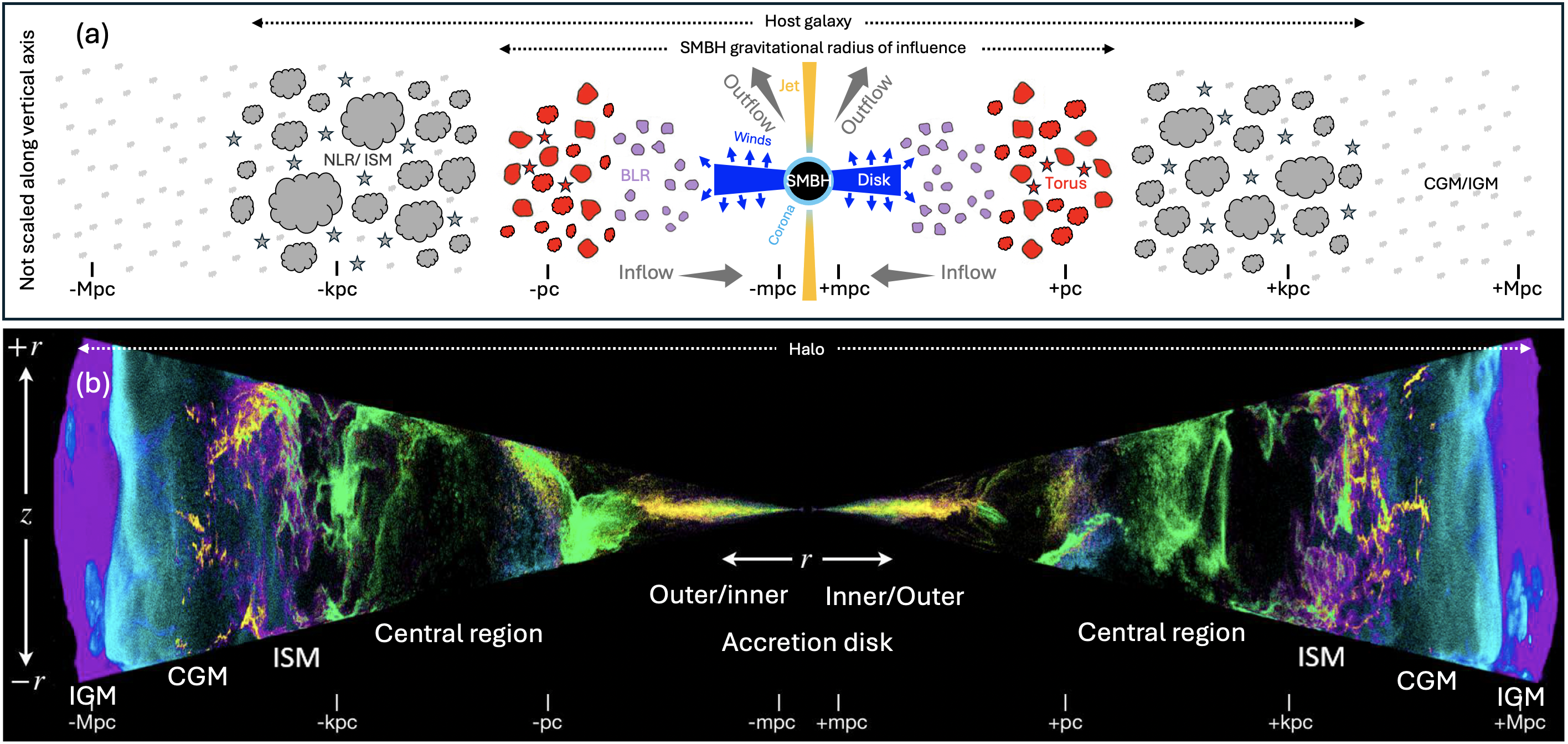}	
	\caption{(a) Simple schematic representation of an AGN within the broader host-galaxy and halo environment over (logarithmic) milli-parsec to mega-parcsec scales in the horizontal plane; note the schematics are not scaled in the vertical plane. Key regions are highlighted including those under the gravitational influence of the SMBH (the AGN: accretion disk, X-ray ``corona", potential winds/jets launched from the accretion disk, broad-line region (BLR), and the dusty molecular ``torus"), the host galaxy, and halo environment. The AGN is aligned with the host galaxy for presentation purposes; in reality since it is under the gravitational influence of the SMBH, it can be randomly orientated with respect to the host galaxy. The general schematic style is inspired by \cite{Ramos-Almeida:17} but using a colour scheme to match the SED components in Fig.~\ref{fig:SED}. (b) Multiphase gas cross-section simulation over (logarithmic) milli-parsec to mega-parcsec scales. The $r$ coordinate is defined from the centre of the SMBH and the cross section is displayed to cover the range $-r$ to $+r$ in the $z$ direction. Colours denote the different gas-phase temperatures: T $<$ 10$^3$ K (green), 10$^3$ $<$ T $<$ 10$^4$ K (yellow), 10$^4$ $<$ T $<$ 10$^5$ K (magenta), 10$^5$ $<$ T $<$ 10$^6$ K (purple), T $>$ 10$^6$ K (cyan). The various regions have been highlighted: IGM (intergalactic medium), CGM (circumgalactic medium), ISM (interstellar medium), central kpc region of the galaxy, and the outer and inner regions of the accretion disk. See \cite{Harrison2024} for a complementary schematic representation of the gas conditions over the same size scales. {\it Source:} panel b adapted from Fig.~9 of \cite{Hopkins2024}.}
	\label{fig:agn_model}
\end{figure*}

\begin{figure}[h]
	\centering 
	\includegraphics[width=0.5\textwidth, angle=0]{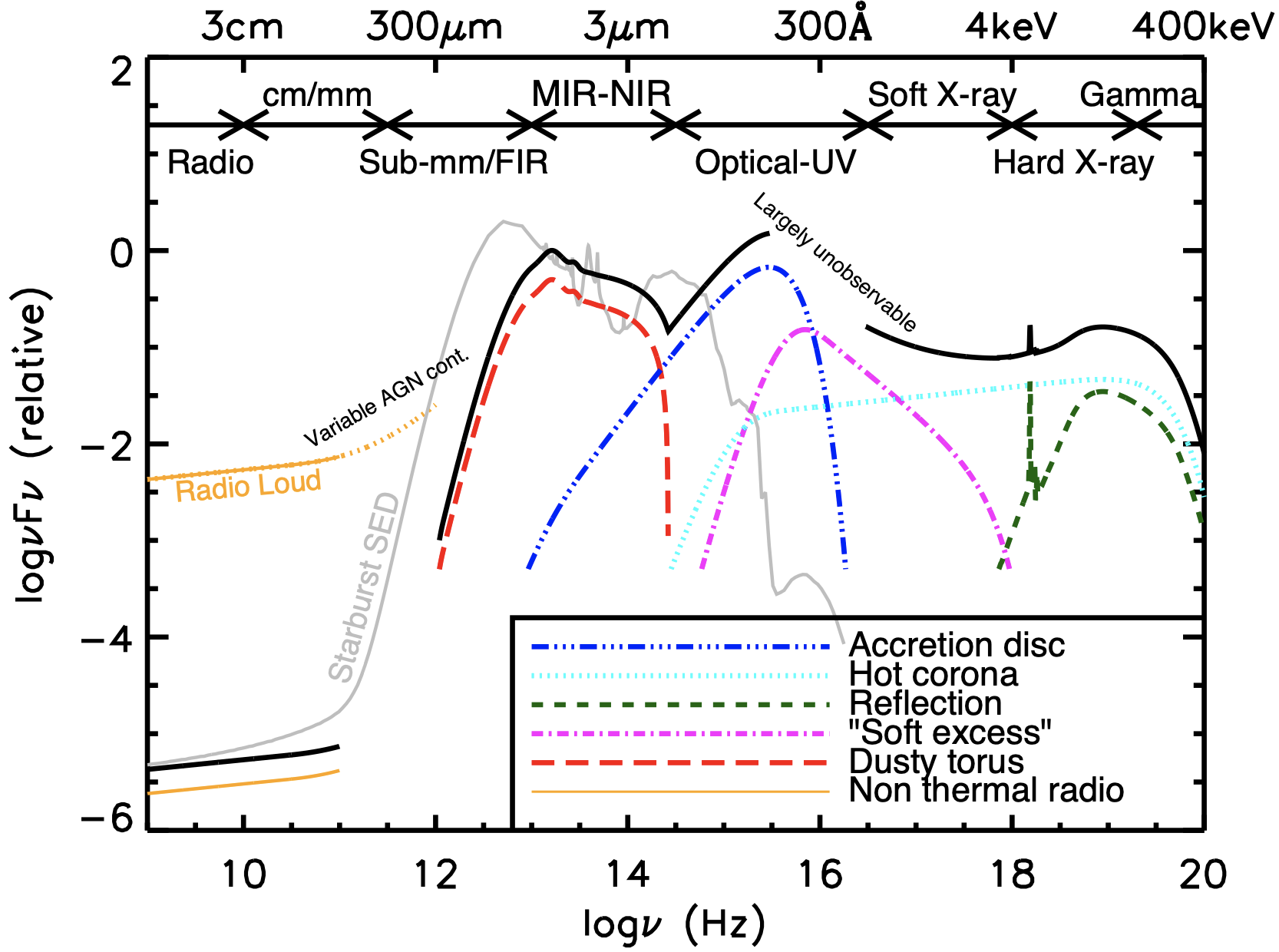}	
	\caption{Broad-band SED of an unobscured AGN (black curve) across the broad gamma ray--radio waveband: the frequency and wavelength ranges for individual wavebands are highlighted. Individual continuum-emitting components are coloured and labelled including emission from a star-forming component (grey curve) and the brighter radio emission from a radio-loud AGN (labelled yellow curve); key emission regions are schematically represented in Fig.~\ref{fig:agn_model}a. {\it Source:} reproduced from Fig.~1.2 of \cite{Harrison:PhD}.} 
	\label{fig:SED}
\end{figure}

The majority of the gas that gives rise to this AGN activity is, at least initially, distributed over the broader host-galaxy environment, which is orders of magnitude larger than the SMBH. For example, for a non-rotating $10^{8}$~$M_{\odot}$ SMBH, the event horizon ($R_{\rm S}$; $\approx$~3~$\times$~10$^{-5}$~pc) and the radius of gravitational influence ($R_{\rm RoI}$; $\approx$~10~pc) are $\approx$~9 and $\approx$~3 orders of magnitude smaller than the host galaxy, respectively.\footnote{The radius of the event horizon of a non-rotating SMBH is calculated as $R_{\rm S} = 2 G M_{\rm BH}/c^2$ (i.e.,\ the Schwarzschild radius) and the radius of the gravitational influence of a SMBH is calculated as $R_{\rm RoI} = G M_{\rm BH}/{\sigma}^2$, where $M_{\rm BH}$ is the SMBH mass and $\sigma$ is the velocity dispersion of the host galaxy.\label{foot:SMBHrad}} To overcome this vast difference in size scale the gas needs to lose angular momentum through gravitational torques from stars either within the galaxy or external to the galaxy \citep[e.g.,\ from galaxy mergers and interactions;][]{Jogee2006,Hopkins2010_gas_feed}. Once the gas reaches $R_{\rm RoI}$, the SMBH dominates the gravitational potential and, along with radiation and magnetic pressure, gas inflow and outflow, defines the key features of the standard AGN model \citep[e.g.,\ the dusty ``torus"; broad-line region; accretion disk;][]{Netzer2015,Ramos-Almeida:17,hickox_obscured_2018}. See Fig.~\ref{fig:agn_model} for a schematic representation of the key components of an AGN within the host galaxy environment and Fig.~\ref{fig:SED} for colour-coded broad-band SEDs of these components and the radio--gamma ray waveband definitions. 

Despite this angular-momentum challenge, we know that AGN ``events" occur regularly in many (potentially all) galaxies. Key evidence for this view is (1) the identification (or inference) of a dormant SMBH at the centres of most galaxies, and (2) the remarkable relationship found between the mass of the SMBH and the galaxy bulge in nearby galaxies, which provides both the archaeological evidence for past phases of AGN activity and infers a connection between the growth of the SMBH and the host galaxy \citep[e.g.,][]{AH12,Kormendy:13,Heckman:14}.

The aims of this article are twofold. The first, and primary, objective is to review our understanding of AGN to address the very broad question: what drives the growth of black holes? This article provides an update of our previous \cite{AH12} review (hereafter AH12) and describes the exciting scientific discoveries and progress made over the last decade. The second objective is to investigate the drivers of this scientific progress, using the AH12 review as a ``snapshot'' of the research field from over a decade ago. This second objective provides interesting insight on how science progresses within a rapidly advancing research discipline.  To enable the closest tracking of scientific progress in this article, we have used the same scientific sections as those adopted in AH12.

From a very broad perspective, the last decade has witnessed major discoveries in the field of AGN and black holes. The Event Horizon Telescope \citep[EHT;][]{EHT:19_facility} has used a global (world wide) very long baseline interferometer to take the first ``direct'' image of a SMBH at millimetre wavelengths in two systems: the nearby massive galaxy M87 and Sgr~A$*$ at the centre of our Milky Way galaxy \citep{EHT:19_M87,EHT:22_Sgr}. The SMBHs are identified from the silhouette or shadow due to them lying in the foreground of synchrotron emission produced in the vicinity of the SMBH. From the shape and size of the SMBH shadow, key fundamental measurements can be inferred (e.g.,\ the mass and spin of the SMBH), even testing general relativity, and complementing other on-going experiments 
\citep[see][for a review]{Genzel:24}. 

The whole of physics, not just astronomy, has been transformed by the first significant detections of gravitational waves \citep[][]{LIGO:15,Abbott:16}. The gravitational waves are produced from the energy losses of two compact objects (i.e.,\ black holes or neutron stars) in a tightly bound orbit and peak just prior to the objects coalescing \citep[e.g.,][]{Blanchet2014_gravwave}. The current gravitational-wave observatories are most sensitive to stellar-mass objects and have detected gravitational waves from $\approx$~100 merging systems with masses up-to $\approx$~150~$M_{\odot}$ \citep{Abbott:20,Abbott:23}. Pulsar timing arrays (PTAs) are providing the first gravitational-wave constraints from SMBH mergers and have identified a gravitational-wave background believed to be produced by merging SMBHs \citep[e.g.][]{EPTA2023,Nanograv2023,Reardon2023_GWB,Xu2023_GWB}. The first direct identification of merging SMBHs is expected in the next decade with launch of LISA \citep[Laser Inteferometer Space Antenna;][]{LISA2017}.

Large field of view and all-sky monitoring observatories \citep[e.g.,\ the {\it Swift} high-energy observatory;][]{gehrels_swift_2004}, allied to rapid response follow-up facilities, have greatly opened up the parameter space to identifying exceptional short-lived phenomena such as tidal disruption events \citep[TDEs; e.g.,][]{Komossa:15,Gezari_2021}, which are thought to be due to the accretion of stars in the immediate vicinity of a SMBH. The unrelenting progress in instrumentation and facility development over the last decade with larger field of view instruments, greater sensitivity, higher spatial, spectral, and temporal resolution, and greater multi-plexing has also provided us with a deeper panchromatic view of AGN and richer astrophysical insight. The last decade has also witnessed a huge increase in computing power allowing for more sophisticated and detailed simulations with greater spatial, temporal, and mass resolution over larger volumes with refined physical prescriptions.

A major motivation for this review was our ``What Drives the Growth of Black Holes: A Decade of Reflection'' workshop held in Reykjavík, Iceland, over 26$^{\rm th}$--30$^{\rm th}$ September 2022. This workshop was the culmination of a series of meetings organised over the last decade (see \ref{workshops} for details). The sessions of the Iceland workshop were arranged to be identical to those of our original ``What Drives the Growth of Black Holes?'' meeting from 2010, which was the inspiration for the AH12 review, to facilitate a more direct assessment of the growth of scientific progress. Briefly, the four scientific sessions of the workshop addressed the following key questions:

\begin{itemize}

\item How does the gas accrete onto black holes, from
kilo-parsec to sub-parsec scales?

\item What are the links between black-hole growth and
their host galaxies and large-scale environments?

\item What fuels the rapid growth of the most massive
(and also the first) black holes?

\item What is the detailed nature of AGN feedback and
its effects on black-hole fuelling and star formation?
    
\end{itemize}

In this article we first describe the approach we take in this review, including a brief primer on how we can measure scientific progress (\S\ref{sec:approach}). In the subsequent four sections (\S\ref{sec:section3}--\S\ref{star-formation}) we then address each of the key questions above, providing an overview of the research field and highlighting the significant areas of progress over the last decade. In \S\ref{sec:discprogress} we then briefly reflect on the progress made in each of these research fields, utilising citation network analyses to quantify the relationships between different sub fields, and comment on expected progress over the next decade. In \S\ref{sec:conclusion} we provide a brief summary and conclusions. As with all review articles, we are unable to assess every relevant paper in the scientific literature. Our aim is more to provide a clear overview of this broad research field and to provide a starting point for more detailed investigations.

\vspace{0.2cm}
\section{Approach taken in this review}\label{sec:approach}

In this section we first provide an overview of the broad black-hole (BH) research field (\S\ref{context}) and show where the AH12 review (and, by association, this review) sits within that broader research framework. We then ask the question what exactly \textit{is} scientific progress (\S\ref{progress}), and define practical categories that we can use to track and identify scientific progress along with examples of each category (\S\ref{tracking}; see Table~\ref{Table1}).

\subsection{Scientific context: the broader black-hole research field}\label{context}

To place this review and the AH12 review into context, we first consider the much broader field of BH research, covering not only AGN and growing SMBHs but also more fundamental theoretical BH research. To demonstrate this, we use an approach called citation network analysis, which applies graph theory and network science to bibliographic data to identify clusters of papers that are densely connected through citation links. As has been shown in prior research \citep{Klavans2017_CNA,Leng2021}, these clusters correspond to specific sub-fields and topics, reflecting the tendency for papers on a particular topic to cite other works in the same research area more frequently than those addressing different questions.

To construct the citation network covering black-hole research, we selected all peer-reviewed articles in indexed astrophysics journals from the Web of Science that contained terms indicating a focus on BHs or AGN/quasars, and which were published between 2003 and the first quarter of 2024. After parsing out the bibliographies of these papers, we constructed a network containing 60,924 papers connected together by 1,378,057 citation links. We then applied modularity maximisation to this network using the Leiden algorithm to detect clusters of densely interconnected papers \citep{Traag2019}: overall 13 clusters are identified. See \ref{CNA} for more details about the construction of the citation network.

\begin{figure}
	\centering 
	\includegraphics[width=0.5\textwidth, angle=0]{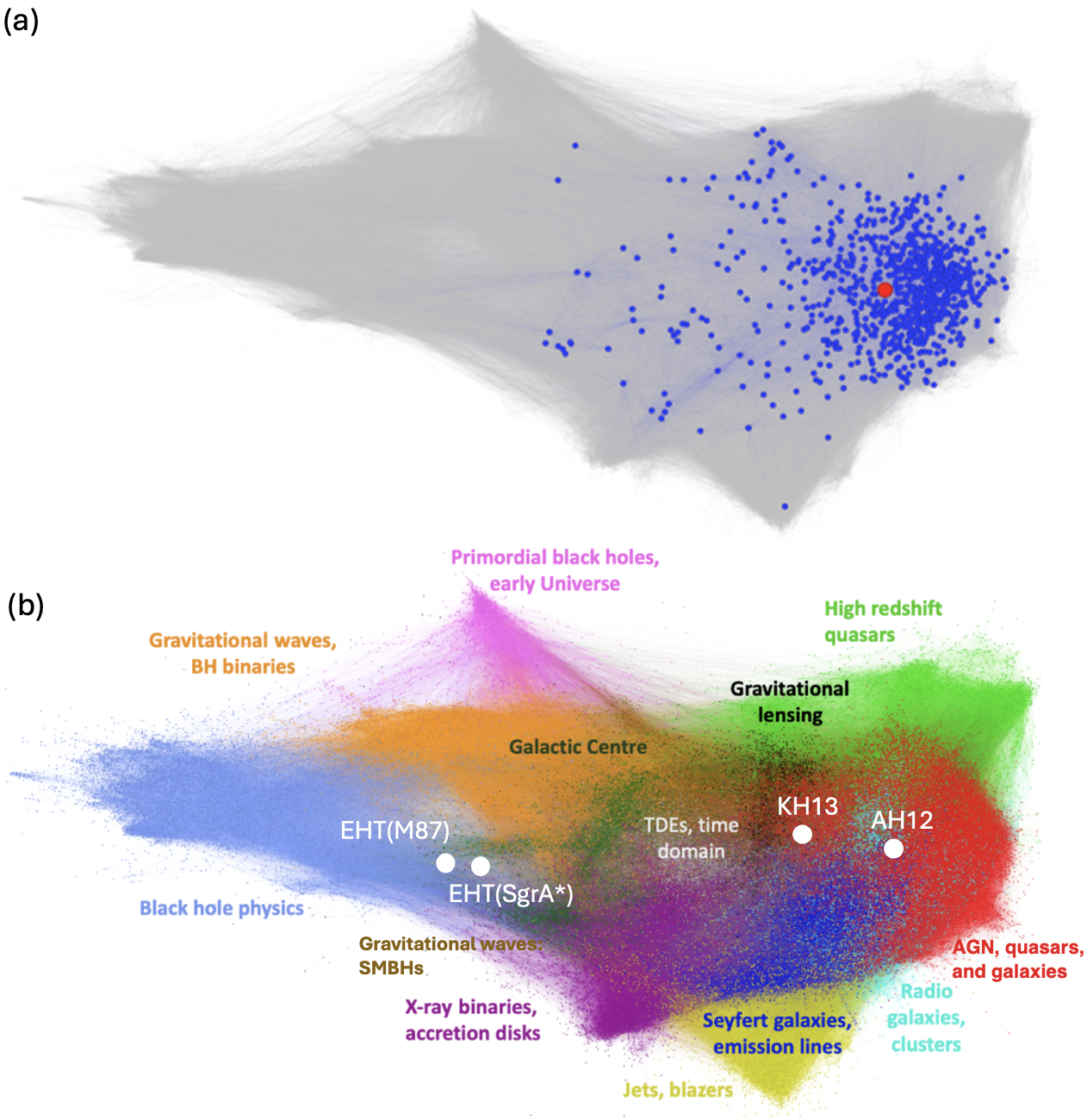}	
	\caption{Citation network diagram (CND) for (a) papers related to black-hole research with connections between papers plotted in grey. The location of AH12 across the CND is indicated by a red circle and the location of papers that either cited or referenced by AH12 are plotted using blue circles. (b) the 13 identified clusters across the CND plotted in different colours. The most closely associated cluster to AH12 is ``AGN, quasars, and galaxies", although it also has good association to the ``Seyfert galaxies, emission lines", and ``Radio galaxies, clusters" clusters indicated in red, blue, and cyan, respectively. The location of three influential studies across the CND are also plotted as white circles, along with AH12: \cite{Kormendy:13,EHT:19_M87,EHT:22_Sgr}.} 
	\label{fig:network}
\end{figure}

Fig.~\ref{fig:network} shows the full citation network for BH research, where nodes represent papers coloured by membership of particular clusters that relate to specific topics and sub fields. The relative locations of these nodes to one another is based on their citation relationships - nodes that are interconnected by a citation are pulled together, with those without a citation link pushed apart \citep{Jacomy2014}.\footnote{A node refers to individual connections for a given paper while a cluster refers to all research identified to be associated to a specific sub field using citation network analyses.} As can be seen from the top panel, the AH12 review is towards one side of this citation network and tightly associated with the second largest sub-field cluster which is related to research focused on ``AGN, quasars, and galaxies"; however, it also connects more sparsely to many other clusters. For context, the sub field on the opposite side of the citation network to AH12, the largest cluster, focuses on the theoretical fundamental properties of BHs including general relativity, quantum gravity, and Hawking radiation. Interestingly, the cluster at the centre of the citation network is research related to studying the SMBH at the centre of our Galaxy, a relatively small and diffuse sub field whose results informs both fundamental BH research and also the broader AGN/quasar research community. The location of three influential articles focused on measuring the properties of SMBHs are indicated on the bottom panel of Fig.~\ref{fig:network}, showing their relative ``domains" in comparison to AH12: \cite{Kormendy:13,EHT:19_M87,EHT:22_Sgr}. More details about each cluster, including their sizes and associations to other clusters are given in Tables~\ref{Table_CNA} and \ref{Table_contingency}.

\begin{table*}
\begin{tabular}{l} 
 \hline
Scientific-progress category definitions\\ 
 \hline
{\bf 1.\ Facilities:} technological advances due to new facilities; e.g., instruments, telescopes, observatories, super computers.\\
{\bf 2.\ Data:} increase in the amount of available data from existing facilities; e.g., larger (or more complete) data samples.\\
{\bf 3.\ Techniques:} new data analysis approaches leading to new insight; e.g., machine learning, Bayesian methods.\\
{\bf 4.\ Community:} group sociology, interactions, sizes, facilitated workshops and discussions, and potential for impact and funding.\\
{\bf 5.\ Conceptual:} new ideas from existing data; e.g., cross disciplinary research, inspiration, focused discussions (e.g., at workshops).\\
 \hline
\end{tabular}
\caption{Short description of the categories used in this article to indicate the drivers of scientific progress. The category labels are highlighted in bold.}
\label{Table1}
\end{table*}

\subsection{What is scientific progress?}\label{progress}

Before considering our pragmatic approach to measuring scientific progress in this review we first consider the more fundamental questions: what is scientific progress, how can we recognize when it has occurred, how (if at all) can we quantify it, and what are the drivers of scientific progress?

In the philosophy of science, scientific progress has been characterised in a range of different ways. These needn't be thought of as competitor characterisations \citep{TV17} since when science progresses, it can so in different ways at different times. Two prominent dimensions of scientific progress concern \textit{epistemic} progress (increase in knowledge), and \textit{functional} progress \citep[e.g.,\ solving scientific problems, or development of better conceptual frameworks;][]{bird}. Whilst these two aspects of progress often go hand in hand, they can also come apart: advances in scientific knowledge needn't be functional advances (e.g.,\ knowledge of a new exoplanet), and functional advances needn't be advances in scientific knowledge \citep[e.g.,\ the introduction of Dirac's delta function;][]{Dirac1927}.

When we are considering an \textit{increase} in something (e.g.,\ knowledge, or problem solving), the question naturally arises to what extent can we identify and measure that increase? We can describe new discoveries by concrete tokens; e.g.,\ the first black-hole merger detected through gravitational waves \citep[][]{LIGO:15,Abbott:16}, or the much more gradual discovery that SMBHs play a significant role in the formation and evolution of galaxies, first postulated $\approx$~20--30 years ago \citep[e.g.,][]{Silk:98,Fabian_1999,Bower:06,Croton_2006}. We could in principle list all such discoveries and add them up in an effort to determine how much knowledge we have gained in the last ten years. But there are two problems with using such a method for measuring progress.

First, there is a description problem: there are many ways of articulating the new knowledge that has come out of, e.g.,\ a single experiment, so that under one description we have ``progress = 1'' because we have only a single concrete token of new knowledge, but for another description we could have ``progress = 50'' if we broke down the information in that concrete token and separated it into fifty, smaller pieces of knowledge. This is especially obvious with discoveries that are more gradual, taking many years, since typically there are numerous mini discoveries that contribute to the broader discovery.

Second, and perhaps more importantly, not all progress is equal. Some progress can be characterised as incremental while other areas of progress can be considered ``revolutionary", which itself could be sub-categorised into ``major" and ``minor" varieties. We might also think that one conceptual revolution is somehow \emph{more progress} than a thousand incremental pieces of new knowledge. The same could be said of progress understood as problem solving, or conceptual developments. Such distinctions and subtleties are important, since the same piece of research could be packaged in radically different ways {\it vis-a-vis} its contribution to scientific progress, and biases could influence the choice of packaging.

Viewed from a distant standpoint, and learning from the history of science, some cases of apparent scientific progress actually turn out not to be progress at all, or at least, only in a thin sense. For example, the phlogiston paradigm of combustion, the caloric paradigm of heat, pre-evolutionary biology, the `fixist' paradigm prior to acceptance of continental drift, and the steady state cosmological model prior to the detection of the cosmic microwave background (CMB): these are all instances of apparent scientific progress that is now largely forgotten, considered irrelevant or misguided ever since a subsequent scientific revolution. Thus, the question arises to what extent, if at all, can we distinguish apparent progress from genuine progress? We may also effectively think that the major questions in some fields are addressed and all that remains is to tidy up the details, but history has taught us that revolutions in apparently mature fields can happen \citep{kuhn}.

However, the idea that all scientific progress should only be considered `apparent', on the grounds that we cannot rule out a future scientific revolution that completely reshapes the field, is quite extreme. Arguably, there are many examples of `solid' scientific knowledge, where major revolutions are not only unexpected, but rather implausible \cite[p.~98]{Godfrey-Smith03}; \cite{vickers22}. Even in BH research - one of the purest examples of ``remote science" - much is now known with a high degree of certainty. For example, today it is an established scientific fact, supported by overwhelming evidence, that the energy from an AGN is produced by material being accreted onto a SMBH but it took many decades to firmly establish this fact.

\subsection{Scientific progress: approach adopted in this review}\label{tracking}

After we accept that we \textit{can} identify instances of scientific progress, the question remains: how can we practically measure or quantify the amount of progress, both across the broad research field, and also within specific topics and sub fields?

In the absence of any widely accepted approach to quantify or otherwise measure scientific progress, in this review we adopt a ``know it when you see it" approach. That is to say, we rely on specialists working in the field, as well-placed to identify both areas where there has been significant progress and areas where there has been a notable lack of progress, using the AH12 review as a snapshot of the field back in 2012 to provide a convenient baseline from a decade ago. In each of the subsequent four sections (\S\ref{sec:section3}--\ref{star-formation}) we attempt to highlight where significant progress has, and has not, been made over the last decade. In cases where we identify significant progress, we also consider the likely drivers of this progress; in this way, we can also develop hypotheses concerning how these drivers might potentially be manipulated, in order to optimise the possibility of progress or to increase the efficiency of progress in the context of future research (see \S\ref{sec:progress_future}).

While the drivers of scientific progress are neither discrete nor completely enumerable, we have postulated five main types of drivers of progress within scientific research (see Table \ref{Table1}). Some of these drivers are easier to recognise than others, and should lead to demonstrable scientific progress: for instance new facilities (category 1), the compilation of larger and more comprehensive datasets (category 2), and the development of new data-analysis techniques (category 3). Additionally, we suggest that developments in community engagement (category 4) and conceptual shifts (category 5) can be important drivers of progress, but often in more subtle ways than categories 1-3. They can be harder to quantify precisely because they are the places where amorphous phenomena such as ``serendipity" occur: you cannot predict when two people having a conversation over coffee will stumble upon a novel concept. However, one proxy for community engagement is citation networks, which we introduced in \S\ref{context}, and will utilise again in \S\ref{cna}.

To flesh out these progress-driver categories (Table~\ref{Table1}), we list some examples below.

\begin{itemize}
\item {\bf Category 1 (facilities):} this category relates to significant progress enabled by new facilities that have greatly opened up new regions of parameter space through orders of magnitude greater resolution (spatial, spectral, temporal), spatial coverage, and sensitivity. A clear example of this is {\it JWST} \citep{Gardner2023} which within a couple of years of operations is already driving huge advances in our understanding of AGN activity across all redshifts, particularly at high redshift, due to unprecedented sensitivity and unattenuated coverage at near-IR--mid-IR wavelengths.
\item {\bf Category 2 (data):} this category relates to significant progress enabled by significant increases in data from existing facilities. An example of this category is the increase in the number of quasars spectroscopically identified from the Sloan Digital Sky Survey (SDSS): $\approx$~106~k around the time of the AH12 review \citep{Schneider2010} to $\approx$~750~k from the latest quasar data release \citep[][]{Lyke2020} and the $\approx$~3~M quasars expected from the DESI survey \citep{Chaussidon2023}. 
\item {\bf Category 3 (techniques):} this category relates to significant progress enabled by developments in data-analysis approaches. This category is quite often closely related to category 2 since a huge increase in data volume requires the adoption of techniques to efficiently analyse such large data sets. A good example is machine learning and neural network based techniques: over an order of magnitude more published astronomy papers have adopted these techniques in the last decade.\footnote{On the basis of a search for the word ``machine learning'' in the abstract field using NASA ADS: $\approx$~1,400 refereed papers in 2012 and $\approx$~18,400 refereed papers in 2023.\label{foot:ML}} 
\item {\bf Category 4 (community):} this category relates to developments facilitated by community engagement, which can be powerful drivers of progress. This is especially obvious when research groups, previously largely isolated from each other (e.g.,\ working in different fields or disciplines: inter disciplinary), start to interact at regular focused meetings, with novel research materialising as a direct result. An example is the \href{https://www.lorentzcenter.nl/index.php?pntType=ConPagina&id=638&conBestandId=674&pntHandler=DownloadAction}{Lorentz Center workshop} in Leiden 16--20 October 2017, which led to a series of influential articles explaining the myths and reality of AGN feedback \citep[e.g.,][]{Cicone2018,Cresci2018,Harrison2018,Wylezalek2018}.
\item {\bf Category 5 (conceptual):} this category is more nuanced. Conceptual progress is often easy to identify; for example, seeing AGNs as typical ``events" in the lifetime of galaxies, rather than as rare or unusual phenomena, represents progress in the form of a conceptual shift. But there is likely to be another key \textit{driver} of this progress relating to any of the other categories; e.g.,\ new facilities, more data, improved analysis techniques, facilitated community engagement possibly through inter-disciplinary research. However, the resultant new conceptual framework can subsequently be defined as a driver of progress.
\end{itemize}


\section{How does matter accrete onto black holes from kilo-parsec to sub-parsec scales?}
\label{sec:section3}

AGN activity requires a supply of gas. In this first scientific section we consider how the gas is driven from the kilo-parsec scale of the host galaxy to the SMBH radius of influence and onto the accretion disk to be accreted  by the SMBH.

The fuelling of the SMBH is confronted with a huge angular momentum problem, regardless of the abundance of the available gas reservoir. All scales are concerned by this angular momentum problem, from the outer galaxy disk (10~kpc scale), to the sphere of influence of the SMBH (pc scale), down to the accretion disk scale (0.01~pc); see Fig.~\ref{fig:agn_model}.  Low-mass black holes may accrete stars in their close vicinity, especially those with low angular momentum; i.e.\ in almost radial orbits. Stars that have a low angular momentum, and can interact strongly with the black hole, come from a region in phase space known as the loss-cone \citep{Frank1976}, since they will be swallowed by the black hole. When these stars are depleted, the relaxation time to refill the loss-cone is long. The relaxation time is the time-scale necessary for a star to lose its energy through encounters with others. In a galaxy disk, the relaxation time is much larger than the Hubble time, although it could be shorter in dense regions, like globular clusters, or nuclear regions. To fuel the black hole, it is more efficient to rely on the gas component, which can cool and dissipate energy. It may lose angular momentum through gravity torques from the non-axisymmetric potential, due to stellar bars and spirals.

\begin{figure}
\includegraphics[clip,width=0.48\textwidth,angle=0]{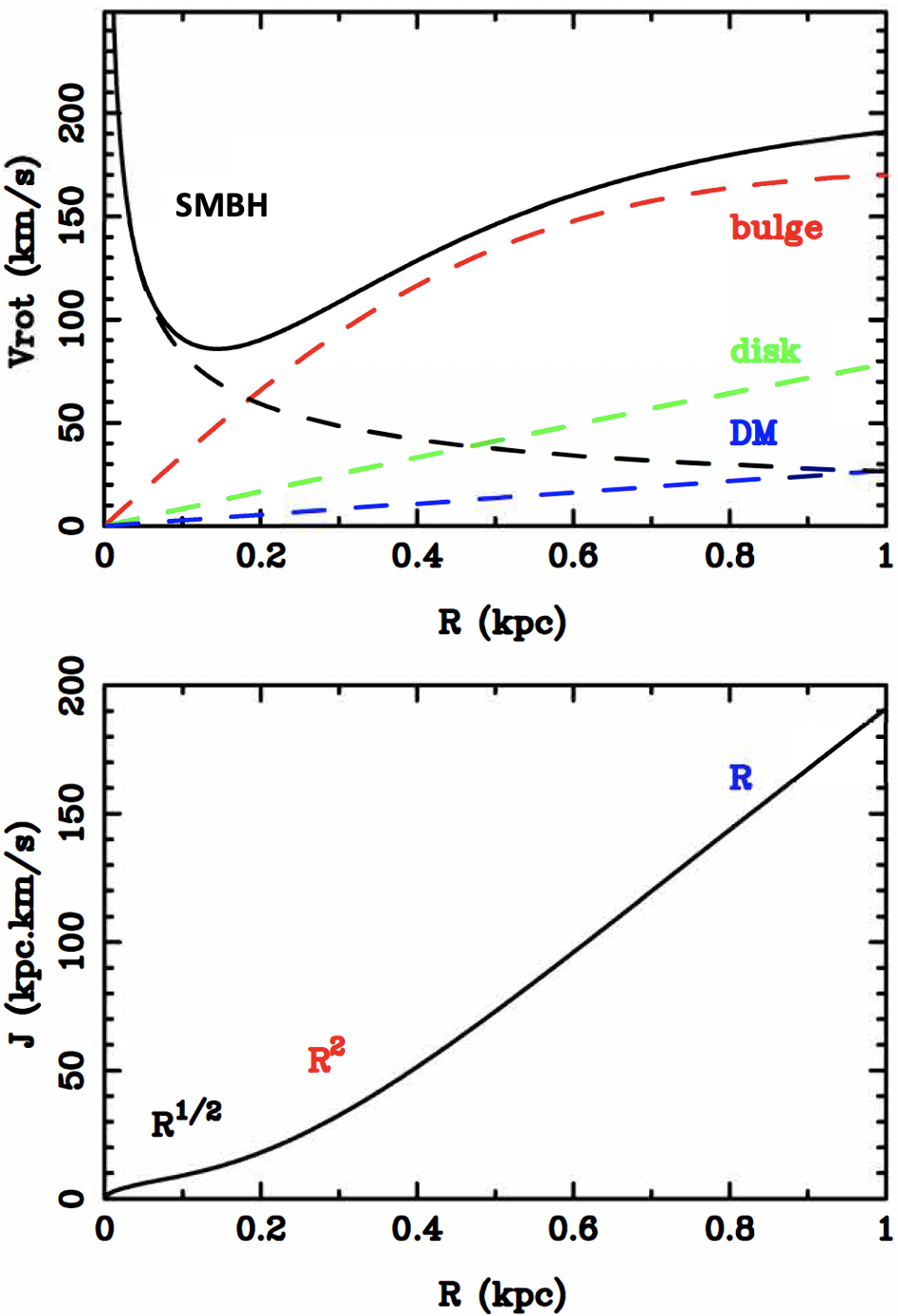}
\caption{Rotational velocity and angular momentum versus radius around a SMBH. Top panel: the SMBH dominates the inner rotation curve (Keplerian potential), which is decreasing; then the curve rises due to the 
        bulge, and eventually reaches a plateau due to the conspiracy of the gravitational potential of the disk and the
        dark matter halo (DM). The different colors mark these various contributions; 
     Bottom panel:
        the corresponding angular momentum per unit mass J, increases as power-laws
        of the radius, with the indicated slopes: 1/2 (Keplerian), 2, and 1 for the three zones.
        }
\label{fig:vrotJ}
\end{figure}

The angular momentum of matter in rotating galaxy disks is proportional, per unit mass, to the circular velocity V$_c$ and to the radius $R$. A typical rotation curve in a spiral galaxy can be decomposed into roughly three regions: a central SMBH-dominated zone ($R_{\rm RoI}$; see Footnote~\ref{foot:SMBHrad}), where the potential is Keplerian with V$_c\propto R^{-0.5}$, a rising velocity zone dominated by the bulge V$_c\propto R$, and finally a flat rotation curve zone V$_c$=V$_{flat}$, due to the conspiracy of the stellar disk and dark matter halo potentials. The corresponding specific angular momentum (J = R$\times$ V) always increases with radius with a power-law dependence with slopes of 0.5, 2 and 1 for these 3 zones, respectively; see Fig. \ref{fig:vrotJ}.

For the material within the central kpc region of the galaxy, it will need to lose a factor of at least 1,000 in specific angular momentum to reach the accretion disk at a radius of $\approx$~10$^{-2}$ pc. The amount of gas inflow (dM/dt), to account for the AGN luminosity, assuming a radiative efficiency of 10\% and a luminosity L = 0.1 dM/dt c$^2$, would need to vary from 10$^{-3}$~M$_\odot$~yr$^{-1}$ to  2~M$_\odot$~yr$^{-1}$ for the most luminous quasars of 10$^{13}$ L$_\odot$. Assuming a typical duty cycle of the order of 100~Myrs, the SMBH would therefore have to accrete about 2~$\times$~10$^8$~$M_{\odot}$ during its ``active" period, and the gas has to inflow within just a few dynamical time-scales, since the rotation period at 1~kpc radius is $\approx$~30~Myr. Very efficient dynamic mechanisms are required to exchange the angular momentum at those scales in the galaxy. Normal gas viscosity is negligible in this region, given the low density of clouds \citep[e.g.][]{Lin1987}. If viscous torques are negligible, only gravity torques can drive the gas inwards, and make it lose angular momentum. These torques are due to the tangential forces, produced by non-axisymmetric disturbances, such as bars. Stellar bars create a robust and long-lived bi-symmetric potential, that is able to drive the gas inwards. The stars will then gain the angular momentum, and the bars will weaken \citep{Bournaud2002}. From optical images, two thirds of spiral galaxies are classified as barred, and from near-infrared images, more nuclear bars are detected, that were not seen before \citep[e.g.][]{Salo2015}. Given that bars can disappear and form again, it is also likely that bars have developed in every galaxy at some time during their evolution.


To better understand how the gas can be driven inwards to fuel the central SMBH, we describe the main dynamical features of a barred spiral galaxy, the orbital structure, the resonances, and the gravity torques between stars and gas in \S\ref{sec:dynamics}. Some fraction of SMBH growth can also be due to the accretion of stars, which is detailed in \S\ref{sec:TDE}. A large part of the progress in the last decade in this research field has been facilitated by more sensitive observations and more extensive data, essentially through increased spatial resolution, brought by new facilities and interferometers. This will be described in \S\ref{sec:observations}, with ALMA (Atacama Large Millimeter Array) discoveries of molecular tori, kinematically and morphologically decoupled from the main galaxy disk, and with VLTI (Very Large Telescope Interferometer) mid-IR interferometric images, where the main dust emission near the AGN traces out a polar cone geometry. These observations are progressively changing our concepts and paradigm about the dusty torus and the AGN unification model. At even higher spatial resolution (sub-milliarcsec scales probing $\sim$~0.05 pc), the radio very long baseline interferometry, and the EHT, are now revealing several features of the accretion disk. Our knowledge of the SMBH growth mechanism on these scales is reviewed in \S\ref{sec:accretion-disks}. The small sizes of the AGN emission regions imply that they can vary on day--year time scales. The stronger variations may result from a complete transformation from an optical type 1 AGN (with prominent broad emission lines) to a type 2 AGN (lacking broad emission lines),
and these changing-look AGN, extreme variability events, and quasar periodic eruptions teach us a lot about the SMBH accretion mechanisms, as described in  \S\ref{sec:variability} and \S\ref{sec:QPE}. Numerical simulations and new techniques using ever more powerful computers have
enlightened these mechanisms in the last decade, as shown in \S\ref{sec:simulations}. Finally, \S\ref{sec:progress} summarizes the progress made over the last decade, noting that the future is bright both in the high-spatial resolution and the time domain. 

\begin{figure}
	\centering 
	\includegraphics[width=0.48\textwidth, angle=0]{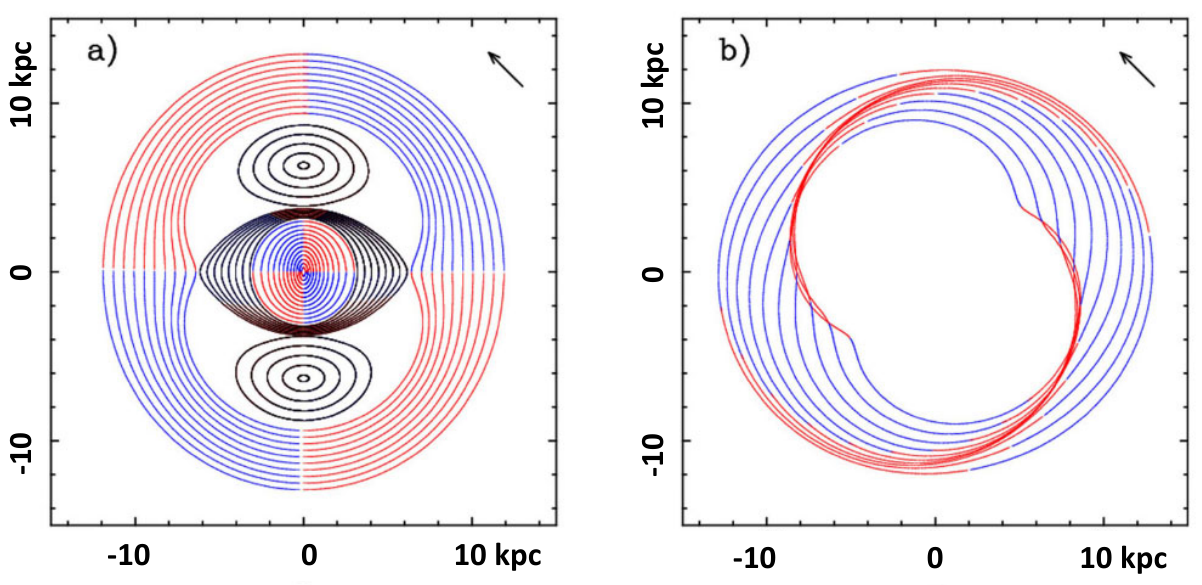}	
	\caption{Left: Main stellar orbits in the bar rotating frame. The bar is along the horizontal axis. Note the x1 orbits, parallel to the bar, until $x$~=~6~kpc. At the corotation radius, the orbits are restricted to epicycles around the Lagrangian points. Bar torques are positive in the red quadrants, and negative in the blue ones, with respect to the galactic rotation, indicated by the arrow in the top right corner.
           Right: Gas-cloud orbits, and spiral formation, triggered by the bar. The cloud-cloud collisions tilt the orbits, that turn gradually by 90$^\circ$ at each resonance. The elongated orbits of the left panel, with their respective colour, are now precessed to form a logarithmic spiral. } 
	\label{fig:Orbits}
\end{figure}

\subsection{Gas infall due to bars and spirals} \label{sec:dynamics}

From a theoretical stand point, most of our understanding of gas inflow through bars and spirals was developed in the decades prior to AH12, with the main progress over the last decade coming from observations, which confirmed some aspects of the theoretical predictions that were not previously testable. The non-axisymmetric disk potentials due to bars and spirals produce gravity torques able to drive the gas from their corotation inwards. The actual SMBH fuelling requires several steps, depending on the spatial scale. A primary bar of typical radius 5~kpc, can drive in a first step the gas from its corotation near 5~kpc to the inner Lindblad resonance (ILR) at 100 pc scale, where the gas accumulates in a ring. Being then aligned with the bar, the gas does not suffer any torque, and the gas inflow may be stalled. In a second step, a secondary bar may decouple from the primary bar, and continue its influence from 100~pc to 10~pc scales. The various non-axisymmetric features are characterized by their Fourier components: the surface density, or the implied potential, is decomposed in a series A(r)  = A$_0$(r) + $\Sigma$ A$_m$(r) cos[$m(\theta$ - $\theta_m$(r))], where typically $m$ is considered up to the 6th harmonic. Bars have $m=2$ and a constant $\theta_m$(r), while the latter gives the winding shape for a spiral. There can exist several embedded structures, mostly of $m=2$ morphology; however, close to the nucleus, when the SMBH Keplerian gravitational potential begins to dominate, $m=1$ (lopsided) features may contribute.

The sign of the torques can be derived from the main characteristics of stellar orbits in a bar potential (see Fig. \ref{fig:Orbits}, left). The stellar orbits are structured around a skeleton of periodic orbits, playing the role of building blocks, trapping orbits around them. The most important periodic orbits are the x1 family, parallel to the bar inside corotation, and the x2 family, perpendicular to the bar in between the two ILR, if they exist  \citep[see the review by][]{Contopoulos1989}.  The existence of two ILRs within the epicyclic approximation might not be sufficient for the x2 family to appear. When the bar is strong enough, the x2 orbits disappear. Then the elongations of the orbits change by 90$^\circ$ at each resonance, so that outside corotation the orbits are perpendicular to the bar until the outer Lindblad resonance (OLR). They do not support the bar, which ends at about its corotation radius.

\begin{figure}
	\centering 
	\includegraphics[width=0.48\textwidth, angle=0]{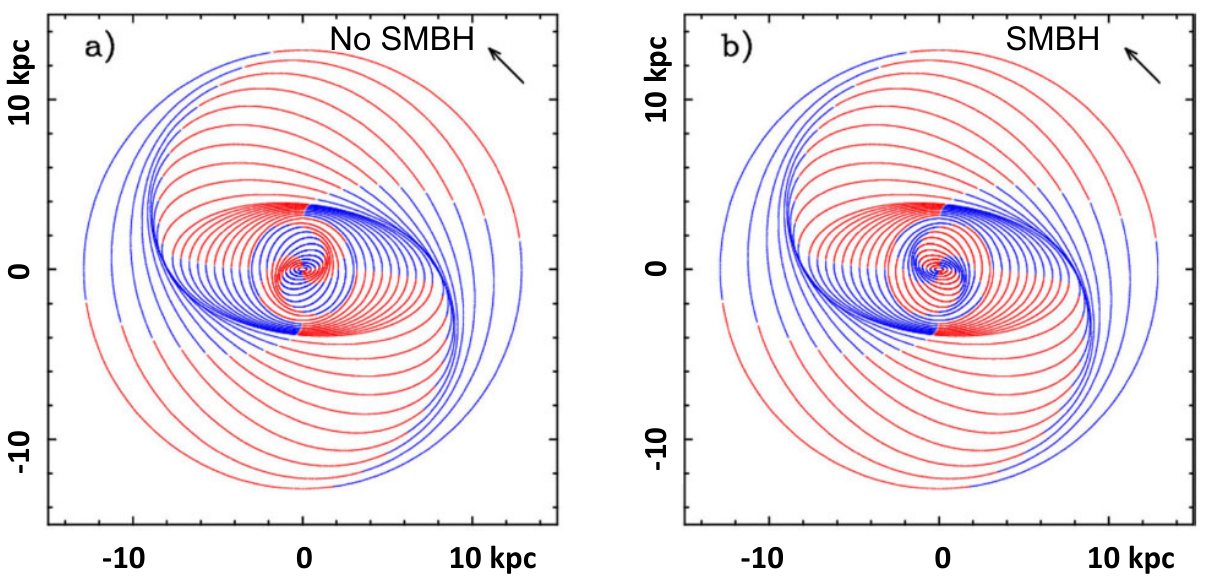}	
	\caption{Left: in the absence of a central SMBH, a leading spiral is expected inside the ILR.
           Right: in the radius of gravitational influence of the SMBH, all frequencies, and in particular the $\Omega-\kappa/2$ precessing frequency, are increasing with decreasing radius and the spiral inside ILR becomes trailing. } 
	\label{fig:Leading-trailing}
\end{figure}

While stars are a collisionless component, and perpendicular orbits can coexist, the gas has a different behavior, because of dissipation. Gas clouds following periodic orbits collide with each other, which tilts their orbits; they also change by 90$^\circ$ at each resonance, but gradually (see Fig. \ref{fig:Orbits}, right). The crowding of their stream lines produces a spiral morphology \citep{Sanders1976}. The spiral is relatively open, and can wind at maximum by 180-360$^\circ$, according to the presence or not of 2 Inner Lindblad Resonances (ILR), corotation (CR) and Outer Lindblad Resonance (OLR).

Since gas and stars are not aligned, stellar bars exert a torque on the gas, except at resonances, where the gas configuration falls back into symmetry with the bar. The bar splits the plane into four quadrants, where the gravity torques change sign, with respect to the sense of rotation of the galaxy (see Fig. \ref{fig:Orbits}, left). Between corotation and ILR, the gas spiral arms gather in the quadrants of negative torques, and gas is driven inwards, increasing the depth of the potential well. Then all frequencies, including $\Omega$ and $\Omega-\kappa/2$, the orbit precessing rates, increase towards the center and trigger the existence of two ILRs. The vertical resonance with the bar develops a peanut-shape bulge, weakening the bar further. This triggers the
decoupling of a secondary bar, an embedded nuclear bar, rotating faster than the primary bar \citep[e.g.][]{Buta1996}.

When the dissipative gas is driven inwards, the precessing rate of its elongated orbits $\Omega-\kappa/2$ first increases, while the radius decreases. This leads to a trailing spiral structure (see Fig. \ref{fig:Leading-trailing}). In between the two ILR, the precessing rate then decreases, after having reached a maximum, and a leading spiral is expected. However, when entering the $R_{\rm RoI}$ of the SMBH, all frequencies will increase again, since the potential becomes Keplerian,
see Fig. \ref{fig:vrotJ}. In this region, a trailing spiral is then expected, as shown in the right panel of Fig. \ref{fig:Leading-trailing}. The torque, which was positive on the leading arm, and maintained the gas in the ring, then becomes negative again, providing the possibility to fuel the AGN.

\subsection{Black hole growth through the accretion of stars} \label{sec:TDE}

A small amount of SMBH fuelling, especially for gas-poor early-type galaxies, can come from stars in the vicinity of the nucleus. The capture and accretion of nearby stars will give rise to a flaring event, called a TDE, for Tidal Disruption Event \citep{Komossa:15, Velzen2020,Gezari_2021}. Due to the rapid progress of time-domain astronomy over the last decade, transient event monitoring facilities like the Zwicky Transient Facility \citep[ZTF;][]{Kochanek2017_ASASSN}, or the All-Sky Automated Survey for Supernovae \citep[ASAS-SN;][]{Bellm_2019ASASSN}, have allowed for many TDEs to have been detected, which bring new and precious knowledge of the AGN accretion disks. In the near future the Vera Rubin Telescope will yield a wealth of such events \citep{Hambleton2023}, likely leading to significant growth in the size of the TDE community (see \S\ref{cna} for the growth of this community over the last decade).

When a star (r$_*$, M$_*$) with low angular momentum reaches the tidal radius of a central SMBH R$_T$ =  r$_*$ (M$_{BH}$/ M$_*$)$^{1/3}$, the star can be disrupted by tidal forces, and a large fraction of its gas becomes captured by the accretion disk. Assuming for the average stellar density, $\rho_*$= 1.4 g.cm$^{-3}$ the average density of solar mass stars, 

\begin{equation}
R_T (cm) =  6\times10^{13} M_8^{1/3}  \rho_*^{-1/3},
\end{equation}

\noindent where M$_8$ is the black hole mass in units of 10$^8$~M$_\odot$.
As the SMBH event-horizon radius R$_{\rm S}$ = 2~G~M$_{\rm BH}$/c$^2$~=~3~$\times$~10$^{13}$~cm~M$_8$,
grows faster with M$_{\rm BH}$ than the tidal radius,
there is a limit, when M$_8$ $\sim$ 3, above which the star disruption occurs inside
the black hole, and there is no gaseous release or AGN activity (but the SMBH grows even more rapidly). This is the critical Hills limit \citep{Hills1975}. For SMBH masses lower than this limit, these capture events can be observed as X-ray flares, and/or optical-UV flares, during typically one year, their signature being a characteristic light curve decreasing with time with a -5/3 power-law slope \citep{Gezari2012}. These TDEs are expected to occur at the frequency of one every 10,000 years in the Milky-Way galaxy. Some TDEs have been detected in external galaxies, where their temperature appears lower than expected, and decreasing less rapidly in time. This could be due to feedback effects in the form of a disk wind, when the accretion rate is of the order of the Eddington limit \citep{Miller2015}. But soon, the depletion and loss cone effect dries up this fuelling source, unless galaxy interactions reshuffle the stellar distribution, and creates nuclear star clusters, in a nuclear starburst.

Several dozens of TDE have now been observed, and discovered not only in the X-ray, but also in the UV--optical \citep{Velzen2020}. TDEs observed in both wavebands are rare. It has been advocated that the optical-UV emission is due to the reprocessing of the X-ray photons by material around the accretion disk. It could even be an orientation effect, if this material preferentially resides within the equatorial plane, as in the unified model \citep{Dai2018}. Most TDEs occur in post-starburst or quiescent galaxies \citep{French2020}. They are not detected in AGN hosts, which may be due to the difficulty in distinguishing a TDE; i.e.,\ the accretion of a single star would likely be lost within the emission produced by an AGN accreting a large amount of gas. The TDE might be confused with AGN flares, or there might be too much dust extinction to detect the TDE. Therefore, AGN hosts suffer selection biases against the detection of TDEs \citep{French2020}.

For the various TDEs already observed in detail, the luminosity decreases with time as a power law, as expected, although sometimes it depends on the determination of the time of the emission peak, which is not always observed. The peak intensity and the duration, from a few months to one year, reveals a large diversity. There is a good correlation between the fall-back time and the SMBH mass (estimated from the M-$\sigma$ relation), which means that the TDE observation can help to determine M$_{\rm BH}$
\citep{Velzen2020}. Optical spectra have helped to distinguish TDEs from AGN flares and from supernovae. Lines of hydrogen and helium are observed with different ratios varying with time, due to radiative transfer and opacity issues. The width of the emission lines vary with time, from initially $v\approx$~10$^4$ km~s$^{-1}$ but becoming narrower on the timescale of a month. The broad width of the emission lines is interpreted as scattering by hot electrons, and the decrease with time as the decrease in the density and optical depth of the emitting region.  The existence of outflows associated with TDEs is particularly revealed by broad absorption lines (BALs) in the UV spectrum, in particular in the ionized lines (C IV, Si IV), similar to what appears to be occuring in BAL quasars (see \S\ref{sec:QSOaccretion}). 

The TDE host galaxies have a higher central stellar concentration than other galaxies of the same morphological type, which is not unexpected \citep{Hammerstein2021}. However, the current small sample sizes do not exclude the possibility of observational biases. Observations until now allow for the estimation of the amount of mass accreted by the SMBH through TDEs, and this is at best a few percent of the total SMBH mass. Other sources of fuelling are therefore also required, in the form of gas accreted from the host galaxy, feeding normal AGN \citep{Yuan2014}. Observations of TDEs are, however, key for probing otherwise dormant SMBHs, explore general relativistic effects in strong gravity, and the investigation of magneto-hydrodynamic gas streams near black holes. They are also an opportunity for reverberation mapping \citep{Velzen2021}.

\begin{figure}
	\centering 
	\includegraphics[width=0.48\textwidth, angle=0]{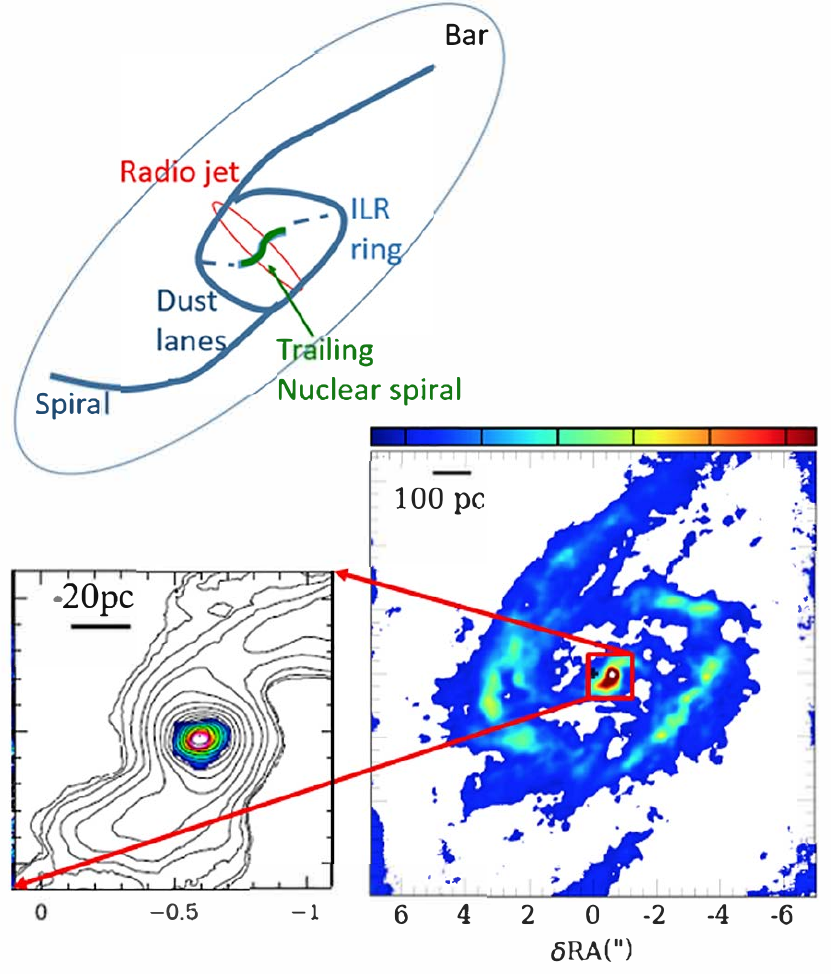}	
	\caption{Top: Schematic representation of the main morphological features observed in
 NGC~613. The dust lanes are leading the
large scale bar. The filaments are shown in dashed lines between the
ILR star-forming nuclear ring and the nuclear trailing spiral. 
           Bottom: the molecular torus (zoomed at left, in colour) is observed inside the nuclear trailing spiral traced by the black contours of ALMA CO(3-2) emission. {\it Source:} adapted from Figs~4, 5, \& 16 of \cite{Audibert2019}. } 
	\label{fig:N613-torus}
\end{figure}

\subsection{Fuelling the AGN: a review of recent observations \label{sec:observations}}

\begin{figure}
	\centering 
	\includegraphics[width=0.48\textwidth, angle=0]{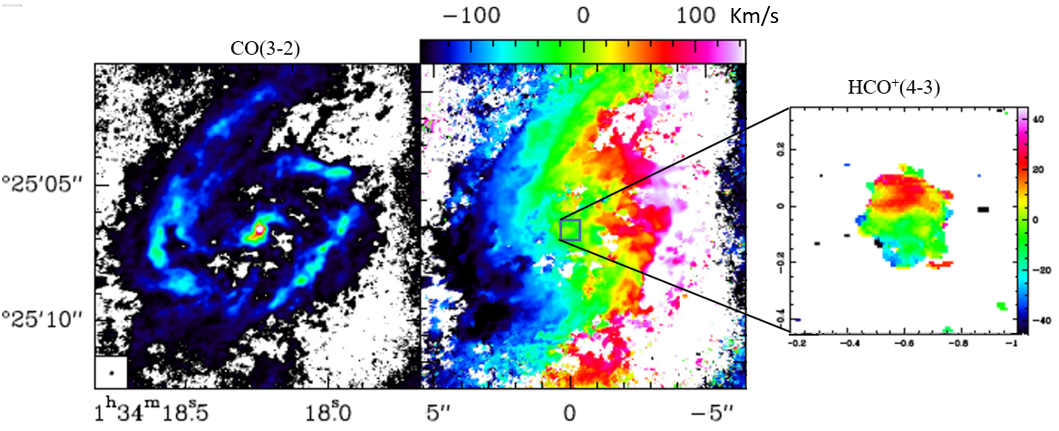}	
	\caption{Left: Molecular gas distribution in NGC~613: inside the ILR ring, the CO(3-2) emission reveals a nuclear spiral. Middle: Corresponding velocity field, with a major axis oriented almost East-West.
           Right: zoom into the molecular torus, inside the nuclear trailing spiral. The HCO$^+$(4-3) velocity field has a major axis oriented almost North-South. This HCO$^+$ torus has the same extent as the zoomed emission in colour of the left bottom panel of Fig. \ref{fig:N613-torus}. The scales are in arcsec (1'' = 83~pc), and km/s. {\it Source:} adapted from Figs 2 \& 7 of \cite{Combes2019} and \cite{Audibert2019}.} 
	\label{fig:N613-kinematic}
\end{figure}

The key to progress in our understanding of the fuelling of SMBH growth is tracing the cold molecular gas and dust at high spatial resolution in the central region of galaxies. This has been made possible thanks to recent large interferometric facilities, in both infrared (IR) and millimeter wavelengths; see Fig.~\ref{fig:SED} for waveband definitions. At millimeter wavelengths, it is possible to reach a resolution of $\sim$~13~mas with ALMA, and much more, $\sim$~40~$\mu$as with the very long baseline of the EHT \citep{EHT:19_facility}. In the near-IR waveband, the maximum resolution is 3~mas, corresponding to just $\sim$ 0.22~pc, at a typical distance of D~=~15~Mpc, for nearby spiral galaxies, sufficient to map out the emission from the hot dust in the vicinity of the AGN.  

The advent of ALMA in 2011 has been instrumental in advancing our empirical understanding of gas inflows down to small ($<$~10~pc) scales over the last decade \citep{Burillo2014, Burillo2021}.
 It is now possible to trace molecular gas not only in the primary and nuclear bar regions, as expected by the dynamical theory presented in \S\ref{sec:dynamics}, but also inside the nuclear spiral structures, which has allowed for the discovery of additional molecular disk components. The latter has been found to be randomly oriented with respect to the rest of the disk, and is kinematically decoupled: we associate this structure to the molecular torus, which is central to the unified AGN model \citep{Urry1995}. The gas fuelling of these molecular tori is thought to be the result of secular evolution, with the nuclear spiral being the last step in the fuelling process (cf \S\ref{sec:dynamics}). 
 The molecular tori is within the region where the SMBH dominates the gravitational potential (i.e.,\ at $R_{\rm RoI}$, see Footnote~\ref{foot:SMBHrad} and Fig.~\ref{fig:agn_model}).
 The accretion time scale onto the molecular tori within this region is quite short ($<$~10~Myr), and nuclear bars are more ephemeral than primary bars. However the decoupling of an embedded bar weakens the primary bar, and also the positive torques, outside corotation. This allows the replenishment of the gas disk, so that the molecular torus may be fuelled intermittently.

Many barred Seyfert galaxies, like NGC~613, have been found to be in a feeding phase, fuelling the molecular torus (cf.\ Fig. \ref{fig:N613-torus}), with a nuclear trailing spiral inside the ILR ring, as expected when the gas enters the radius of gravitational influence of the SMBH, as in Fig. \ref{fig:Leading-trailing}. The computation of the torques has confirmed the evidence of fuelling the molecular torus in NGC~1566 \citep{Combes2014}, NGC~613 \citep{Audibert2019}, and NGC~1808 \citep{Audibert2021}.

NGC~613 is one of the prototypical cases of a strong barred galaxy, with a star forming gas ring at ILR. The first ALMA observations, with moderate spatial resolution, could see only the ring, without resolving the internal structure \citep{Miyamoto2017}. The computation of the torques indicated that the gas can lose all of its angular momentum in only one rotation, when inside a $\approx$~50~pc radius. The observed molecular gas velocities show that the molecular torus, at the center of the nuclear spiral, has decoupled kinematics; i.e., the kinematic major axis is not aligned with that of the large-scale disk, see Fig. \ref{fig:N613-kinematic}.

\begin{figure}
	\centering 
	\includegraphics[width=0.48\textwidth, angle=0]{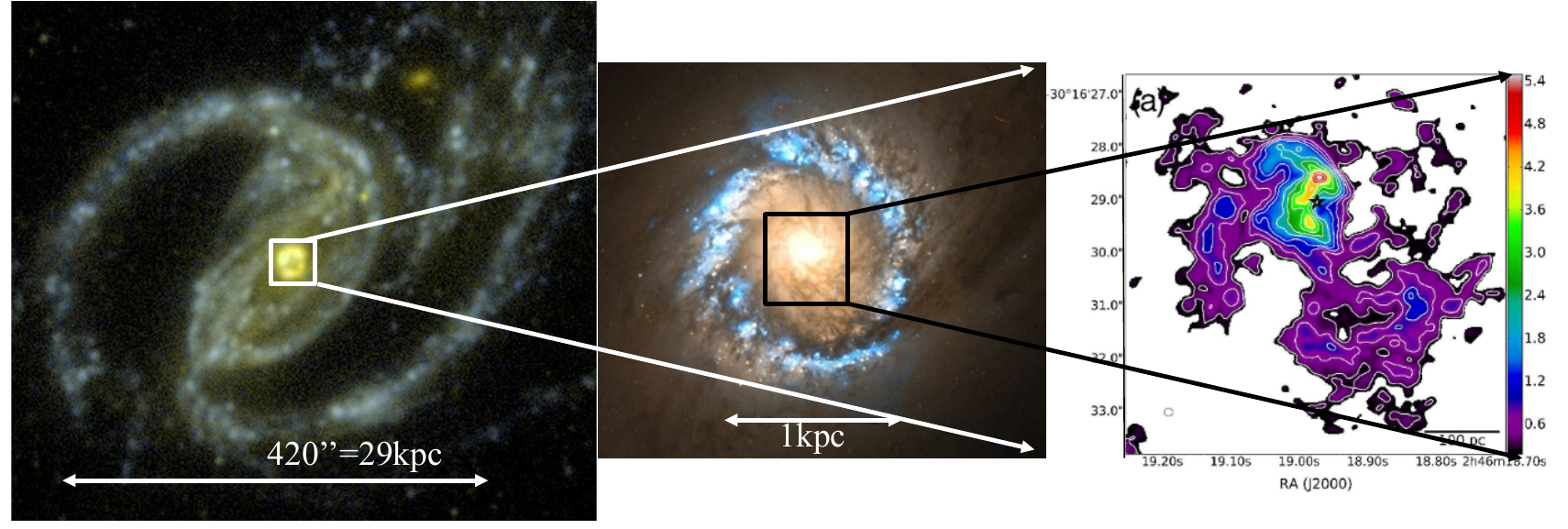}	
	\caption{Zoom-in on the nuclear spiral observed with ALMA inside the ILR ring of NGC 1097. The left image is a large-scale UV image of the galaxy with {\it Galex}, the middle is an HST image of the star forming ring at ILR \citep{Fathi2006}, and at right is the CO(3-2) map from \cite{Izumi2017}. {\it Source:} adapted from Fig.~3 of \cite{Izumi2017}.} 
	\label{fig:N1097-zoom}
\end{figure}

Conversely, Fig. \ref{fig:N1097-zoom} shows an example of a nuclear spiral, but without the molecular torus in NGC~1097; the latter could have been suppressed by AGN feedback,
see \S\ref{star-formation}. Indeed, it has been shown that the cold molecular gas concentration, at scales below 50~pc, is quite high compared to the 200~pc-scale molecular surface density but only below a critical X-ray luminosity \citep{Burillo2024}. Above this critical luminosity, the concentration falls by 4 orders of magnitude. This may be attributed to AGN feedback, which depletes the central gas. This result has been obtained for a sample of 64 nearby (D~=~7--45~Mpc) disk galaxies including 45 AGN and 19 non-AGN \citep{Burillo2024}, where there is sufficient spatial resolution to identify a molecular torus, when present. Additional data is being acquired to improve the source statistics.

\begin{figure}
	\centering 
	\includegraphics[width=0.5\textwidth, angle=0]{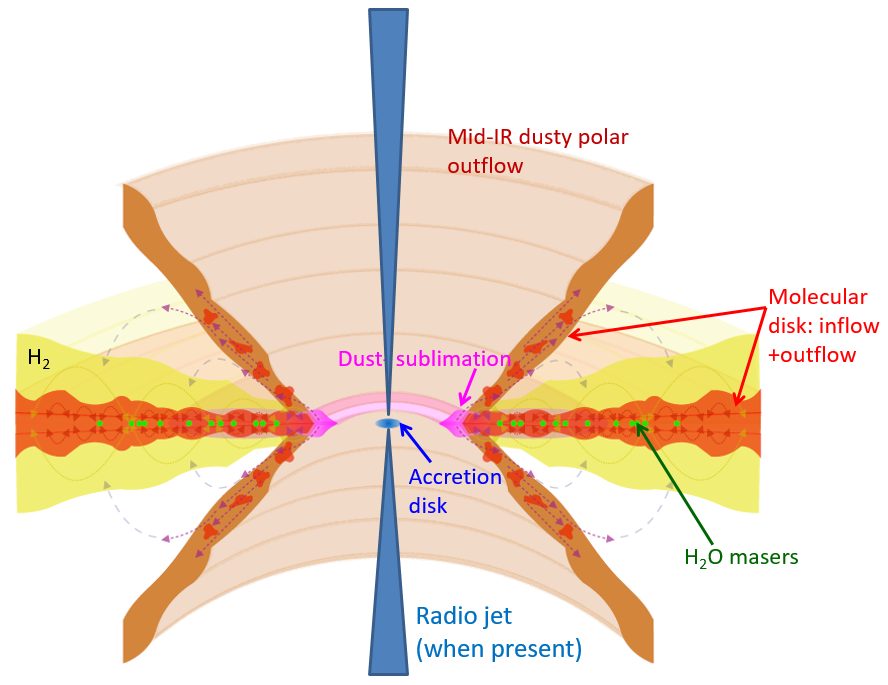}	
	\caption{Schematic view of the circum-nuclear environment of the SMBH, with the radio jet, the molecular circum-nuclear disk, or molecular torus, the hollow dusty polar cone shaped by the AGN wind, and molecular outflows tracing the cone. In some nearby AGN, the circum-nuclear disk contains H$_2$O maser emission. The density of the disk is expected to decrease exponentially with radius and height. {\it Source:} adapted from Fig.~4 of \cite{Hoenig2019}. } 
	\label{fig:AGN-schema}
\end{figure}

The original AGN unification paradigm was based on the existence of a dusty torus whereby the broad line region (BLR) and accretion disk of a Type 1 AGN was obscured if the dusty torus lay along the line of sight towards the observer, such that only the Type 2 AGN features are identified. However, high-resolution infrared observations, including ground-based interferometers (in particular VLTI-MATISSE) have shown that the dust in the central parts of AGNs can often be detected in the polar direction, instead of being aligned in the same plane as the putative dusty torus \citep{Asmus2016, Asmus2019, Leftley2021, Leftley2024}. In a revised AGN unification scenario, a conceptual advance driven by new facilities, the obscuration can come both from a dense and geometrically thin dusty disk as well as the base of the polar cone, which is quite clumpy \citep{Isbell2022, Isbell2023}, with the geometrical thickness from the combination of these obscuring regions corresponding to the relative abundance of Type 1 and Type 2 AGN \citep{Alonso_Herrero_2011,Ramos2011,Elitzur_2012,Ramos-Almeida:17}. The polar dust component is most likely ejected (or even formed) within an AGN-driven wind \citep{Alonso-Herrero2021}, starting from the dust sublimation radius (pc size), and forming the border of a hollow cone \citep{Hoenig2019}. The gas is therefore initially predominantly in infall within a geometrically thin disk identified as the molecular torus, but can later be in outflowing in a direction perpendicular to the torus and the accretion disk. However, polar dust is not found in all AGN \citep[for example, observations of two high-Eddington ratio sources show no evidence for polar dust;][]{Drewes_2025} and it is expected to be most commonly identified in sources with accretion properties that are conducive to a long-lived extended outflow \citep[e.g.,][]{Venanzi_2020,Garcia_Bernete_2022}. Molecular outflows are more difficult to detect along the colder molecular gas, and/or could be dragged by radio jets, which are usually oriented  parallel to the wind direction,
as sketched in Fig. \ref{fig:AGN-schema}.

Further key progress has been facilitated in this research field by Integral Field Units (IFU) like MUSE on the VLT, which can map the kinematics of the ionized gas across the galaxy \citep{Husemann2019,Bernete2021,Juneau_2022}. These kinematical studies have been of paramount importance to distinguish between rotational motions of the gas and outflows, that could produce efficient feedback, as discussed in more detail in \S\ref{star-formation},  Outflows accelerated by AGN are commonly observed in the form of coherent, mildly collimated high-velocity gas directed along the AGN ionisation cones, and kinetically powerful jets \citep{Bernete2021}. The outflow is comprised of gas in different phases (i.e.,\ it is multi phase), with the highest velocity gas being ionized but where the bulk of the mass in the outflow is in the molecular phase \citep{Fiore2017}.
If a jet component is present and lies perpendicular to the disk of the galaxy then it's impact will be at a maximum, due to a stronger coupling with the ISM gas, even more so if the jet has a low power (i.e.,\ non relativistic) since it can interact with the ISM for a longer duration. This scenario appears to be occurring in NGC~1068, the prototypical nearby Seyfert 2 \citep{Burillo2016,Burillo2019} where the jet-ISM interaction drives a cocoon in the galaxy disk, and produces lateral velocity perturbation perpendicular to the jet direction \citep{Venturi2021,Girdhar2022}. The gas excitation and broadening of the emission lines is predominantly due to shock ionization. An ionized collimated outflow with symmetric blueshift and redshift may exist in the direction parallel to the jet and wind. Complex situations may occur when the jet and wind are completely aligned with the galaxy disk, entraining ionized and molecular gas outflows, while an oblique ionization cone can also be identified where the ISM is elevated in a fountain by the star formation feedback and illuminated by the central AGN \citep{Husemann2019}. For each outflow ``event", the AGN feedback is not sufficiently energetic or efficient to quench star formation on the galaxy scale, and would have a more local impact. However, when repeated outflows occur, their cumulative effect may well be crucial to slowly quenching star formation; see \S\ref{star-formation}.

High resolution near-IR AGN imaging of NGC 1068 has also been obtained with VLTI-GRAVITY \citep{GRAVITY2020}. The interpretation of the results in terms of a circum-nuclear ring has been debated by \cite{Gamez2022}. Finally, combining GRAVITY and MATISSE data has resulted in a coherent multi-wavelength model, where the parsec-scale region is composed of a dust disk plus a polar wind \citep{Leftley2024}.
Breakthrough results were also obtained for M~87 using the Global Millimeter VLBI Array (GMVA) complemented with ALMA, and Greenland Telescope (GLT) observations to not only image the shadow of the SMBH (see \S\ref{introduction}), but also the radio jet at the base where it is formed and launched, showing that it produces edge-brightened structures connected to the accretion flow and the gravitationally lensed rings \citep{Lu2023}. In Centaurus~A, it was also possible to reveal the launching and initial collimation region of the radio jet, down to 10 gravitational radii \citep{Janssen2021}. As in M~87, the radio jet is found to be edge brightened, and its structure is remarkably comparable, when scaled by SMBH mass, leading to the potential identification of a universal profile for SMBHs over a range of scales.

To further progress our understanding of the AGN feeding and growth of SMBH, there are at least two clear approaches to follow: first, to obtain high resolution data of ionized, molecular gas and dust on a larger sample of systems allowing to then calculate timescales and duty cycles, in comparison with simulations of the circum-nuclear regions, and in particular producing the torus; and second to obtain much higher resolution observations, namely with VLBI revealing water maser emission like in Fig~\ref{fig:AGN-schema}, closer to accretion disks, where warping and differential precession dominate \citep{Greenhill1997, Herrnstein1999, Gallimore2004, Gallimore2024}.
 
\subsection{The accretion disk}  \label{sec:accretion-disks}

\begin{figure}
	\centering 
	\includegraphics[width=0.5\textwidth, angle=0]{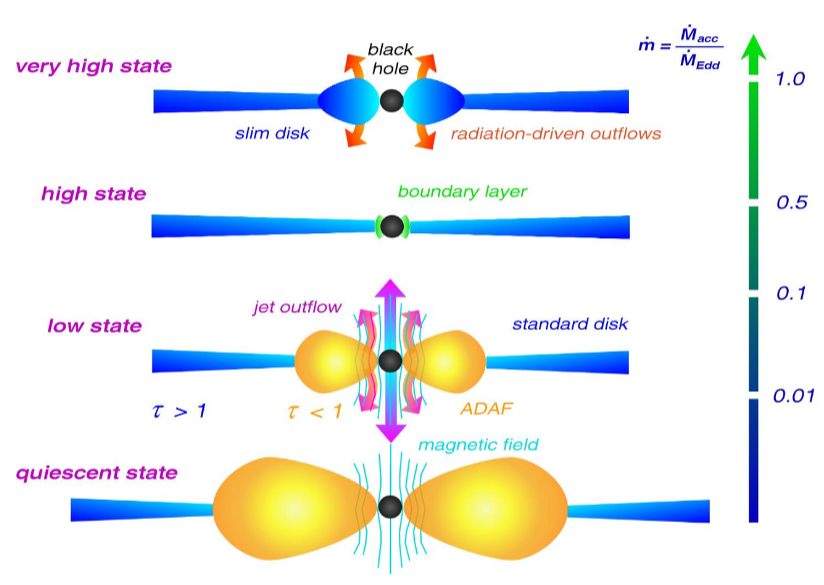}	
	\caption{Accretion flows can have different configurations, according to the Eddington-normalized
accretion rate, indicated in the right vertical line. At the bottom, with low accretion rate, the density of the flow is so low that its cooling is not efficient: the flow heats and becomes a geometrically thick cloud. At higher accretion rate, a cold and geometrically thin disk begins to form in the outer parts, while the center is an ADAF, with a radio jet. At quite high accretion rate, the cold thin disk extends down to the last stable orbit, and is efficiently radiating. Above this accretion rate, the luminosity is so high that the radiation pressure triggers a disk wind. {\it Source:} adapted from \cite{Esin1997} and redrawn and presented in Fig.~4.4 of \cite{Muller2004}.} 
	\label{fig:Accretion-flows}
\end{figure}

How is the gas accreted towards the SMBH at very small, sub-parsec scales? Theoretically we would expect the structure and geometry of the accretion disk to depend on the accretion rate, with respect to the Eddington rate. The latter corresponds to the Eddington luminosity, which is proportional to the mass of the black hole, i.e. L$_{\rm Edd}$/L$_\odot$ = 3.2 $\times$ 10$^4$ M$_{\rm BH}$/M$_\odot$. This is the maximum luminosity reached by an AGN, where radiation pressure from the accretion disk balances the gravitational force, assuming a spherical geometry. In a typical highly accreting AGN, where the accretion rate is just below the Eddington limit (typically expressed as the Eddington ratio $\lambda_{\rm Edd}$= L/L$_{\rm Edd}$ between 0.01 and 1), we expect accretion to occur in a classical geometrically thin rotating accretion disk, where viscous torques transport angular momentum away. The disk is optically thick, relatively cold (i.e.,\ its temperature between 10$^3$ and 10$^4$K in the outer parts, is below the virial temperature, of order 10$^9$K), and radiating efficiently, like a black-body. When the accretion rate is above the Eddington limit, then the disk is expected to become ``puffy" (i.e.,\ geometrically thick) and cannot radiate efficiently, and some of the energy is accreted into the SMBH by advection. This is often referred to as a slim disk \citep[e.g.][]{Kubota2019}; see Fig. \ref{fig:Accretion-flows}. This accretion regime may explain in particular Narrow Line Seyfert 1s (NLS1s), which have broad emission lines that are narrower than typical Seyfert 1s \citep{Wang2003}. Note that the different states of accretion flows have been established first for X-ray binaries, where they are more easy to establish due to much shorter timescales \citep[e.g.][]{Fender2004}.

At the other extreme, if the accretion rate is low, $\lambda_{\rm Edd}<0.01$, then the disk is less optically thick and can even become optically thin, and again can no longer radiate efficiently, becoming a RIAF (Radiatively Inefficient Accretion Flow) \citep[e.g.][]{Yuan2014}. The energy instead heats the gas rather than being radiated, and towards the SMBH the medium becomes hot and geometrically thick. If an AGN transitions from a high accretion state to a low accretion rate then the thin disk is expected to be destroyed, initially near the SMBH and then moving out towards the edge of the accretion disk. In this low accretion rate state, energy can also be advected directly into the SMBH, and this regime is called an Advection Dominated Accretion Flow (ADAF) \citep[e.g.][]{Lasota2016}. The radial velocity of the accretion is higher, since it will be proportional to the scale height of the medium, and also the rotational velocity will be less due to the gas pressure. This regime is included within the broader concept of ``hot accretion flows", which also host gaseous outflows and convection \citep[e.g.][]{Yuan2014}. The hot medium or hot ``corona", of size $\approx$~3--10 gravitational radii that forms as a by-product of the accretion process \citep{Fabian2015}, is responsible for the emission of hard X-rays through inverse Compton emission, as shown in Fig. \ref{fig:Accretion-flows}. The hard X-ray spectrum is characteristic of a low-luminosity regime. This is shown in \citet{Kang2024} and \citet{Hagen2024}, with observational evidence of changes in accretion disk SED with Eddington ratio, in particular see Fig. 5 of \citet{Hagen2024}.

The physics of AGN feedback is expected to be very different across these accretion regimes (cf \S\ref{star-formation}). At high Eddington ratio, radiation pressure on the ionized gas, close to the SMBH will drive very high velocity winds, observable as broad absorption lines in X-rays \citep{Tombesi2010} and in the UV/optical \citep{Borguet2013}. At lower-luminosity ($\lambda_{\rm Edd}<0.01$), lower-velocity winds are expected to be launched from the accretion disk at larger radii by radiation pressure on dust \citep{ricci_radiative_2017}. Kinetic feedback can also be produced via collimated outflows commonly observed as magnetically driven radio jets, which drag out atomic and molecular gas at lower velocities but which contains most of the mass of the outflow. This may happen for any Eddington ratio, provided that a radio jet exists.  Radio jets are most frequently seen in early-type galaxies, and are related to the SMBH spin \citep{Chiaberge2011, Kormendy:13}. Several regimes can be identified, corresponding to the large diversity of optical and UV observations, especially when considering line driven accretion disks winds, when most of the opacity is occurring in the spectral lines \citep{Giustini2019}. This can explain inactive nuclei, low-luminosity AGN (the most frequent in nearby galaxies), Seyfert galaxies, NLS1s and up to the broad absorption line quasars (BALQSOs); see also \S\ref{sec:QSOaccretion}.

Another important phenomenon to take into account is the spin of the SMBH. Measurements through X-ray lines and reflection spectroscopy have shown that the spin is frequently quite high, and close to the maximum in AGN \citep{Reynolds2013}. In the future, the identification of photon rings or the last stable gas circular orbit with high resolution (EHT or VLTI-GRAVITY, \citet{EHT:19_M87, GRAVITY2018}) also have the potential to directly measure spin. This is important to confirm, since one of the main mechanisms to launch a radio jet is a rapidly spinning SMBH \citep{Blandford1977}. The gravity of a rotating black hole exerts a relativistic ``frame-dragging" effect on the surrounding gas, causing a warp and Lense-Thirring precession. The gas infalling from the galaxy disk is likely to have a random orientation with respect to the black hole spin, and will be submitted to the torque of the SMBH, which will have the effect of aligning it within a disk perpendicular to its spin \citep{Bardeen1975}. The inner and outer disk are no longer aligned, and the disk develops a warp out to $\approx$~100~$R_{\rm S}$ (see Footnote~\ref{foot:SMBHrad}). This warping is a characteristic feature of the water masers detected in nearby AGN accretion disks \citep{Kartje1999, Gallimore2004, Caproni2006}.
The high spatial resolution brought by VLBI has revealed polarised emission of the water masers in the NGC~1068 accretion disk \citep{Gallimore2024}. The derived orientation of the magnetic field with respect to the accretion disk suggests that the large-scale outflow may be triggered in the accretion disk, through hydromagnetic instabilities.

If the viscosity of the disk is insufficient, instead of warping it may split into several rings, with different inclinations and precessing rates. When the various rings, with differential precession, collide, they will more rapidly drive the gas inwards toward the SMBH \citep{Nealon2015, Pounds2018}. General relativistic magnetohydrodynamic simulations of very thin disks around rotating black holes reveal the tearing of the disk. The torques from the frame dragging wins over the disk viscosity, due to magnetized turbulence, fragments the disk into a rapidly precessing inner part and slowly precessing outer part \citep{Liska2021}. The inner disk aligns perpendicular to the SMBH spin, while the outer disk keeps the ``souvenir" of its unaligned angular momentum.

\subsection{Transient phenomena: changing look AGN}  \label{sec:variability}
AGN have been known for a long time to be variable over timescales of days to months and years, in particular in the optical/UV and X-ray wavebands, where the emission regions around the SMBH are small. X-ray spectra show absorption and emission features, mainly from reflection, which helps to derive sizes \citep{Turner2009}. However, in the last decade, more variability has been discovered, with greater amplitude, and strongly variable objects are called changing-look AGN \citep[e.g.,][]{Ricci2023}. These objects show dramatic flux and spectral changes, in both X-ray and optical/ultraviolet bands, changing optical class from Type 1 to Type 2 (or vice versa). These objects are challenging the unification AGN model; their number has been growing rapidly over the past decade, from initially a few examples to now a few hundred \citep{Ricci2023,Guo2024_CLAGN}, With the advent of new facilities and surveys focused on the transient universe, like the ZTF, ASAS-SN and multi-epoch wide-area X-ray surveys like {\it e-ROSITA} \citep{Merloni2020} combined with extensive optical spectroscopy of AGN and quasars over at least two epochs. In the future, the time domain will be further developed with new instruments, in particular the Vera Rubin telescope \citep{Hambleton2023}, the {\it Einstein Probe} \citep{Yuan2015}, {\it ULTRASAT} \citep{Ben-ami2022}, and these variations will bring new insights in the physics of active nuclei, and in particular on the unification scenario for AGN. 

\begin{figure}
	\centering 
	\includegraphics[width=0.48\textwidth, angle=0]{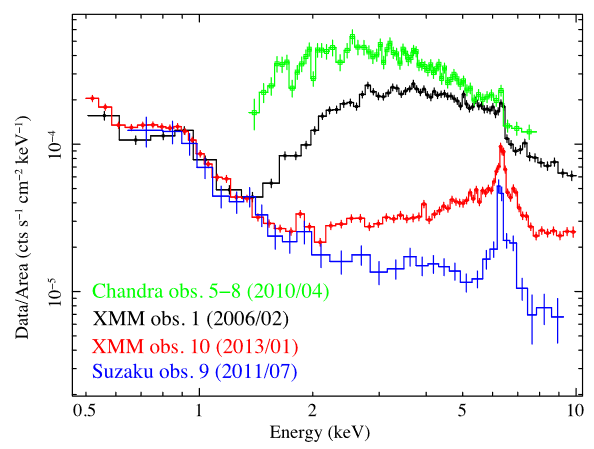}	
	\caption{X-ray spectra of ESO 323–G77 showing significant time-variable column density variations. Data from {\it Chandra} (2010), {\it XMM–Newton} (2006, 2013), and {\it Suzaku} (2011). When the obscuration is high, the Fe~K$\alpha$ 
 line at 6.4~keV is most prominent. The column density varied between 10$^{22}$ cm$^{-2}$ to 1.5 $\times$ 10$^{24}$ cm$^{-2}$. {\it Source:} reproduced from Fig.~1 of \cite{Miniutti2014}.} 
	\label{fig:Dust-variations}
\end{figure}
 
There exist two main mechanisms to explain the changing-look phenomena: one is changing obscuration, where the line-of-sight column density of gas and dust varies on short time scales due to the clumpiness of the obscuring medium, and the other is an intrinsic changing accretion state, due to variable accretion. In the first scenario, the obscuring clouds are expected to have a large velocity and could be associated with outflows transiting over the core of the AGN \citep[e.g.][]{King:15}, as well as the outer regions of the torus. Sensitive X-ray monitoring campaigns have revealed that changes in the absorbing column density towards individual AGN are common over timescales of days to years \citep[e.g.,][]{Risaliti_2002,Markowitz_2014,Pizzetti_2025}. However, significant absorption changes corresponding to a unobscured--obscured or obscured--unobscured transition are far less common \citep{Ricci2023}; a significant changing-obscuration event in the Seyfert 1 galaxy ESO 323-G77 by \cite{Miniutti2014} is shown in Fig. \ref{fig:Dust-variations}. In the second scenario, where no obscuration is detected in X-ray, or when the shape of the UV-optical continuum slope implies no dust obscuration, the changing look may be explained by a changing accretion rate. Both the continuum emission and broad emission lines may appear and disappear, and the AGN may change from a type 1 to a type 2 and reverse back on the timescale of years--decades. This implies a significant change in the accretion rate. 

\begin{figure}
	\centering 
	\includegraphics[width=0.48\textwidth, angle=0]{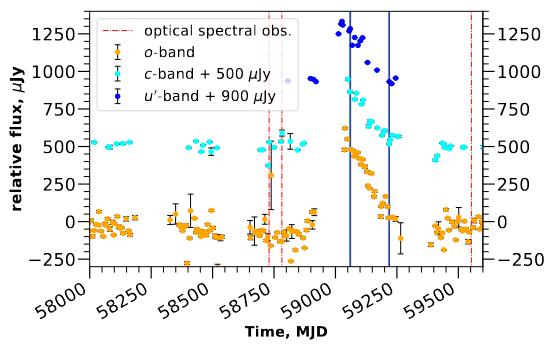}	
	\caption{An outburst in the changing-look AGN Mrk~1018 shown at three wavelengths, the u'-band (blue: 300-400~nm), c-band (cyan: 420-650~nm), and o-band (orange: 560-820~nm). The u' and c-band data have been shifted vertically for visualisation purposes. The magnitude of the outburst is independent of the colour, implying that dust obscuration is not the cause. {\it Source:} reproduced from Fig.~9 of \cite{Brogan2023}. } 
	\label{fig:Mrk1018-burst}
\end{figure}

One prototypical example of a changing look AGN is Mrk~1018. When discovered, Mrk~1018 was classified as a Type 1.9 Seyfert galaxy, although it transitioned to a Type 1 Seyfert galaxy a few years later before returning to its initial classification as a Type 1.9 Seyfert galaxy after 30 years \citep{Husemann2016}. It is in this phase, that a strong outburst was observed in 2020, as shown in Fig. \ref{fig:Mrk1018-burst} \citep{Brogan2023}. The Eddington ratio increased by a factor $\approx$~13 from 0.004 to 0.052 during the outburst. The shape of the decline over the following $\approx$~200 days is incompatible with a TDE, and therefore the event is most likely caused by a drastic, short-term increase in the accretion rate. Since Mrk~1018 is undergoing a galaxy merger, it is possible that these events are due to a binary SMBH or a recoiling SMBH.
The first changing-look AGN ``events" to be identified occurred in Seyfert galaxies \citep[e.g.][]{McElroy2016}, but the first changing-look quasar was identified about decade ago \citep{Lamassa2015}. Through an SDSS and Pan-STARRS-1 survey, \citet{MacLeod2019} found 17 new changing-look quasars, yielding a changing-look rate larger than 20\% among variable quasars (changes in optical continuum magnitudes by $\approx$~0.5--1~mag), while the combination of the SDSS and DESI quasar surveys is now allowing for the systematic identification of changing-look quasars in sample sizes of $>500$ \citep{Guo2024_CLAGN}. 

\begin{figure*}
	\centering 
	\includegraphics[width=0.8\textwidth, angle=0]{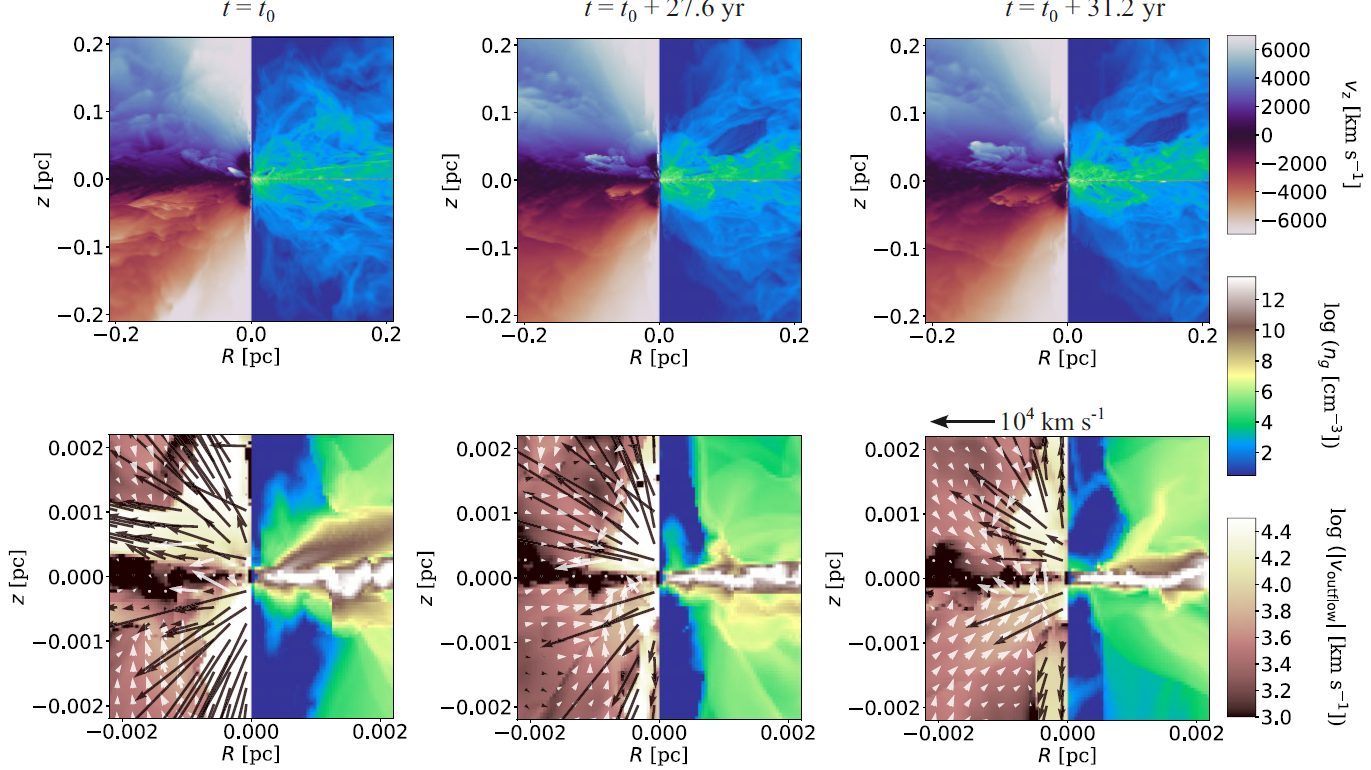}
	\caption{Hydrodynamical simulations of gas accretion onto an SMBH, illustrated at 3~epochs separated by just 30~years. The strong variability illustrates the intrinsic changing-look AGN phenomenon, through accretion rate variations. Top and bottom plots focus on the central 0.2~pc and 0.002~pc regions, respectively. The left-half panels show V$_z$ (top) and V$_{outflow}$ (bottom). The right-half panels show the gas number density. {\it Source:} adapted from Fig.~2 of \cite{Wada2023}.} 
\label{fig:Wada-2023}%
\end{figure*}

Simulations of changing-look AGN, have explained the observations through accretion-rate variations, but also
from the stochastic character of the gas dynamics in the sub-pc region, see Fig.~\ref{fig:Wada-2023}. 
Since in numerical simulations, there exist frequent variations due to the intermittent outflows around an AGN, \citet{Wada2023} propose that the time-scale of $\sim$ 10~years is too short for a change of the accretion disk, but the changing look could come from changes in the structures of the emission line regions. Even without a direct change of accretion rate, it is possible that the stochasticity of the outflow phenomena produces significant changes in the spectrum of the ionized gas, in particular the Balmer emission lines. In the simulation, the equivalent width of H$\alpha$ and H$\beta$ changed by a factor of 3 with the emission lines disappearing over 30~yr. These variations come from the gas dynamics within the dust sublimation radius. The AGN radiation pushes the dust and drives the gas outflow intermittently in the sub-pc region of the disk.

\subsection{Extreme events and quasi periodic eruptions}  \label{sec:QPE}
 Extreme quasar variability has been studied as flares superposed on their normal stochastic variability \citep{Graham2017}. These flares occur in 5~$\times$~10$^{-5}$ of the million quasars considered and last typically 3 years. Micro-lensing by stars in foreground galaxies could be one explanation  \citep{Lawrence2016}. But the majority could be due to a changing-look event, and explosive stellar-related activity in the accretion disk: super-luminous supernovae, TDEs, and mergers of stellar mass black holes. 
 They can also be due to the
 interaction of a supermassive black hole with a lower mass companion in an extreme mass-ratio inspiral (EMRI)
 \citep{Chakraborty2024}, or with a single star \citep{Linial2023}.
 TDEs can produce indeed spectacular variation, as in the Seyfert 2 galaxy GSN~069, where the X-ray emission increased by more than a factor of $\approx$~240 in 16~years \citep{Miniutti2019}. After that, the decrease was slow over 8~years, with only a thermal spectrum. However, there were then X-ray quasi-periodic eruptions (QPE), with jumps by two orders of magnitude, over the course of 54 days on time scales of hours. Since GSN~069 is a dwarf galaxy with a 4~$\times$~10$^{5}~$M$_\odot$ SMBH, the class of AGN with more massive SMBHs could have the same eruptions on time scales of months, given the relative relevant sizes. We should note that X-ray bursts are quite common in stellar-mass systems \citep{Galloway2008}, but difficult to identify in SMBHs, although they should also be quite common. A handful of QPE have been detected since 2019 in low-mass AGN-hosts \citep{Arcodia2024}. They are thought to be due to 
gas accretion from a white dwarf in a very eccentric orbit about the central massive black hole.
The orbit of the white dwarf is progressively reduced in radius and excentricity, due to radiation of gravitational waves \citep{King-QPE2023}. The energy liberated at each passage is then much lower, while the phenomenon becomes a quasi-periodic oscillation, hardy detectable. The constraint of a closely orbiting star with high excentricity favors the low-mass SMBH \citep{King-why2023}.

To represent the observations of AGN X-ray and optical variability, \citet{Kubota2018} built a model of AGN including a standard thin disk in the outer parts, truncated towards the inner part, to contain a warm region to produce the soft X-ray excess and a hot corona, the source of hard X-rays. The size of the hard X-ray corona is equal to the radius where there is sufficient accretion energy to power the observed X-ray emission. Such a model succeeds in producing a harder X-ray spectrum towards lower Eddington ratios, reproduces the X-ray/UV correlation, and an increasing optical variability with decreasing Eddington ratio.

A changing look event can be dramatic, as shown by \cite{Ricci2020}. In this example the Type 2 AGN 1ES~1927+654 experienced drastic changes in the X-ray band, with variations up to $\sim$4~dex over $\sim$~100~days, with short time-scale variability up to $\sim$2~dex in $\sim$~8~hrs. This was accompanied by a complete disappearance of the hot corona, and depletion of the whole inner disk. After several months the disk and hot corona were recreated. These extreme AGN variability events with factors of 2--10 optical/UV/X-ray luminosity changes over 1--10 yr time scales suggest a change in our view of accretion disks. Indeed, with the standard thin accretion disk, the propagation time-scale is too long, of the order of 100~years, for a 10$^6$ M$_\odot$ black hole; however, in a geometrically thick disk, the accretion flow will be shorter, on par with the variability timescale \citep{Dexter2019}.

Variations are not only seen at X-ray/UV/optical wavelengths, but also in the mid-IR waveband, in more than a hundred sources selected from the {\it WISE} survey \citep{Jiang2021}. Only four galaxies were found to be radio loud, indicating that synchrotron radiation from relativistic jets is not the main cause of the variability. It is likely that these mid-IR outbursts are dominated by the dust echoes of transient accretion onto SMBH, either through TDEs or changing-look AGN. This is supported by the compatibility of the mid-IR peak luminosity with X-ray and optical TDE phenomena at the high end.

\subsection{Simulations of cosmic SMBH growth}  \label{sec:simulations}

Over the last decade, there has been considerable progress in simulations of SMBH growth, in particular within the broader cosmological context \citep{Vogelsberger:2014,Sijacki:15, 2014MNRAS.444.1453D, Volonteri2016, Weinberger:2017, Dave:2019}, adopting special recipes to take into account of sub-grid physics, such as the Bondi accretion rate, or the repositioning of the SMBH, to oppose its artificial wandering. The repositioning is applied in simulations that lack a subgrid model for dynamical friction, which applies to the majority of existing simulations today. In some simulations, a drag force is implemented, exerted on the SMBH by the gas and/or stars and/or dark matter \citep{Sijacki2011, Dubois2015,2017MNRAS.470.1121T,2022MNRAS.510..531C,2022MNRAS.513..670N}. Fully modelling dynamical friction requires a spatial resolution finer than 10 pc \citep{2019MNRAS.486..101P}, which is beyond the capacities of most cosmological simulations.
In parallel, at much smaller scales, it has been realized that the standard model of thin disks, for Eddington ratios over 0.01--0.2, were thermally unstable, and incompatible with observations \cite[e.g.][]{Davis2020}. The advent of 3D-magneto hydrodynamical (MHD) simulations, taking into account of radiation and general relativity, have made great progress towards solving this problem. With high spatial and temporal resolution many of these simulations have been performed at the individual object scale, to follow the accretion flow close to the SMBH, including general relativity (or pseudo-Newtonian) and MHD, with radiative transfer. For sub-Eddington ratios, of 0.07--0.2, \cite{Jiang2019} finds departures from the standard model of a thin accretion disk, depending on the amount of magnetic pressure, which is key to thickening the accretion disk. 

Several 3D-MHD simulations have shown that magnetic-pressure-dominated thin, sub-Eddington accretion disks can maintain thermal equilibrium, in contrast to thermally unstable disks dominated by radiation pressure \citep{Sadowski2016}. The magnetically supported thin disks are thicker than standard accretion disks at the same radius and accretion rate. These thicker disks are thought to be more compliant with observations, in particular when the observed X-ray/UV/optical variability is concerned. As discussed in the previous sub-sections, the observed variations, when due to intrinsic variations of accretion rate, are too rapid to be explained within the standard model of a thin accretion disk, although thicker accretion disks may better explain the observations \citep{Dexter2019}. \cite{Mishra2022} demonstrated that magnetically-dominated disks are thermally stable and that the magnetic fields can both ensure the stability of somewhat thicker disks, and drive accretion onto the central SMBH through magneto-rotational instability (MRI). For that to occur the best configuration of the magnetic field is vertical, perpendicular to the accretion disk, where it is possible to generate a toroidal field near the mid plane of the disk.
 
While at lower Eddington ratios (0.01--0.07), the accretion flow is mainly through the optically thin corona, for higher accretion rate systems, the accretion disk surface density increases with radius. The turbulence driven by the MRI which transports the angular momentum away, fostering a rapid accretion flow. Super-Eddington accretion ($\lambda_{\rm Edd}$ = 20) onto a maximally rotating SMBH has been simulated by \cite{McKinney2014}, resulting in the formation of a persistent radio jet through the Blandford-Znajek mechanism, and driving an accretion-disk wind. The energy loss from the wind reduced the accretion efficiency down to 20\%, and the total radiative efficiency dropped down to 1\%, implying an observed AGN luminosity equal to that expected for the Eddington limit.\footnote{For the definition of efficiencies, let us note that not all of the mass accretion serves to grow the SMBH, first a component (typically $\approx$~10\%) is radiated away. At low Eddington ratios ($<$~0.01), part of the mass is advected into the SMBH, without radiation (increasing the accretion efficiency but lowering the radiation efficiency). Also part of the mass is lost to driving an outflow (i.e.,\ a jet and/or a wind), further reducing the accretion efficiency.} The accretion disk is quite geometrically thick, corresponding to the slim disk regime. This shows that it might be possible in the early Universe, to grow SMBHs quickly through super-Eddington accretion (see \S\ref{sec:QSOhighz} for potential super-Eddington accretion at high redshift).

It is quite important to simulate also the galactic environment, to better understand how the gas replenishment occurs and allows for more or less continuous fuelling of the AGN. The required range of scales in such a simulation are huge, and hydrodynamical codes with refinement, and even hyper-refinement are required. Starting from zoom-in simulations, identifying a given halo in a cosmological context, \citet{Angles-Alcazar2021} have then run an hydrodynamical simulation refined down to a resolution of 0.1~pc around the SMBH. They chose three epochs corresponding to a pre-AGN, a full luminous AGN (i.e.,\ a quasar) and a late-time AGN, differing in gas mass, star formation rate (SFR), and SMBH accretion rate at z$\sim$~2. They found a quasi-steady state is reached, where the gas inflow at 0.1~pc scale is at least 6~M$_\odot$~$yr^{-1}$, in the full luminous AGN phase. This active phase is short ($\approx$~2~Myrs) and the duty cycle is brief, separated by long inter-cycles. Star formation consumes the bulk of the gas inflow, although supernovae feedback is also taken into account in the simulation, but not the impact of AGN feedback. The mechanisms driving the gas inflow are the $m=2$ non-axisymmetries, primary and secondary, and the gas at sub-pc scales remains unaligned with respect to the large-scale disk (see Fig. \ref{fig:Leading-trailing} and Fig. \ref{fig:N613-kinematic}).

Following a cosmological halo identified at $z=6$, \citet{Dubois2014} studied the evolution and growth of a central SMBH, through gas accretion and galaxy mergers. Since the physics of SMBH accretion needs to be at the sub-grid scale in the simulation, they used semi-analytical recipes for AGN feeding and feedback. They found the SMBH can rapidly change its mass and spin throughout its lifetime of gas accretion and mergers. In the high-redshift universe, the gas is quite abundant in cosmic filaments and gas accretion dominates the SMBH growth, while galaxy mergers, and subsequent SMBH mergers, dominate at lower redshift. At high redshift, gas accretion has a more coherent angular momentum, which allows the spin of the SMBH to grow, while at lower redshifts, mergers of galaxies and their SMBHs have more random orientations, implying a reduced spin. Radiative efficiency is also higher early in the evolution, since it follows the spin amplitude. The resulting alignment between the SMBH spin and angular momentum of the host galaxy is then never completely achieved, but it is higher for intermediate mass galaxies; note, SMBH accretion disks and their galaxy hosts are randomly oriented in massive galaxies.

Large-scale cosmological simulations are essential for capturing the growth of the entire SMBH population, from those residing in dwarf galaxies to the largest ones, over cosmic time. These simulations, with box side lengths ranging from tens to hundreds of Mpc \citep{2014MNRAS.444.1453D,Hirschmann2014,2014MNRAS.445..175G,Vogelsberger:2014,Sijacki:15,2015MNRAS.450.1349K,Volonteri2016,2018MNRAS.473.4077P,Weinberger:2017,Dave:2019,2022MNRAS.513..670N}, are primarily calibrated to reproduce galaxy properties (e.g., stellar mass function, fraction of quiescent galaxies, sizes) and the SMBH to stellar mass ratios of local massive galaxies with $M_{\star}\geqslant 10^{10.5}\, \rm M_{\odot}$ \citep[][for an overview]{habouzit_supermassive_2021,2022MNRAS.509.3015H}.  Since these simulations are not calibrated on other SMBH or AGN properties, they enable meaningful comparisons with observations. In particular, they produce a relatively good agreement with the SMBH mass function at $z=0$, 
and produce the AGN downsizing, or anti-hierarchical AGN distribution with time; i.e.,\ the number density of luminous AGN peaks at higher redshifts than those of faint AGN \citep{Hirschmann2014}. This is attributed to the higher gas abundance at high redshift (see \S\ref{sec:rapid-growth}). The mean of the logarithm of the Eddington ratios varies across simulations; however, all show an increase with redshift \citep{2022MNRAS.509.3015H}, consistent with observations \citep{kelly_demographics_2013}.
The simulations also approximately reproduce the AGN luminosity function at $z=0$ but encounter issues at the faint end for $z\geqslant 1$ where most of the simulations overpredict the number of AGN with $L_{\rm bol}\leqslant 10^{45}\, \rm erg/s$. The faint end is driven by a combination of the subgrid physics related to SMBH formation (i.e., seed mass), feedback from SNe and AGN, and the accretion of SMBHs, all of which are designed to address the limitations in resolution that hinder the self-consistent capture of these processes. The above subgrid models and their variations across simulations lead to different SMBH-galaxy mass correlations, particularly in low-mass galaxies. The lack of stochasticity in these models often results in tighter relations than observed in the local Universe \citep[][and Fig.~\ref{fig:mbh_mstar_diagram}]{reines_relations_2015}.
Some simulations produce a population of SMBHs that are overmassive relative to the mass of their host galaxies or halos; all due to tidal stripping of their hosts \citep{Volonteri2016,2016MNRAS.460.1147B}. In simulations without SMBH repositioning, the high-mass end of haloes may even host several SMBHs, and when two are gas fuelled and radiate, this produces dual AGN hosted within the same galaxies (see also \S\ref{sec4:mergers}). This quite rare occurrence corresponds to the observations at low redshift.

AGN feedback via AGN-driven outflows is thought to be important in regulating the galaxy mass and producing the tight M-$\sigma$ relation, but also in preventing the formation of galaxies that are over massive compared to those observed. Abundance matching across a range of cosmic epochs shows that the star formation in massive galaxies peaks early, before being quenched, and that this quenching is related to their SMBH growth \citep{Moster2013}. The right equilibrium between runaway star formation in massive galaxies and the ejection of most of the gas outside of galaxies is hard to fine tune.  
Towards this goal, \citet{Weinberger:2017} introduce a new model for the AGN 
feedback. For high-accretion rates, above a given threshold depending on SMBH mass and the Eddington ratio, a fraction of the mass energy of the accretion heats the gas, preventing star formation (see also \S\ref{star-formation}). 
In the case of low accretion rates, they employed a kinetic feedback model designed to be more effective. This model pushes gas in random orientations, generating shock waves that heat the gas and subsequently quench star formation.
This heating is more uniformly distributed, and has more impact on galaxies. 
Their cosmological simulations succeeded in producing the red and dead massive elliptical galaxies \citep{2017arXiv170703395N}, although with a population of SMBHs not fully aligned with observational constraints. 

Zoom-in simulations, able to manage a large range of scales, are now common practice in the cosmology and galaxy domain. Care must be taken to adapt the approximations made to simulating a particular scale, to the next zoom-in scale of the simulation. In particular, the physics of the gas in the AGN circum-nuclear region is quite different from the gaseous disks forming stars at kpc scales. Feedback recipes are different, and the impact of the magnetic field or the radiation transfer have to be adapted to the scale. In their ambitious multi-scale enterprise, \citet{Hopkins2024} modified their simulations to modify the physics from cosmological scales to radiation-magnetohydrodynamic (RMHD) computations of accretion flow near a central SMBH. Their simulation spans scales from $\sim$~100~Mpc down to $\sim$~100~au to feed a bright quasar, with accretion rates as high as 10--100~M$_\odot$~yr$^{-1}$. The gas infall is driven by gravitational torques from stellar bars and spirals down to 1~pc scales. The physics at sub-pc scales changes, with a high fraction of ionized gas and strong magnetic fields. In the accretion disk, the gas is turbulent, no longer self-gravitating, and subject to the MRI instability. The star formation, is also efficiently suppressed. An aspect of the multiphase gas over this huge range of scales (accretion disk, BLR, torus, host galaxy, and halo) is displayed in Fig.~\ref{fig:agn_model}.

\subsection{Summary of the key drivers of progress} \label{sec:progress}

Looking back over the last decade since AH12, huge progress has been made in the domain of AGN fuelling and SMBH growth, due both to new observational facilities, great advances in the telescope sensitivity and spatial resolution, and also theoretical and conceptual advances. On the observational data side, the last decade has seen the advent of ALMA, yielding for nearby AGN about 10~mas resolution, resolving the radius of gravitational influence of the SMBH ($R_{\rm RoI}$). Since 2018, the VLTI with MATISSE and GRAVITY have provided 5--10~mas resolution in the near and mid IR wavebands, and a conceptual breakthrough has been made in mapping circum-nuclear regions of AGN, identifying polar dust cones. Thanks to the high resolution of the VLT interferometer, \citet{GRAVITY2024} has measured the BLR sizes in some AGN, and found them compatible to that determined through reverberation mapping; see also \S\ref{sec:BHmass}. The EHT has achieved a record-breaking 20~$\mu$as in resolution at millimetre wavelengths, allowing us to witness the shadow of two SMBHs, and the launching of the radio jets. Since 2022, {\it JWST} has brought a wealth of data on the dust and gas surrounding the SMBH, with a huge amount of excited lines, together with the H$_2$ rotational ladder, and the PAH. As described in \S\ref{sec:QSOhighz}, one of the main surprises from {\it JWST} is the potential over-massive SMBHs from candidate AGN identified in the early universe, with respect to their host stellar mass. Super Eddington accretion \citep[e.g.][]{Madau2025} may help reduce the black hole mass estimates, which are quite uncertain \citep{2024A&A...686A.256L, 2024A&A...689A.128L}.

On the theoretical side, numerical simulations have been flourishing, with cosmological runs following the SMBH growth across cosmic times, and GR-MHD codes, with radiative transfer in 3D, which are transforming our vision of the accretion disks, and their different regimes, as a function of the Eddington ratio. The huge advances were due to several factors: not only the large increase in computing power, allowing higher resolution and larger volumes and, therefore, greater statistical power, but also more detailed and refined physics, with new algorithms and techniques.

The last decade has also seen great advances in monitoring AGN variability thanks to new facilities such as ZTF and ASAS-SN which provide large-area optical photometric imaging to allow for the identification of transient phenomena and to track small-amplitude variations in the output of objects in the time domain. This has allow for the first large-scale searches for TDEs and the monitoring of changing-look AGN. These developments have helped to show that differences between Type 1 and Type 2 AGN cannot be just explained by the unified AGN model, accretion rate is also a key factor in determining the observed properties. In the future, more sensitive time-domain facilities such as the Vera Rubin telescope will further develop this research field, opening up time-domain astronomy to a larger fraction of the research community.


\vspace{0.2cm}
\input{sec4}

\vspace{0.2cm}
\input{sec5}

\vspace{0.2cm}
\section{What impact do AGN have on the black-hole fuelling and star formation?}\label{star-formation}

\input{sec6}

\section{Discussion of the scientific progress}\label{sec:discprogress}

\begin{figure*}[h]
    \centering
    \includegraphics[width=0.85\linewidth]{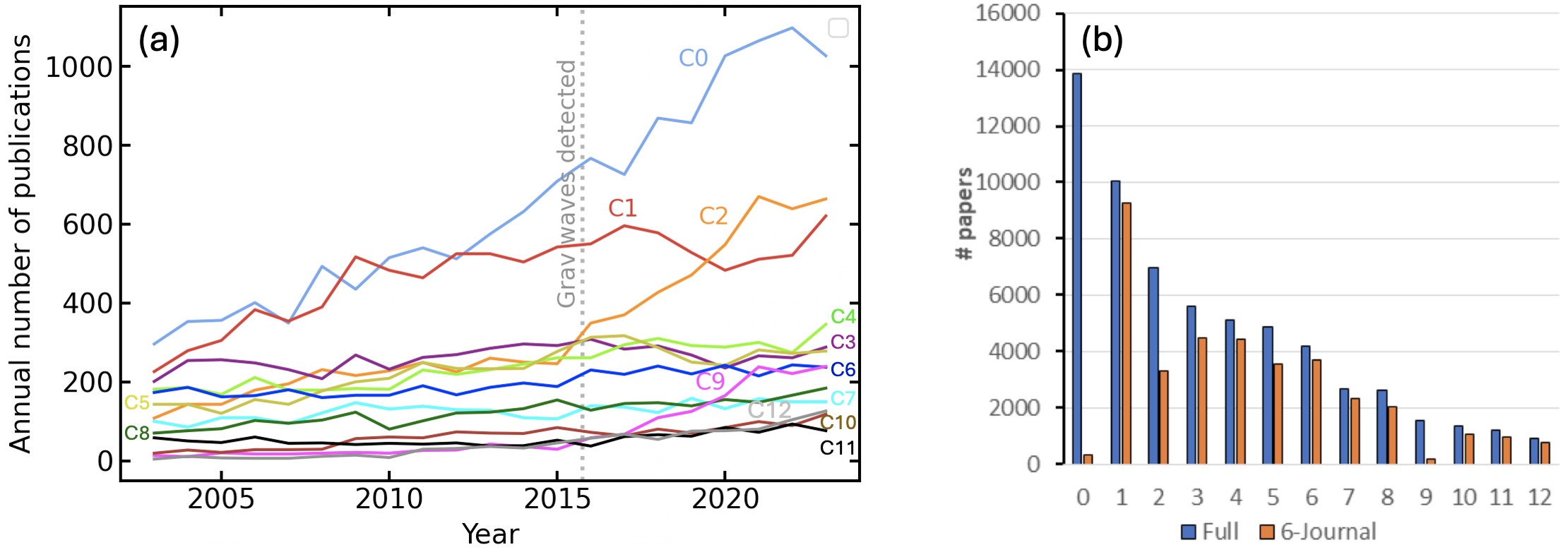}
    \caption{(a) Annual number of publications per year for the 13 research clusters identified in \S\ref{context} over 2003--2023 using the same colour coding as that adopted in Fig.~\ref{fig:network}; see Table~\ref{Table_CNA} for the cluster labels. Several clusters with a strong rise in publication rate since $\approx$~2015 are highlighted along with C1, the cluster most closely associated with the research in this review. (b) Total number of publications for each research cluster (``Full": blue), along with the number of publications in the six selected ``astrophysics" journals (``6-Journal": orange; see Footnote~\ref{foot:astro_journals}).}
    \label{fig:Cluster progression}
\end{figure*}

\subsection{Reflections on the last decade of progress}\label{sec:progress_reflections}


The last decade has witnessed huge advances in our understanding of ``what drives the growth of black holes?", as detailed in the last four sections (see \S\ref{sec:section3}--\S\ref{star-formation}). Here we briefly summarise some of the key areas of significant progress, making direct reference to the progress drivers defined in Table~\ref{Table1}.

New {\it facilities} (category 1) have been a strong driver of progress in each of the scientific areas, allowing many key conceptual advances. For example, new facilities have allowed for significant progress in shaping our understanding of the nature of the obscuring medium in AGN (ALMA; VLTI-GRAVITY; VLTI-MATISSE; see \S\ref{sec:observations}), the time-variable nature of accretion from close photometric and spectroscopic monitoring (ZTF; ASAS-SN; spectroscopic follow up; see \S\ref{sec:TDE} \& \S\ref{sec:variability}), the identification of faint high-redshift candidate AGN ({\it JWST}; see \S\ref{sec:QSOhighz}), the nature and impact of AGN-driven outflows and feedback (LOFAR and many new IFU facilities; see \S\ref{sec:AGNFeedbackObservations}), and the development of higher spatial and temporal resolution simulations with improved sub-grid physics (see \S\ref{sec:simulations} and \S\ref{sec:AgnFeedbackEffects}--\ref{sec:AGNFeedbackImpact}), amongst other discoveries. It is unsurprising that significant progress has been made through the exploitation of new facilities since astronomy is, largely, a facility driven discipline and facilities are defined and approved on the very basis that they will address the key scientific issues in the research field.

Increases in the quantity and quality of {\it data} (category 2) from existing facilities and the development of new {\it techniques} (category 3) have also been significant drivers of scientific progress. These two drivers often go in parallel since the increase in data will often necessitate the employment of new techniques to handle the larger datasets. For example, the increase in the availability of optical spectroscopic data from the SDSS survey, along with developments in the modelling of the physics of accretion disks, has enabled greater insight on how the observed properties of quasars connect to their underlying accretion properties (see \S\ref{sec:QSOaccretion}). Our understanding that AGN are typical ``events" in the broader lifetimes of galaxies has also largely come about from the development of techniques to take account of limitations in the Eddington-ratio parameter space probed by existing data, along with the realisation that the different variability timescales of AGNs and galaxies (itself assisted by improved hydro-dynamical simulations of accretion disks), which showed that different {\it apparent} correlations will be identified (or not) depending on what observed quantities are compared (see \S\ref{sec4:varying_growth} \& \S\ref{sec4:mergers}). The development of simulation tools to directly predict the observed properties of AGN-driven outflows, another example of techniques, also greatly helped the interpretation of outflow and ``AGN feedback" signatures from spatially resolved spectroscopy, and ultimately showed that the impact of outflows on the host-galaxy environment is more localised and less ``immediate" than the community initially assumed (see \S\ref{sec:AGNFeedbackObservations}).

The last two scientific drivers, {\it community}
(category 4) and {\it conceptual} (category 5),
are often more difficult to assess. The definition of {\it community} is broad in this context and essentially refers to approaches that aim to engage and receive feedback from a broader community rather than from an individual research team. A ``pure" example of community driven progress is the Galaxy-Zoo project \citep[][]{Lintott2008}, which engaged the general public (``Citizen Science") to assist in the classification of the morphology of galaxies, which led to the identification of unusual systems such as the ``Voorwerpje", now understood to be the light echoes of past AGN episodes (see \S\ref{sec4:echoes}). A more typical example of community driven progress is the organisation of scientific workshops \citep[occasionally multi or inter disciplinary;][]{Burns_2025} that bring together groups of scientists with complimentary skills and scientific backgrounds to facilitate broad discussion along with the development of new approaches to analyse and intrepret data. As mentioned in \S\ref{tracking}, the \href{https://www.lorentzcenter.nl}{Lorentz Center workshops} are good examples of this type of community engagement, as are the \href{https://aspenphys.org/summer-workshops/}{Aspen workshops} which require participants to commit to attending for an extended duration to allow for more in-depth discussion; community engagement was also a major driver behind the development of our series of black-hole workshops (see \ref{workshops}). Many examples of key conceptual advances have been highlighted above. Perhaps the key factor with this progress driver is that these advances are typically achieved in association with one of the other four drivers rather than being a ``Eureka" moment of pure inspiration without any other input. Indeed, the examples highlighted above represent conceptual advances but were driven by new {\it facilities}, increases in existing {\it data}, the development of new {\it techniques}, or facilitated through broader {\it community} engagement.

\begin{figure}[h]
    \centering
    \includegraphics[width=1\linewidth]{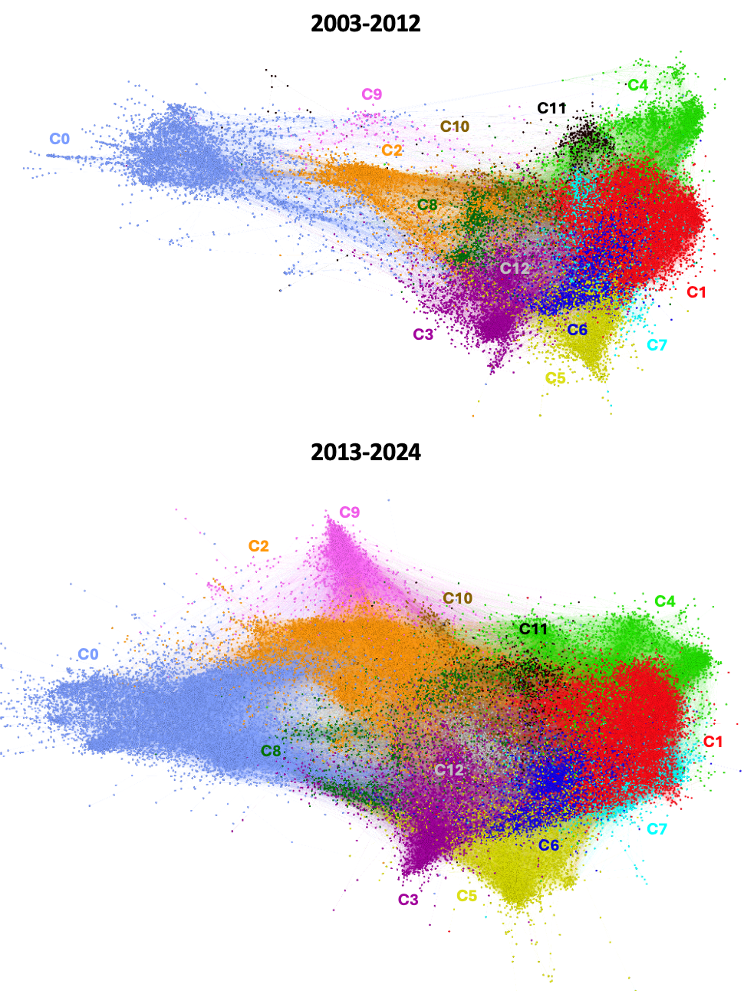}
    \caption{The temporal evolution of the citation network diagram presented in Fig.~\ref{fig:network}, split into the last decade (2013--2024: bottom) and the prior decade (2003--2012: top), with the 13 identified clusters colour coded and highlighted; see Table~\ref{Table_CNA} for the cluster labels.}
    \label{fig:CNA_evolution}
\end{figure}

\subsection{A view from citation network analyses}\label{cna}

Citation networks are powerful tools to track scientific progress through the identification of research clusters, such as those defined in Fig.~\ref{fig:network}. These clusters are essentially research ``communities" (i.e.,\ the community with which an individual researcher would likely identify). However, these clusters are not static but change over time as new papers are published, citation patterns shift, and new research connections are made. We can employ the citation network analysis tools introduced in \S\ref{context} to track how the clusters shown in Fig.~\ref{fig:network} have progressed and evolved with time, a measure of community progress.

In Fig.~\ref{fig:Cluster progression}a we show the changes in the number of publications per year for the 13 research clusters over the 2003--2023 period. A number of features are evident from this figure. First, over decadal-long timescales, the number of publications within each research cluster have increased: the typical increase from the 2003--2012 period to the 2013--2023 period is a factor of $\approx$~1.5. Three clusters (C0, C2, and C9) have clearly undergone strong growth over the last decade with particularly rapid increases in the number of papers since 2015. Each of these clusters are tightly connected to fundamental BH research and the identification of the first gravitational waves, and the consequential need to interpret and understand these data are strong drivers of the impressive growth in these clusters. However, it is notable that the majority of the papers in these clusters are not published in typical ``astrophysical" journals,\footnote{In this analysis we used the following six journals: \textit{Mon Not R Astron Soc}; \textit{ApJ}; \textit{ApJL}; \textit{Astron. J}; \textit{Astron. Astrophys}; and \textit{New Astron. Rev}.\label{foot:astro_journals}} as shown in Fig.~\ref{fig:Cluster progression}b: the researchers in these clusters probably feel more associated to theoretical physics and particle physics than astronomy and astrophysics. Although less clear from  Fig.~\ref{fig:Cluster progression}a, C12, which focuses on transient and TDE-related research, has also grown rapidly over the last decade almost certainly due to the increased availability of transient photometric data from the PTF, ZTF, and ASAS-SN facilities. The C1 cluster, that is most closely associated with this review, has grown at the median rate over the last decade. However, it underwent much more dramatic growth in the previous decade likely driven by the discovery of black-hole--galaxy scaling relations in the mid-late 1990's and the demonstration that ``AGN feedback" is required in cosmological simulations to explain the observed properties of galaxies in the early-mid 2000's.

In Fig.~\ref{fig:CNA_evolution} we have sub divided the CND from Fig.~\ref{fig:network} into two decadal-long periods: 2003--2012 and 2013--2024. The larger clusters (C0--C6) were already well defined in the first decade and have kept a similar morphology (a measure of clustering strength) through the last decade. However, it is notable that C0 was originally much more disparate and offset from the other clusters but has become better connected over the last decade. This is largely due to the detection of gravitational waves and the EHT constraints on the physical properties of the SMBHs in Sgr~A$*$ and M~87, which bridge interests in both theoretical physics and astronomy; indeed, see the relative ``domains" of the EHT papers on Sgr~A$*$ and M~87 in Fig.~\ref{fig:network}. C9 has become much better defined over the last decade, reflecting it's growth from originally a small cluster to now a medium size cluster (see Fig.~\ref{fig:Cluster progression}a). Most of the other smaller clusters are more diffuse and filamentary but reside in approximately the same positions across both CNDs. The exception is C8, which has a distinct new extension towards C0 due to the SMBH constraints on Sgr~A$*$ from the EHT.

We finish this brief analysis with a couple of related notes. First, as shown by analyses of the publication output across Physics over the last century, increases in the number of published papers are driven by increases in the number of researchers working within a given cluster, rather than individual researchers becoming more productive \citep{Sinatra2015_century,Fanelli2016_publication_rate}. Of course, there may be some  variations between sub fields and clusters, although overall any significant rises in publication output over the median trend will be driven by researchers ``flocking" to these research fields, possibly due to greater funding opportunities. Second, we should be cautious about forecasting future growth based on any dramatic increases in publication within a cluster as they are unlikely to be long lived and may even decline with time as other areas of research become more appealing to both researchers and funding agencies.

\subsection{Anticipated progress over the next decade}\label{sec:progress_future}

Identifying the drivers of scientific progress over the last decade not only helps us retrospectively understand where we have been able to make progress and why, it also helps us to provide guidance for generating future progress more efficiently. Facilities are a major driver of the scientific progress in astronomy and the largest new facilities are decided and defined a decade prior to starting operations so we already have a good idea of what the next decade will bring in terms of facilities. 

A non-exhaustive list of expected major new facilities includes several extremely large ground-based optical--infrared telescopes with $>30$~m diameter primary mirrors (e.g.,\ the European ELT; the Thirty Meter Telescope), the square-kilometer radio array (SKA), the {\it Euclid} and {\it Roman} near-IR space telescopes, the {\it Athena} and {\it AXIS} X-ray telescopes, the Cherenkov Telescope Array (CTA) Gamma-ray facility, the {\it LISA} and {\it Einstein} gravitational-wave observatories, ground-based large dedicated multi-object spectroscopic telescopes (DESI; MOONS; PSF; 4MOST; WEAVE), and the LSST survey with the Vera Rubin telescope. These are just some of the major new facilities and doesn't include the current facilities which are already making great advances and yet still maturing such as the {\it LOFAR} radio telescope, the {\it eROSITA} observatory, and the {\it JWST} space telescope. In terms of computing facilities, graphics processing units (GPUs) are likely to become more prevalent over the next decade offering much greater performance than central processing units (CPUs) particularly for parallel processing. Artificial intelligence (AI)-based processors will also likely become available, along with improvements in the capacity and performance of data storage and memory.

We can expect great discoveries and potential conceptual advances from all of these facilites, which open up new parameter space across the broad frequency--sensitivity--area--source statistic plane. Given the prominent growth rates of clusters C0, C2, C9, and C12 we highlight {\it LISA} which will provide the first sensitive constraints on gravitational waves from SMBHs and the Vera Rubin telescope which will fully open up the astronomical community to unprecedented large-scale sensitive multi-band optical photometric monitoring and the identification of transient phenomena. On the simulation side, developments in computing architecture over the next decade should allow for the first simulations on exascale supercomputers ($>10^{18}$~double-precision operations/s) and provide at least an order of magnitude improvement in the box-size--simulation resolution ratio over current state-of-the-art simulations. 
Improvements in sub-grid modelling will be possible thanks to enhanced numerical resolution and to closer, cross-scale connections to ever more comprehensive and physically-accurate small-scale simulations. 
Such work is already underway \citep{Costa:2020, Rennehan:2024, Koudmani:2024, Husko:2024} and future large-scale models will certainly tap into predictions from small-scale radiative transfer and general-relativistic-magnetohydrodynamic simulations \citep{Ricarte:2023, Kaaz:2024}, simulations performed with more accurate gravity solvers \citep{Genina:2024}, and more \emph{ab initio} simulations of BH seed formation \citep{Regan:2024}.
Innovative machine learning approaches to sub-grid modelling, currently in their infancy, may come to play an important role in porting small-scale simulation results and predictions to large-scale, coarse-resolution simulations \citep[e.g.][]{Stachenfeld:2021}.   

It is also undeniable that astronomy has become more data rich over the last decade with the current generation of facilities already providing such vast quantities of data that it is becoming impractical for any individual to store all of their data \citep[e.g.,\ LOFAR generates tens--hundreds of terrabytes of data per day;][]{Manzano2024}. In some cases it has even become impractical for the observatory to retain all of the raw data as it is more effective to simply re-take the observations. This increase in data will drive further development of data-analysis techniques capable of handling such large datasets. Over the last decade, we have seen a large increase in the adoption of machine-learning approaches to analyse large datasets (see Footnote~\ref{foot:ML}), which will surely become even more prevalent over the next decade. Given the massive rise in the ability of generative artifical intelligence (gAI) over just the last few years, it also seems inevitable that gAI will play a more dominant role in data analyses over the next decade.


A great, largely untapped, potential for significant scientific progress over the next decade is improved interactions between different research communities, potentially through multi or inter disciplinary workshops \citep{Burns_2025}. The CNDs shown in Figs.~\ref{fig:network} \& \ref{fig:CNA_evolution} show connections between different ``black hole" research communities, some of which are comparatively isolated and could benefit from having broader scientific discussions with other research communities. This can be easily achieved through the facilitation of focused workshops, aimed to address key questions by bringing together experts from different communities (even beyond the astronomical community) who can bring new knowledge, skills, and approaches to address key questions. Compared to the large budgets and long timescales required to develop new facilities, focused workshops provide extremely cost effective and immediate solutions to making significant scientific progress.

\section{Conclusions}\label{sec:conclusion}

In this review we have provided an overview of our understanding of the growth of SMBHs to address the broad question: what drives the growth of black holes? This question is somewhat rhetorical as, has been established for several decades, the growth of black holes is the result of the accretion of gas onto the black hole. However, the broader implication of this question is {\it how} does the gas reach the vicinity of the black hole, {\it what} was the earlier origin of that gas, {\it are} there are specific environments more conducive to black-hole growth than others, and {\it what} influence does the black hole have on the evolution of the host galaxy? We have shown that huge progress has been made in addressing each of these questions over the last decade. 

We now have a much greater physical understanding of the accretion process, driven by both theoretical and observational advances, and have opened up the time-domain axis in AGN astronomy, identifying instances of changing accretion states. Our understanding of the obscuring medium has undergone a revolution and led to the development of more physical models of the AGN ``torus", largely driven by high spatial resolution observations as well as a better understanding of the role of galaxy scale structures in AGN obscuration. We have a better understanding of the connection between the AGN and its hosting galaxy and larger-scale environment and have demonstrated that AGN are ``events" within the broader life cycle of galaxies which can significantly impact the long-term evolution of the galaxies through energetic ``outflows" (accretion driven winds and jets). We also have a significantly greater census of AGN activity across the luminosity--obscuration--redshift plane, with sample sizes now numbering in the millions, and we are starting to witness the very earliest phases of SMBH growth.

We have also evaluated the drivers of this scientific progress, using the AH12 review as a ``snapshot" of the field from a decade ago. Inevitably much of the scientific progress is driven by new facilities and more abundant data across the whole of the broad population in luminosity, redshift, and wavelength space, which has required the development of new techniques to analyse the ever increasing data volume. This has led to several conceptual advances such as the examples given in the previous paragraph. We have also shown how scientific progress can be driven through community developments, both in citizen science and the professional research community, and utilised citation network analyses to demonstrate the full breadth of the black-hole research network. A key component of community engagement is scientific workshops such as our Durham--Dartmouth black-hole workshops, which form the backbone of this (and the original AH12) review. With that final concluding thought we thank you for reading this review and we hope to see you at the next black-hole workshop focused on understanding ``What drives the growth of black holes?".




\section*{Acknowledgements}

We dedicate this review to the memory of Richard Bower who sadly passed away on 6th January 2023. Richard was not only a brilliant scientist and polymath, making significant contributions to theory, observation, and instrumentation in astronomy and beyond, he was also a warm and entertaining colleague and friend. Richard was also one of the greatest patrons of our black-hole workshop series, missing just two meetings, lighting up a session with his knowledge, passion, and personality. His influence was particularly great in our 2012 workshop on ``What is the role of AGN in the evolution of galaxies?" at Dartmouth College where he became the {\it de facto} arbiter on the outcome of scientific debates, giving his thumbs up or thumbs down in agreement or not ({\it pollice verso})!

We thank the 187 participants of our Iceland meeting in 2022 for the great contributions to the discussions and debates, which reinvogorated our passion for scientific conferences following 2 years of online meetings during the Covid-19 pandemic. We also want to thank everyone who has attended at least one of our Durham--Dartmouth black hole meetings over the last decade, particularly those who contributed to the great discussions which have helped to guide the field and lead to scientific progress. A special thanks to Roger Blandford FRS, our surprise attendee to the Iceland meeting, for presenting the awards to our long-serving patrons at the Iceland meeting and for the great after-dinner speech on the drivers and progress over the last 50 years of black-hole research.

We thank the referee for constructive and positive feedback and the following people for commenting on earlier drafts of this review and/or providing data and figures: Alastair Basden, Xiaohui Fan, Vicky Fawcett, Claire Greenwell, Stephanie Juneau, Dale Kocevski, Rohit Kondapally, Karen Leighly, Rusen Lu, James Matthews, John Paice, Krisztina Perger, Cristina Ramos-Almeida, Amy Rankine, Jan-Torge Schindler, Jan Scholtz, Matthew Temple, Carolin Villforth, and Sarah White. We extend a special thanks to Matthew Temple for producing Figure 27. DMA thanks the Science Technology Facilities Council (STFC) for support through the Durham consolidated grant (ST/T000244/1).  
JA, CMH, and LKM acknowledge funding from United Kingdom Research and Innovation (UKRI) grants MR/T020989/1, MR/V022830/1, and MR/Y020405/1, respectively. MH acknowledges funding from the Swiss SNSF Starting Grant (218032). RCH acknowledges support from NASA via grant numbers 80NSSC22K0862 and 80NSSC23K0485, and is grateful to the Durham Department of Physics and Center for Extragalactic Astronomy for hosting a very enjoyable and productive sabbatical visit. We thank the Matariki Network of Universities, of which both Dartmouth College and Durham University are partners, for financial support of our institutional meetings. 


\appendix

\section{Durham--Dartmouth black-hole workshops}
\label{workshops}

Over the last decade we have held six black-hole growth workshops, our Durham--Dartmouth black-hole workshop series. Brief details on each workshop, including the number of participants, are given in Table~\ref{workshop_table}. Overall, 512 participants have attended at least one of our workshops. A small group of elite participants have attended four or more workshops: James Aird, Richard Bower, Ric Davies, Chris Harrison, Dale Kocevski, James Mullaney, David Rosario, Carolin Villforth, and Belinda Wilkes. Five of these, our greatest workshop patrons, also attended the Iceland conference and received a ``long-service'' award, presented by our special guest, Roger Blandford. 

\begin{figure*}
	\centering 
	\includegraphics[width=0.98\textwidth, angle=0]{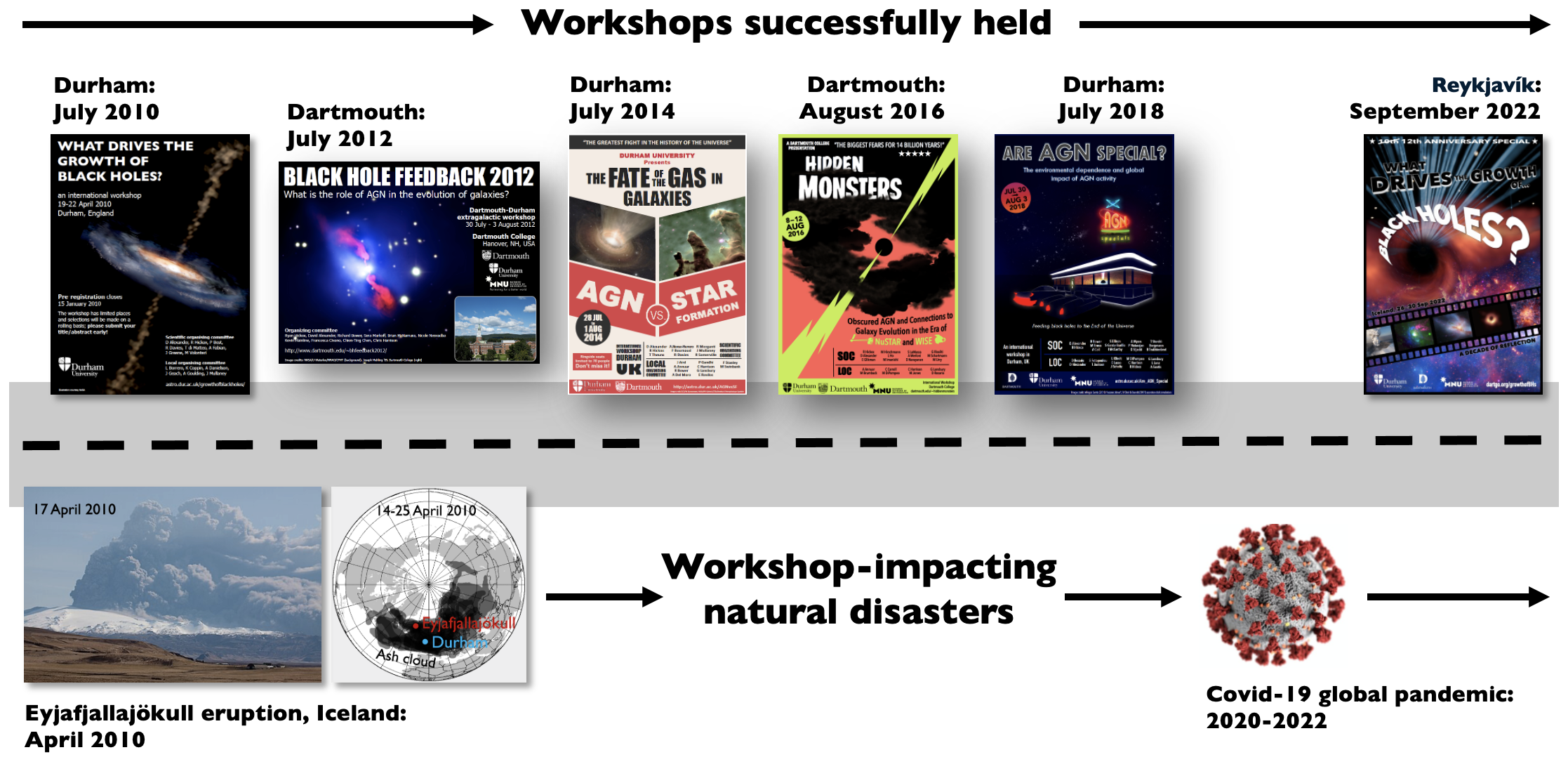}	
	\caption{The long and interesting road to the first decade of our black-hole growth workshops, illustrated by the workshop posters (top) and punctuated with natural disasters (bottom). The 2014, 2016, and 2022 workshop posters were designed by George Lansbury and inspired by a 1950's boxing promotion poster, a horror B movie, and a disaster movie, respectively. The 2018 workshop poster was designed by Allegra Santis-Rosario, inspired by science fiction and the ``Restaurant at the edge of the Universe" book by Douglas Adams. The natural-disaster images show from left to right: the Iceland volcanic-eruption on Eyjafjallajökull, taken on 17 April 2010 (two days before the original start of the 2010 workshop) by Henrik Thorburn, the ash-cloud distribution over the Northern hemisphere constructed by the Metoffice over 14-25 April 2010 with Eyjafjallajökull (red) and Durham (blue) highlighted and labelled, and a Covid-19 coronavirus schematic by CDC on Unsplash to illustrate the global pandemic which lasted from early 2020 to early 2022.} 
	\label{fig_conf}%
\end{figure*}

\begin{table*}
\begin{tabular}{lllc} 
 \hline
Workshop title & Dates & Location & N\\ 
 \hline
What drives the growth of black holes? & 26--29 July 2010 & Durham & 119\\
What is the role of AGN in the evolution of galaxies? & 30 July--3 August 2012 & Dartmouth & 64\\
The fate of the gas in galaxies: AGN versus star formation & 28 July--1 August 2014 & Durham & 97\\
Hidden monsters: obscured AGN and connections to galaxy evolution & 8--12 August 2016 & Dartmouth & 98\\
Are AGN special? The environmental dependence and global impact of AGN activity & 30 July--3 August 2018 & Durham & 112\\
What drives the growth of black holes? A decade of reflection & 26---30 September 2022 & Reykjavík & 187\\ 
 \hline
\end{tabular}
\caption{Basic details of our series of black-hole workshops along with the workshop dates, locations, and number of participants.}
\label{workshop_table}
\end{table*}

An illustration of these workshops, including the challenging series of natural hazards related to any of our workshops with the words ``What drives'' in the title, is shown in Fig.~\ref{fig_conf}! Our first ``What drives the growth of black holes?" workshop was originally scheduled for 19-22 April 2010 at Durham University. However, the Icelandic volcano Eyjafjallajökull, which had first started erupting one month earlier, drove large quantities of fine ash into the atmosphere which got blown across a large fraction of the Northern hemisphere and, critically, all of the UK; see Fig.~\ref{fig_conf}. This caused European airspace to shut down, forcing us to reschedule the workshop to July, thankfully an eruption-free month. Two determined participants still managed to travel to Durham for the original week in April!



\begin{figure}
	\centering 
	\includegraphics[width=0.47\textwidth, angle=0]{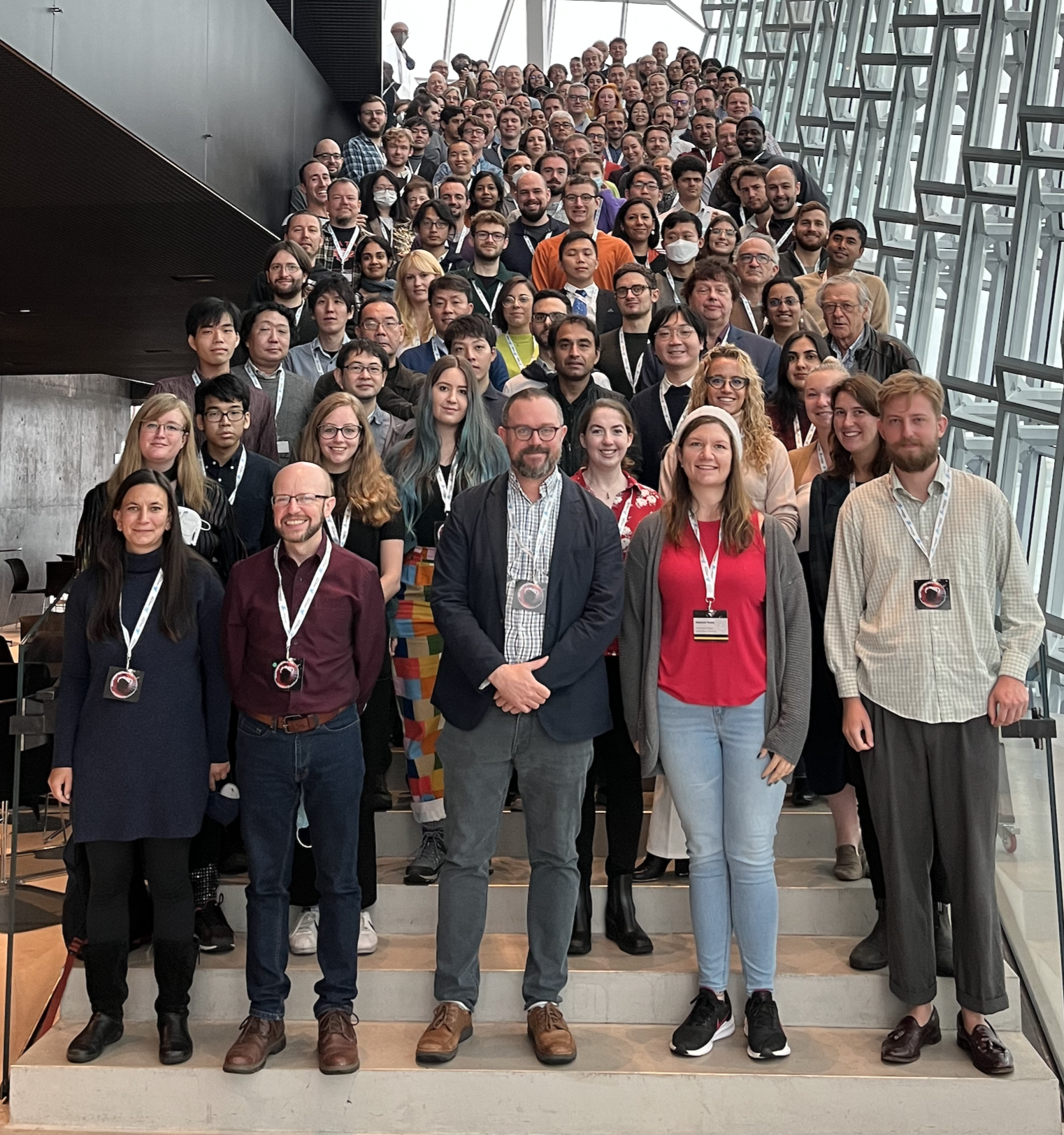}	
	\caption{Iceland conference photo taken inside the Harpa concert and conference venue, Reykjavík, in September 2022.} 
	\label{fig_conf_photo}%
\end{figure}

\begin{figure}
	\centering 
	\includegraphics[width=0.47\textwidth, angle=0]{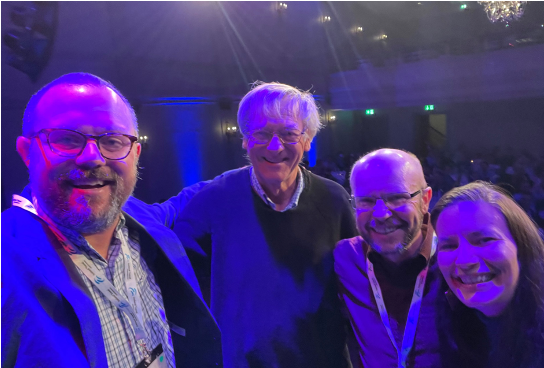}	
	\caption{Iceland conference dinner entertainment team on stage at the Gamla Bíó theatre following the long-service awards ceremony. From left to right: Ryan Hickox, Roger Blandford FRS, David Alexander, and Leah Morabito.} 
	\label{fig_conf_awards}%
\end{figure}

A series of workshops followed on a two-year cycle alternating between Dartmouth College and Durham University. In these meetings we focused on specific subject areas covered in the original ``What drives the growth of black holes?" workshop and AH12 review. Early on in the planning we decided that our sixth workshop, 10 years on from the first workshop, would be a celebration of the scientific progress made over the last decade. There could only be one location for this meeting: Iceland, to face  Eyjafjallajökull and challenge nature to further disrupt our workshop series! Nature, instead, gave us a 2-year global pandemic with the Covid-19 coronavirus! But, eventually, after failed attempts of holding the meeting in 2020 and then 2021, with some angst and trepidation we successfully scheduled and held the sixth workshop in September 2022 at the Harpa Concert and Conference venue in Reykjavík, Iceland. This meeting was eponymously named after the original workshop and scheduled to have the same scientific sessions but with the additional focus of investigating the scientific progress made over the last decade: ``What drives the growth of black holes? A decade of reflection". The conference photo from this meeting is shown in \ref{fig_conf_photo} and a photo taken following the long-service awards at the conference-dinner venue, the Gamla Bíó theatre, is shown in \ref{fig_conf_awards}.

\section{Citation Network Analysis}
\label{CNA}

Citation network analysis is the application of graph theory and network science to bibliographic data. Citation networks consist of nodes that represent individual papers, with directed edges between nodes representing citations \textit{from} a citing \textit{to} a cited paper. These networks are used to analyse how the scientific literature is structured into specific disciplines, fields, and topics. These structures are self-organising: papers on the same or similar topics cluster together into densely interacting literatures due to the tendency of authors to cite other papers relevant to their own \citep{Price1965}. To understand how a literature clusters into different areas, modularity maximisation \citep{Newman2004} via the Leiden Algorithm\citep{Traag2019} is a common and well validated approach \citep{Klavans2017_CNA}. By these methods, citation networks have been used to produce ‘maps’ of literatures in a number of different areas, including sustainability sciences \citep{Kajikawa2014}, oxytocin research \citep{Leng2021,Leng2022}, and energy storage \citep{Mejia2020}). Beyond its use in literature survey, it is also used to test for citation bias and error propagation in evidence-bases informing the evaluation of hypotheses \citep{Greenberg2009,Leng2018}. 

\begin{table}
\begin{tabular}{cccl} 
 \hline
\# & Papers & Avg. & Label\\ 
 & & year\\
 \hline
C0 & 13872 & 2015.5 & Black-hole physics\\
C1 & 10043 & 2014.3 & AGN, quasars, and galaxies\\
C2 & 6984 & 2016.2 & Gravitational waves: binary BHs\\
C3 & 5574 & 2013.5 & X-ray binaries, accretion disks\\
C4 & 5104 & 2014.4 & High-redshift quasars, reionisation\\
C5 & 4846 & 2014.5 & Jets, blazars\\
C6 & 4195 & 2013.9 & Seyfert galaxies, emission lines\\
C7 & 2686 & 2013.9 & Radio galaxies, clusters\\
C8 & 2613 & 2014.6 & Galactic centre\\
C9 & 1566 & 2018.7 & Primordial BHs, early universe\\
C10 & 1332 & 2015.7 & Gravitational waves: SMBHs\\
C11 & 1186 & 2014.3 & Gravitational lensing\\
C12 & 923 & 2018.0 & TDEs, time domain\\
\hline
\end{tabular}
\caption{Details of the citation-network clusters shown in Fig.~\ref{fig:network}: cluster number, number of papers in cluster, average year of publication, and cluster label.}
\label{Table_CNA}
\end{table}

To construct the citation network discussed in \S\ref{context}, we retrieved a large literature from the Web of Science (WoS) Core Collection by searching for the following terms in the title, abstract, and associated keywords of indexed papers in prominent astrophysics journals: TS = (``black hole" OR ``black holes" OR ``active galactic nucleus" OR ``active galactic nuclei" OR ``galaxies: nuclei” OR ``quasar” OR “quasars”).

This search query returned a total of 61,036 articles and reviews published between January 2003 and March 2024. All metadata associated with these were retrieved in March 2024. These data were restructured in the R statistical environment into a: ‘Node-attribute list’ that contains (i) a unique ID for each unique reference string; (ii) WoS unique identifier number; (iii) Digital Object Identifier (DOI), if available; (iv) name of all authors; (v) title of paper; (vi) journal of publication; (vii) year of publication; and (viii) total number of citations recorded by WoS). An ‘edge list’ was also built from the bibliographies of retrieved papers containing two columns: Source and Target, which record the unique identifiers of each citing and cited paper, respectively.

\begin{table*}
\centering
 Proportion of within-network citations from within a given cluster and between clusters\\
 \begin{tabular}{c|c|c|c|c|c|c|c|c|c|c|c|c|c}
 \hline
 & {\bf C0} & {\bf C1} & {\bf C2} & {\bf C3} & {\bf C4} & {\bf C5} & {\bf C6} & {\bf C7} & {\bf C8} & {\bf C9} & {\bf C10} & {\bf C11} & {\bf C12}\\ 
 \hline
C0 & {\bf 93.3\%} & 0.2\% & 4.9\% & 2.6\% & 0.4\% & 0.8\% & 0.8\% & 0.2\% & 7.4\% & 2.6\% & 1.2\% & 1.4\% & 0.5\% \\
C1 & 0.1\% & {\bf 77.3\%} & 0.7\% & 3.1\% & 9.9\% & 2.3\% & 13.4\% & 18.2\% & 4.5\% & 0.2\% & 7.8\% & 2.8\% & 2.1\% \\
C2 & 3.2\% & 0.8\% & {\bf 85.1\%} & 4.4\% & 1.3\% & 0.8\% & 0.4\% & 0.4\% & 5.8\% & 4.5\% & 8.2\% & 1.7\% & 5.9\% \\
C3 & 0.6\% & 1.0\% & 1.9\% & {\bf 75.1\%} & 0.6\% & 2.4\% & 4.7\% & 2.2\% & 4.2\% & 0.1\% & 1.6\% & 0.7\% & 2.2\% \\
C4 & 0.1\% & 6.1\% & 1.0\% & 0.8\% & {\bf 79.5\%} & 1.0\% & 3.8\% & 2.1\% & 1.0\% & 0.7\% & 2.4\% & 3.1\% & 0.6\% \\
C5 & 0.2\% & 1.2\% & 0.3\% & 2.9\% & 0.8\% & {\bf 83.1\%} & 3.7\% & 6.2\% & 2.7\% & 0.1\% & 2.0\% & 1.2\% & 1.2\% \\
C6 & 0.2\% & 4.8\% & 0.2\% & 4.8\% & 2.9\% & 2.5\% & {\bf 66.7\%} & 2.8\% & 0.8\% & 0.0\% & 2.2\% & 7.1\% & 3.7\% \\
C7 & 0.0\% & 4.4\% & 0.1\% & 1.7\% & 1.2\% & 3.4\% & 1.5\% & {\bf 65.4\%} & 0.7\% & 0.1\% & 1.5\% & 0.5\% & 0.6\% \\
C8 & 1.2\% & 1.0\% & 1.4\% & 2.3\% & 0.5\% & 1.9\% & 0.4\% & 0.6\% & {\bf 68.6\%} & 0.4\% & 2.4\% & 0.3\% & 2.5\% \\
C9 & 0.8\% & 0.1\% & 1.8\% & 0.2\% & 0.7\% & 0.2\% & 0.0\% & 0.1\% & 0.6\% & {\bf 90.1\%} & 2.4\% & 1.2\% & 0.1\% \\
C10 & 0.1\% & 1.9\% & 1.8\% & 0.7\% & 1.0\% & 0.8\% & 1.5\% & 1.0\% & 1.7\% & 0.9\%	& {\bf 66.1\%} & 1.4\% & 1.5\% \\
C11 & 0.1\% & 0.5\% & 0.2\% & 0.1\% & 0.9\% & 0.2\% & 1.5\% & 0.3\% & 0.1\% & 0.4\%	& 0.3\% & {\bf 78.6\%} & 0.1\% \\
C12 & 0.1\% & 0.8\% & 0.8\% & 1.4\% & 0.2\% & 0.5\% & 1.6\% & 0.4\% & 1.9\% & 0.1\%	& 1.9\% & 0.2\%	& {\bf 78.8\%} \\
 \hline
\end{tabular}
\\
 Proportion of within-network references from within a given cluster and between clusters\\
 \begin{tabular}{cccccccccccccc}
 \hline
 & C0 & C1 & C2 & C3 & C4 & C5 & C6 & C7 & C8 & C9 & C10 & C11 & C12\\ 
 \hline
{\bf C0} & {\bf 88.8\%} & 0.3\% & 4.6\% & 1.6\% & 0.3\% & 0.4\% & 0.6\% & 0.1\% & 2.4\% & 0.5\% & 0.2\% & 0.1\% & 0.1\% \\
 \hline
{\bf C1} & 0.0\% & {\bf 82.5\%} & 0.4\% & 1.2\% & 4.0\% & 0.6\% & 5.7\% & 3.4\% & 0.9\% & 0.0\% & 0.9\% & 0.2\% & 0.2\% \\
 \hline
{\bf C2} & 3.3\% & 1.5\% & {\bf 84.6\%} & 3.0\% & 1.0\% & 0.4\% & 0.3\% & 0.1\% & 2.0\% & 0.9\% & 1.7\% & 0.2\% & 0.9\% \\
 \hline
{\bf C3} & 0.9\% & 3.2\% & 2.9\% & {\bf 80.1\%} & 0.6\% & 2.0\% & 5.6\% & 1.2\% & 2.3\% & 0.0\% & 0.5\% & 0.1\% & 0.5\% \\
 \hline
{\bf C4} & 0.1\% & 15.1\% & 1.2\% & 0.7\% & {\bf 75.7\%} & 0.7\% & 3.8\% & 0.9\% & 0.4\% & 0.2\% & 0.6\% & 0.4\% & 0.1\% \\
 \hline
{\bf C5} & 0.4\% & 4.1\% & 0.6\% & 3.6\% & 1.1\% & {\bf 78.1\%} & 5.2\% & 3.9\% & 1.7\% & 0.0\% & 0.7\% & 0.2\% & 0.3\% \\
 \hline
{\bf C6} & 0.3\% & 13.3\% & 0.2\% & 4.6\% & 3.0\% & 1.8\% & {\bf 72.6\%} & 1.4\% & 0.4\% & 0.0\% & 0.7\% & 1.1\% & 0.8\% \\
 \hline
{\bf C7} & 0.1\% & 23.1\% & 0.3\% & 3.0\% & 2.5\% & 4.8\% & 3.1\% & {\bf 61.0\%} & 0.7\% & 0.0\% & 0.9\% & 0.1\% & 0.3\% \\
 \hline
{\bf C8} & 3.7\% & 5.7\% & 4.2\% & 4.7\% & 1.1\% & 3.0\% & 0.9\% & 0.7\% & {\bf 72.9\%} & 0.3\% & 1.5\% & 0.1\% & 1.2\% \\
 \hline
{\bf C9} & 3.5\% & 0.9\% & 7.6\% & 0.6\% & 2.3\% & 0.4\% & 0.1\% & 0.1\% & 1.0\% & {\bf 80.7\%} & 2.2\% & 0.5\% & 0.1\% \\
 \hline
{\bf C10} & 0.6\% & 15.8\% & 7.3\% & 2.0\% & 3.1\% & 1.7\% & 4.9\% & 1.5\% & 2.4\% & 0.7\% & {\bf 58.4\%} & 0.6\% & 1.0\% \\
 \hline
{\bf C11} & 1.0\%	& 7.7\% & 1.4\% & 0.8\% & 5.7\% & 1.0\% & 9.5\% & 0.8\% & 0.4\% & 0.7\% & 0.5\% & {\bf 70.2\%} & 0.2\% \\
 \hline
{\bf C12} & 0.3\%	& 8.5\%	& 4.5\% & 5.0\%	& 0.8\% & 1.5\% & 6.9\% & 0.7\% & 3.6\% & 0.1\% & 2.1\% & 0.1\% & {\bf 65.9\%} \\
 \hline
\end{tabular}
\caption{Contingency table showing the degree of cross citation or cross referencing between clusters. The top section shows the percentage of the total within-network citations (in degree) contributed by papers within and between the clusters; the data is read in column format, as indicated. The bottom section shows the percentage of the total within-network references (out degree) directed within and between clusters; ; the data is read in row format, as indicated.}
\label{Table_contingency}
\end{table*}

We restrict our analysis to the retrieved papers with full bibliographic data, and focused only on the main component; i.e.,\ the largest connected network in which an undirected path of edges connects any two randomly chosen nodes together. This leaves a network containing 60,924 papers connected together by 1,378,057 citation links. To cluster this network, modularity maximisation \citep{Newman2004} via the Leiden algorithm \citep[][]{Traag2019} was performed in R using the ‘igraph’ package resulting in 13 clusters at $Q=$~0.67 (see Eqn~\ref{eqn:modularity} below) with the following parameters: resolution 1, iterations 100, repetitions 100. Modularity maximization is a method used to identify communities within a network by optimizing the modularity value $Q$. It measures the strength of division of a network into clusters, with higher values indicating denser connections within clusters compared to between them. Due to the directed acyclic structure of citation networks, we converted the network into an undirected variant following \cite[][]{Speidel2015}. For an undirected graph, modularity is calculated by:

\begin{equation}
    Q={\frac{1}{2m}{\sum_{i,j}\left(A_{ij}-{\frac{d_i d_j}{2m}}\right)}}={\delta(c_i,c_j)},
    \label{eqn:modularity}
\end{equation}

\noindent where $m$ is the total number of edges, $A_{\rm ij}$ is set to 1 if there is an edge between nodes $i$ and $j$ or otherwise is set to 0, $\frac{d_i d_j}{2m}$ is the probability of an edge between two nodes proportional to their degree, the delta function ($\delta$) is set to 1 if $i$ and $j$ are in cluster $c$ or otherwise is set to 0.

We report the contingency matrices for the cluster network in Table~\ref{Table_contingency} which show the percentage of papers within-network citations (top) and within-network references (bottom) directed within and between clusters. For example, the diagnonal lines indicate how ``insular" a given sub field (i.e.,\ research community) is in terms of citations (top section) and references (bottom section): C0 and C9 (``black-hole physics" and ``primordial BHs and the early universe") tend to reference (and are cited by) other studies within their cluster while C7 and C10 (``radio galaxies, clusters" and ``Gravitational waves: SMBHs") are much more open in their referencing and citations, although they are much smaller clusters than C0.



\bibliographystyle{elsarticle-harv} 
\bibliography{example}






\end{document}

%% file: sec4.tex
\section{What properties of the host galaxies or larger-scale environment affect black-hole growth?}
\label{sec4:hosts}
\newcommand{\ja}[1]{\textcolor{purple}{JA: \textbf{#1}}}
\newcommand{\rch}[1]{\textcolor{green}{RCH: \textbf{#1}}}



The previous section reviewed progress in our understanding of the mechanisms that brings gas to the centres of galaxies to fuel the accretion process and provided an overview of accretion onto SMBHs. In this section we discuss what we have learnt about whether these processes preferentially occur---and thus trigger AGN---within galaxies with certain properties or within certain large-scale environments. Such studies can help us understand how SMBH growth and the growth of galaxies proceed together over cosmic time as well as revealing whether AGN phases and the huge energy they release in different forms has an \emph{impact} on the properties and evolution of galaxies or their larger scale environment (such ``feedback'' processes are discussed further in \S\ref{star-formation}). 

\begin{figure*}
	\centering 
    \includegraphics[width=0.9\textwidth, angle=0]{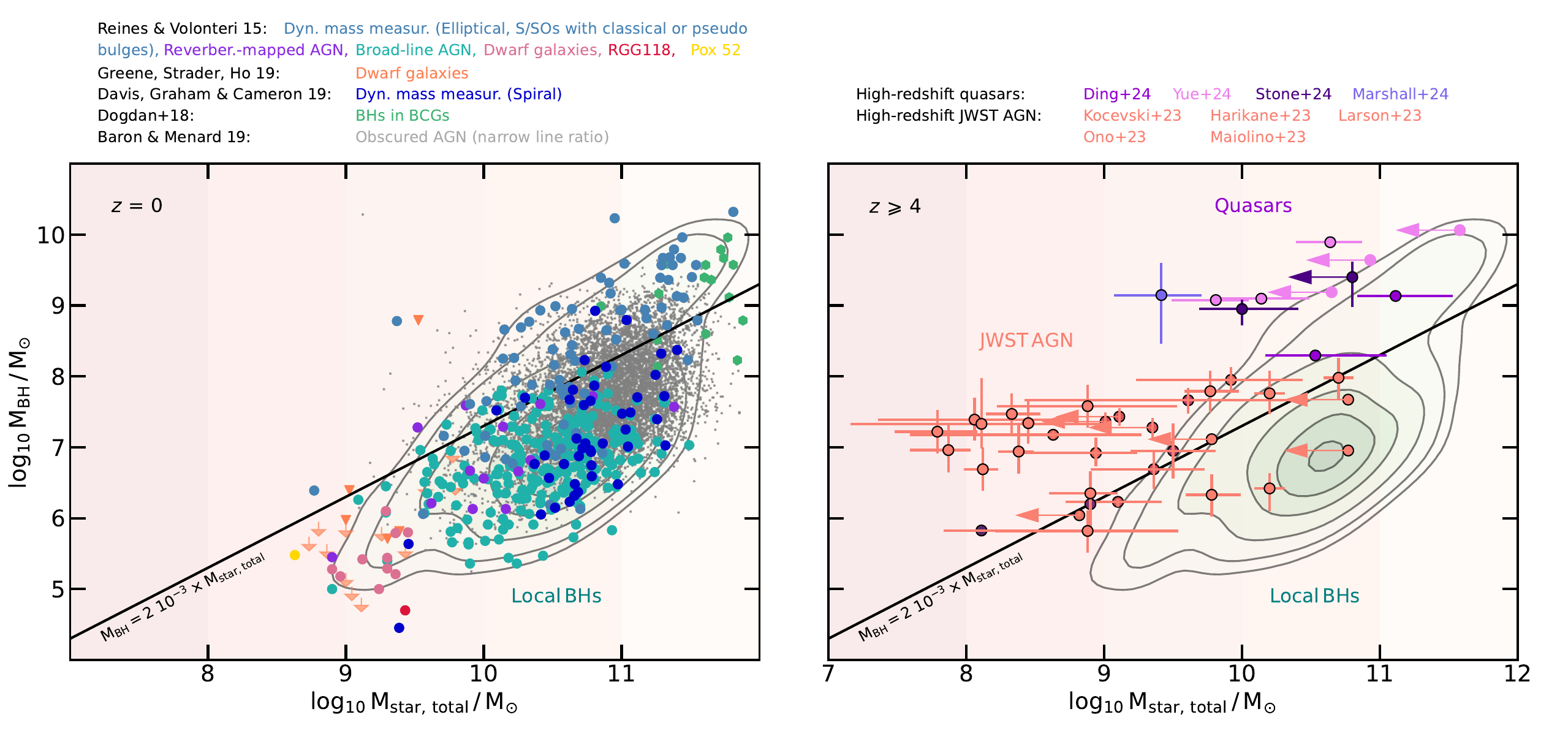}
	\caption{{\it Left panel:} SMBH mass as a function of the total stellar mass of their host galaxies in the local Universe, demonstrating the overall connection between the growth of galaxies and their central SMBHs. The sources are categorized into several groups: SMBHs with dynamical mass measurements in large galaxies (light blue) and spiral galaxies (dark blue), reverberation-mapped AGN (purple), and broad-line AGN (teal), all presented in \citet{reines_relations_2015,2018ApJ...869..113D}, and SMBHs in low-mass and dwarf galaxies \citep[pink to red symbols, including several upper limits,][]{reines_relations_2015,greene_intermediateMass_2020}, and SMBHs in BCG galaxies \citep[green,][]{2018ApJ...852..131B}. The contours represent the combined populations of these groups. Additionally, we include a large sample of obscured AGN from SDSS in the background \citep[$z\leqslant0.3$, grey,][]{baron_black_2019}. 
    The solid black line shows $M_\mathrm{BH}=0.002 M_\mathrm{star,total}$ \citep{marconi_2003} that is widely adopted in definitions of specific SMBH accretion rates or kinetic power (see footnotes \ref{footnote:sBHAR} and \ref{footnote:sBHKP}).
    {\it Right panel:} Sources at high redshift ($z\geqslant 4$) discussed in \S\ref{sec:highzqso}. Quasars characterized with {\it JWST} observations, i.e. with both SMBH mass and total stellar mass measurements, are represented by pink and purple symbols \citep{2023Natur.621...51D,2023ApJ...953..180S,2024ApJ...966..176Y,2024arXiv241011035M}. {\it JWST}-discovered AGN candidates with stellar mass estimates are shown in orange \citep{2023arXiv230200012K,2023arXiv230311946H,2023arXiv230801230M,2023ApJ...951...72O,2023arXiv230308918L}. The contours represent systems in the local Universe and are shown to guide the eye. 
    }
	\label{fig:mbh_mstar_diagram}
\end{figure*}

Two key results that indicate the existence of an \emph{overall} connection between the growth of galaxies and the black holes that lie at their centres have been well established now for a number of decades.
First, the SFR density and the black hole accretion rate density are found to follow a similar pattern over cosmic time, both peaking when the Universe was around 2--6~Gyr in age ($z\sim3$ to $z\sim1$, referred to as ``cosmic noon'')\footnote{We note that {\it JWST} discoveries of surprisingly large populations of candidate AGN at higher redshifts (see \S\ref{sec:highzqso}) indicate substantial BH mass assembly may occur at early cosmic times and \emph{precede} the bulk of galaxy stellar assembly \citep[see e.g.][]{yang_ceers_2023,2024arXiv240610341A}.}
and dropping by a factor of $\sim10$ thereafter \citep[e.g.,][]{boyle_cosmological_1998, madau_cosmic_2014, aird_evolution_2015}. 
Thus the total amount of galaxy stellar growth and the total black hole growth via accretion appear to be correlated, suggesting a common origin.  
Second, the masses of \emph{inactive} black holes at the centres of galaxies in the local ($z\sim0$) Universe are found to be correlated with a number of different, larger-scale galaxy properties---most tightly with the velocity dispersion of the central galaxy bulge, producing the famous ``$M_\mathrm{BH}-\sigma$'' relation \citep{ferrarese_fundamental_2000, gebhardt_relationship_2000, tremaine_slope_2002}.
Correlations are also found with the bulge luminosity or mass \citep[e.g.][]{magorrian_demography_1998,marconi_2003,haring_2004} and the total stellar mass \citep[e.g.][see Fig.~\ref{fig:mbh_mstar_diagram}, left panel]{cisternas_secular_2011, reines_relations_2015}, albeit with larger scatter and possible dependencies on other galaxy properties such as SFR \citep[e.g.,][]{terrazas_supermassive_2017, greene_intermediateMass_2020}.
Direct, dynamical SMBH mass measurements only remain possible for a small number of nearby \emph{inactive} galaxies, where the SMBH sphere of influence can be resolved \citep[see review by][]{Kormendy:13}.
However, extensive efforts over the last decade have extended SMBH scaling relations to include both broad-line AGN (see \S\ref{sec:BHmass} for further discussion) and obscured AGN \citep[e.g.][]{baron_black_2019} and to push constraints to lower SMBH and galaxy masses in the dwarf galaxy regime \citep[see reviews by][]{reines_observational_2016,greene_intermediateMass_2020}.
The exact form of these scaling relations and the origin/implications of the scatter remain an area of active research \citep[e.g.][]{shankar_scaling_2019,habouzit_supermassive_2021,terrazas_diverse_2024}.
Nonetheless, these relations indicate a broad connection between the build up of the black hole mass (i.e. the result of past AGN activity within a galaxy) and the build up of the stellar component, showing that a connection between black hole and galaxy assembly may apply to \emph{individual} galaxies. 

Much research effort over the last decade has focused on trying to track this relationship \emph{as the build up of both galaxies and their SMBHs occurs} across different epochs of cosmic time. 
The last decade has seen an advance in the \emph{data} available to identify AGN and their hosts due to both the culmination of major survey efforts with the premier facilities (e.g. \textit{Chandra}, \textit{XMM-Newton}, \textit{GALEX}, \textit{HST}, \textit{Spitzer}) and a range of new facilities that provide unprecedented sensitivity or sky coverage at key wavelengths (e.g. \textit{NuSTAR} at hard X-ray energies, VISTA in the near-IR, 
\textit{WISE} in the mid-IR, \textit{Herschel} in the far-IR, LOFAR at radio wavelengths).
Alongside, there have been crucial developments in the \emph{techniques} to disentangle the emission from AGN and their host galaxies and thus enable accurate measurements of the galaxy properties. 
We provide a succinct review of the progress due to both these elements in \S\ref{sec4:multiwavelength} below. 
A result of such work has been a major \emph{conceptual} development in our understanding of SMBH growth as a transient, varying process throughout the lifetime of a galaxy, which is discussed in \S\ref{sec4:varying_growth}. 
In \S\ref{sec4:jetted} we focus on a different mode of SMBH activity---associated with strong relativistic jets that are most clearly revealed by radio surveys---that is found to be distributed differently across the galaxy population and appears to play a role in capping both the growth of a galaxy and its central SMBH.
An interesting development that has followed from our new understanding of how AGN (of differing types) occupy galaxies has been to revise our understanding of the relationship between large-scale environment and AGN activity, indicating (for the most part) that any such connection may be secondary to the more direct connection with the host galaxy properties (see \S\ref{sec4:environment}). 
In \S\ref{sec4:mergers} we aim to clarify an issue of much debate over the past decade: the role that mergers between galaxies play in the triggering of AGN activity while in
\S\ref{sec4:dwarfs} we discuss how a new field of research has emerged over the last decade focusing on active massive black holes in dwarf galaxies (i.e., galaxies with low stellar masses).
Finally, in \S\ref{sec4:keydrivers} we reflect on the key drivers that have led to progress in this field. 

\subsection{Identification of AGN and measurements of galaxy properties}
\label{sec4:multiwavelength}

A primary feature of AGN is emission across the electromagnetic spectrum, and thus a wide variety of identification techniques at different wavelengths are well established \citep[see][for systematic reviews]{padovani_active_2017,hickox_obscured_2018}. 
All methods of identifying AGN produce incomplete samples due to both the diversity of intrinsic AGN properties (e.g., accretion rate, black hole mass, obscuration/absorption by different components, strength of any collimated jet) and the relative strength of light from the host galaxy that can make it difficult to distinguish emission due to an AGN from that produced by other processes (e.g., ongoing star formation, existing stellar populations, hot interstellar gas). Ultimately, all methods of AGN selection require choices to be made and will thus be incomplete in terms of their sampling of the underlying \emph{SMBH} population.\footnote{A related difficulty when selecting samples is determining when an active SMBH should be consider an AGN. 
A threshold may be chosen in terms of luminosity (at a given wavelength).
Alternatively an Eddington rate threshold (i.e. relative to the SMBH mass) may correspond to a more fundamental parameter, reflecting the relative growth rate of the SMBH mass,
but is difficult to measure (requiring accurate SMBH mass measurements) and introduces an issue for lower mass systems as a given Eddington rate will produce a lower absolute luminosity that may preclude detection. 
Another route is to consider the luminosity produced by the AGN relative to the emission due to host galaxy processes (in particular star formation) at the same wavelength. While such a selection can produce a robust sample of definitive AGN, such choices will exclude intrinsically weaker AGN as well as AGN that lie in more luminous galaxies (e.g. due to higher SFRs) which overwhelm the AGN signatures. 
}
One of the major advancements over the last decade has been to improve our understanding of these incompletenesses, 
primarily by adopting multiwavelength approaches to determine the limitations of different AGN selection techniques
\citep[e.g.][]{goulding_tracing_2014,mendez_primus_2013,azadi_mosdef_2017,padovani_active_2017,hickox_obscured_2018, 
lamassa_sdssiv_2019,georgakakis_forward_2020}. %
Here we describe some of the most widely adopted approaches to identify AGN and the techniques used to measure their host galaxy properties, with a focus on new developments over the past decade.

\subsubsection{Optical / UV / emission-line selection}
\label{sec4:opt-uv-selection}

Selection of AGN based on their optical/UV properties provides arguably the most direct tracer of SMBH growth, 
as the accretion disk emission peaks at UV wavelengths and has a blue continuum that extends into the optical regime (see Fig.~\ref{fig:SED}).
However, such selection is particularly sensitive to the impact of host galaxy dilution -- weaker AGN may easily be overwhelmed by the emission from stars at these wavelengths -- as well as obscuration effects as relatively small amounts of dust or gas in the line-of-sight (within the host galaxy or associated with the AGN system itself) will drastically redden or completely extinguish the AGN light. 
Nonetheless, selection on the basis of blue optical colours combined with spectroscopic confirmation (primarily via the presence of broad emission lines, verifying the presence of a ``Type-1'' AGN) has been widely exploited as a means of identifying large samples 
of luminous and unobscured quasars.
These quasars evolve strongly over cosmic time, peaking in space density at $z\sim2$, although recent studies find only a weak ``downsizing'' whereby the space density of lower luminosity sources peaks later in cosmic time (i.e. at lower redshift) than higher luminoisity sources \citep[e.g.][cf. the stronger downsizing signature seen for X-ray selected AGN discussed in \S\ref{sec4:xray-selection} below]{ross_quasar_2013,kulkarni_evolution_2019,schindler_extremely_2019}.
The properties of the most massive and most rapidly accreting examples of such quasars are explored in more detail in \S\ref{sec:rapid-growth}.
The presence of broad emission lines may be used more generally as an indicator of AGN in blanket spectroscopic follow-up programmes \citep[e.g.][]{driver_gama_2009,hahn_desi_2023} including in sources that lack a blue optical continuum (e.g. due to dust redddening) but are selected by other means
\citep[e.g.][see also \S\ref{sec:highzqso} for discussion of the spectroscopic identification of high-redshift AGN candidates in ``little red dots'' using \textit{JWST}]{glikman_first_2004,Banerji2012}.

A more recent area of development has been to use optical variability as a means to identify the presence of low-luminosity AGN even when the bulk of the optical light is dominated by the host galaxy \citep[e.g.][]{de-cicco_variability_2015,de-cicco_optically_2019,sanchez-saez_quest_2019, pouliasis_robust_2019}, addressing some of the issues with host galaxy dilution. Such techniques are likely to be revolutionized in the near future by the LSST with the Vera Rubin Observatory \citep{ivezic_lsst_2019}. 
Isolating blue, quasar-like light from galactic centres using high spatial resolution imaging (e.g. with \textit{JWST} or \textit{Euclid}) may also provide a route for the identification of weak optical/UV AGN within luminous host galaxies \citep[e.g.][]{Margalef-Bentabol_euclid_first_2025,stevens_euclid_active_2025}.

The strong UV emission produced by an accretion disk also photo-ionises gas clouds at distances up to $\sim$kpc scales, resulting in characteristic narrow emission lines that are seen even when the direct line-of-sight to the central engine is blocked by obscuring material, producing a ``Type-2'' AGN. 
Emission-line ratio diagnostics that use the strongest optical emission lines (e.g. H$\alpha$, H$\beta$, [O~III], [N~II], [S~II]) to identify AGN and distinguish them from star-forming systems are well established and had already been widely applied to 
provide a detailed picture of AGN activity within the local ($z\lesssim0.3$) galaxy population at the time of the AH12 review \citep[see section 3.2 therein, see also][]{Heckman:14}.
However, the last decade has seen a greater understanding of the contribution of stellar processes  to these emission lines that can dilute the signatures of lower luminosity AGN, particularly in star-forming galaxies, and has led to a re-assessment and re-interpretation of many of these results \citep[e.g.][see \S\ref{sec4:varying_growth} below]{trump_biases_2015,jones_intrinsic_2016,agostino_crossing_2019}. 
More extensive IFU spectroscopy has also enabled a spatially resolved picture that can aid in the separation of galaxy-wide star formation and central AGN activity \citep[e.g.][]{wylezalek_manga_2018,comerford_toward_2022}.
Furthermore, new facilities that enable extensive multi-object spectroscopy in the near-to-mid IR (e.g. Keck-MOSFIRE, VLT-KMOS, \textit{JWST}/NIRSpec) have allowed these rest-frame optical diagnostics to be applied to statistical samples of distant, higher redshift galaxies \citep[e.g.][]{coil_mosdef_2015,harrison_kmos_2016}.
While such studies are effective in revealing AGN activity at these earlier cosmic epochs (including AGN that may not be identifiable via other techniques), the higher SFRs, lower metallicities and younger stellar populations of high-redshift galaxies can also produce strong emission lines that leads to greater dilution of the AGN signatures by the host galaxy light -- an issue that appears to impact \textit{JWST} spectroscopic studies that are now enabling rest-frame optical/UV emission-line diagnostics out to even higher redshifts \citep[e.g.][see \S\ref{sec:QSOhighz}]{scholtz_jades_2023}.
Star formation dilution may be mitigated by using fainter, higher ionisation potential lines found at rest-frame UV, optical or IR wavelengths that require the strong ionising continuum of an AGN \citep[e.g.][]{Feltre_2016, Calabro_2023}. IR emission lines have the added benefit of being less affected by dust extinction; however, such lines are typically weak and current IR spectrographs have much more limited multi-plexing than optical spectrographs, making it more challenging to obtain sensitive IR spectroscopy for large samples.

\subsubsection{X-ray selection}
\label{sec4:xray-selection}

X-ray emission, primarily produced by inverse Compton-scattering of UV photons from the accretion disk by a hot, compact corona \citep[extending to $\lesssim$ a few $r_s$, e.g.][]{Fabian2015}, provides an alternative tracer of the AGN accretion power and an efficient means of identifying AGN \citep[see][for reviews]{Brandt_2015,Kara_2025}; see Figs.~\ref{fig:agn_model} \& \ref{fig:SED}. A major advantage of X-ray selection of AGN is that other galactic sources of X-rays (such as the population of X-ray binaries throughout the galaxy or hot gas) are relatively weak, resulting in a strong host--to--AGN contrast. Furthermore, higher energy X-rays ($\gtrsim10$~keV) are relatively immune to photoelectric absorption (up to equivalent hydrogen column densities, $N_\mathrm{H}\sim10^{24}$~cm$^{-2}$) while the absorption of softer X-rays produces a clear spectral signature that enables \emph{measurements} of the column density. 
At higher column densities ($N_\mathrm{H}\gtrsim1.5 \times 10^{24}$~cm$^{-2}$), corresponding to the Compton-thick regime, X-rays of all energies are no longer able to penetrate without being Compton scattered, which substantially suppresses the directly transmitted spectrum, while some photons may be scattered back into the line of the sight with an altered spectrum with a characteristic ``Compton hump'' peaking at rest-frame energies $\sim30$~keV (see Fig.~\ref{fig:SED}).
The need for a substantial population of Compton-thick AGN ($\sim30-50$\% of moderate-luminosity AGN) to explain the peak in the overall cosmic X-ray background (CXB) at $\sim20-40$~keV has been well established \citep[e.g.][]{comastri_contribution_1995,gilli_synthesis_2007,ananna_accretion_2020}.
However, an area of significant progress has been the development of models to describe the absorption, line fluorescence, and scattering of X-rays by a realistic obscurer, expanding earlier work adopting toroidal geometries \citep[e.g.][]{murphy_xray_2009a,ikeda_study_2009,brightman_xmmnewton_2011} to now include clumpy, varying and dynamic distributions \citep[e.g.][]{tanimoto_application_2020,buchner_xray_2019,buchner_physically_2021} that align with our modern picture of accreting SMBH systems (see \S\ref{sec:section3}).

\begin{figure*}
    \centering
    \includegraphics[width=1.1\columnwidth,trim=0 0.7cm 0 1cm]{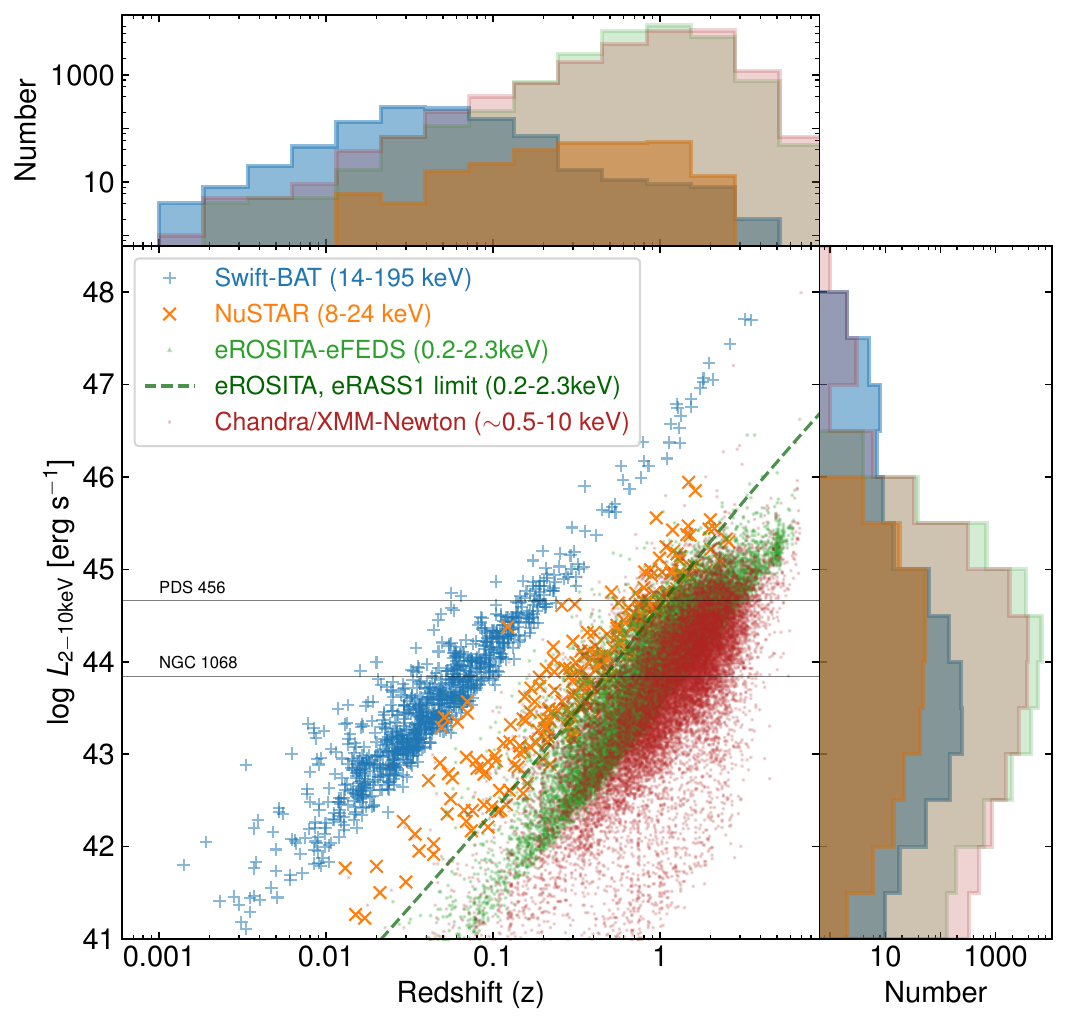}
\caption{
Coverage of the rest-frame 2-10~keV luminosity--redshift plane obtained through X-ray selection of AGN with different observatories that are sensitive to different observed-frame energy bands (as indicated in legend).
Blue crosses show (non-beamed) sources from the 
\textit{Swift}-BAT 158-month sample \citep{lien_swiftbat_2023} that provides an all-sky survey at 14-195~keV.
Orange crosses indicate sources selected at 8--24~keV compiled from a range of \textit{NuSTAR} surveys \citep{mullaney_nustar_2015,civano_nustar_2015,masini_nustar_2018,zhao_pearls_2024,greenwell_nustar_2024}.
Red points compile data from a number of deep and wide \textit{Chandra} and \textit{XMM-Newton} surveys  \citep{luo_cdfs_2017,xue_chandra_2016,nandra_aegis_2015,marchesi_cosmoslegacy_2016,masini_cdwfs_2020,chen_xmmservs_2018,ni_xmmservs_2021,peca_stripe_2024}.
Small green points show the sampling achieved in the $\sim140$~deg$^2$ \textit{eROSITA} Final Equatorial Depth Survey \citep[eFEDS][]{brunner_efeds_2022,salvato_efeds_2022} which already provides sample sizes comparable to prior efforts (primarily probing soft energies, 0.2--2.3~keV), while the green dashed line corresponds to the flux limit of the first \textit{eROSITA} all-sky survey \citep[eRASS1,][]{merloni_erass1_2024} that promises unprecedented statistics for future studies of the X-ray AGN population.
The horizontal grey lines indicate the intrinsic rest-frame 2--10~keV luminosities of two nearby AGN, the quasar PDS~456 and the heavily obscured AGN NGC~1068.
}
\label{fig:lx_z}
\end{figure*}

Given the relative immunity to obscuration, selection using hard ($\gtrsim$10~keV) X-rays can provide a comprehensive sampling of AGN.  
The \textit{Neil Gehrels Swift Observatory} \citep[\textit{Swift}, launched 2004,][]{gehrels_swift_2004} and the \textit{International Gamma-Ray Astrophysics Laboratory} \citep[\textit{INTEGRAL}, launched 2002,][]{winkler_integral_2003} were both primarily designed to identify and study Gamma-ray bursts; 
however, both contain wide-field hard X-ray instruments that provide all-sky survey coverage at $>15$~keV energies (\textit{INTEGRAL}-IBIS, \citealt{ubertini_ibis_2003}; \textit{Swift}-BAT, \citealt{Barthelmy_BAT_2005}) enabling surveys to find AGN.
The ongoing missions have enabled progressively deeper surveys to be built up over the last decade and have identified samples of AGN (primarily $z\lesssim 0.2$) of sufficient size ($\gtrsim 1000$ sources) for robust statistical analysis of the population \citep[see Fig.~\ref{fig:lx_z}]{krivonos_integralibis_2022,oh_swiftbat_2018,lien_swiftbat_2023}. 
A major effort to obtain optical spectroscopy for the \textit{Swift}-BAT sources over the past decade \citep[the BAT AGN Spectroscopic Survey, BASS:][]{koss_bat_2017,koss_bass_2022,oh_bass_2022} as well as other supporting multiwavelength data has enabled a robust characterization of the physical properties of the local AGN population, including placing constraints on the overall $N_\mathrm{H}$ distribution and Compton-thick fraction in the local Universe \citep{ricci_bat_xray_2017}, constraining the distribution of Eddington ratios \citep{ananna_bass_xxx_2022}, revising our understanding of the impact of radiation pressure on the structure and covering factors of the putative AGN ``torus'' \citep{ricci_radiative_2017,ricci_feedback_2022,ricci_covering_factor_2023,ichikawa_covering_2019,ananna_structure_2022}, and setting a local baseline for the connection between galaxy mergers and luminous AGN activity \citep[see also \S\ref{sec4:mergers}]{koss_mergers_2018}. 

Launched in 2012, the \textit{Nuclear Spectroscopic Telescope Array} \citep[\textit{NuSTAR},][]{harrison_nustar_2013} has provided $\sim$2 orders of magnitude improvement in sensitivity at hard ($\gtrsim10$~keV) X-ray energies that enables the identification of AGN out to $z\sim3$ \citep[e.g.][]{aird_nustar_2015,greenwell_nustar_2024}, drastically expanding our sampling of the X-ray AGN population compared to the non-focusing hard X-ray telescopes discussed above (see Fig.~\ref{fig:lx_z}). 
\textit{NuSTAR} has confirmed the increasing importance of obscured AGN populations out to early cosmic times \citep{alexander_nustar_2013,aird_nustar_2015,delmoro_nustar_2017,lansbury_hunting_2017,zappacosta_nustar_2018,zhao_pearls_2024} and directly resolved $\sim$35\% of the emission of the CXB at 8--24~keV (close to the peak) into individual AGN \citep{harrison_nustar_2016}. 
However, \textit{NuSTAR} surveys still remain sensitive to only the brightest sources and the sample sizes remain relatively small \citep[$\lesssim$2000 sources, with the vast majority identified serendipitously when targetting known sources:][]
{lansbury_nustar_2017,greenwell_nustar_2024}.

\begin{figure}
    \centering
    \includegraphics[width=\columnwidth]{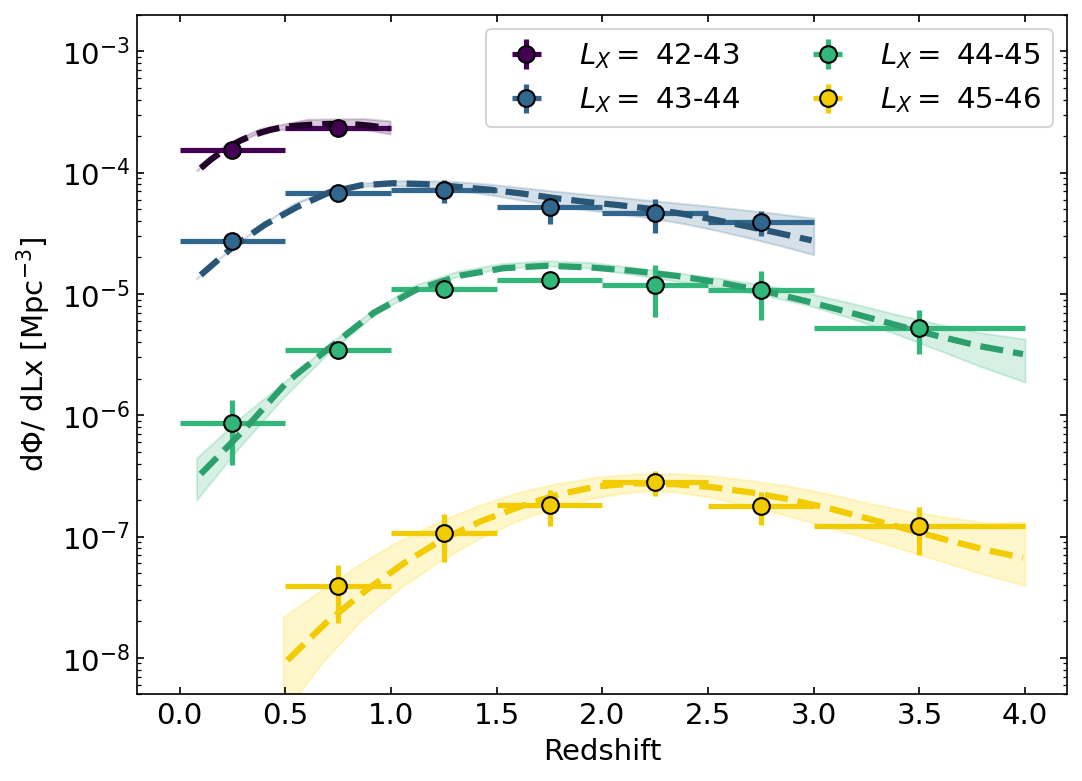}
    \caption{Measurements of the space density of X-ray selected AGN of different ranges of luminosity (as indicated by the different colours), demonstrating the ``downsizing'' evolutionary behavior whereby lower luminosity AGN peak in space density later in cosmic time (i.e. at lower redshift) compared to high luminosity AGN. {\it Source:} reproduced from Fig.~18 of \citet{peca_evolution_2023}.
    }
    \label{fig:spacedens_vs_z}
\end{figure}

Our most extensive census of X-ray AGN has instead been achieved with surveys at lower energies ($\sim$0.5--10~keV) thanks to the culmination of major deep and wide survey efforts with the flagship observatories, \textit{Chandra} and \textit{XMM-Newton}, that have the sensitivity to  efficiently identify and characterize lower luminosity and obscured AGN populations out to $z\gtrsim1$ and thus track the bulk of the AGN population \citep[e.g.][see Fig.~\ref{fig:lx_z}]{nandra_aegis_2015,civano_cosmos_2016,pierre_xxl_2016,lamassa_stripe82_2016,luo_cdfs_2017,chen_xmmservs_2018,masini_cdwfs_2020,ni_xmmservs_2021}. Combined with extensive multiwavelength imaging and spectroscopic follow-up campaigns to obtain redshifts, these surveys confirm a strong ``downsizing'' 
evolution, whereby the space density of the most luminous X-ray AGN is found to peak at $z\sim2$ whereas lower luminosity AGN peak in space density at progressively lower redshifts (see Fig.~\ref{fig:spacedens_vs_z}) 
as well as revealing a luminosity-dependent fraction of absorbed AGN that increases toward higher redshifts \footnote{Our current understanding of the underlying causes of both the downsizing behaviour and evolution of the absorbed fraction are discussed further in \S\ref{sec4:varying_growth}.} \citep[e.g.][]{ueda_evolution_2014,aird_evolution_2015,buchner_evolution_2015,peca_evolution_2023}.

While \textit{Chandra} and \textit{XMM-Newton} have been effective in covering small-to-moderate sky areas ($\sim$0.1-100~deg$^2$) and yielding the highest AGN sky densities ($\gtrsim10^4$~deg$^{-2}$), \textit{eROSITA} (launched in 2019) has recently provided a new all-sky X-ray survey, yielding $\sim$900,000 point sources (primarily associated with AGN) in the first half-sky data release \citep{merloni_erass1_2024}, offering unparalleled statistics for future studies of the X-ray AGN population. 
The vast majority of these sources are detected at soft energies (0.2--2.3~keV) and are associated with luminous and unobscured AGN populations \citep{liu_efeds_2022} spanning redshifts out to $z\sim6$ \citep{wolf_efeds_2021,wolf_erass_2024}. 
In contrast, the sub-sample of $\sim5500$ AGN that are detected in the harder, 2.3--5~keV band reaches $\gtrsim$an order of magnitude fainter in X-ray luminosity than the \textit{Swift}-BAT samples \citep{waddell_srg_2024,nandra_efeds_2024}, contains a higher proportion of obscured sources than the soft-selected samples and is helping reveal the complex structure of the obscuring material across the AGN population \citep{waddell_efeds_2024}.



\subsubsection{Infrared selection}
\label{sec4:ir-selection}

The dusty, molecular gas that surrounds accreting SMBHs will absorb optical/UV light, which is then re-emitted at IR wavelengths with the emission typically peaking at around 15--60~\textmu m \citep{mullaney_2011}, reflecting the typical temperature of the dust ($\sim$100~K).
Detailed studies of nearby AGN systems over the last decade, exploiting the spatial information provided by high-resolution mid-IR observations using large ground-based 8--10m telescopes \citep[e.g.][]{Hoenig2013,Asmus2016,Asmus2019,Leftley2021,Garcia_Bernete_2022}, have shown that dust with a range of temperatures is typically present in a single AGN, with hotter dust likely associated with an outflowing wind extending in the polar direction, while cooler dust is likely associated with an inflowing, clumpy molecular gas disk that is also seen at sub-mm wavelengths (see \S\ref{sec:observations} and Fig.~\ref{fig:AGN-schema} above).

The identification of AGN on the basis of a rising continuum toward longer wavelengths in the mid-IR, tracing the heated dusty material, was well established based on extensive surveys with the \textit{Spitzer} telescope \citep[e.g.][]{lacy_2004, stern_2005, donley_2005, donley_2008, donley_2012, alonso-herrero_2006}.
The launch of \textit{WISE} in 2009 \citep{wright_wise_2010} and subsequent release of all-sky imaging at 3.4, 4.6, 12, and 22\textmu m \citep[][]{cutri_wise_2013,schlafly_unwise_2019} has enabled such approaches to be extended to much larger datasets, revealing up to $\sim$3 million candidate AGN \citep[e.g.][]{assef_ndwfs_2013,assef_wiseagn_2018,stern_midinfrared_2012,mateos_xmmwise_2012}.
\textit{WISE}-selected AGN have been particularly useful when very large samples are required for statistical measurements (e.g. clustering analyses, see \S\ref{sec4:environment} below) or to provide reliable identification of counterparts and SED constraints for AGN identified at other wavelengths \citep[e.g.][]{salvato_nway_2018,salvato_efeds_2022,Auge2023}.

IR identification is immune to obscuration effects (tracing emission from the obscuring dust itself) and as such is able to identify heavily obscured populations that are missed by other selections \citep[e.g.][]{DelMoro2016,carroll_largepopulation_2021,Andonie2022}.
It can also provide a reliable tracer of the accretion power \citep[e.g.][]{Gandhi_resolving_2009A&A...502..457G,mullaney_2011}. However, dust heated by recent star formation or circumstellar dust around older stellar populations can produce substantial mid-IR emission that can dilute the signal of intrinsically weaker AGN. 
Careful, cross-community work over the last decade comparing IR selections with AGN identified at other wavelengths has shown that IR selection is effective at selecting AGN that are highly luminous relative to their hosts but is biased against the selection of lower luminosity AGN in galaxies with high SFRs or with the highest stellar masses \citep[][]{mendez_primus_2013,azadi_mosdef_2017,ji_agn_selection_2022}.
Studies of the demographics also find a downsizing behaviour \citep[i.e. lower luminosity AGN peaking in space density at lower redshift than higher luminosity sources; see e.g.][]{assef_mid-ir_2011,lacy_spitzer_demographics_2015} that is broadly consistent with findings at X-ray wavelengths (cf. Fig.~\ref{fig:spacedens_vs_z} and \S\ref{sec4:xray-selection}).

More recently, new imaging obtained with the Mid-Infrared Instrument (MIRI) on {\it JWST} \citep{rieke_mid-infrared_2015} provides more than an order of magnitude improvement in mid-IR sensitivity compared to \textit{Spitzer}, albeit over a small field-of-view (74~arcsec × 113~arcsec) that has meant overall survey coverage is limited to $\lesssim$300~arcmin$^2$ \citep[e.g.][]{yang_2023ApJ...956L..12Y,alberts_2024ApJ...976..224A, leung_2024arXiv241112005L}. 
Initial studies have used SED-fitting to reveal populations of AGN including low-luminosity sources (in correspondingly faint/low-mass host galaxies) and likely obscured AGN that were unidentifiable with prior facilities \citep[e.g.][]{yang_ceers_2023,chien_finding_2024, kim_cosmic_2024, li_epochs_2024, lyu_active_2024}. Selection biases related to the distinction between host galaxy and AGN emission are if anything more acute in this regime compared to prior \textit{Spitzer} or \textit{WISE} studies \citep{kirkpatrick_ceers_2023}. 
These fainter/obscured populations do not significantly increase estimates of the total black hole mass accretion density, at least to $z\sim3$, and thus do not drastically alter our understanding of AGN demographics\footnote{{\it JWST} discoveries of substantial AGN populations at $z\gtrsim3$ with unexpectedly high space densities, including the ``little red dots'' found based on red near-IR colours, are discussed further in \S\ref{sec:QSOhighz}.} \citep{yang_ceers_2023}.
Mid-IR \emph{spectroscopic} diagnostics, which have been shown with \textit{Spitzer} to be highly effective as a means of identifying highly complete AGN samples and mitigating both obscuration and host galaxy biases \citep[e.g.][see \citealt{Sajina_past_2022} for a review]{goulding_towards_2009,Alonso-Herrero_local_2012,DelMoro2016,feltre_optical_2023,Bierschenk_bass_2024}, 
are only just starting to be exploited with \textit{JWST} \citep[e.g.][]{armus_goals_2023,goold_ReveaLLAGN_2024}. 

\subsubsection{Radio selection}
\label{sec4:radio-selection}

For deep extragalactic surveys, the prime means for identifying AGN in the radio band is synchrotron emission from relativistic particles accelerating in magnetic fields. When referring to radio AGN, often researchers focus on the canonical Fanaroff-Riley type 1 / type 2 sources \citep{1974MNRAS.167P..31F}, where the synchrotron emission is produced in powerful radio jets that range from kpc to Mpc in scale, far beyond the extent of their host galaxies. These types of AGN were historically defined as `radio-loud' as they could be easily identified with a ratio of 5 GHz to B-band luminosity $>10$ \citep{1989AJ.....98.1195K}; those with a ratio $<10$ were classified as `radio-quiet' and the radio emission was thought to be dominated by star formation. With more sensitive radio surveys it has become clear that, while more than one population is required to make up the complete range of observed sources, there is a large overlap region between `radio-quiet' and `radio-loud' AGN \citep[Fig. 15 in][]{2007ApJ...654...99W}. 

To pick out ``clean" samples of AGN hosting radio jets, often a single cut in radio loudness, or a luminosity cut that gives effectively the same selection (i.e.,\ samples selected from flux-limited surveys with a small redshift range and similar host galaxy B-band luminosities), is used. However, while a high enough cut in radio loudness can provide a clean jetted-AGN sample, it also provides a biased picture. The fraction of galaxies containing radio-loud AGN
increases steeply as a function of stellar mass \citep{sabater_lotss_2019}.  Radio AGN also tend to be strongly clustered, living in dense large-scale environments
\citep{2004MNRAS.350.1485M,hickox_host_2009,2009MNRAS.393..377M}. The radio selection of AGN also often picks up radiatively inefficient AGN, which are thought to be fed by advection dominated accretion flows \citep[ADAFs; e.g.][]{1995ApJ...452..710N} rather than the standard radiatively efficient accretion disk \citep[see also \S\ref{sec:accretion-disks} and Fig.~\ref{fig:Accretion-flows}]{Shakura1973}. These `hot mode' accretors show no evidence for an optically thick accretion disk \citep{2007MNRAS.376.1849H} and the bulk of their energy is contained in radio-loud jets \citep[e.g.][]{2007MNRAS.381..589M}. They will therefore not be selected as AGN at any other waveband. 
While initial studies demonstrated that radiatively inefficient accretion was linked to lower accretion rates \citep[e.g.][]{2012MNRAS.421.1569B}, newer radio surveys find a broader range of accretion rates associated with the different AGN classifications
\citep{whittam_mightee_2022}. 

In galaxies where it is not clear that a powerful radio jet is present (generally referred to as `radio-quiet'), the source of radio emission is still a topic of debate. A recent review succinctly lays out the different options \citep{2019NatAs...3..387P}: star formation, shocks from AGN winds, low-power jets, and coronal effects. Coronal emission, while possible, is only likely to be fruitfully investigated at mm frequencies \citep{2015MNRAS.451..517B} and have only a small impact in the regime where synchrotron radiation dominates ($\lesssim 10\,$GHz). The debate of star-formation vs. AGN activity as the origin of radio emission has supporters in both camps, and arguments usually rely on defining sources as \textit{radio excess} above what would be expected for star formation as measured from another waveband, usually at far-IR wavelengths \citep{delvecchio_vla-cosmos_2017,best_lofar_2023}. 

It is difficult to distinguish between synchrotron emission from star formation and AGN activity. Brightness temperature measurements can pick out AGN cores independent of using a separate measure of radio excess, given high enough resolution \citep{1991ApJ...378...65C}. Using milli-arcsec resolution with wide-field VLBI techniques at GHz frequencies is now relatively developed  \citep{2001A&A...366L...5G,2017A&A...607A.132H,2018A&A...619A..48R}, but still limited in field of view and number of sources, which makes it difficult to place these radio-selected AGN within the context of the general AGN population. Recent work using brightness temperature measurements from high-resolution LOFAR imaging, which increases field of view and sample sizes by over two orders of magnitude, has shown that even in non-radio excess sources, there can be a significant fraction (up to 50\% more than expected) of radio emission contribution from AGN activity \citep{2022MNRAS.515.5758M,2025MNRAS.536L..32M}. This also becomes evident when forward-modelling the radio population in the LoTSS Deep Fields assuming a two-component model: star formation + AGN activity. \cite{2021MNRAS.506.5888M} demonstrated that this two component model, where the AGN population is drawn from a power-law distribution down to low radio power, can completely reproduce the source counts and radio-loud fraction in the observations. What is also revealed is that there is an entire population of sources which would be classified as `radio-quiet' but have more than half of their radio emission generated by AGN activity \citep[][see their Fig. 12]{White_2017,2021MNRAS.506.5888M}. The models are agnostic to whether the AGN radio emission is generated by a jet (as assumed by this single component model) as opposed to other radio AGN emitting processes, which allows for a new way to explore the radio properties of different AGN sub samples \citep{Yue2024}.

\subsubsection{Galaxy properties}
\label{sec4:gal-properties}

Studying the links between the growth of SMBHs and their host galaxies has also required progress in the techniques to measure galaxy physical properties in the presence of an AGN.

AH12 presented some of the early results from \textit{Herschel}, which launched in 2009 \citep{Pilbratt_herschel_2010} and provided unprecedented sensitivity to far-IR wavelengths (55--671$\mu$m).
These wavelengths probe the black-body emission associated with dust that is heated through star-formation process and as such can provide a robust tracer of the total SFR of galaxies that is not impacted by dust obscuration \citep[e.g.][]{Kennicutt_1998ARA&A..36..189K,Lutz_2014ARA&A..52..373L}.
The emission from the hotter dust heated by an AGN peaks instead at mid-IR wavelengths (see \S\ref{sec4:ir-selection}) and is relatively weak at far-IR wavelengths. 
As such, over the last decade \textit{Herschel} has been vital as a means of verifying other AGN selection approaches, when the lack of strong far-IR emission can exclude a star-formation origin and confirm that a galaxy is an AGN \citep[e.g.][]{DelMoro_2013A&A...549A..59D,Andonie2022}.
\textit{Herschel} has also provided ``clean'' measurements of SFRs for moderate-to-high luminosity AGN in cases when the AGN light dominates at other wavelengths \citep[e.g.][]{rosario_mean_2012,stanley_remarkably_2015}.
The possible contribution of cooler, extended \emph{but AGN-heated} dust at far-IR wavelengths remains a topic of debate \citep[e.g.][]{symeonidis_cooler_2016,lyu_intrinsic_2017ApJ...841...76L,xu_revisiting_2020ApJ...894...21X,symeonidis_2022}; at the highest AGN luminosities the AGN light may again dominate and prevent accurate measurements of the SFR \citep[e.g.][]{symeonidis_2018}. 
\textit{Herschel}'s sensitivity limits also means that it is only able to detect the highest SFR galaxies at a given redshift, providing biased samples of galaxies and an incomplete sampling of AGN hosts.

Measuring the SEDs of galaxies---the intensity of the observed light over a wide range in wavelengths---and fitting to models of the underlying stellar populations thus remains the primary tool to determine SFRs, stellar masses, and other galaxy properties \citep[see][for a pedagogic overview of the SED-fitting process]{Iyer_SED_review_2025}. 
In studies of galaxy evolution, recent advances have included improvements to stellar population models \citep[e.g.][]{stanway_reevaluating_2018MNRAS.479...75S,conroy_metal-rich_2018ApJ...854..139C,gotberg_spectral_2018A&A...615A..78G}, more flexible methods to provide accurate descriptions of the diversity/burstiness of star formation histories \citep[e.g.][]{leja_how_2019ApJ...876....3L,robotham_prospect_2020MNRAS.495..905R}, and the use of spectroscopic data alongside photometry \citep[e.g.][]{chevallard_modelling_2016MNRAS.462.1415C,carnall_bagpipes_2018MNRAS.480.4379C}.
However, the main challenge when studying the host galaxies of AGN (identified using \emph{any} of the methods described in \S\ref{sec4:opt-uv-selection}--\ref{sec4:radio-selection}) is to separate the light produced by the galaxy from that produced by the AGN. 
To this end, many SED fitting codes now allow an AGN component to be included at UV/optical/IR wavelengths
\citep[e.g.][]{calisto_rivera_agnfitter_2016ApJ...833...98C,boquien_cigale_2019A&A...622A.103B,Rosario_2019,buchner_genuine_2024A&A...692A.161B}.\footnote{See \citet{pacifici_art_2023} for an overview and comparison of contemporary SED-fitting codes.}
More recent advances have included extending SED fitting into the X-ray and radio regimes, which can provide further constraints on the AGN contribution and thus improve constraints on the underlying host galaxy properties \citep[e.g.][]{yang_xcigale_2020,yang_fitting_2022,thorne_long_2023,azadi_disentangling_2023}.
High-resolution optical and near-IR imaging can also enable the central AGN light to be \emph{spatially} distinguished from the wider host galaxy, improving constraints on both galaxy and AGN properties.
Such approaches are now being applied to {\it JWST} imaging as well as high-quality ground-based datasets from new facilities such as HyperSuprimeCam (HSC) on Subaru, allowing host-galaxy measurements to be extended to bright unobscured quasars where the AGN dominates the integrated SED 
\citep[e.g.][]{li_sizes_2021ApJ...918...22L,2023Natur.621...51D}.

While only able to carry out relatively targetted studies due to the limited field-of-view, ALMA provides exquisite angular resolution and is uniquely capable of measuring certain galaxy properties, such as the gas content via either the sub-mm continuum emission or through molecular emission lines \citep{Combes_2018A&ARv..26....5C}. ALMA is also able to constrain the gas \emph{dynamics}, and thus track in detail the inflow or outflow of molecular gas, as well as measuring dynamical masses -- one of the few accessible probes of host galaxy mass when considering the most luminous/highest redshift AGN (see \S\ref{sec:highz_demographics}). 

These new facilities and approaches have vastly improved our ability to measure host-galaxy properties, even when a bright AGN is present. 
However, we cannot completely mitigate the impact of the host-galaxy emission on the various AGN selection strategies described in \S\ref{sec4:opt-uv-selection}-\S\ref{sec4:radio-selection}. New \emph{conceptual} approaches have thus also been a key driver of progress over the last decade that we explore next.

\subsection{
The varying growth of black holes alongside the galaxy population
} 
\label{sec4:varying_growth}

Over the last decade, our understanding of the links between galaxies and the growth of their SMBHs has undergone a \emph{conceptual} advance: recognizing AGN as \emph{events} within the lifecycles of galaxies. 
During such events, which may correspond to relatively short phases ($<100$~Myr) compared to the overall lifetime of a galaxy ($\sim$Gyr), the SMBH may accrete gas, generate emission at different wavelengths (producing the various signatures used to identify AGN described in \S\ref{sec4:multiwavelength}), launch relativistic jets, or drive winds and outflows of gas, all \emph{at varying rates}.\footnote{Variations also occur in AGN on much shorter timescales, including variations in emission properties on $\sim$days to $\sim$years, while substantial changes in the accretion rate are observed on decadal timescales in some changing-look AGN (see \S\ref{sec:variability}).}
A given galaxy may experience multiple AGN events throughout its lifetime, and thus experiences multiple such AGN phases, returning in-between to a phase when the central SMBH is relatively inactive.\footnote{The potential to still produce powerful jets even when the accretion rate is low---leading to a distinct jet-dominated AGN phase---is discussed in \S\ref{sec4:jetted}.}
As such, the rate of SMBH growth can vary substantially as it proceeds alongside the evolution of the galaxy population. 
In the following, we first discuss how this picture emerged over the last decade from early simulation work and observations,
before describing the current state-of-the-art, the related (re-)interpretation of AGN downsizing, the connection to obscuration properties, and the discovery of ``echoes'' of this varying SMBH growth. 

\subsubsection{Early theoretical and observational developments}
\label{sec4:early-developments}

Patterns of variable SMBH growth were predicted in early simulation work, which showed that the rate at which gas is driven from kpc scales to the sub-pc scales of the SMBH can vary substantially over galaxy timescales. 
Initial studies focussed on SMBH fuelling within elliptical galaxies \citep[e.g.][]{ciotti09,novak11} and showed how---even in galaxies that are initially devoid of gas and have passively evolving stellar populations---that gas (primarily produced by stellar winds) may be driven onto the central SMBH leading to intermittent phases of AGN activity over galaxy timescales \citep[see also][]{yuan_elliptical_2018,choi_origins_2023}.
Variations in the rate at which cold gas is delivered to the galactic centre are also expected, whether purely from stochastic accretion of distinct molecular clouds \citep[e.g.][]{hopkins_lowlevel_2006} or variations in the flow rate of gas that is driven into the centres of galaxies by secular processes (see \S\ref{sec:section3}) or galaxy mergers (see \S\ref{sec4:mergers}).
The fuelling rate will also be modified by localised feedback processes from the resulting AGN that may lead to additional variation in the instantaneous rate of accretion onto the SMBH \citep[e.g.][]{negri_subgrid_2017}. 


The bulk of early observational studies linking AGN and their host galaxies used either narrow emission line or X-ray selection (for studies in the local and more distant Universe, respectively), which trace the radiatively efficient AGN phase (and thus the bulk of SMBH mass growth through accretion) and leave the host galaxy optical SED relatively uncontaminated (see \S\ref{sec4:multiwavelength}).
Early studies found that most of the AGN hosts in these samples lie in either the ``red sequence" or, primarily, the ``green valley'' of the galaxy colour-magnitude diagram \citep[e.g.][]{nandra_aegis_colourmag_2007,silverman_hosts_2008,hickox_host_2009,schawinski_moderateluminosity_2009, schawinski_observational_2007}. 
However, by the time of AH12 it was realised such colours are in fact typical of moderate-to-high stellar mass galaxies ($M_*\gtrsim 10^{10}M_\odot$) that are found to host AGN \citep[e.g.][]{silverman_onging_2009,cardamone_bimodality_2010,xue_colourmag_2010}.
The predominance of high-mass galaxies in AGN samples  was initially thought to indicate that a massive host is a strong requirement to drive SMBH growth.

This viewpoint was challenged by \citet{Aird2012} 
who showed that the incidence of X-ray AGN in galaxies across the full stellar mass range is characterised by a power-law shaped distribution of \emph{specific black hole accretion rates} (sBHAR, denoted $\lambda_\mathrm{sBHAR}\propto L_\mathrm{AGN}/M_*$)\footnote{Specific black hole accretion rates are often converted to Eddington ratio equivalent units, given by $\lambda_\mathrm{sBHAR}=L_\mathrm{AGN}/(1.3 \times 10^{38} \times 0.002 \times M_*/M_\odot)$ where $L_\mathrm{AGN}$ is the bolometric AGN luminosity in erg~s$^{-1}$ \citep[e.g.][]{Aird2012,aird_xrays_2018}. While this choice is practical and accounts for an overall scaling with mass, a direct equivalence with Eddington ratios, $\lambda_\mathrm{Edd}$ (see \S\ref{sec:accretion-disks}), is only true if a universal scaling between $M_\mathrm{BH}$ and $M_*$ holds at all redshifts; recent observations indicate SMBH masses can span $>2$ orders of magnitude at a given $M_*$ at $z\sim0$ (see Fig.~\ref{fig:mbh_mstar_diagram}).
A certain radiative efficiency must also be assumed (typically $\approx0.1$) to convert $\lambda_\mathrm{sBHAR}$ into a tracer of SMBH \emph{mass} growth; however, this quantity may also vary considerably between different AGN phases.\label{footnote:sBHAR}} that likely reflects the varying rates of AGN activity on short timescales compared to the evolution of the host galaxy \citep{aird_primus_2013}.  
The prevalence of high $M_*$ hosts in AGN samples is predominantly a selection bias: 
AGN in more massive galaxies produce higher observable luminosities for a given $\lambda_\mathrm{sBHAR}$ making them easier to detect and leading to their dominance in flux-limited AGN samples.
Subsequent studies have confirmed that this overall selection bias holds over a broader range in redshift \citep[e.g.][]{bongiorno_accreting_2012,georgakakis_observational_2017,aird_xrays_2018,yang_linking_2018,birchall_incidence_2022} and for other AGN selections \citep[e.g.][]{mendez_primus_2013,trump_biases_2015,ji_agn_selection_2022}.
Thus, substantial central SMBH growth appears to be much more widespread across the galaxy population than was previously thought.

Also shortly after the AH12 review, \citet{mullaney_hidden_2012} used new \textit{Herschel} surveys to identify reliable samples of star-forming galaxies (independent of AGN content, see \S\ref{sec4:gal-properties}) and by stacking the X-ray emission demonstrated the existence of a ``hidden AGN main sequence'' whereby the average SMBH growth rate increased with stellar mass in an analogous manner to the increase in the SFRs of galaxies with stellar mass.
The ratio of average SMBH growth \emph{relative to galaxy growth} thus remains roughly constant across the star-forming galaxy population. 
However, the AGN main sequence is ``hidden'' by the broad distribution of $\lambda_\mathrm{sBHAR}$ that reflects the varying \emph{instantaneous} rates of SMBH growth.\footnote{
On a philosophical aside, it is interesting to note that while the studies of \citet{Aird2012} and \citet{mullaney_hidden_2012} did take advantage of new facilities and new data, it was the \emph{limitations} of the available 
galaxy samples that contained only a subset of known X-ray AGN that motivated the need for new analysis \emph{techniques} (measuring the probability distribution of accretion rates or stacking to probe average accretion rates in each case) and ultimately led to the \emph{conceptual} advance to interpret these results in terms of varying growth of SMBHs over galaxy timescales. 
New data is helpful, but sometimes limitations in our data push us to think in new ways!}

Contemporaneous studies focusing on correlations between AGN luminosity and SFR (i.e. tracers of SMBH growth versus galaxy growth) appeared to show contradictory results.
Studies that took samples of AGN and used \emph{Herschel} data (or other probes) to estimate galaxy-wide SFRs found little or no correlation between the observed AGN luminosity and the SFRs of their hosts, at least at low-to-moderate AGN luminosities, although a rise in the typical SFRs toward higher redshift that tracks the overall evolution of the star-forming main sequence was apparent \citep[e.g.][] {shao_starformation_2010,mullaney_goodsherschel_2012,harrison_noclear_2012,santini_enhanced_2012,rosario_mean_2012,stanley_remarkably_2015}. 
In contrast, studies that selected samples of bright, star-forming galaxies detected by \textit{Herschel} and \emph{averaged} the amount of SMBH growth through X-ray stacking analysis \citep[following][]{mullaney_hidden_2012} found a strong correlation whereby the average AGN luminosity increases with SFR \citep[e.g][]{symeonidis_hermes_2011,chen_correlation_2013}.

\begin{figure*}
    \includegraphics[width=\textwidth]{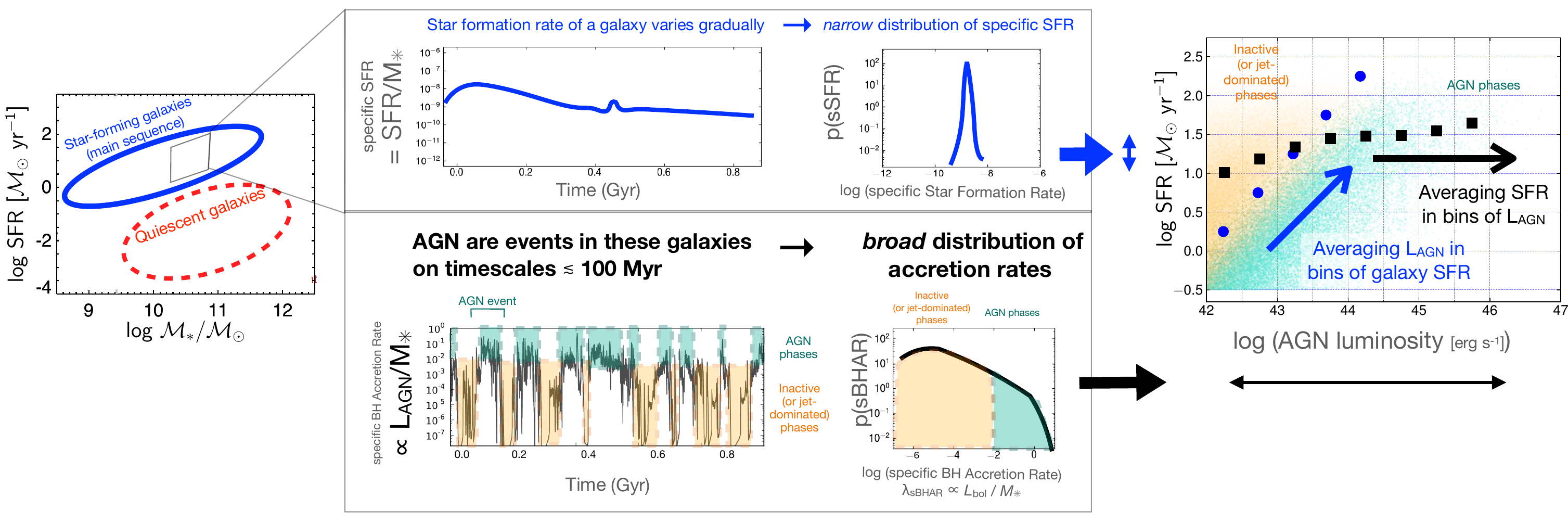}
    \caption{Diagram illustrating the picture established over the last decade that explains the relation between AGN and galaxy SFRs over the bulk of cosmic time (based on results at $z\lesssim3$).
    Star-forming galaxies mostly lie on the ``main sequence of star formation'' such that (at a given redshift) higher stellar mass galaxies have higher SFRs (left panel). The SFR of a galaxy varies relatively slowly on $\sim$Gyr scales, leading to a relatively narrow distribution of specific SFRs (central upper panels). 
    AGN may be viewed as multiple, relatively short events (lasting $\lesssim 100$~Myr) within the lifetime of a galaxy when the rate of accretion onto the central SMBH exceeds a certain level (highlighted in green in the lower central panels). 
    The accretion rate may also vary substantially on shorter timescales during AGN and inactive phases, leading to a broad distribution of accretion rates.
    Strong jets can still be produced even during phases when the accretion rate is low (highlighted in orange) and other AGN signatures are absent, corresponding to a ``jet-dominated'' or ``radiatively inefficient'' AGN phase (see Figs.~\ref{fig:AGNphases} and \ref{fig:radio-agn-fraction} and discussion in \S\ref{sec4:jetted}). 
    The broad distribution of accretion rates, due to this short-timescale variability, blurs any direct correlation between the total (radiative) AGN luminosity and the SFR (green and orange dots in right panel). 
    A similar average SFR is found at different AGN luminosities, corresponding to main sequence of moderate-mass galaxies at a given redshift (black squares). In contrast, the average AGN luminosity for galaxies of a given SFR reveals a correlation (blue circles) showing that---once short-term variations in AGN luminosity are averaged out---higher SFRs are associated with increased growth rates of central SMBHs.
    }
    \label{fig:agn_vs_SF}
\end{figure*}

\subsubsection{Variability on mega-year timescales explains the AGN--star formation connection}
\label{sec4:agn-sf}

\citet{hickox_variability_2014}, building on an idea proposed in the AH12 review, demonstrated how the  results discussed above could be reconciled by considering the variability of AGN activity over galaxy timescales. 
They proposed that the galaxy-wide SFR is correlated with the \emph{average} SMBH accretion rate but the \emph{instantaneous} SMBH accretion is drawn from a very broad distribution that reflects \emph{variability} on $\sim$Myr timescales.

Fig.~\ref{fig:agn_vs_SF} summarizes this 
conceptual picture. 
At a given redshift, galaxies can be divided into two populations (left panels): star-forming galaxies, which mostly follow a reasonably tight correlation between total stellar mass, $M_*$, and SFR \citep[the ``galaxy main sequence of star formation'', see e.g.][]{noeske_main_sequence_2007,elbaz_goods_2011,popesso_main_sequence_2023}; and quiescent galaxies where star formation has quenched and SFRs are much lower, which dominate the galaxy number densities at the highest $M_*$ (at $z\lesssim 1$). 
Most radiatively efficient AGN
\footnote{We refer here to radiatively efficient AGN as those where a standard accretion disk is present that ensures an efficient production of radiation (primarily at UV/optical wavelengths but also producing the X-ray and IR emisson, as discussed in \S\ref{sec4:multiwavelength}); see Fig.~\ref{fig:Accretion-flows}. Such an accretion mode is assumed to be operational when the SMBH is accreting at Eddington ratios $\gtrsim 1$\%, or the corresponding $\lambda_\mathrm{sBHAR}$. However, radiatively efficient AGN may still be identified with $\lambda_\mathrm{sBHAR}<0.01$, in part due to the large scatter in the scaling between $M_\mathrm{BH}$ and $M_*$ (see Footnote~\ref{footnote:sBHAR}). 
}
over the bulk of cosmic time (i.e. at $z\lesssim3$) are found within main-sequence star-forming galaxies \citep[e.g.][]{rosario_nuclear_2013,aird_xrays_2018}. 
The distributions of specific SFRs (sSFRs) for these galaxies will be comparatively narrow \citep[top central panels of Fig.~\ref{fig:agn_vs_SF}, e.g.][]{mullaney_hidden_2012,Schreiber2015}. 
In contrast, the instantaneous rate of accretion onto the central SMBH is highly variable and can span a wide range (bottom central panels of Fig.~\ref{fig:agn_vs_SF}).
Over \emph{galaxy evolutionary timescales} ($\sim$Gyr) numerous AGN events (typically lasting $<$100~Myr) may occur when the SMBH switches between inactive phases when the rate of mass accretion is relatively low (highlighted in orange in Fig.~\ref{fig:agn_vs_SF}), and AGN-dominated phases where the SMBH is accreting at a high rate (highlighted in green).
Thus, within a sample of star-forming galaxies with a relatively narrow range of sSFRs, a broad distribution of sBHARs is found, 
blurring any direct correlation between SFR and AGN luminosity 
(right panel of Fig.~\ref{fig:agn_vs_SF}).
Selecting galaxies of a given SFR and averaging their AGN luminosity (e.g. through X-ray stacking), thus averaging over the variable sBHAR distribution, reveals the underlying correlation (blue circles in right panel of Fig.~\ref{fig:agn_vs_SF}). In contrast, AGN selected at different luminosities all have a wide range of SFRs with an average consistent with the main sequence of star formation at the corresponding redshift but no correlation is found \citep[except possibly at the highest AGN luminosities, which require high mass host galaxies that have higher SFRs, producing a weak correlation, e.g.][]{rosario_mean_2012,dai_relationship_2018}.

\subsubsection{Our current picture of SMBH growth as a function of stellar mass, SFR and galaxy structure}
\label{sec4:across-galaxy-population}

A result of this conceptual shift is the realisation that studying the properties of an \emph{individual} AGN is insufficient to determine how host properties affect SMBH growth as the instantaneous AGN luminosity is a poor indicator of the overall accretion rate on cosmic timescales.
Instead, new observational techniques have been developed to quantify the level of AGN activity in terms of both the occurrence of AGN events and the overall SMBH mass growth in different galaxy populations, primarily through measurements of the probability distribution of $\lambda_\mathrm{sBHAR}$,\footnote{These probability distributions, that we refer to as $p(\lambda_\mathrm{sBHAR})$, are defined as the probability density of an AGN of a given $\lambda_\mathrm{sBHAR}$ occuring in a sample of \emph{galaxies} of a certain type/range of properties (e.g. stellar mass).} and derived quantities such as the fraction of galaxies with an AGN (above specified limits in luminosity or $\lambda_\mathrm{sBHAR}$) or sample-averaged accretion rates for different galaxy populations \citep[e.g.][]{bongiorno_agn_2016,yang_stellar_2017,georgakakis_observational_2017,yang_linking_2018,zou_mapping_2024}. 

\begin{figure*}
\centering
    \includegraphics[width=1.5\columnwidth,trim=0 2cm 0 1cm]{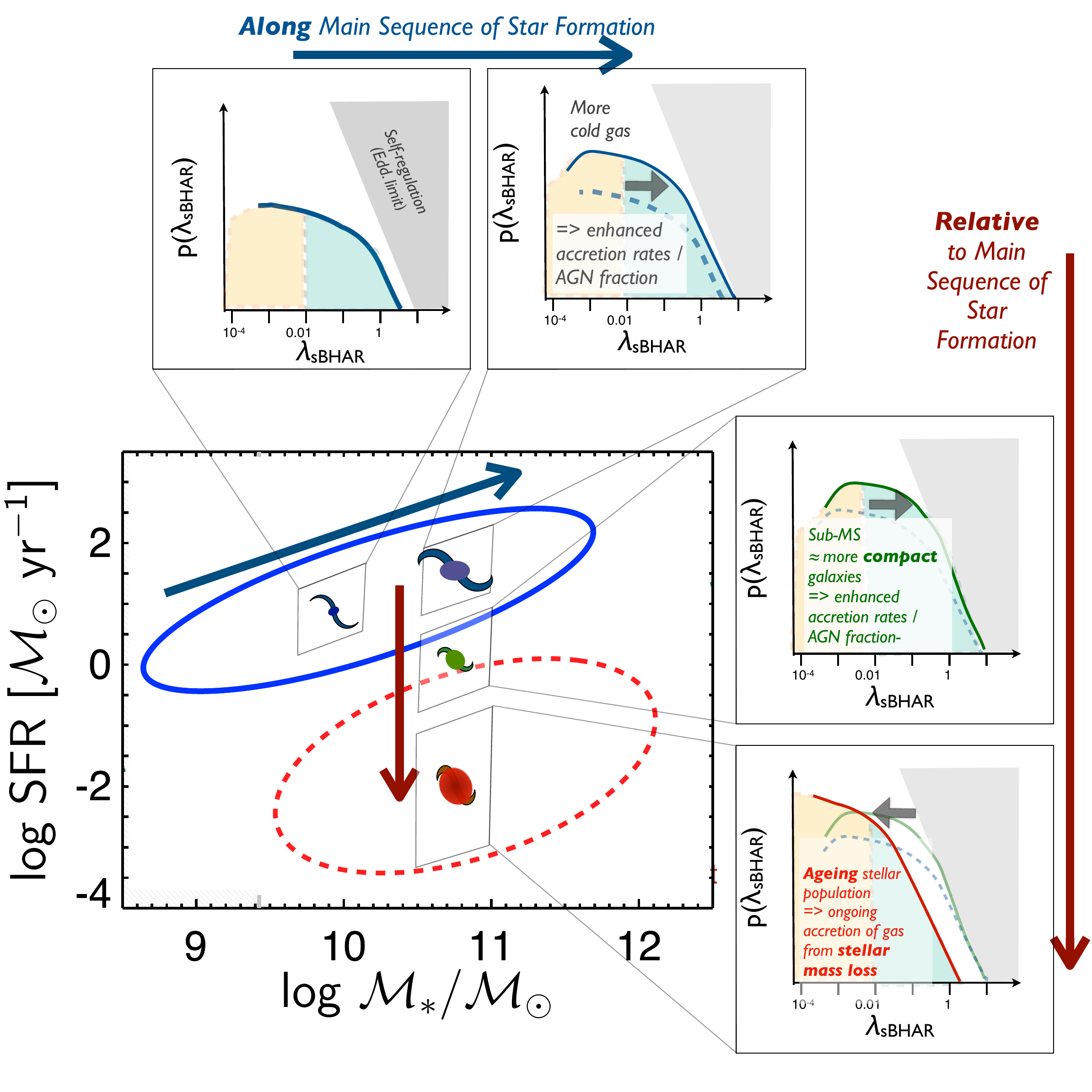}
    \caption{Schematic showing how the incidence of AGN, quantified by the probability distribution of specific SMBH accretion rates, $p$($\lambda_\mathrm{sBHAR}$), changes depending on the SFRs and stellar masses (indicated by the axes of the main panel) as well as the \emph{structure} of galaxies (illustrated by the cartoon galaxies). 
    Galaxies further along the main sequence (i.e. more massive/higher SFR/physical larger star-forming galaxies) are associated with a shift of $p$($\lambda_\mathrm{sBHAR}$) toward higher $\lambda_\mathrm{sBHAR}$ that is likely related to increases in the availability of cold gas (top panels).
    An enhancement in the \emph{fraction} of galaxies that host AGN is also found in sub--main-sequence galaxies (moving relative to the main sequence at fixed $M_*$, right panels)
    that appears to be related to their more compact structure (i.e.,\ smaller physical sizes, as illustrated). 
    AGN are also found in galaxies with lower SFRs (i.e.,\ in quiescent, bulge-dominated galaxies) but typically with lower $\lambda_\mathrm{sBHAR}$, likely fuelled by stellar mass loss from a passively evolving stellar population and there may be an \emph{anti}-correlation with size; see Footnote~\ref{footnote:size_vs_age}. 
    Radiatively \emph{inefficient} AGN with low $\lambda_\mathrm{sBHAR}$ but significant radio jets are also common in such galaxies 
    (see Fig.~\ref{fig:radio-agn-fraction} \& 
    \S\ref{sec4:jetted}).}
    \label{fig:agn_alongacrossMS}
\end{figure*}

Fig.~\ref{fig:agn_alongacrossMS} summarises the emerging picture of how such measurements depend on a number of key galaxy properties.
In general, $p(\lambda_\mathrm{sBHAR})$ has a broad shape with a break at high $\lambda_\mathrm{sBHAR}$ corresponding to $\sim$the Eddington limit and thus reflecting the (likely) self-regulation of SMBH growth through localised feedback mechanisms
\citep[e.g.][]{aird_primus_2013,bongiorno_agn_2016,georgakakis_observational_2017,zou_mapping_2024}.  
AGN are found throughout the star-forming main sequence, but $p(\lambda_\mathrm{sBHAR})$ shifts toward higher values at higher $M_*$, producing both a slight increase in AGN fraction (indicating a higher rate of triggering of AGN events) and higher average accretion rates  \citep[e.g.][]{aird_xrays_2018,yang_linking_2018}.
The total galaxy stellar mass may be the primary determinant of the \emph{absolute} level of SMBH mass growth in such galaxies \citep[e.g.][]{yang_stellar_2017}. However, \citet{aird_xrays_2019} suggest that---once an overall scaling with stellar mass is accounted for via measurements of $\lambda_\mathrm{sBHAR}$---the absolute SFR provides a better predictor of AGN activity along the main sequence while also accounting for the strong increase in AGN fraction toward higher redshift i.e. from $z\sim0$ to $z\sim3$, where the normalisation of the main sequence and thus typical SFRs for galaxies of given $M_*$ also increase
\citep[see also][]{Mountrichas_star_formation_2022A&A...661A.108M}.
Such scalings with SFR suggest that it is the overall availability of cold gas---whether stochastically delivered to the central regions of a galaxy or indirectly as the fuel to form stars that drive strong stellar winds---that determines both the rate of AGN events and the overall amount of SMBH mass growth that takes place in a galaxy on cosmic timescales. 
However, determining whether stellar mass or SFR is the primary determinant of AGN activity remains difficult for main-sequence star-forming galaxies where these quantities are correlated.

An additional determining factor in SMBH growth may be the prominence of a central stellar bulge. 
Indeed, higher mass star-forming galaxies tend to be more bulge-dominated and thus bulge mass, rather than total stellar mass, may be more closely related to AGN activity \citep[e.g.][]{yang_evident_2019}. 
Considering instead galaxies with different SFRs \emph{relative} to the main sequence (at a given $M_*$), the subset of galaxies with slightly lower SFRs (``sub--main-sequence'' galaxies) show a higher incidence of AGN \citep[e.g.][]{shimizu_decreased_2015,aird_xrays_2019,ji_agn_selection_2022} that may correspond to their typically more centrally concentrated stellar densities (i.e. more compact sizes) which also appears to be an important determinant of AGN fraction and sample-averaged accretion rates \citep[e.g.][]{rangel_evidence_2014,kocevski_candels_2017,ni_does_2019,ni_revealing_2021,aird_agn_2022}. 
However, it is important to note that a central stellar bulge is not a \emph{requirement} for a galaxy to grow a central SMBH or host an AGN \citep[e.g.][]{simmons_galaxy_zoo_2013}.
Starburst galaxies, which have enhanced SFRs relative to their stellar masses and thus lie above the main sequence, also exhibit an enhancement in AGN fraction \citep[e.g][]{Bernhard_enhanced_2016MNRAS.460..902B,aird_xrays_2019}.\footnote{The enhancement of AGN activity in starburst galaxies, however, may not be as strong as the enhancement in the SFRs, see e.g. \citet{Rodighiero_relationship_2015ApJ...800L..10R}.}
Gas-rich major mergers may be an important route to trigger this important sub-set of the AGN population (see \S\ref{sec4:mergers} and \S\ref{sec:QSOphases} for further discussion). 


Finally, we note that AGN are fairly widespread within the quiescent galaxy population and while they typically have lower $\lambda_\mathrm{sBHAR}$ there may still be substantial SMBH growth occurring in such galaxies \citep[e.g.][]{wang_agn_2017,aird_xrays_2018,birchall_relationship_2023}.
These galaxies, by definition, have little ongoing star formation and lack a ready supply of cold gas that is thought to be the main driver of SMBH growth in star-forming galaxies. 
Several studies have found that the AGN incidence in such galaxies anti-correlates with direct tracers of stellar age \citep[determined from spectroscopic tracers or through modeling of stellar populations, e.g.][]{kauffmann_feast_2009,hernan-caballero_higher_2014,mountrichas_histories_2022,georgantopoulos_comparing_2023,ni_incidence_2023,Almaini_evidence_2025}
or tracks another property that is known to correlate with age (e.g. size,\footnote{Physically larger quiescent galaxies are found to have younger stellar populations \citep[e.g.][]{vanderWel_size_2009ApJ...698.1232V,Hamadouche_vandels_2022MNRAS.512.1262H}, resulting in enhanced AGN fractions in larger / more extended quiescent galaxies (cf. star-forming galaxies where AGN activity appears to be boosted in more compact galaxies).\label{footnote:size_vs_age}}
see \citealt{rangel_evidence_2014,aird_agn_2022}).
Such findings suggest that recycled gas---produced by the mass loss through supernovae and stellar winds as a stellar population evolves---may be a crucial source of fuel for AGN in such galaxies, as predicted in theoretical works discussed above \citep[e.g.][]{ciotti09,choi_origins_2023}.
In the most massive quiescent galaxies, the hot gas reservoir may be sufficient to fuel frequent AGN events at relatively low accretion rates, either directly \citep[e.g.][]{Yuan2014} or via a fraction that condenses into cold gas and is accreted stochastically by the SMBH \citep[e.g.][]{gaspari_chaotic_2015}, further enhancing the AGN incidence at low sBHARs in such galaxies \citep[e.g.][]{ni_incidence_2023}.

\subsubsection{Re-assessing AGN downsizing}
\label{sec4:downsizing}

\begin{figure}[t]
    \centering
    \includegraphics[width=\columnwidth]{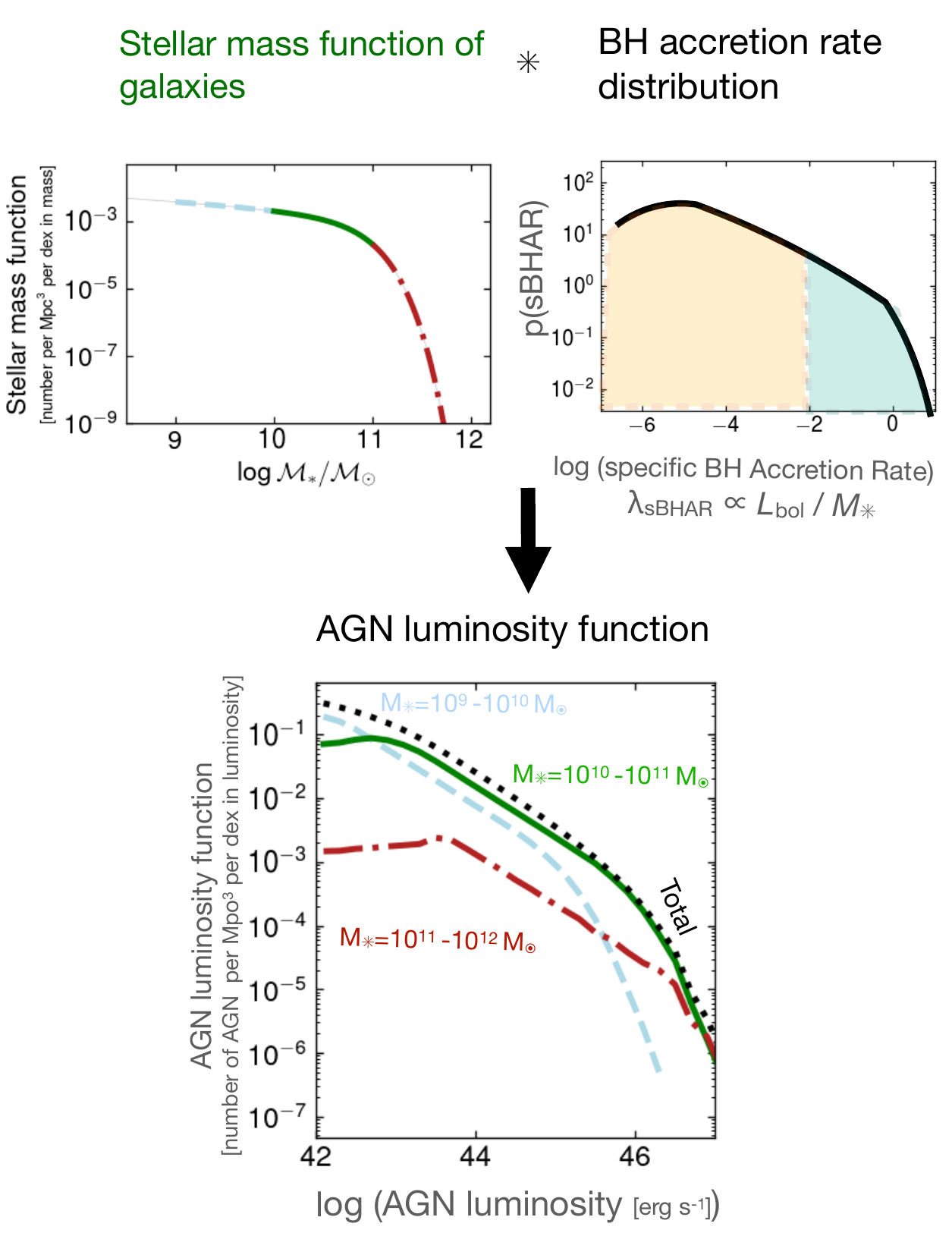}
    \caption{Illustration of how the AGN luminosity function (space density of AGN as a function of luminosity) at moderate redshifts ($z\lesssim3$) is built up from the galaxy stellar mass function (space density of galaxies with different stellar masses, top left panel) and the broad probability distribution of specific SMBH accretion rates, $p(\lambda_\mathrm{sBHAR})$ (top right panel). 
    Due to the broad range in $\lambda_\mathrm{sBHAR}$, galaxies of a given stellar mass (indicated by colours/linestyles in the top left panel) 
    produce AGN with a wide range of luminosities (corresponding colours/linestyles in bottom panel). However, the steepness of $p(\lambda_\mathrm{sBHAR})$ relative to the stellar mass function means that host galaxies with $M_*\sim10^{10}-10^{11} M_\odot$ (solid green line) dominate the total AGN luminosity function except at the very highest and very lowest luminosities.}
    \label{fig:agn_lf}
\end{figure}

The discovery of the broad distribution of specific SMBH accretion rates, and the associated development of our understanding of how host galaxy properties affect SMBH growth, also feeds into our interpretation of the space density of AGN of different luminosities (i.e. the AGN luminosity function) and the observed ``downsizing'' pattern with redshift (see \S\ref{sec4:xray-selection}, Fig.~
\ref{fig:spacedens_vs_z}).
The total space density of AGN at a given luminosity is the result of a convolution of the space density of galaxies of a given stellar mass (i.e. their stellar mass function, SMF) and the broad distribution of $\lambda_\mathrm{sBHAR}$ (see Fig.~\ref{fig:agn_lf}).
Due to the relative steepness of $p(\lambda_\mathrm{sBHAR})$ compared to the SMF, the space density of AGN over a wide range in luminosities ($L_\mathrm{AGN}\approx 10^{43}-10^{46}$~erg~s$^{-1}$) is dominated by host galaxies with masses of $M_*\sim10^{10}-10^{11} M_\odot$ \citep[lower panel of Fig.~\ref{fig:agn_lf};][]{aird_primus_2013,caplar_agn_2015,georgakakis_forward_2020}, without requiring any preferential mass range for AGN activity.
Higher mass galaxies will produce higher luminosity AGN (for the same distribution of $\lambda_\mathrm{sBHAR}$), but as such galaxies are much rarer they only make a significant contribution to the total space density of AGN at the highest luminosities. 
Lower mass galaxies, on the other hand, are more common but produce lower luminosity AGN events. 
To produce a \emph{given} AGN luminosity for a lower mass galaxy requires a higher $\lambda_\mathrm{sBHAR}$, and such phases are rarer than at lower $\lambda_\mathrm{sBHAR}$~(see \S\ref{sec4:dwarfs}).

The interplay between the evolution of the galaxy SMF (in particular the increase in normalisation due to the build up of the galaxy population between $z\sim4$ and $z\sim1$) and the evolution of $p(\lambda_\mathrm{sBHAR})$ (which shifts toward lower values between $z\sim2$ and $z\sim0$, likely due to the decrease in the availability of cold gas) is able to produce the observed ``downsizing'' in the evolution of the space densities of AGN of different luminosities, as shown in Fig.~\ref{fig:spacedens_vs_z} \citep{caplar_agn_2015}. 
Differential growth of SMBHs of different masses is \emph{not} required to produce this pattern.
However, independent observations---in particular for bright sub-samples where SMBH masses may be measured directly (see \S\ref{sec:BHmass})---indicate a true downsizing whereby more massive SMBHs appear to be more active at earlier cosmic times than their lower mass counterparts \citep[e.g. ][see discussion in AH12 \S3.3.1]{heckman_present-day_2004,kelly_demographics_2013}. 
Some of this discrepancy may be due to the broad range of SMBH masses for galaxies of similar stellar mass, allowing for SMBH-mass-dependent growth to occur without a strong dependence on host stellar mass. 
Indeed, recent observations suggest that the scaling between SMBH and stellar mass depends on host galaxy SFR and has a large scatter \citep[e.g.][see also Fig.~\ref{fig:mbh_mstar_diagram}]{terrazas_supermassive_2017,greene_intermediateMass_2020} that may require a range of SMBH growth pathways even in galaxies of similar stellar mass \citep[e.g.][see also \S\ref{sec:QSOhighz}]{guetzoyan_bootes_2024,terrazas_diverse_2024}. 

To fully track the growth of the masses of central SMBHs over cosmic time and reconcile with the assembly of their host galaxies requires detailed modelling either through data-driven techniques that combine multiple observation constraints \citep[e.g.][]{shankar_accretion-drive_2013,georgakakis_forward_2020,zhang_trinity_2023} or from physically-motivated hydrodynamical simulations \citep[e.g.][]{habouzit_supermassive_2021}.
However, a consensus view is yet to emerge.

\subsubsection{Obscuration properties and their connection to varying SMBH growth rates}
\label{sec4:obscuration}

A remaining open question is how the obscuration properties of AGN fit into this scheme of distinct AGN events with varying levels of SMBH growth throughout a galaxy lifetime.
Line-of-sight obscuration may occur due to gas within the wider ISM of the host galaxy \citep[e.g.][]{Buchner_2017MNRAS.464.4545B,Gilli2022} and as such be associated with the same reservoir of gas responsible for fuelling AGN events. 
As such, the level of obscuration may differ between different galaxy types due to the distinct physical conditions and AGN triggering mechanisms at play across the galaxy population \citep[e.g.][]{rangel_evidence_2014,georgantopoulos_comparing_2023,mountrichas_populations_2024}.
The parsec-scale material associated with the ``torus''  will also be a major contributor, with the line-of-sight column density dependent on viewing angle \citep[][]{Netzer2015,Buchner2017}.
Winds, driven by the central accretion disk, can also introduce variations in line-of-sight obscuration that depend on the viewing angle. 
Distinct phases that form a sequence within a single AGN event, as gas is funnelled into the central regions, fuels accretion onto the central SMBH and the associated AGN, and is subsequently expelled from the central regions due to feedback from the AGN itself, can also lead to variations in the level of obscuration (see \S\ref{sec:QSOphases}). 
Indeed, the obscuration may be due to the same material that is ultimately accreted by the SMBH (see \S\ref{sec:section3}) and is eventually disrupted by localised feedback processes \citep[e.g. radiation pressure, see][]{ricci_radiative_2017,Laloux_2024MNRAS.532.3459L} and as such closely linked to a cycle of fuelling and feedback that determines the length of AGN events.


\subsubsection{Echoes of past AGN activity}
\label{sec4:echoes}

\begin{figure*}
\centering
    \includegraphics[width=1.5\columnwidth,trim=2cm 2cm 0 0]{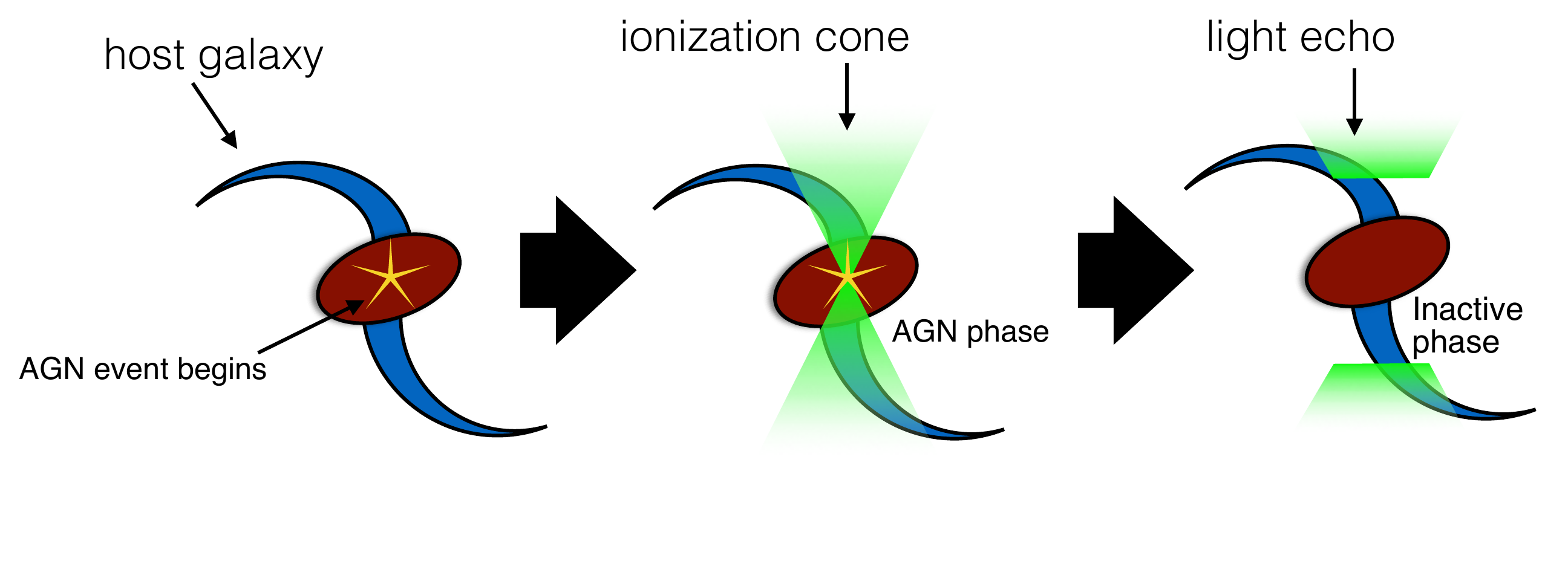}
    \caption{Schematic illustration of how an AGN event occurring in a galaxy leads to the production of extended emission line regions (the ionization cone) during the AGN phase and subsequently leaves a short-lived ``echo'' of prior activity when the SMBH returns to inactive phase, revealed via the discovery of Voorwerpe around a fraction of nearby galaxies. \textit{Source:} reproduced from Fig.~1 of \citet{schawinski_flicker_2015}.
    }
    \label{fig:voorwerps}
\end{figure*}

Given the emerging view of AGN as events in galaxies, we might expect to find evidence of \emph{past} AGN in galaxies that are currently in an inactive phase. Indeed, a new sub-field has emerged over the past decade that involves observational searches for such evidence \citep[e.g.][]{keel_fading_2017,Lansbury2018}. 
One of the first and clearest indicators of such past activity was discovered thanks to the engagement of a broader community of \emph{citizen scientists} through the Galaxy Zoo project \citep{Lintott2008,Lintott_2011MNRAS.410..166L}, which provided images of galaxies from the SDSS to the general public to classify based on their visual morphologies, a task that was insurmountable for professional astronomers alone. 
This process also enabled the identification of rare and unusual sources. 
One such finding was ``Hanny's Voorwerp'',\footnote{Discovered by Dutch schoolteacher Hanny van Arkel, a citizen scientist participant in Galaxy Zoo. The name ``Voorwerp'' is Dutch for object or ``thingy'' but has now been adopted by the professional astronomy community as the generic term for these objects associated with AGN light echoes.}
a highly irregularly shaped object that appears extremely green in the SDSS images.
This object has subsequently been attributed to an ionisation ``echo'' related to past AGN activity in the centre of the nearby galaxy, IC 2497 \citep{lintott_2009MNRAS.399..129L} analagous to the extended emission line regions of known AGN \citep[e.g.][]{Stockton_2006NewAR..50..694S,Keel_2022MNRAS.510.4608K} but lacking the central power source. 
The discovery has prompted wider searches for similar ionisation echos, leading to the discovery of a number of ``Voorwerpje'' with similar properties \citep[e.g.][]{keel_2012MNRAS.420..878K}. 
\citet{schawinski_flicker_2015} used the incidence of such Voorwerpje to estimate the length of AGN events and proposed an evolutionary sequence (reproduced in Fig.~\ref{fig:voorwerps}) linking the start of an AGN event, the production of extended ionisation cones during the AGN phase, and the remnant light echoes seen as the Voorwerpje when the central SMBH drops in accretion rate and returns to an inactive phase. 
Evidence of the impact of past AGN jets (see \S\ref{sec4:radio-selection} and \ref{sec4:jetted}) is also seen in remnant radio sources across a range of systems \citep[see][and references therein]{Mahatma_2023Galax..11...74M}, including in Hanny's Voorwerp \citep{smith_relic_2022}. 

There is also evidence that the SMBH at the centre of the Milky Way, Sagitarius A*, has undergone multiple past AGN phases, including the presence of bubbles of hot gas seen at Gamma-ray wavelengths \citep[with \textit{Fermi},][]{Su_2010ApJ...724.1044S} and at X-ray wavelengths 
\citep[most clearly with new all-sky data from \textit{eROSITA},][]{Predehl_2020Natur.588..227P} that extend several kiloparsecs above and below the Galactic plane, although the possibility of a non-AGN origin is still under debate \citep[e.g.][]{Yang_2022NatAs...6..584Y,Liu_2024ApJ...967L..27L,Churazov_2024A&A...691L..22C,Sarkar_2024A&ARv..32....1S}.


\subsection{Jetted AGN -- capping the growth of galaxies and their black holes}
\label{sec4:jetted}

The discussion in \S\ref{sec4:varying_growth} above has focussed on AGN that are identified by their radiative output: most effectively identified at optical/UV, X-ray, or IR wavelengths. 
However, the last decade has seen a drastic increase in our observational capabilities at radio wavelengths that enable us to trace another form in which SMBHs send energy out into the wider environment: highly collimated, powerful jets. 
While most clearly observed at radio wavelengths (see \S\ref{sec4:radio-selection}), the bulk of the power in such jets is carried by the \emph{kinetic} energy of the outflowing material rather than emerging as electromagnetic radiation.
In this section, we discuss how the presence of jets relates to the broader AGN population, the different phases of activity, and our understanding of AGN as ``events''.
We also discuss recent studies of how such jetted AGN populate galaxies and highlight the role that jetted AGN are thought to play in capping the ultimate growth of the stellar mass of their host galaxies and the mass growth of SMBHs themselves. 

\begin{figure*}
\centering
    \includegraphics[width=0.9\textwidth,trim=0cm 1cm 0 0]{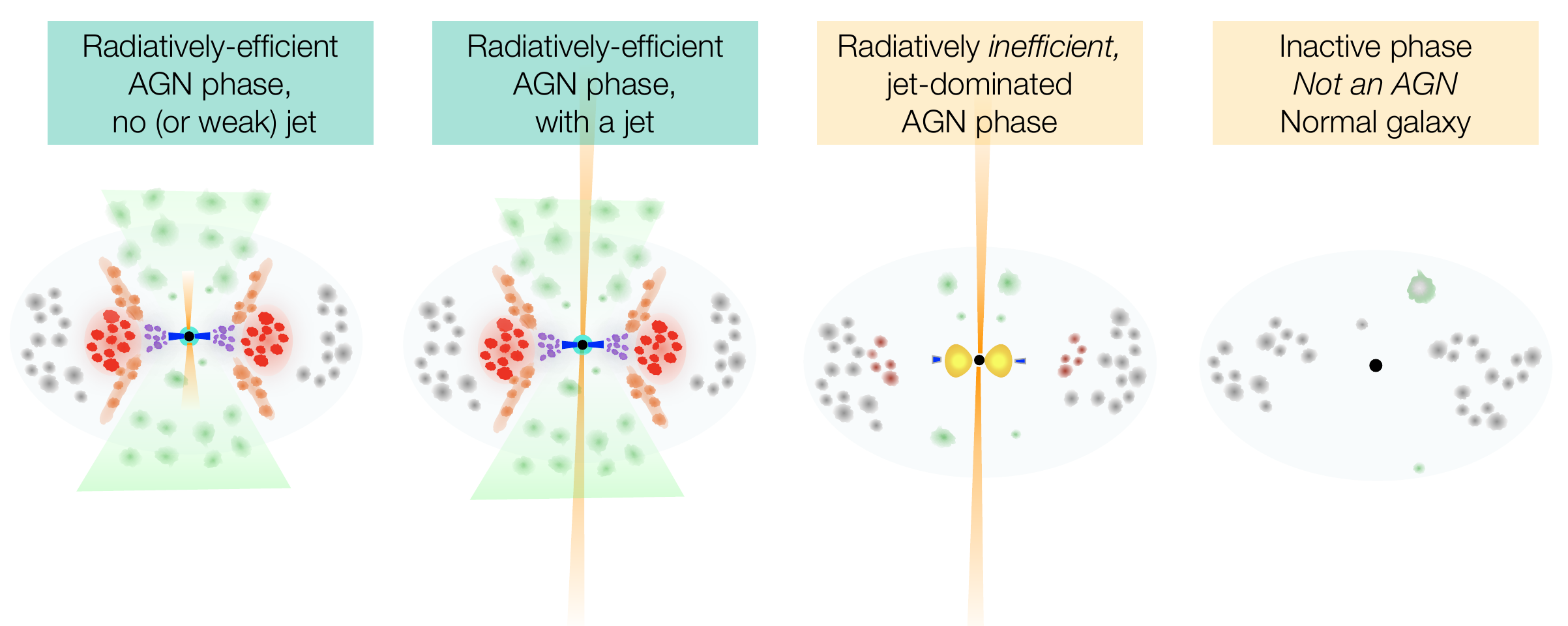}
    \caption{Illustration of different phases of a central SMBH (cf. Fig.~\ref{fig:agn_model} which shows size scales), including radiatively-efficient phases (left two diagrams, highlighted in green) corresponding to $\lambda_\mathrm{Edd}\gtrsim1$\% when a standard accretion disk (shown in blue) and broad-line region (purple), narrow-line region (green), X-ray corona (cyan) and warm dusty obscurer (red, with extended outflows of hotter dust in the polar direction shown in brown as in Fig.~\ref{fig:AGN-schema}) are all present and a radio jet (orange) \emph{may} be present but may be weak/small (first panel) or strong and extend up to $\sim$Mpc scales (second panel). 
    At lower accretion rates (corresponding to $\lambda_\mathrm{Edd}\lesssim 1$\%, right two diagrams highlighted in yellow) a strong radio jet may still be produced, corresponding to a radiatively inefficient AGN phases when the accretion is in the form of a RIAF/ADAF (shown in yellow in third panel, see \S\ref{sec:accretion-disks} and Fig.~\ref{fig:Accretion-flows}) or the SMBH may be inactive and neither growing in mass nor powering substantial radiation (i.e. not an AGN, fourth panel). 
    We note that these phases all likely form a continuum; they are \emph{not} four completely distinct phases. 
    Whether such phases can occur in a sequence as part of a single AGN ``event''---and if so, their order, how long each phase lasts, and the persistence of different structures across phases---or whether they only occur as part of distinct events remains a key open question (see also \S\ref{sec:QSOphases} for discussion of additional phases that may occur as part of an evolutionary sequence for the brightest quasars). 
    }
    \label{fig:AGNphases}
\end{figure*}

First, it is important to note that strong, extended jets are associated with at least some fraction of radiatively efficient AGN that also possess a standard accretion disk and the many other features of this phase of activity \citep[e.g.][]{whittam_stripe_2018,kondapally_lofar_2025}. 
Whether a jet is \emph{always} present in a radiatively efficient AGN---but may be weaker/less extended in some cases---is still unclear, in part due to the difficulty in distinguishing such jets from other radio-emitting processes (e.g. stellar processes, AGN driven-winds; see discussion in \S\ref{sec4:radio-selection}).

Jets are also produced when the SMBH is in a \emph{radiatively inefficient} accretion mode (i.e. a RIAF/ADAF, see \S\ref{sec:accretion-disks} and Fig.~\ref{fig:Accretion-flows}), typically corresponding to $\lambda_\mathrm{Edd}\lesssim$1\%, or the equivalent $\lambda_\mathrm{sBHAR}$, 
and when the mass accretion rate is usually relatively low. 
This dichotomy is reflected in the observational classification of radio-selected AGN into high-excitation radio galaxies (HERGs) and low-excitation radio galaxies (LERGs)\footnote{The original classification of radio galaxies as HERGs or LERGs was based on the presence or absence, respectively, of high-excitation emission lines in their optical spectra \citep[e.g.][]{2012MNRAS.421.1569B} but as deeper radio surveys have become available the definitions have expanded to refer more generally to sources where the radio emission is dominated by the AGN and a radiatively efficient AGN is identified in the broader multiwavelength SED (i.e.~the detection of an X-ray AGN or an optical-to-IR SED that indicates the presence of a quasar or obscured AGN) in the case of a HERG, or that these other signatures of a radiative AGN are absent and only the radio emission indicates the presence of an AGN in the case of a LERG \citep[e.g.][]{best_lofar_2023,kondapally_lofar_2025}.}  
that broadly correspond to radiatively efficient and inefficient regimes, respectively.
Fig.~\ref{fig:AGNphases} illustrates these different phases of activity of a SMBH (as well as completely inactive phases), where the green and yellow highlighting is chosen to connect to the phases of high and low \emph{radiative} output (in terms of $\lambda_\mathrm{sBHAR}\approx\lambda_\mathrm{Edd}$) as in Figs.~\ref{fig:agn_vs_SF}--\ref{fig:agn_lf}.\footnote{We note that discrepancies between $\lambda_\mathrm{Edd}$ and $\lambda_\mathrm{sBHAR}$, given the range of $M_\mathrm{BH}$ in galaxies of a given $M_*$ (see Footnote~\ref{footnote:sBHAR} and Fig.~\ref{fig:mbh_mstar_diagram}) result in an imperfect correspondence between radiatively efficient ($\lambda_\mathrm{Edd}>0.01$) and inefficient/jet-dominated ($\lambda_\mathrm{Edd}<0.01$) AGN phases and different $\lambda_\mathrm{sBHAR}$ values. It is also important to note that the SMBH \emph{mass} accretion rate may still be relatively high in low-$\lambda_\mathrm{Edd}$ but radiatively inefficient phases.}
We stress that the different phases of SMBH activity illustrated in Fig.~\ref{fig:AGNphases}---including additional, intermediate phases where the various emission components range in strength---may correspond to discrete AGN ``events'', and thus can occur in distinct host galaxies with differing duty cycles and distinct triggering/fuelling mechanisms.
Whether an evolutionary sequence between some of these phases within a single AGN event occurs, at least in some cases, remains an important open question.

\begin{figure}[t]
\centering
    \includegraphics[width=0.9\columnwidth,trim=0cm 1.5cm 0 0]{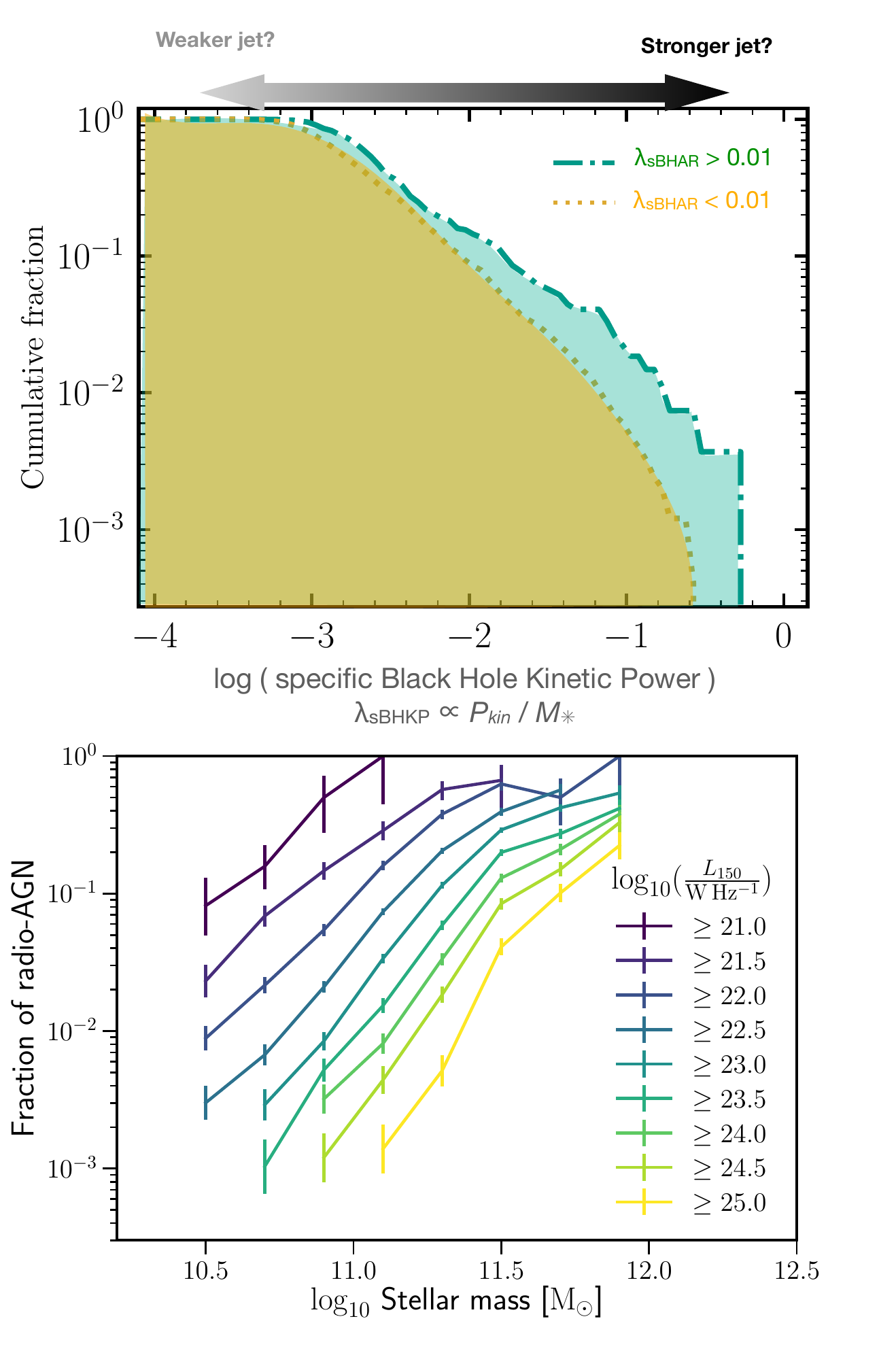}
    \caption{\emph{Top:} Cumulative fraction of radio-selected  AGN with specific Black Hole Kinetic Power ($\lambda_\mathrm{sBHKP}\propto P_\mathrm{kin}/M_*$) exceeds a given value, providing a tracer of the relative power carried by the jet. 
    We have divided the radio-selected AGN into those with high radiative power ($\lambda_\mathrm{sBHAR}>0.01$) and low radiative power ($\lambda_\mathrm{sBHAR}<0.01$), which should roughly correspond to radiatively efficient and inefficient AGN phases, respectively. 
    Both AGN phases are associated with a broad range of jet power, although the strongest jets are mostly associated with radiatively efficient AGN.
    \emph{Source:} produced using results from \citet{kondapally_lofar_2025}.
    \emph{Bottom}: Fraction of galaxies with a radio AGN above different radio luminosity thresholds, showing that the most massive galaxies always produce some level of radio emission and thus indicating that their SMBHs are ``always switched on'' i.e. producing radio jets.
    \emph{Source:} reproduced from Fig.~5 of \citet{sabater_lotss_2019}.
    }
    \label{fig:radio-agn-fraction}
\end{figure}

In the local ($z\lesssim0.3$) universe, a strong rise in the fraction of galaxies with a radio-emitting AGN (predominantly radiatively inefficient, jet-dominated sources) with increasing stellar mass is well established \citep[see AH12 \S3.2.1;][]{best_host_2005,Heckman:14}. 
Recent large-area, sensitive surveys with LOFAR confirm this pattern and indicate that at high stellar masses ($M_* >10^{11} M_\odot$) a radio-emitting, albeit low-luminosity jet is \emph{always} present and revealed in sufficiently deep imaging \citep[e.g.][see Fig.~\ref{fig:radio-agn-fraction}, bottom panel]{sabater_lotss_2019}. 
Low-frequency radio emission may persist up to timescales $\sim10^8$~yr, and thus some of the radio detections likely correspond to remnants of recent activity rather than indicating the \emph{current} activity of the central SMBH (cf. the echoes of past radiative AGN accretion discussed in \S\ref{sec4:echoes}).
Indeed, work to identify not only remnant radio sources but also concurrent evidence of re-starting radio jets is emerging as a major field given the sensitivity, frequency range and spatial resolution of the latest radio surveys, enabling detailed insights into the lifecycles of massive galaxies and their central SMBHs (e.g. \citealt{shabala_duty_2020,jurlin_life_2020}, see \citealt{morganti_what_2024} for a detailed recent review).
The high radio AGN fractions in low-redshift massive galaxies require repeated and near-continual fuelling of the central SMBH, albeit at low levels.
Such galaxies are predominantly quiescent in terms of star formation and generally lack a ready supply of cold gas but contain large reservoirs of hot gas. Gradual cooling of this hot gas is likely sufficient to provide the low-but-continual accretion rates needed to sustain the high duty cycle of jetted AGN.
The cooling hot gas may condense along filaments and could thus in fact fuel the SMBH through a stochastic supply of \emph{cold} gas \citep[e.g.][]{gaspari_chaotic_2015} that triggers the initial AGN event and leads to brief, radiatively efficient AGN phases (see \S\ref{sec4:across-galaxy-population}) before the SMBH enters a more sustained, radiatively-inefficient jet-dominated phase. 

The substantial increase in the incidence of jetted AGN in galaxies with stellar mass persists even if an overall scaling of jet power with stellar mass is accounted for
 (e.g. \citealt{igo_lofar_2024}, although see also \citealt{Hardcastle2019} for discussion of the impact of differing source lifetimes across the mass scale). 
The total kinetic power carried by jets, $P_\mathrm{kin}$, may be estimated from the cavities they appear to inflate 
in the hot IGM of massive clusters or via assumptions about the synchrotron emitting plasma, as these are found to be in balance \citep{2004ApJ...607..800B,2006ApJ...652..216R,2022A&A...668A..65T}. $P_\mathrm{kin}$ is found to scale (albeit with large scatter) with the jet radio 
luminosity\footnote{The relation from a compilation of cavity-based measurements by \citet{Heckman:14}, giving $P_\mathrm{kin} = 2.8\times10^{37} \left(\frac{L_\mathrm{1.4GHz}}{10^{25}\;\mathrm{W\;Hz^{-1}}}\right)^{0.68}$~W, is widely adopted. The same quantity may be referred to as $Q$ or as the ``mechanical luminosity'', $L_\mathrm{mech}$.}
and extrapolating this relationship allows estimates of the jet power more generally across the radio AGN population. 
We consider here the \emph{specific black hole kinetic power}, sBHKP, $\lambda_\mathrm{sBHKP}\propto P_\mathrm{kin} / M_*$ where $M_*$ is the host galaxy stellar mass, as a mass-normalised tracer of the jet power in a manner analogous to the adoption of $\lambda_\mathrm{sBHAR}$ that provides a normalised measure of radiative power \footnote{As with $\lambda_\mathrm{sBHAR}$ (see Footnote~\ref{footnote:sBHAR}), the specific black hole kinetic power can be given in ``Eddington-ratio-equivalent'' units if a single scaling relation between $M_*$ and $M_\mathrm{BH}$ is assumed, i.e. $\lambda_\mathrm{sBHKP}=P_\mathrm{kin}/(1.3\times10^{38}\times 0.002 \times M_*/M_\odot)$ where $P_\mathrm{kin}$ is the kinetic power in erg~s$^{-1}$.
Such units provide a useful scale to assess how much kinetic power is being released relative to the theoretical maximum \emph{radiative} power from accretion onto a SMBH.
\label{footnote:sBHKP}
}
\citep[e.g.][]{igo_lofar_2024}.
The top panel of Fig.~\ref{fig:radio-agn-fraction} shows the distribution of $\lambda_\mathrm{sBHKP}$ for radio-selected AGN (i.e. sources where the radio emission is predominantly due to AGN) that are identified in deep LOFAR surveys \citep{kondapally_lofar_2025}, divided into those with higher radiative power ($\lambda_\mathrm{sBHAR}>0.01$, where the radiative AGN luminosity, $L_\mathrm{AGN}$, has been estimated via SED fitting) and lower radiative power ($\lambda_\mathrm{sBHAR}<0.01$), which should broadly correspond to radiatively efficient and inefficient phases.\footnote{As discussed throughout this section, the imperfect correspondence between $\lambda_\mathrm{sBHAR}$ and $\lambda_\mathrm{Edd}$ is an important caveat when considering the true division of AGN based on their radiative power. Furthermore, recent studies indicate there is not a simple correspondence between higher and lower $\lambda_\mathrm{Edd}$ and the classification of sources as either HERGs or LERGs \citep[e.g.][]{whittam_mightee_2022}.} 
It is clear that jets spanning a broad range of $\lambda_\mathrm{sBHKP}$ are found across both AGN phases and weaker jets dominate in both cases. However, the highest jet power sources are associated with high radiative powers (i.e. high $\lambda_\mathrm{sBHAR}$). 
Thus, the power carried by jets appears to have a weak---but still significant---relationship to the \emph{current} growth rate of the central SMBH.

Over the last decade, new facilities have allowed radio surveys to reach sufficient sensitivity to probe the demographics of the radio source population outside the local universe, reaching to $z\sim4$ \citep[e.g.][]{delvecchio_vla-cosmos_2017,delvecchio_smbh_2018,kondapally_cosmic_2022,Whittam_counterparts_2024MNRAS.527.3231W}.
The overall evolution of the radio luminosity function of the two broad classes of jetted AGN follow somewhat different patterns. 
Radiatively-efficient, jetted AGN (HERGs) increase in number density toward higher redshift in a similar manner to the broader population of ``non-jetted'' radiatively-efficient AGN \citep[e.g.][]{best_cosmic_2014,ceraj_vla-cosmos_2018}.
In contrast, the number densities of radiatively-inefficient jet-dominated AGN (LERGs) appear to undergo relatively little evolution in space density between $z\sim2$ and $z\sim0$ \citep[e.g.][]{williams_lofar-bootes_2018,kondapally_cosmic_2022}. 
However, this weak overall evolution in space density masks an important transition in their host galaxies from predominantly quiescent in the local universe to predominantly star-forming host galaxies by $z\sim1$\footnote{\citet{kondapally_lofar_2025} show that the \emph{fraction} of quiescent galaxies (of given $M_*$) that host radiatively inefficient AGN (LERGs) in fact remains roughly constant from $z\sim0$ to $z\sim1$, indicating that the same fuelling from hot halo gas operates in quiescent galaxies regardless of redshift. However, the space density of such quiescent galaxies drops rapidly over the same redshift interval, whereas star-forming galaxies become more prevalent and the fraction that host LERGs also increases, leading to the dominance of ``star-forming LERGs'' at high redshift.}  \citep[e.g.][]{whittam_stripe_2018,whittam_mightee_2022,kondapally_cosmic_2022,kondapally_lofar_2025}.  
Thus, there appears to be a population of radiatively-inefficient, jet-dominated AGN that appear to be fuelled by the same cold gas supply that is thought to trigger radiatively-efficient AGN in star-forming galaxies, lending support to the idea that both AGN types may in fact be phases within a \emph{single} AGN event. 

The prevalence of jetted AGN in high-$M_*$ galaxies and the fact that (at least by $z\sim0$) such galaxies tend to have quenched their star formation suggests that jets play a vital role in \emph{capping} the growth of the most massive galaxies -- providing a crucial feedback mechanism that maintains the large-scale hot gas reservoir in their host halos. 
By limiting---if not completely halting---the rate of gas cooling and the associated flow of \emph{cold} gas into the central regions, these jets may also cap the mass growth of SMBHs themselves. 
Our developing knowledge of this delicate feedback cycle is discussed in \S\ref{star-formation}.

\subsection{Do AGN preferentially occur in certain large-scale environments?}
\label{sec4:environment}

A major driver of the evolution of galaxies and black holes is their large-scale cosmic environment, which can be parameterized by the mass of the host dark matter halo or the properties of the parent galaxy group, cluster, supercluster, or filament. Galaxy evolution is connected to environment via a range of processes including tidal effects, galaxy mergers, gas inflows and outflows, and interactions with the IGM \citep[e.g.,][]{mo10galform, peng10sfgal}. In general, galaxies residing in denser environments or more massive halos are more likely to have quenched star formation and older stellar populations, and there may be a corresponding impact on the fuelling of SMBH accretion. Different fuelling processes, for example condensation from a hot atmosphere as opposed to torques from galaxy mergers, may be more likely to occur in different environments, making studies of environment a common observational tool for understanding SMBH growth.

The effects of environment on AGN activity have so far been less clear than those for galaxy evolution, owing in part to more limited statistics and the stochastic behaviour of SMBH accretion. Studies of AGN environments have generally followed one of two approaches: (1) direct studies of the populations of AGN among member galaxies in groups, clusters, filaments, and voids, and (2) statistical studies of the spatial correlations of AGN populations, which enabled inferences as to how the AGN populate dark matter haloes. Overall, these studies find significant connections between AGN activity and environment, but any {\em additional} impact beyond the relationships seen for galaxies is relatively small for most populations of AGN.

\subsubsection{Direct measurements of AGN incidence in groups and clusters}

Over the past decade, direct measurements of the AGN populations in galaxy groups and clusters have extended to lower luminosities and higher redshifts thanks to improvements in observational sensitivity. For X-ray selected AGN, the AGN fraction in clusters (defined as the fraction of galaxies with AGN above some $L_X$ threshold) is found to decrease with increasing cluster mass, and decrease with radius toward the centers of clusters \citep[e.g.,][]{ehle14clustagn, ehle15clustagn, noor20agnenv}, with cluster outskirts reaching values approximately that of more typical ``field'' environments 
\citep[e.g.,][]{ehle14clustagn, koul24clustagn}. The X-ray AGN population in clusters also evolves rapidly with redshift, with the AGN fraction increasing by more than an order of magnitude to $z\sim$~1.5 \citep[e.g.,][]{mart13clustagn}, and optical- and IR-selected AGN in clusters show similar behavior \citep[e.g.,][]{albe16clustagn, lope17agnenv}. There has been some evidence presented for a potential excess of AGN outside the cluster virial radius, indicating fuelling of SMBH growth during galaxy infall \citep[e.g.,][]{fass12agninfall, koul19outskirts}, but these effects are not seen in all studies and may simply be owing to statistical fluctuations and projection effects \citep[e.g.,][]{muno24agnoutskirts}. 

Recent years have also seen direct measurements of how AGN populate galaxy groups. Massive galaxy clusters ($M\gtrsim10^{14} M_\odot$) provide large samples of galaxies and clear identifications with large-scale structures, but the bulk of the mass in stars and SMBHs resides in halos of galaxy group scale ($10^{12}\lesssim M \lesssim 10^{13} M_\odot$). Increasingly sensitive optical and X-ray surveys have enabled identification of galaxy groups to $z\sim$~1 and beyond, with corresponding measurements of how AGN occupy these systems \citep[e.g.,][]{alle12groupagn, oh14groupagn, gord18groupagn}. Generally groups are found to have higher fractions of AGN than clusters, extending the mass trends observed for clusters, with the incidence similarly increasing with redshift.  

Identification of AGN with groups and clusters has also enabled direct measurements of the halo occupation distribution (HOD), defined as the number of AGN hosted in central and satellite galaxies as a function of group or cluster mass. These studies find that the incidence of AGN has a relatively weak but increasing dependence on halo mass, with a characteristic $M_{\rm halo}$ of $\sim10^{13}$ $M_\odot$ and the increasing numbers of AGN at higher $M_{\rm halo}$ corresponding to more AGN in satellite galaxies \citep[e.g.][]{chat13agnhod, chak18qsohod}.

Overall, the trends of AGN fractions increasing with clustocentric radius and redshift, and decreasing with cluster group mass are also observed for star formation in galaxies, with the corresponding behaviour in galaxy HODs \citep[e.g.,][]{brod13clustsf}. This indicates that the processes that drive the presence of cold gas and ongoing star formation also drive SMBH growth, as observed for field galaxies (see \S\ref{sec4:varying_growth}).

Recently, a number of studies have explored the AGN population of  protoclusters and comparable large-scale structures at high redshift $(z>2)$. These works generally find substantial numbers of AGN in high-redshift systems, with evidence for a (sometimes modest) enhancement in the AGN fraction compared to similar field galaxies, with corresponding enhanced SF \citep[e.g.,][]{lehm09proto, lehm13z2, tozz22spiderweb, mons23sa22, vito24protoagn, shah24agndense}. This represents a reversal of the relationship between AGN and SF activity for dense regions in the local Universe and continues a trend previously found at lower redshifts
\citep[e.g.,][]{mart13clustagn}. An open question remains the extent to which the virialisation of large-scale structures is important for suppressing SMBH growth in the central regions, since high-redshift protoclusters are typically less relaxed and gravitationally bound compared to lower-redshift systems of similar total mass \citep[e.g.,][]{over16proto}. Nonetheless, the increasingly deep protocluster observations indicate that high-density structures are important regions for SMBH growth in the early Universe.

\begin{figure*}[t]
\centering
    \includegraphics[width=0.9\textwidth,trim=0cm 1.5cm 0 0]{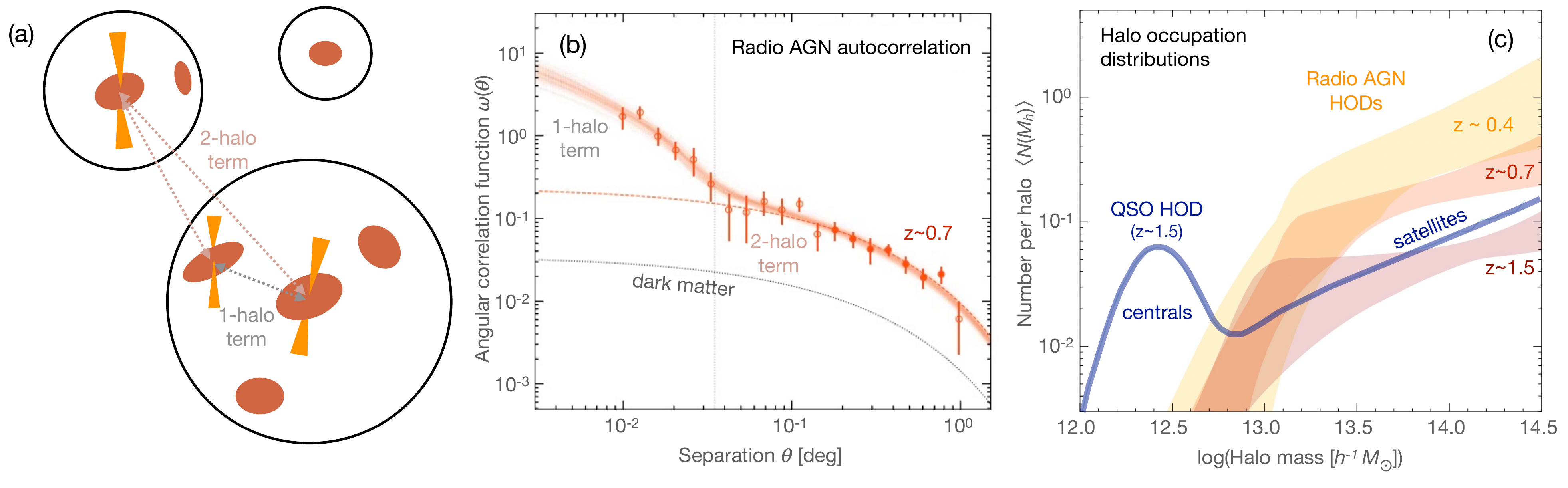}
    \caption{Illustration of the determination of the HOD from spatial clustering measurements. (a) Schematic of an autocorrelation measurement for radio AGN; the large circles represent dark matter halos, the ovals galaxies, and those with orange ``jets'' indicate radio AGN. The correlation function includes separations between AGN in separate halos (the two-halo term) and those within the same halo (the one-halo term). (b) Measurement of the angular autocorrelation function $\omega(\theta)$ for radio AGN at $z\sim0.7$ from \citet{pett24radio}.  The theoretical autocorrelation function for dark matter is shown by the lower dotted curve; the upper dotted curve is a  scaling (by the square of the linear bias) of the dark matter $\omega(\theta)$. The orange shaded region shows a fit to a full HOD model, including the one-halo term on small scales. (c) Halo occupation distributions for radio AGN (orange and red shaded regions) illustrating their evolution with redshift, and compared to the HOD for optical quasars \citep{prad23arxiv}. Each HOD consists of a term for central galaxies plus a power-law function rising to higher halo mass, representing satellites. The HODs clearly demonstrate the preference for higher halo masses for radio AGN compared to optically-selected quasars.  
    \emph{Source:} adapted from Figs~4 \& 9 of \citet{pett24radio}.
    }
    \label{fig:clustering-schem}
\end{figure*}

\subsubsection{Spatial correlation measurements}
\label{sec4:clustering}

In addition to direct measurements of the incidence of AGN in galaxy groups and clusters, there has been a great deal of work using two-point correlation functions to measure the spatial distribution of AGN and to infer their HODs; see Fig.~\ref{fig:clustering-schem} for an illustration. These analyses can take the form of autocorrelation of AGN samples (directly measuring their spatial correlations; e.g., \citealt{myer06clust, efte15qsoclust, comp23clustagn}) or cross-correlation with galaxies, which has the advantage of much larger statistical samples, but requires an understanding of the clustering of the comparison galaxies \citep[e.g.,][]{coil09xclust, hickox_host_2009}. 

The basic technique for measuring the halo mass from clustering involves the computation of the galaxy or AGN {\em bias}; the square of the bias is found by the relative amplitude of the correlation function or power spectrum of galaxies, compared to that for the underlying dark matter distribution. Using numerical simulations of cluster formation and analytic theories for halo formation via statistical collapse, formulae have been developed to convert a measurement of bias to a characteristic $M_{\rm halo}$ \citep[e.g.,][]{shet01halo, tin10}.

The simplest, but physically unrealistic, analysis assumes that all objects in a clustering sample have the same $M_{\rm halo}$, while others adopt a minimum $M_{\rm halo}$ with the assumption that the AGN occupy all haloes above this mass with a constant occupation fraction.  More sophisticated analyses involve modeling the full HOD, which is a measure of the number of central and satellite galaxies in the sample as a function of parent halo mass 
\citep{berl02hod}. In constraining the HOD, the ``two-halo''  term (observed on scales larger than the size of a single halo, or $\gtrsim 1$ comoving Mpc) sets the overall halo mass scale, while the ``one-halo'' term (on scales $\lesssim$ 1 Mpc) determines the fraction of satellite galaxies. The full HOD is described as the average number of central and satellite galaxies as a function of halo mass. 

The variety of ways to model the clustering of different halo masses have led to challenges in comparing different measurements of the bias or halo mass of AGN, so a measure of caution is required in directly comparing results from different studies \citep[e.g.,][]{aird21agnclust}. However, the {\em relative} clustering and halo masses of different populations within a given analysis technique tend to be robust. 

Another related method of determining AGN halo masses comes from weak lensing measurements. Lensing gives a measurement of the integrated mass along the line of sight, and so cross-correlation with AGN (or galaxy) sky positions, along with knowledge of the source redshift distribution, can yield a measurement of bias and halo mass that has largely independent systematics compared to AGN and galaxy auto- or cross-correlation functions \citep[e.g.,][]{bart01lens, lewi06cmblens}. This method has emerged in the past decade due to the availability of high spatial resolution surveys of galaxy shapes, for galaxy weak lensing, as well as high-precision lensing maps of the CMB with ground-based CMB experiments as well as the {\it Planck} satellite. Weak lensing using galaxies and the CMB has been used for AGN samples identified in a range of wavebands including X-rays, optical, mid-IR, and radio \citep[e.g.,][]{sher12qsocmb, geac13qsocmb, dipo15qsocmb, comp23clustagn, pett24radio}.  

Two foundational results in AGN clustering set the stage for progress in the past decade. First, optically-selected quasars show weak, if any dependence in halo mass on luminosity and redshift, being consistently found in halos of the small group scale ($M_{\rm halo}\sim 3 \times 10^{12}$ $M_\odot$; e.g., \citealt{croo05, myer07clust1, ross09qsoclust}). Second, AGN selected in different wavebands demonstrate significantly different clustering amplitudes, with increasing halo masses for AGN selected in the mid-IR, optical and X-ray, and radio, respectively, with X-ray and radio AGN found in halos of $>10^{13} M_\odot$ \citep[e.g.,][]{hickox_host_2009}. These results contributed to a picture in which SMBHs grew, on average, in a radiatively efficient mode up until halos reached the group scale, after which various processes (e.g.,\ merger activity, AGN feedback, emergence of a hot virialized atmosphere) led to a decline in star formation and SMBH growth \citep[e.g.,][]{hickox_host_2009,goulding_tracing_2014}.

In the past decade, a number of studies have computed clustering of AGN selected in different wavebands and using a variety of auto- and cross-correlation techniques as well as local environmental density measures. These studies estimated both the absolute halo masses of AGN, as well as bias relative to ``inactive'' galaxies with similar properties such as stellar mass, SFR, and redshift. On the whole, these studies indicate no obvious difference between AGN hosts and that of comparable ``inactive'' galaxies, with results that are consistent across a wide range of redshifts and AGN selection techniques \citep{mend16agnclust, yang18agnenv, viit19xagnclust, kris20xclust}. The typical halo mass of radio-selected AGN is consistently higher than for radiatively-efficient AGN identified in X-rays, optical, or mid-IR \citep[e.g.,][]{dono10clust, deve19radiocmb, pett24radio}; however, this in broad correspondence with the halos of radio AGN host galaxies, which are typically high-mass, passive systems (see \S\ref{sec4:jetted}) and is consistent with a picture in which mechanical feedback from the AGN emerges over cosmic time along with the passive galaxy population. Robustly measuring the full AGN HOD remains challenging due to degeneracy in HOD parameters and selection effects 
\citep[e.g.,][]{powe24agnhod}; various studies indicate a range of satellite fractions (from $<$10\% up to $\sim$40\%) for AGN in typical group-scale haloes \citep[e.g.,][]{rich12qsohod, alam21qsohod, comp23clustagn, pett24radio}.

A related result is the measurement of clustering as a function of SMBH mass as determined through virial SMBH mass estimators (see \S\ref{sec:BHmass}). These measurements suggest that while there is a weak relation between clustering and $L_{\rm AGN}$ as seen in many previous studies, there is a strong, roughly linear correlation between $M_{\rm halo}$ and $M_{\rm BH}$ \citep[e.g.,][]{krum15xclust, krum23agnclust}, with no clear dependence on Eddington ratio. This indicates that the primary relationship of AGN to their host halos is via SMBH mass, as may be expected given the relations between $M_{\rm BH}$ and galaxy mass, and galaxy mass and $M_{\rm halo}$. This interpretation is consistent with the picture presented in \S\ref{sec4:varying_growth} in which AGN are short-lived and largely stochastic events in the lifecycle of their hosts.

An interesting but somewhat puzzling set of measurements relate the strength of clustering with AGN obscuration. In the simplest ``unified'' model of AGN obscuration, obscuration depends only on small (torus) scale orientation which should introduce no connection between obscuration and large-scale environment. However, measurements using mid-IR selected AGN (initially with {\it Spitzer}-IRAC, then with larger {\it WISE} samples) provided evidence that obscured AGN reside in more massive halos than their unobscured counterparts. These studies have used both spatial autocorrelations as well as CMB lensing measurements, with increasingly higher-significance results with larger and more precise data sets \citep[e.g.,][]{hick11qsoclust,Donoso2014,dipo17qsoclust,mitr18qsohod,Petter2023}. Some studies of X-ray selected AGN in the low-redshift ($z<0.5$) have found correspondingly stronger clustering for obscured sources \citep{Powell2018,krum18agnclust}. These results immediately suggest that processes on scales larger than the unified model ``torus'' are linked to AGN obscuration, as predicted by some theoretical models of AGN fuelling \citep[e.g.,][]{Blecha2018}. Empirical modelling of various physical scenarios and  selection effects suggest that an evolutionary scenario is needed to reproduce the clustering results \citep{dipo17model, Whalen2020}; see \S\ref{sec:QSOphases}. 

However, other recent studies using X-ray, and optical, and mid-IR selected AGN have found no significant difference between obscured and unobscured AGN \citep[e.g.,][]{viit23agnclust} or even {\em stronger} clustering for the unobscured sources \citep[e.g.,][]{alle14xclust, Cordova2024}. This indicates a range of potential physical processes and/or substantial selection effects that are, as yet, not well understood, and opens up the possibility of new insights with the next generation of large spectroscopic galaxy and AGN surveys.

\subsection{The complex connection (or lack thereof) between mergers and AGN activity}
\label{sec4:mergers}

A great deal of research has focused on the potential for galaxy mergers to provide the drivers for rapid SMBH growth. This is largely motivated by early observational results from the local Universe suggesting that the most powerful starbursts and AGN are associated with major galaxy mergers, and may be part of a sequence in which rapid obscured star formation is followed by obscured and then unobscured quasar activity \citep[e.g.,][]{Sanders1988}. This merger-AGN connection was then extended to a broader theoretical framework to explain the co-evolution of SMBHs and galaxy bulges \citep[e.g.,][]{Hopkins2008} and provided compelling reasons to search for connections between these processes.

\subsubsection{Techniques for measuring the merger-AGN connection}

In light of the long relevant timescales and complexities of the observational signatures involved, a range of different techniques have been utilized to test the AGN--merger connection:

\begin{enumerate}[(a)]

\item Perhaps the most widely used and well-known method is to determine the fraction of AGN host galaxies that are in mergers and compare to ``control'' galaxies that do not show AGN signatures but are otherwise matched in properties such as redshift, stellar mass, and SFR \citep[e.g.,][]{cist11agn, vill17merge}. These analyses involve a number of different challenges, including the detectability of tidal features and other merger signatures at low surface brightness (particularly in the presence of a bright nuclear source), difficulties in creating appropriately matched galaxy samples, and statistical constraints, especially for the AGN samples which are often limited in size. Identification of merger signatures can use visual classification, often dividing into merger subclasses such as major and minor mergers and interacting systems. Quantitative techniques include measurements of asymmetry or clumpiness in the galaxy images \citep[e.g.,][]{lotz04morph}, or more recently the use of advanced machine learning algorithms \citep[e.g.,][]{pear19mlmerge}.

\item An alternative method is to consider all galaxies, separated into mergers and non-mergers (or subclasses therein) and to determine the fraction of each class that show AGN signatures or the corresponding average AGN luminosity \citep[e.g.,][]{Goulding2018, come24merge}. This approach uses similar techniques to overcome the challenges of identifying mergers and control galaxies, but has the advantage of better statistical precision in utilizing the full galaxy population.

\item Finally, several studies have expanded on technique (b) by focusing on the AGN fraction in interacting galaxies as a function of their separation distance, thus tracing the incidence of AGN over the course of the merger process \citep[e.g.,][]{elli08pair, li08pair, scot14pair, Ellison2019}. These can use spectroscopically identified kinematic pairs, or simply close separations in photometric samples. These studies are sometimes augmented by samples of late-stage mergers that can only be identified on the basis of disturbed morphologies rather than close pairs of distinct galaxies \citep[e.g.,][]{elli13merge}. Recently, the {\em post-merger} evolution (i.e.,\ after coalescence) has been explored by using dating of stellar populations in post-starburst galaxies, finding that the AGN activity declines after a peak around coalescence \citep{elli24agncoal}. 

\end{enumerate}

\subsubsection{Overview of results for merger fuelling}

Using the technique (a) above, a broad range of studies collectively find no clear difference in merger fractions between AGN and ``inactive'' galaxies that are matched in key properties, leading to the conclusion that the bulk of AGN activity is {\em not} associated with mergers \citep[e.g.,][]{cist11agn, vill23merger}. There is evidence that mergers become more prevalent at the highest luminosities or for dust-reddened systems \citep[e.g.,][]{trei12merge, Glikman2015} although these results have been contested \citep[e.g.,][]{vill23merger}.

Alternatively, techniques (b) and particularly (c) have yielded clear positive associations of mergers with AGN activity. These results are especially pronounced for close separations and/or advanced stages of mergers \citep[e.g.,][]{elli13merge, Ellison2019}. This is especially true for obscured AGN (identified for example in the mid-IR or X-rays), for which the enhancement seems increasingly clear at all redshifts \citep[e.g,.][]{saty14wisepairs, Kocevski2015, Goulding2018, Ricci2021}; see also \S\ref{sec:QSOphases}. The particular connection of obscured AGN activity with mergers has been predicted by a number of theoretical works \citep[e.g.,][]{Hopkins2008,Blecha2018}.

The ultimate conclusion from these works is that {\em some} SMBH growth is connected to galaxy mergers and interactions, and it may be especially important for fuelling of the most powerful and rapid growth (e.g.,\ when the AGN luminosity is driven above the ``knee'' of the luminosity function for a substantial period of time). However, mergers are probably {\em not} associated with the {\em bulk} of the AGN population and corresponding SMBH growth, which is likely driven by secular processes (see \S\ref{sec:section3}). Overall, any physical process that drives gas within the potential of the SMBH is likely to lead to AGN activity.

\subsubsection{Dual AGN}

An exciting new area of investigation into the connections between mergers and AGN activity over the last decade is the identification of {\em dual} AGN. 
While dual AGN (or binaries) likely provide direct insights into SMBHs in post-merger systems, current observational facilities primarily enable the detection of dual AGNs in separate galaxies (especially at high redshift), prior to any potential galaxy merger.
Recent discoveries by {\it JWST} have confirmed the presence of approximately 
30 
candidate dual and triple AGNs with separations ranging from a few to tens of kpc, extending out to $z\sim 5$ \citep[e.g.,][]{2023A&A...679A..89P,2023arXiv231003067P,2024arXiv240514980L}; these observational constraints can constrain the subgrid modeling of large-scale cosmological simulations \citep{2019MNRAS.483.2712R,2022MNRAS.514..640V,2023MNRAS.522.1895C,2025MNRAS.536.3016P}. This is a growing field of research and new techniques such as varstrometry with {\it Gaia} data or the {\it Gaia} multi-peak method combined with follow-up observations using {\it HST} or MUSE adaptive-optics spectroscopy have been successful to confirm multiple sources \citep[][]{2022ApJ...925..162C,2022NatAs...6.1185M,2023arXiv230507396M,2023A&A...671L...4C,2024A&A...690A..57S}.
While these systems are crucial for understanding the hierarchical growth of SMBHs through mergers, dual AGNs with separations of less than a kpc are even more interesting for SMBH fuelling. Even more difficult to confirm, observed dual systems with separations of hundreds of parsecs have been detected \citep[e.g.,][]{2015ApJ...813..103M,2023ApJ...942L..24K}. 
Searching for pc or even sub-pc scale binary systems is feasible through observations with very long baseline interferometric (VLBI) facilities, such as the Very Long Baseline Array \citep[e.g.,][]{2016ApJ...826..106G} and European VLBI Network \citep[e.g.,][]{evn20}. These arrays can achieve angular resolution of $<0.5$ mas at $\sim$cm wavelengths. This resolution corresponds to $<1$ pc at $z=0.1$, allowing VLBI to separate jet or accretion flow emission from two AGN even in extremely close binary orbits, and more modest orbits have already been observed \citep[e.g.,][]{2014Natur.511...57D}. In the future, higher-frequency observations with the Next Generation Event Horizon Telescope \citep{ngeht23} could achieve even higher resolution of $\sim$0.15 $\mu$as, corresponding to spatial resolution of $<$13 pc across {\em all} redshifts. 

In addition to directly resolving binary AGN, it is possible to identify them  via the analysis of specific emission lines, periodic light curves, shifts in the radial velocity of broad emission lines, or variability \citep[e.g.,][]{come13dualagn,grah15dualagn}. Capturing the gas dynamics, inflows, and outflows at small scales in these systems, both theoretically and observationally, will be key to understanding AGN fuelling and the role of the galactic environment and mergers \citep[][for a review]{2019NewAR..8601525D}.

Finally, it is worth noting that a brand new probe of dual SMBHs has emerged in the past few years with the detection of the stochastic gravitational wave background from pulsar timing observations \citep[][]{{EPTA2023,Nanograv2023,Reardon2023_GWB,Xu2023_GWB}}. These results provide a fascinating {\em integrated} constraint on the population of massive binary SMBHs over cosmic volumes; the large amplitude of the stochastic background signal points toward a tension with the most straightforward models of SMBH-galaxy relationships and merger rates \citep{2018ApJ...856...42S,2024A&A...685A..94E}. 

\subsection{Black holes and AGN in dwarf galaxies}
\label{sec4:dwarfs}

Alongside the work showing that SMBH growth is prevalent across a wide range of host galaxy properties and large-scale environments, discussed in \S\ref{sec4:varying_growth}--\ref{sec4:mergers} above, the last decade has seen concerted efforts to find evidence of lower-mass BHs in low-mass galaxies, with stellar masses of $M_{\star}\leqslant 10^{9.5}~\rm M_{\odot}$ (hereafter ``dwarf galaxies'').
Dwarf galaxies are likely to have experienced a less dramatic growth and merger history compared to more massive galaxies. As a result, they may offer indirect constraints on BH formation through ``archeology''. Specifically, the mass of their BHs may have remained close to their initial mass at birth, and thus the number of dwarf galaxies hosting a BH (i.e., the occupation fraction) serves as a crucial constraint on the formation efficiency of BHs. Furthermore, probing lower mass BHs \citep[reaching the elusive ``intermediate-mass" BH regime of $10$--$10^5$~$M_{\odot}$;][]{greene_intermediateMass_2020} and constraining the extent of their growth is vital to understand the assembly of the SMBHs that appear ubiquitous in the higher mass galaxy population. Given their importance---and the availability of powerful new facilities to push into this low-mass regime---the last decade has seen studies of BHs and AGN in dwarf galaxies emerging as a distinct research field. 

Currently, there is an ongoing systematic search for the presence of BHs in 
dwarf galaxies in the local Universe \citep{Greene2008,reines_relations_2015,reines_observational_2016,2016ApJ...817...20M,2017IJMPD..2630021M,2018ApJ...863....1C,greene_intermediateMass_2020,2020ApJ...888...36R}. 
The least massive BHs were found in RGG118, with a mass of $5\times 10^{4}\, \rm M_{\odot}$ \citep{2015ApJ...809L..14B}, and in NGC4395 with a mass estimate of $9100{-}4\times 10^{5}\, \rm M_{\odot}$ \citep{2019NatAs...3..755W}.
The identification and confirmation of these low-mass BHs rely on multi-wavelength observations including optical spectroscopy, observations in X-ray, radio, and IR \citep[e.g.,][]{2004ApJ...610..722G,Greene2008,2012ApJ...755..167D,2018ApJ...863....1C,2018ApJS..235...40L,2018MNRAS.478.2576M,2019MNRAS.488..685M,2019MNRAS.489L..12K}, and remain challenging due to potential contamination from various sources, such as young starburst regions in low-metallicity environment (reassembling AGN), winds from both supernovae (SN) and massive stars, and emission from X-ray binaries \citep[e.g.,][]{2008ApJ...680..154G,2010ApJ...714...25G,reines2011actively,2016ApJ...817...20M,2016ApJ...831..203P,2017ApJ...842..131S,2018MNRAS.478.2576M,2018ApJ...865...43F,2020arXiv201102501S}. Recent observations revealed that, in the low-mass regime, the presence of BHs is not confined to the centres of galaxies; e.g.,\ \citet{2020ApJ...888...36R} discovered that half a dozen dwarf galaxies were hosting offset radio sources that could be AGN, and faint off-centre AGN signatures were detected in dwarf galaxies through MaNGA IFU observations \citep{2020ApJ...898L..30M}.

Most constraints indicate an AGN occupation fraction in dwarf galaxies, across all luminosity ranges, of $\leqslant 0.03$, with significant variations spanning more than an order of magnitude
\citep{2008ApJ...688..794S,Aird2012,schramm2013black,reines2013dwarf,2015ApJ...799...98M,lemons2015x,2016ApJ...831..203P,aird_xrays_2018,2018MNRAS.478.2576M,2019MNRAS.489L..12K,2020ApJ...898L..30M,2020MNRAS.492.2268B,birchall_incidence_2022}.
Several studies have derived constraints on the BH occupation fraction, requiring corrections for inactive BHs missed in observational samples. These studies
suggest that over half of galaxies with stellar masses of $\leqslant 10^{10}\, \rm M_{\odot}$ may harbor a central BH \citep{2015ApJ...799...98M,trump_biases_2015,2019ApJ...872..104N}. In large-scale cosmological simulations, the BH occupation fraction is of almost unity in low-redshift low-mass galaxies. However, results for the AGN occupation fraction vary significantly, spanning several orders of magnitude due to differences in the subgrid modeling of BH seeding, growth, feedback and galaxy processes among simulations \citep{2022MNRAS.514.4912H}.

\subsection{Summary of the key drivers of progress}
\label{sec4:keydrivers}

The results highlighted above demonstrate the exciting progress that has been made in the past decade on the connection between SMBH growth and their host galaxies and environments. Interestingly, the bulk of these advances have been driven by new conceptual insights as well as novel data analysis and theoretical frameworks, rather than the availability of new data. In the {\em previous} decade, the explosion of extragalactic multiwavelength surveys provided observational foundations but also raised significant puzzles that were only resolved in subsequent years. The progress in this past decade was enabled by a deeper understanding of AGN and galaxy evolution timescales and their impact on observables, along with a more sophisticated handling of selection effects via forward modelling and advanced statistical techniques. Theoretical insights have come from simple ``pen-and-paper'' analytical thinking as well as adapting these ideas to numerical simulations. Much of this progress happened collectively, with advances catalysed by collaborations, conferences, and any discussions. The upshot is a clearer and more comprehensive SMBH and galaxy evolution framework that has been widely adopted by the field.

Of course, as always with astronomy, new faciliites have played a significant role in this progress; new sensitive wide-field sky surveys (for example with {\it WISE} and LOFAR for obscured and jet-emitting AGN, respectively) have enabled greater statistics for clustering and host galaxy measurements, while {\it JWST} has provided a number of leaps forward including studies of mergers and dwarf AGN.

Looking forward, exciting progress will be driven by another revolution in data volume, through new sensitive wide-area survey instruments. In the optical and near-infrared, facilities such as Rubin LSST \citep{ivezic_lsst_2019}, {\it Roman Space Telescope} \citep{Akeson_2019arXiv190205569A}, {\it SphereX} \citep{Crill_2020SPIE11443E..0IC}, and the culmination or initiation of a number of ground-based spectroscopic surveys (e.g. DESI: \citealt{DESICollaboration_2024AJ....168...58D}; SDSS-V: \citealt{Kollmeier_2017arXiv171103234K}; WEAVE: \citealt{Jin_2024WEAVE}, 4MOST: \citealt{deJong_2019Msngr.175....3D}; MOONS: \citealt{Taylor_2018SPIE10702E..1GT}), will increase the number of AGN and host galaxy measurements by an order of magnitude or more.  In addition, large sky surveys at other wavelengths, for example with planned or proposed facilities in the radio \citep[Square Kilometer Array:][]{Braun_2015aska.confE.174B}, IR \citep[PRIMA:][]{Moullet_2023arXiv231020572M}, and X-rays ({\it NewAthena}: \citealt{cruise_newathena_2024}; {\it AXIS}: \citealt{2023SPIE12678E..1ER}) will further add to multiwavelength AGN samples and provide vital constraints on their properties.

Taking full advantage of these huge new data volumes will require new techniques in the analysis of observations. There has already been significant progress in machine learning techniques to identify AGN from multiwavelength imaging, spectroscopy, and timing \citep[e.g.,][]{geac12map, fais19mlagn, carv23mlradio, savi23lsst, hvid24agn} and these will become especially valuable as sample sizes increase. 
However, the last decade has shown us the power of conceptual advances that are driven not just by new larger datasets, but communities of scientists from different viewpoints coming together to assess the findings and draw a new consensus.

In addition to large sky surveys, new smaller-field instruments will provide deeper and high-resolution views of AGN and their hosts. Ongoing studies with {\it JWST} continue to push the boundaries in studying the host galaxies and environments of AGN \citep[e.g.,][]{saxe24jwstagn,bona25jwstagn}, while the E-ELT \citep{gilm07eelt} and other future extremely large ground-based telescopes will produce unprecedented sensitivity and spatial resolution for separating and characterizing the nuclear and stellar components of galaxies. Of particular interest are IFU observations that will yield spatially-resolved maps of velocities, ionized gas properties, and stellar populations to increasingly high redshifts.


Finally, our understanding of SMBH populations, and in particular mergers, will be revolutionized with the advent of space-based gravitational wave astronomy with {\it LISA} \citep{2023LRR....26....2A,2024arXiv240207571C}. {\it LISA} will build on the integrated constraints from PTAs (which probe relatively widely separate SMBH binaries) by directly observing the {\em coalescence} of individual merging-SMBH systems. This will yield corresponding insights into AGN fuelling via the rates of galaxy and SMBH mergers measured across all of cosmic time.



%% file: sec5.tex
\section{What fuels the rapid growth of the most massive (and also the first) black holes?}
\label{sec:rapid-growth}

The previous section focused on the demographics and properties of the overall AGN population, building on the prior section which looked at how gas is accreted onto the SMBH. In this section we take a more detailed look at the most rapidly growing black holes, systems typically referred to as quasars, where the accretion physics and host-galaxy/SMBH environments may be different to the more general AGN population.

\begin{figure*}
	\centering 
	\includegraphics[width=0.8\textwidth, angle=0]{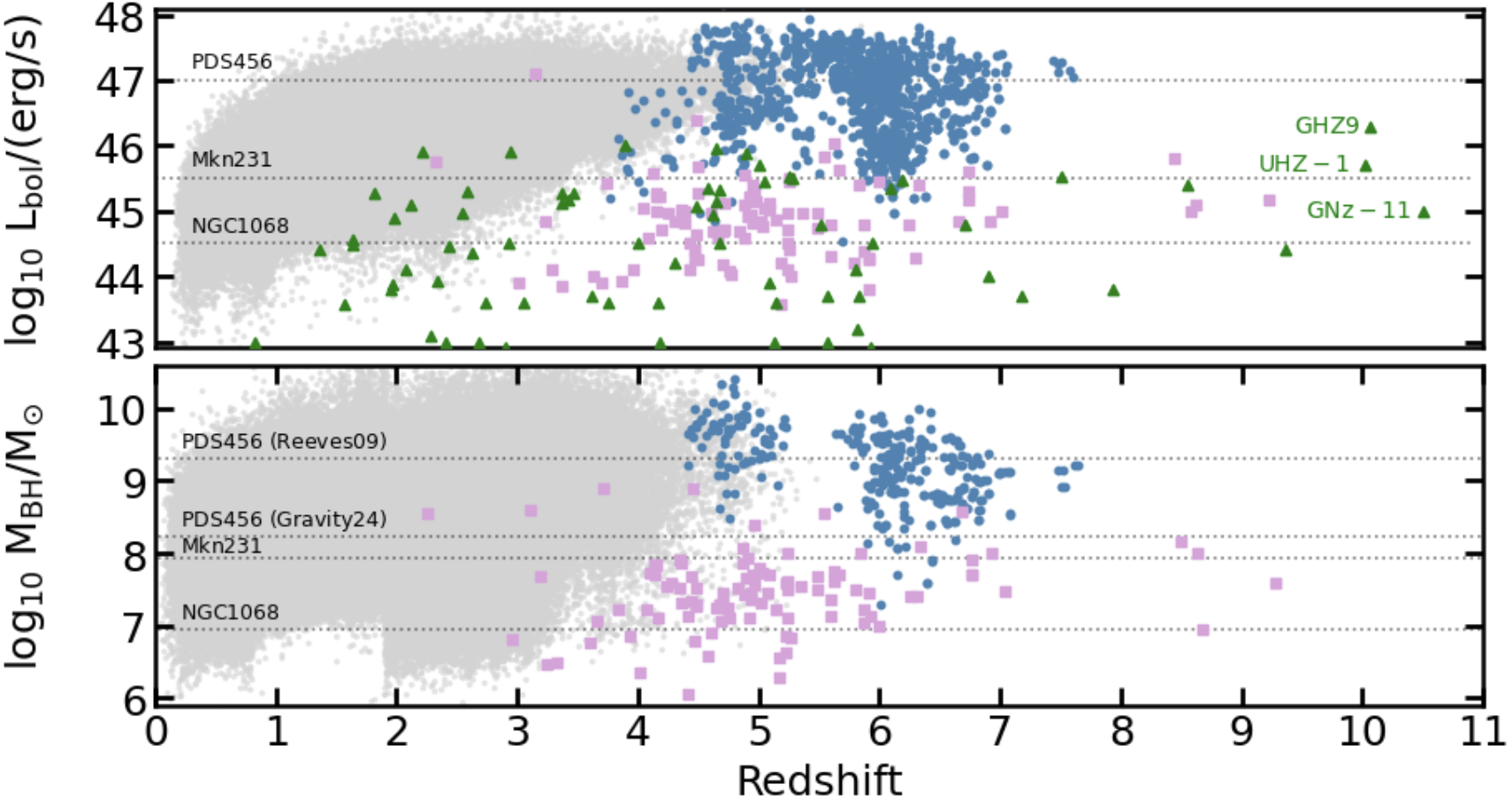}
	\caption{Bolometric luminosity (top) and black-hole mass (bottom) versus spectroscopic redshift for quasars from SDSS DR14 (grey dots) and dedicated $z>$~4 quasar identification studies (dark blue circles), compared to candidate high-redshift AGN identified by {\it JWST}-NIRSpec with either broad Balmer emission lines (magenta squares) or just narrow emission lines (green triangles); the three highest-redshift {\it JWST} AGN plotted in Fig.~\ref{fig_highz_QSO_spectra} are labelled. The bolometric luminosities and black-hole masses of three nearby moderately luminous to luminous AGN (NGC~1068, Mkn~231, PDS~456) are plotted to provide context; PDS~456 is plotted for two different black-hole mass measurements, from virial scaling relationships \citep[][]{Reeves2009} and from GRAVITY+ spatially resolved measurements \citep[][]{GRAVITY2024}. Primary sources: 
    \cite{Glikman2010},
    \cite{Willott2010},
    \cite{McGreer2013,McGreer2018}, \cite{Giallongo2015}, \cite{Yang2017,Yang2019,Yang2021_MBH},
    \cite{Matsuoka2018,Matsuoka2022},
    \cite{Onoue2019},
    \cite{Shen:19},
    \cite{Rakshit2020},
    \cite{Fan2023},
    \cite{Goulding2023},
    \cite{2023arXiv230905714G},
    \cite{2023arXiv230311946H},
    \cite{2023arXiv230200012K,Kocevski2024_LRD_sample}, 
    \cite{2023arXiv230308918L},
    \cite{2023arXiv230801230M,2024Natur.627...59M},
    \cite{scholtz_jades_2023}.
    \cite{2024NatAs...8..126B},
    \cite{2024Natur.628...57F},
    \cite{Juodzbalis2024},
    \cite{Kokorev2024},
    \cite{Lai2024},
    \cite{Lin2024_JWST},
    \cite{2023arXiv230605448M},
    \cite{Mazzolari2024_NLAGN},
    \cite{Napolitano2024}, 
    \cite{Perger2024},
    \cite{Takahashi2024},
    \cite{Treiber2024},
    \cite{Tripodi2024},    \cite{Juodvzbalis_2025},
    \cite{Taylor_2025_z9p3_LRD}.}
	\label{fig_QSO_parameter_space}%
\end{figure*}

Quasars were originally identified as systems where the AGN strongly outshines the host galaxy at optical wavelengths, although more recently quasars are often defined as an AGN with a luminosity above a set threshold, usually the knee in the luminosity function (e.g.,\ a bolometric luminosity of $>10^{45}$~erg~s$^{-1}$, which broadly corresponds to $L_{\rm X}>10^{44}$~erg~s$^{-1}$ and an absolute magnitude of $M_{\rm B}<$-23). While adopting a specific quasar definition is essential for individual studies, since there is not a single universal definition of a quasar, it is more problematic to stick rigidly to a single definition in a review article without leaving out many important studies. Therefore, in this review we adopt a pragmatic quasar definition: those systems in which the authors referred to them as quasars.

The majority of quasar research has focused on the optically bright unobscured subset of the population, where at least a component of the accretion disk is seen in the optical waveband, allowing for direct constraints on the mass-accretion and SMBH properties: we will mostly focus on this  population here. Over the last decade spectroscopic samples of unobscured quasars have expanded by an order of magnitude from $\approx$~100~k \citep[][]{Schneider2010} to of-order a million \citep[e.g.,][]{Lyke2020,Flesch_2023O}, providing sufficient source statistics for more-nuanced SMBH accretion investigations: over the next 2 years the DESI quasar survey alone will increase the quasar demographics to over $\approx$~3~M, pushing down to lower-luminosity systems \citep[e.g.,][]{Alexander2023,Chaussidon2023}. Large advances have also been made in the identification of the highest-redshift quasars: a decade ago the first $z\approx$~7 quasar had just been identified \citep{2011Natur.474..616M} while dedicated deep searches over the last decade have revealed an order of magnitude more $z>7$ quasars \citep{Fan2023}. The recently launched {\it JWST} telescope has now identified lower luminosity, but less secure, candidate AGN out to $z\approx$~10 \citep[e.g.,][]{Adamo_review_2024}. When possible we will also refer to the (theoretically) equally populous obscured quasar population but note that, due to challenges in spectroscopically identifying obscured quasars, their samples number in the thousands.\footnote{There are two key challenges in spectroscopically identifying obscured quasars (1) the difficulty in reliably selecting a complete sample of obscured quasars since the optical-near-IR emission will (typically) be dominated by the host galaxy), and (2) their optical faintness in the optical waveband compared to unobscured quasars. We expect this inequity to be addressed over the coming decade with sensitive large-scale multi-object spectrographs (DESI; WEAVE; 4MOST) with surveys targeting X-ray, infrared, and radio AGN \citep[e.g.,][]{Smith_2016sf2a.conf..271S,Merloni_2019Msngr,Duncan_2023Msngr,Andonie_20254MOST} to construct a more complete census of the AGN and quasar population; see \S\ref{sec4:multiwavelength} for the merits of different multi-wavelength AGN selection approaches.\label{foot:MOSobscured}}

Fig.~\ref{fig_QSO_parameter_space} shows the bolometric luminosity--redshift and SMBH mass--redshift plane for spectroscopically identified quasars and high-redshift {\it JWST}-identified AGN. Dedicated selection and follow-up studies of $z>4$ quasars has allowed for the identification of ``faint" quasars out to $z\approx$~7 with properties which overlap the parameter space explored by {\it JWST}. However, thanks to the high spectroscopic sensitivity of NIRSpec, {\it JWST} is now starting to identify the majority of the AGN population (lower luminosities; less-massive SMBHs) out to $z>8$.

In the following sub sections we first discuss approaches and progress in measuring the masses of SMBHs in distant quasars, a key parameter to determine the relative SMBH growth rate (\S\ref{sec:BHmass}) before investigating the physics of accretion and the surprising overall uniformity in the central-engine properties of unobscured quasars out to $z\approx$~7 (\S\ref{sec:QSOaccretion}). We then review the evidence for different quasar populations representing distinct phases in the overall lifetime of quasars, investigating the well-known but still largely unconfirmed quasar-evolutionary model (\S\ref{sec:QSOphases}). We finally look at the advances in identifying and measuring the properties of the highest-redshift quasars and AGN ($z\approx$~4--10) over the last decade, systems which can place constraints on the origins of the first seed black holes (\S\ref{sec:QSOhighz}), including the latest remarkable discoveries made by {\it JWST}.

\subsection{Black-hole mass measurements of distant quasars}\label{sec:BHmass}

A key advantage in the identification of unobscured quasars is that some of the physical properties of the accretion process can be {\it inferred} due to the direct detection of emission from the accretion disk \citep[see \S4 of][for a brief review]{padovani_active_2017}. We highlight {\it inferred} because the accretion-disk emission from quasars typically peaks in the far-UV \citep[$\approx$~10--100~nm; e.g.,][]{Jin2012,Mitchell2023} where absorption from gas within our own Galaxy precludes direct measurements of the peak of the accretion-disk emission. Consequently, as discussed in \S4.1 of AH12, it is necessary to apply uncertain bolometric corrections \citep[e.g.,][]{2020A&A...636A..73D,2020MNRAS.495.3252S}
to the observed quasar emission to estimate the overall accretion-disk luminosity and infer the mass-accretion rate, which will be dependent on the accretion-disk structure and Eddington ratio \citep[e.g.,][see \S\ref{sec:accretion-disks}]{Hagen2024,Kang2024}. We note, in the absence of obscuration and host-galaxy dilution affects, the most accurate bolometric corrections will be at wavelengths closest to the accretion-disk emission peak.

For the gas in the vicinity of the SMBH ($<10$~pc for a SMBH of $10^8$~M$_\odot$; see Footnote~\ref{foot:SMBHrad} and Fig.~\ref{fig:agn_model}), the kinematics are dominated by the SMBH rather than the host galaxy, providing the potential to estimate the mass of the SMBH. For nearby systems where the spatial resolution is high and the targets are bright, it is possible to directly measure the gas kinematics as a function of distance from the SMBH, allowing for an accurate model of the gravitational potential and hence the SMBH mass \citep[e.g.,][for reviews]{Kormendy1995,Kormendy:13}. However, for the majority of the more distant quasars, we need to resort to mass-scaling relations, which provide an estimate of the SMBH mass based on the widths of the broad emission lines and an estimate of the distance to the broad-line region \citep[BLR; see][for a review]{Shen2013}. These mass-scaling relations are referred to as virial SMBH mass estimators as they assume the BLR gas is in virial equilibrium with the SMBH (see Eqn.~8 from AH12). As the BLR radius is typically too small to be directly measured from imaging data, a technique called ``reverberation mapping'' is used to measure the time lag between accretion disk and broad-line luminosity changes \citep[e.g.,][]{Blandford1982,Peterson1993,Shen2013}, which provides a direct measurement of the accretion disk--BLR distance. 

The foundations of the virial SMBH mass technique were established in the decade prior to AH12 \citep[e.g.,][]{Peterson2004,Bentz2009} from reverberation campaigns of tens of local AGN, where a tight relationship was found between the luminosity of the AGN and the radius of the BLR.\footnote{It is important to note that these relationships are defined for populations and the BLR properties (e.g.,\ geometry; orientation; ionisation; metallicity) are captured as a single ``fudge factor" ($f$) which can differ significantly between individual quasars.} This important discovery showed that SMBH masses can be estimated from single-epoch optical spectroscopic observations with individual mass uncertainties of $\approx$~0.5~dex \citep{2006ApJ...641..689V,Shen2013}, an aspect exploited to estimate SMBH masses for 100,000s of distant quasars \citep[e.g.,][]{Shen2011,Rakshit2020}. Prior virial SMBH mass estimators had been determined for local AGN, rather than distant luminous quasars, and are typically determined for the H$\beta$ emission line, which becomes inaccessible for optical spectroscopy at $z>0.7$ and, instead, the more-poorly calibrated Mg~II and/or C~IV broad emission lines typically need to be adopted for distant quasars \citep[e.g.,][]{Shen2012,Coatman2016}.\footnote{Note that C~IV-derived SMBH masses require a blueshift-dependent correction to account for the outflow component \citep{Coatman2017}; see \S\ref{sec:QSOaccretion}.}

Reverberation mapping (RM) of distant quasars is more challenging than for local AGN because their larger SMBH masses mean the BLR response time to accretion disk changes will be of order years rather than the days/month timescale for lower-luminosity AGN, and their fainter optical magnitudes necessitate longer exposure times. So the previous approach of undertaking RM on a source by source basis, ``sequentially", is too inefficient. The catalyst to allow for significant advances has been the advent of large field of view multi-object spectrographs over the last decade which have allowed for RM for a significant number of distant quasars to be observed in ``parallel" as part of the larger SDSS and DES survey programmes \citep[e.g., SDSS-RM and OzDES-RM;][]{Shen2015,Yuan2015,Malik2023,Shen2024}. These RM campaigns have found flatter luminosity-radius relationships for distant quasars when tracing broad Mg~II than that found for local AGN using broad H$\beta$ \citep{Homayouni2020,Yu2023}, but they remain broadly consistent with a similar overall scatter. Microlensing of the BLR of distant gravitationally lensed quasars provides a complementary approach to measuring the BLR size. While the sample sizes are small, studies have found results broadly consistent with those obtained from RM, once biases have been taken into account \citep{Guerras2013,Hutsemekers2024}. These key measurements provide reassurance on the reliability of prior SMBH mass estimations.

\begin{figure*}
	\centering 
 	\includegraphics[width=1.0\textwidth, angle=0]{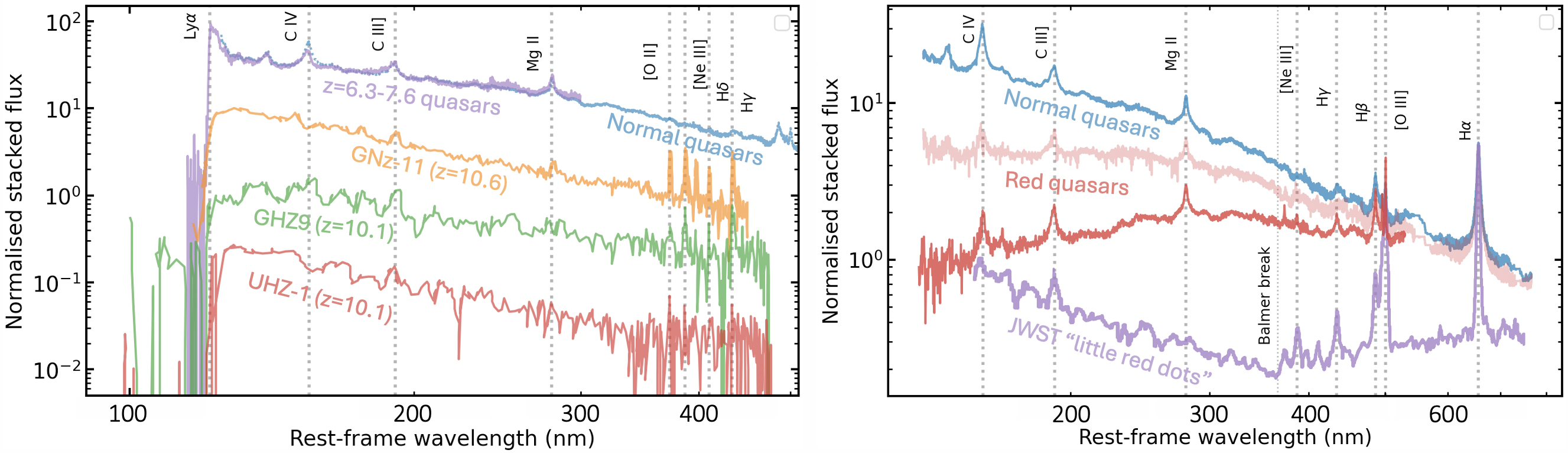}
	\caption{(left): rest-frame UV spectral composite of $z=$~6.3--7.6 quasars (violet) from \cite{Yang2021_MBH} compared to the rest-frame UV--optical spectral composite of $z\approx$~1.5 normal quasars (blue) from \cite{Fawcett2022} and {\it JWST}-NIRSpec spectra of three $z\approx$~10 AGN candidates: GNZ-11 \citep[orange;][]{2024Natur.627...59M}, GHZ9 \citep[green;][]{Napolitano2024} UHZ-1 \citep[red;][]{Goulding2023,2024NatAs...8..126B}; (right): rest-frame UV-optical spectral composite of normal quasars (blue) and red quasars (salmon; red) at $z\approx$~1.5 from \cite{Fawcett2022,Fawcett2023} compared to the spectral composite of {\it JWST}-identified ``little red dots" (violet) at $z=$~2--6 from \cite{Setton2024}. Key emission and absorption lines are indicated and labelled.} 
	\label{fig_highz_QSO_spectra}%
\end{figure*}

The vast majority of distant quasar SMBH mass estimates have relied on the width of broad Mg-II as the proxy for the gas kinematics, which is a less reliable mass tracer than the Balmer broad lines \citep{Shen:19}. However, over the last decade the increase in sensitive near-IR spectroscopic facilities (e.g.,\ Gemini-GNIRS; VLT-XShooter; {\it JWST}-NIRSpec) have allowed for SMBH masses of distant quasars based on the more accurately calibrated H$\beta$ emission line and the empirical relation between the BLR size and rest-frame 5100 \AA \, continuum luminosity \citep{2006ApJ...641..689V}; i.e., the same tracers employed for low-redshift Universe studies. Initial findings indicate that H$\beta$-based mass estimates are, on average, consistent with, or slightly lower than, Mg-II-based masses
\citep{2023Natur.621...51D,2023ApJ...951L...5Y,2024arXiv240414475B,2024ApJ...966..176Y}.

While it is currently impossible to directly measure the BLR radius for the majority of distant quasars due to the need for $\mu$as resolution, the recent GRAVITY+ instrument on the VLT-I interferometer \citep{GRAVITY2022} provides the potential to measure the spatial offset between the blue and red side of the broad emission lines for the brightest quasars hosting the most massive SMBHs. For example, \citet{2024Natur.627..281A} reported the dynamical measurement of the SMBH mass for a $z\approx$~2 quasar with GRAVITY+ finding it was 0.7~dex and 0.4~dex lower than the estimate based on single-epoch H$\beta$ and H$\alpha$, respectively; see also Fig.~\ref{fig_QSO_parameter_space} for virial mass--GRAVITY+ differences for the PDS~456 quasar. The GRAVITY+ wide project \citep{Drescher2022} promises to extend this study to hundreds of $z\approx$~2 quasars, providing important calibrations of the virial SMBH estimators for distant quasars. This will also allow any potential cross calibration issues due to differences in the radius measurement approaches (i.e.,\ responsivity weighted versus emissivity weighted) to be addressed.

\subsection{Accretion environment of distant quasars}\label{sec:QSOaccretion}

The accretion disk is the origin of the high luminosities of quasars and, while neutral hydrogen within our own galaxy prevents us from directly detecting the peak of the emission at far-UV wavelengths, the near-UV--optical waveband traces the cooler Rayleigh-Jeans tail of the accretion-disk emission and provides a plethora of emission lines which yield indirect constraints on the accretion-disk properties \citep[e.g.,][]{Mathews1987,Osterbrock2006}. The emission from quasars at wavelengths shorter and longer than the broad far-UV--optical waveband is typically due to processes associated with the broader accretion-disk environment: thermal IR emission from the dusty obscuring torus heated by the accretion disk, synchrotron radio emission from an AGN-driven jet/wind/shocks, X-ray emission due to inverse Compton scattering from electrons in a high-energy accretion-disk ``corona'', in addition to host-galaxy emission (stellar emission; star formation); see Figs.~\ref{fig:agn_model} and \ref{fig:SED}. We may therefore also expect a clear connection between this broader accretion disk environment and the accretion disk itself. 

Remarkably, the properties of the central engine of optically selected quasars appear to have changed little with cosmic time out to at least $z\approx$~7, despite the huge changes in cosmic environment over this broad time frame.\footnote{We note that most studies have used SDSS-selected quasars which have a specific selection towards blue quasars and so a lack of differences may be partially a selection effect. For example, see \S\ref{sec:QSOhighz} for discussion of the significant differences in the properties of the high-redshift candidate AGN uniquely discovered by {\it JWST}.} The composite rest-frame UV--optical spectrum of $z\approx$~6--7 quasars is largely indistinguishable from that of lower-redshift quasars (see Fig.\ref{fig_highz_QSO_spectra}), at least when matched in luminosity, with essentially identical continuum emission slopes and emission-line properties (i.e.,\ emission line strengths and widths) indicating no significant changes in the overall accretion-disk emission \citep[e.g.,][]{Shen:19,Yang2021_MBH,Dodorico:23}. This result indicates that typical quasars already have ``mature" metal-rich central regions even out to $z\approx$~7 \citep{Lai2022}. The average X-ray properties and the relationship between the soft X-ray and near-UV emission (referred to as the $\alpha_{\rm OX}$ relationship, the equivalent power-law slope between the rest-frame UV emission at 2500~\AA\ and 2~keV in the X-rays), which traces the relationship between the accretion disk and the ``corona'', and the 2500~\AA\ luminosity of quasars, also appears to be {\it largely} unchanged with redshift \citep[e.g.,][]{Salvestrini2019,Vito:22, Wang2021, Rankine_2024}.\footnote{The lack of a strong redshift dependence in the $\alpha_{\rm OX}$--2500\AA\ luminosity relationship allows carefully selected quasar sub samples to be used as standard candles out to at least $z\approx$~3, far exceeding the most distant Type 1a supernovae \citep[e.g.,][]{Risaliti2019,Lusso2020}.} The overall UV--mid-IR SED of quasars also lacks strong redshift evolution, indicating no broad changes in the relationship between the accretion-disk emission and the hot dust at the inner edge of the obscuring ``torus'' \citep[e.g.,][]{Leipski2014,Lyu2022,Auge2023}. In terms of host-galaxy properties, careful subtraction of the quasar emission in rest-frame optical--near-IR imaging (greatly facilitated by {\it JWST} observations), has shown that quasars out to $z\approx$~4 are typically found in massive galaxies ($M_{\rm stellar}>10^{10}$~$M_{\odot}$), and the consequent SMBH--host galaxy mass relationship shows no strong evolution with redshift \citep[e.g.,][]{Suh2020,Mountrichas2023,Tanaka2024,Sun2024}, in contrast to the {\it JWST}-identified high-redshift AGN (see \S\ref{sec:environment}). The average far-IR emission from quasars is found to increase with redshift, tracking the large-scale changes in dust-obscured star formation with cosmic time for typical massive star-forming galaxies \citep[e.g.,][]{Schreiber2015,Stanley2017,Schulze2019,Calistro2021} and implying that the majority of this emission is from the host galaxy rather than from the central engine; see \S\ref{sec4:varying_growth} and \S\ref{sec:environment} for further host galaxy and environmental constraints. 

The essence of these results had been established in the decade prior to AH12, although the increase in data over the last decade have strengthened these conclusions. The fractions of different quasar sub populations also appear to be consistent, with broadly similar fractions of radio-loud (see \S\ref{sec4:radio-selection}) and broad-absorption-line quasars (BALQSOs)\footnote{BALQSOs are quasars where BALs blueward of the broad-emission lines are detected in the rest-frame UV \citep[e.g.,][]{Weymann1981,Weymann1991}. The BALs occur when high-velocity outflowing gas passes across the line of sight between the quasar accretion disk and the observer, causing broad and complex absorption features related to the velocities of the individual absorbing clouds. In the majority of BALQSOs, the BALs are only associated with high-ionisation emission lines such as C~IV, and the system is tyically called a HiBAL, but a smaller fraction of BALQSOs have additional BALs associated with low-ionisation lines such as Mg~II, in which case the system is often referred to as a LoBAL \citep[e.g.,][]{Voit1993,Hall2002}.\label{foot:BALQSO}} identified out to $z\approx$~6, at least for optically selected quasars \citep[e.g.,][]{Banados2018,Liu2021,Yang2021_MBH,Bischetti2023}; we note there is evidence for an increase in the fraction of BALQSOs at $z>6$, not clearly driven by quasar luminosity or accretion rate, suggesting an increase in accretion-driven winds at early epochs.

\begin{figure*}
	\centering 
	\includegraphics[width=1.0\textwidth, angle=0]{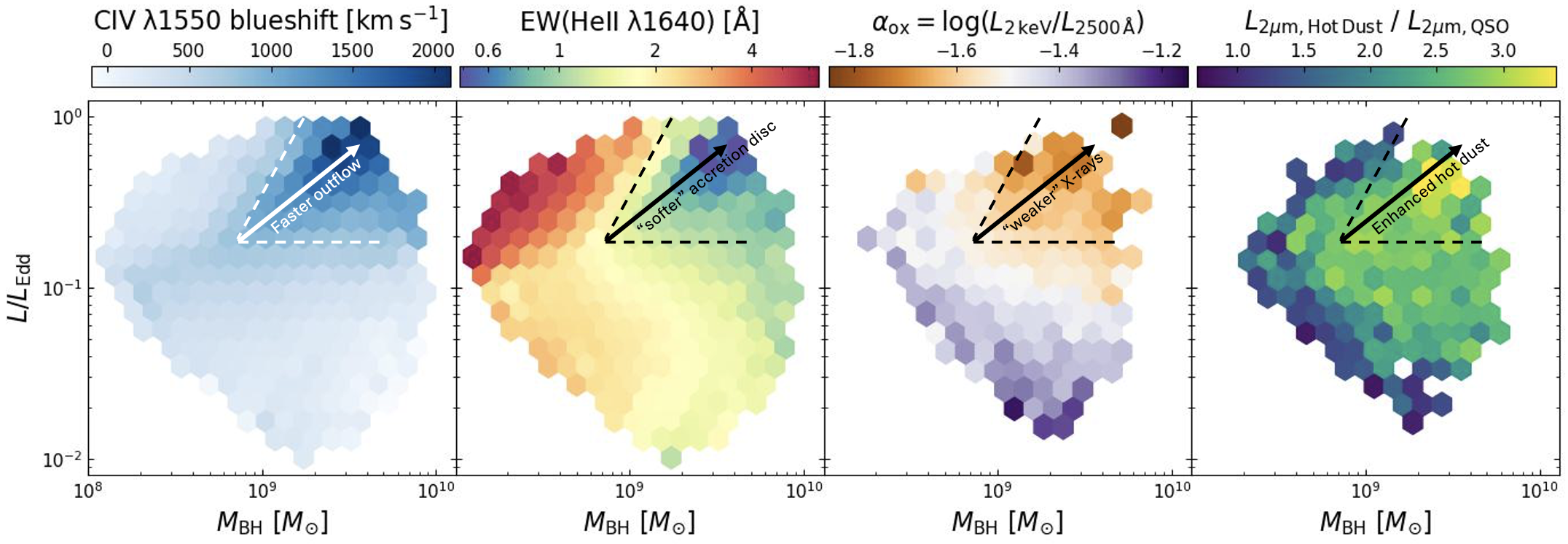}
	\caption{Eddington-ratio--black-hole mass plane for quasars at $z\approx$~2 from the SDSS survey DR16, colour coded to indicate from left to right: C~IV blueshift, He~II EW, $\alpha_{\rm OX}$, and relative excess of hot-dust at 2~$\mu$m. A key region of the left-hand panel highlights the largest C~IV blueshifts, which is repeated in the other panels and shows the connection between C~IV outflowing gas, the accretion-disk SED (as traced by He~II EW and $\alpha_{\rm OX}$), and the connection to excess hot dust. {\it Source:} adapted from \cite[][]{Temple2023} but recalculated and reprojected onto the $L/L_{\rm Edd}-M_{\rm BH}$ plane. The right-hand panel calculates the hot-dust excess following \cite{Temple2021} but using the \cite[][]{Temple2023} sample and projecting the results onto the $L/L_{\rm Edd}-M_{\rm BH}$ plane (M.~Temple, priv. communication).} 
	\label{fig:accretion_outflow_plane}
\end{figure*}

To first order, the lack of change in the overall properties of quasars with cosmic time gives the impression of little diversity in the central engine. However, this is not the case and significant variation is seen between individual systems: indeed, a major focus of quasar research over the last few decades has been identifying the driver(s) of the diversity seen in the spectra of quasars. While several significant empirical correlations have been identified between quasar emission lines and continua \citep[e.g.,][]{Baldwin1977,Boroson1992,Sultentic2000,Richards2011,Shen2014,Marziani2018,Panda2024}, the greatest progress in understanding the physical origin of these relationships has come in the last decade thanks to the order of magnitude increase in the number of sensitive optical spectra for quasars, allied to developments in data-analysis techniques, broader overall sensitive wavelength coverage, and accretion-disk models. 

Foundational early quasar research had previously identified the C~IV$\lambda$1549 broad emission line as a key spectral diagnostic, given the anti correlation of the C~IV equivalent width with the quasar UV continuum luminosity \citep[the famous ``Baldwin" effect;][]{Baldwin1977}. More recently, a connection has been found between the C~IV equivalent width (EW) and the blueshift of C~IV \citep[e.g.,][]{Richards2011,Rivera2022}, indicating that a significant wind component is present in broad C~IV that is not seen in the lower-ionisation broad emission lines. It was found early on that Eddington ratio must be a driver \citep[e.g.,][]{Baskin2005,Shen2014,Sultentic2015} but that property alone could not explain the observed relationships. Greater clarity was made when considering the strength of He~II$\lambda$1640, a sensitive indicator of the unseen far-UV radiation field and, hence, the accretion-disk SED \citep[][]{Rankine2020}.\footnote{He~II$\lambda$1640 provides a sensitive indicator of the $>54$~eV photon rate and, hence, the far-UV continuum. He~II is produced from the recombination of He~III to He~II and so is driven by the ionisation rate of He~II to He~III 
\citep[e.g.,][]{Timlin2021}.} Investigating the He~II$\lambda$1640 and C~IV$\lambda$1549 properties of quasars across the Eddington-ratio--$M_{\rm BH}$ plane, \cite{Temple2023} showed that the C~IV blueshift is tightly connected to {\it both} Eddington ratio and black-hole mass with a distinct region in the Eddington ratio--$M_{\rm BH}$ plane where the strongest outflows are found and He~II is weak. This key result is shown in the first two panels of Fig.~\ref{fig:accretion_outflow_plane}. The weakness of He~II indicates a relatively ``soft'' accretion-disk SED which prevents the gas from becoming over ionised, allowing for the launch of a line-driven wind through radiation pressure \citep[e.g.,][]{Murray1995,Proga2000,Giustini2019} and producing blue-shifted C~IV emission \citep[][]{Yong_2017,Matthews_2023}: quasars with a lower SMBH mass or lower Eddington ratio either overionise the gas or produce insufficient photons to radiatively drive the wind. As can be seen in Fig.~\ref{fig:accretion_outflow_plane}, the same region in the Eddington ratio--$M_{\rm BH}$ plane where strong C~IV blueshifts and weak He~II is found also corresponds to relatively weak X-ray emission (as traced by $\alpha_{\rm OX}$), providing further verification of a ``soft" accretion-disk SED; indeed, \cite{Timlin2021} recently discovered a tight relationship between $\alpha_{\rm OX}$ and He~II equivalent width which demonstrates that $\alpha_{\rm OX}$ provides an indirect indicator of the accretion-disk SED peak wavelength. Overall, these results are in good qualitative agreement with physical accretion-disk models which predict the changes in the accretion-disk SED, X-ray corona, and the strength of line-driven winds with Eddington ratio and $M_{\rm BH}$ mass 
\citep[e.g.,][]{Kubota2018,Giustini2019}; see \S\ref{sec:accretion-disks}. 

Approximately 10--40\% of quasars are referred to as BALQSOs \citep[e.g.,][]{Allen2011}, quasars with powerful outflows observed in absorption (see Footnote~\ref{foot:BALQSO}). An area of considerable uncertainty since their original discovery has been the relationship between BALQSOs and normal ``non-BAL" quasars. Thanks to a suite of complementary multi-wavelength and time-domain studies over the last decade there is growing evidence that most BALQSOs are normal quasars with ``soft" accretion-disk SEDs (either intrinsically ``soft" or obscuration absorbs the hard ionising photons) allowing for the launch of powerful line-driven winds from the accretion disk even as far out as the dusty ``torus" \citep[e.g.,][]{Czerny2011,Giustini2019}. Orientation angle almost certainly plays a role given the connection between dust and BALQSO activity, particularly for LoBALs \citep[e.g.,][]{Hall2002,Gibson2009,Urrutia2009,Matthews_2016}, and the fact that the BAL wind is unlikely to have a 100\% covering factor. Key evidence includes:

\begin{enumerate}[(a)]
    
\item The fraction of BALQSOs within the quasar population increases with decreasing He~II equivalent width, as does the strength of the BAL outflow and dust reddening, with LoBALs (BALQSOs with the most powerful and optically thick outflows) having the weakest He~II emission \citep[e.g.,][]{Baskin2013,Hamann2019}. Indeed, the optical spectral properties for the majority subset (HiBALs) are consistent with those of normal quasars when the emission lines are reconstructed for the impact of the BAL features, with the exception that BALQSOs are not found with strong He~II, large C~IV EW, no C~IV blueshift \citep[e.g.,][]{Baskin2013,Rankine2020}, as expected given Fig.~\ref{fig:accretion_outflow_plane}.
\item BALQSOs are often X-ray weak, producing less X-ray emission than expected from the $\alpha_\mathrm{OX}$--$L_\mathrm{2500}$ relationship, particularly for the most extreme LoBALs \citep[e.g.,][]{Gallagher2006,Luo2014}. The origin of the X-ray weakness often appears to be due to absorption, although some may be intrinsically X-ray weak where the ``corona" is quenched or unable to form \citep[e.g.,][]{Proga2005,Leighly2007,Luo2014,Liu2018}.\footnote{The absorber is speculated to reside between the accretion disk and the BLR \citep[possibly a geometrically and optically thick inner accretion disk, as expected at the highest Eddington ratios; e.g.,][; see Fig.~\ref{fig:Accretion-flows}]{Sadowski2014,Jiang2019}. This absorber further prevents the gas from becoming over ionised and facilitating even more powerful line-driven winds than those seen in normal quasars; see \cite{Higginbottom_2024} for a recent radiative hydrodynamical simulation demonstrating the significant radiative challenges in not over ionising the gas}. These line-driven winds may also be a major contributor to the X-ray obscuration \citep[][]{Matthews_2016,Matthews_2020}.\label{foot:BAL_absorber}
\item BAL features are found to vary on day--year timescales with the strongest BALs demonstrating the weakest variation, likely due to (predominantly) transverse motions of a clumpy absorbing gas \citep[e.g.,][]{Capellupo2011,Filik2013}. In extreme cases the BAL features can disappear altogether and the BALQSO becomes a normal quasar and vice versa \citep[i.e., BAL features can appear in normal quasars;][]{Rogerson2016} with even LoBALs transforming into HiBALs \citep[e.g.,][]{Filik2013,McGraw2017,Yi2021,Yi2022}. The properties of a quasar following a BALQSO transition appear indistinguishable to those of normal quasars \citep{Sameer2019}, although the current samples are small. Based on the fraction of transiting quasars ($\approx$~2\%) and the probed rest-frame timescales ($\approx$~years) the typical BAL lifetime is calculated to be $\approx$~100-1000~years. 
\end{enumerate}

We highlight that differences may occur for the rare FeLoBAL population, where detailed spectral analysis--physical modelling has shown that the launching radius is sometimes on kpc scales rather than in the vicinity of the accretion disk, and where some FeLoBALs appear to accrete at low accretion rates \citep[][]{choi2022,Choi2022_SIMBAL,Leighly2022}.

A subset of the non-BAL quasar population with weak high-ionisation emission lines (C~IV EW~$<$~15\AA; weak-line quasars: WLQs) but normal low-ionisation emission lines \citep[e.g.,][]{Wu2011,Plotkin2015,Chen2024} are also often X-ray weak \citep[e.g.,][]{Pu2020,Wang2022}. 
The X-ray weak fraction (factor $>6$ less X-rays than predicted by the $\alpha_{\rm OX}$--$L_{\rm 2500}$ relationship) appears to increase with decreasing C~IV EW, approaching $\approx$~40\% as compared to $\approx$~6\% for normal quasars \citep[e.g.,][]{tanimoto_application_2020,Pu2020}. The origin of the X-ray weakness appears to be similar to that speculated for X-ray weak BALQSOs \citep[e.g.,][]{Luo2015,Liu2021,Ni2022,Wang2022}; see Footnote~\ref{foot:BAL_absorber}. 

Quasars with strong C~IV blueshifts have also been found to exhibit excess near-IR emission, suggesting that systems with powerful winds have a larger covering factor of hot dust \citep[][]{Temple2021}. In Fig.~\ref{fig:accretion_outflow_plane} (right panel) this result is extended to the full Eddington ratio--$M_{\rm BH}$ plane of \cite{Temple2023} showing that the quasars with the strongest outflows have both the weakest He~II and excess near-IR emission, which may be due to winds produced in the outer part of the accretion disk driving dust away from the equatorial regions and into the polar regions, such as that observed for some local AGN (see \S\ref{sec:observations} and Fig.~\ref{fig:AGN-schema}). Differences in the accretion-disk SED across the Eddington ratio--$M_{\rm BH}$ plane may also help to explain the modest fraction \citep[$\approx$~15\%;][]{Lyu2017} of quasars with relatively weak hot-dust emission, a population that appear to lack a polar-dust component \citep[e.g.,][]{Lyu2018,Lyu2022}.

Good advances in our understanding of the radio emission from quasars have been made over the last decade due to sensitive large field-of-view radio facilities such as LOFAR, allowing for the detection of the weaker ``radio-quiet" emission from the majority of the quasar population. Using deep LOFAR observations, \cite{rankine2021} has shown that radio-detected quasars reside in the same region of the C~IV~EW--blue shift plane as radio-undetected quasars. The radio-detection fraction of quasars increases with C~IV blueshift but the radio-loud fraction {\it decreases} with increasing C~IV blueshift such that the majority of radio-loud quasars have no C~IV blueshift. The radio-loud quasars also have the strongest He~II emission, indicating a strong ionising SED preventing a line-driven wind but allowing a powerful radio-emitting jet to form. BALQSOs and dust-reddened quasars follow this same basic trend, although the presence of a BAL or dust reddening leads to enhanced radio emission \citep[][]{Petley2024}. Dust is a stronger driver of enhanced radio emission than either the C~IV properties or BAL features, with a striking relationship found between the amount of dust reddening towards the quasar and the radio-detection fraction \citep[][]{Fawcett2023,Calistro2023,Petley2024}. This radio--opacity connection (the dust is probably a proxy for general opacity, given dust--gas mixing in the ISM) is likely due to shocks caused by quasar-driven winds or low-power jets interacting with the circumnuclear/ISM environment \citep[e.g.,][]{Zakamska2014,Nims:15}. The identification of excess mid-IR emission from red quasars suggests the dust may be circumnuclear (potentially polar dust; see Fig.~\ref{fig:AGN-schema}) and entrained in a quasar-driven wind \citep{Calistro2021,Zakamska2023}.

\begin{figure*}[t]
	\centering 
	\includegraphics[width=0.99\textwidth, angle=0]{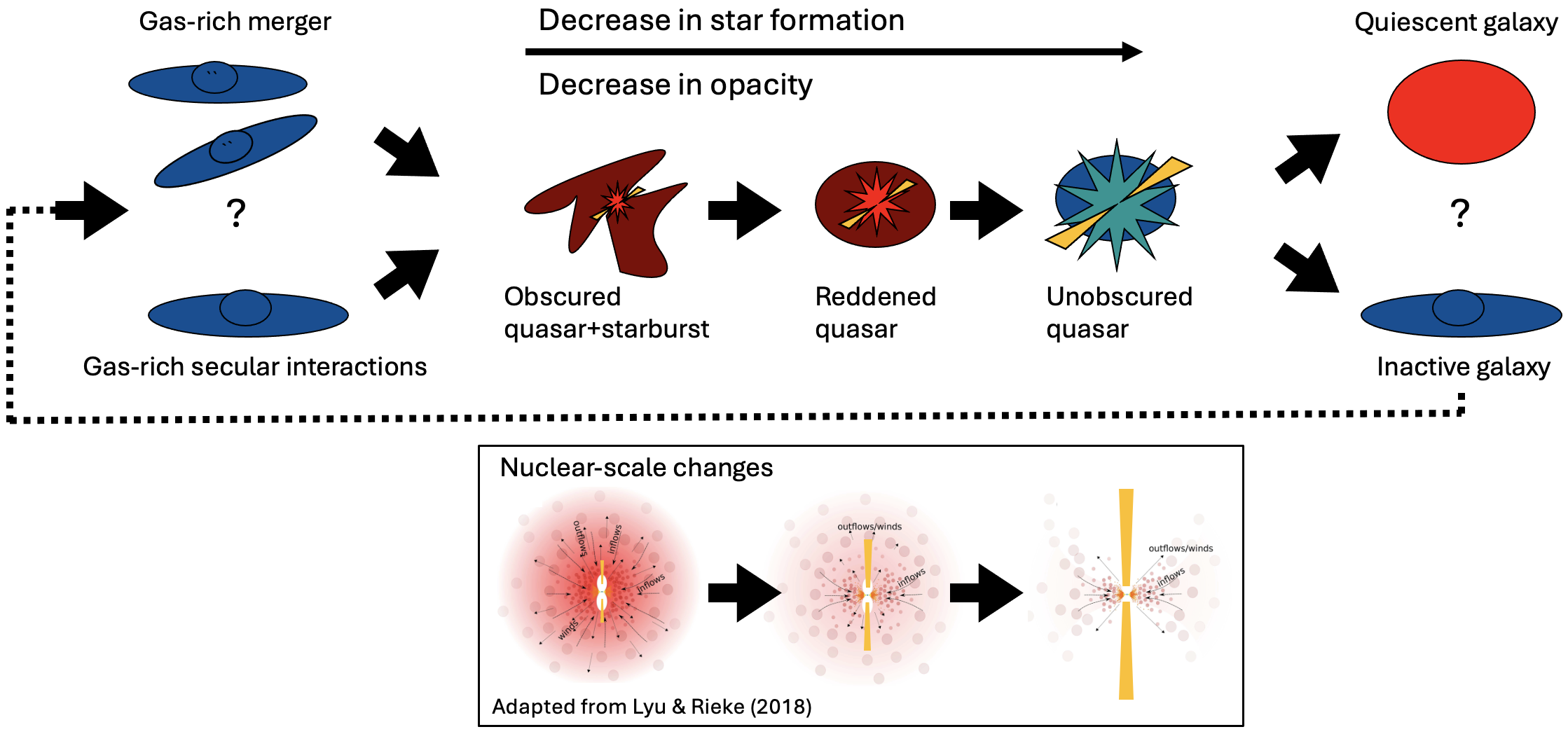}
	\caption{Quasar evolutionary model schematic showing the gas-inflow trigger options (merger or secular), the key quasar phases with differing amounts of obscuration and star formation, and the end result of either a quiescent galaxy or an inactive galaxy where the AGN and star formation have temporarily ceased. The inset panel provides a zoom in on the nuclear regions of the key quasar phases, showing the dust obscuration and indicating gas outflow via winds and jets. {\it Source:} the inset panel is adapted from Fig.~19 of \cite{Lyu2018}.} 
	\label{fig_QSOphases}%
\end{figure*}

\subsection{Evidence for a quasar evolutionary sequence}\label{sec:QSOphases}

A key driver of quasar research over the last few decades has been testing if different quasar populations represent distinct phases within a broader evolutionary sequence. The catalyst in the now ``classical" quasar evolutionary model is the merger of two gas-rich galaxies, which drives gas into the central regions and initiates intense dust-obscured star formation \citep[e.g.,][]{Sanders1988,Canalizo2001,Hopkins2008}. The star formation serves to enhance the gas inflow towards the central SMBH, resulting in quasar activity. The quasar will be initially heavily obscured from both the star formation and the nuclear region. However, due to the interaction between quasar-driven winds (and potentially jets) with the gas and dust (see \S\ref{sec:AGNFeedbackImpact}), the quasar will ultimately drive away the obscuring material, shut down the star formation and reveal an unobscured quasar. We note that the merger only serves to drive gas into the central $\approx$~kpc of the galaxy \citep[e.g.,][]{Hopkins2010_gas_feed}, and other processes (e.g.,\ gas torqued by stars; stellar winds; see \S\ref{sec:dynamics}) are required to drive the gas towards the nucleus. Consequently, given sufficient gas inflow, this evolutionary sequence could also be initiated by secular processes, either from within the galaxy or from interactions with a satellite galaxy. See Fig.~\ref{fig_QSOphases} for a schematic representation of expected changes on the galactic and nuclear scale.

We have already discussed in the previous sub section and \S\ref{sec:variability} that AGN and quasars are not static objects and can vary substantially during their mass-accretion ``event" lifetimes, causing changes to their observed properties (e.g.,\ spectral type; outflow signatures; hot-dust emission). Therefore, how can we distinguish between normal ``expected'' time-variable behaviour and distinct phases within a broader evolutionary sequence? A key difference is that the evolutionary model predicts clear trends over long timescales (e.g.,\ a decrease in SFR with a decrease in nuclear obscuration) while the standard unified AGN model (essentially the ``null hypothesis'' for evolutionary model tests) predicts no significant long-term trends. The most reliable test of an evolutionary model will, therefore, be those based on studying different quasar (and galaxy) populations rather than individual sources. Inherent to any population studies, the selection approaches of the quasar populations must be self consistent, including a ``control'' sample selected in the same basic manner, to ensure any differences are not due to selection biases.\footnote{This is analogous to the apples-to-oranges idiom: don't compare apples with oranges if you want to understand different varieties of apples!}\footnote{We note that, even in the evolutionary model, all quasars should host a dusty/molecular ``torus". Therefore, although the torus covering factor is expected to vary between the evolving populations, unless it is total then at least some unobscured quasars will reside in the earlier more-obscured phase (but aligned so we see the accretion-disk emission) and some obscured quasars will reside in the later more-unobscured phase. However, if different quasar populations predominantly reside in different evolutionary phases then we will see key differences in the average properties and distributions between the populations.} This can be challenging since many quasar populations are rare and faint, particularly when obscured and/or distant. This has often led to research focused on small often heterogenous quasar populations, selected using an approach that differs to that adopted for the control sample, which can lead to apparent conflicting results in the literature on differences (or not) between quasar populations.

\begin{table*}
    \centering
\begin{tabular}{lll} 
 \hline
Extreme source population & Brief selection description & Definition reference \\ 
 \hline
Extremely Red Quasar (ERQ) & Extreme mid-IR-optical slope selection: typically selects & \cite{Ross2015};\\
 & dust-reddened quasars & \cite{Hamann2017}\\
Hot Dust Obscured Galaxy (Hot DOG) & Extreme mid-IR-optical slope and extreme near-IR & \cite{Eisenhardt2012}\\
 & colour selection: typically selects heavily obscured quasars & \\
Little Red Dot (LRD) & Red optical-near-IR colours and compact near-IR morphology: & \cite{2023arXiv230605448M}\\
 & mixture of object types including broad-line systems & \\
 \hline
\end{tabular}
\caption{Selected extreme source populations identified over the last decade, along with brief details including a key reference that defines the population. 
}
\label{Table:extreme_populations}
\end{table*}

Over the last decade a growing number of studies, adopting a range of approaches but utilising well-defined control samples, have found convincing evidence for obscured AGN and quasars residing in statistically different host-galaxy environments to unobscured systems. Broadly, a larger fraction of AGN appear to be obscured in galaxy mergers than found for AGN in isolated galaxies \citep[e.g.,][]{Kocevski2015,DelMoro2016,Donley2018,Goulding2018,Ellison2019}; see also \S\ref{sec4:mergers}. A trend with increasing obscuration has also been found as a function of merger phase and companion galaxy separation. Using sensitive {\it NuSTAR} high-energy observations of nearby luminous and ultra-luminous infrared galaxies, \cite{Ricci2017_mergers,Ricci2021} 
found a significant increase in the fraction of obscured AGN in late merger phases (when the galaxies are close to coalesence; i.e., $<$~10~kpc), when compared to early merger phases.\footnote{While the major merger of gas-rich galaxies is not a requirement to driving significant gas inflow, it does provide an indicator (a ``time stamp") on what point a given galaxy is within the merger sequence (i.e.,\ close separation galaxies are likely later in the sequence than far separation galaxies).} Adopting high-resolution optical/near-IR imaging and the high-energy near obscuration unbiased {\it Swift}-BAT sample, \cite{koss_mergers_2018} showed that the obscured AGN fraction increased yet further for the closest scales of $<3$~kpc. Qualitatively similar results are also found for mid-IR selected AGN samples, again a near obscuration-unbiased selection approach, where the obscured AGN fraction has been found to systematically increase as a function of galaxy separation distance down to host-galaxy scales of at least $\approx$~10~kpc \citep[e.g.,][]{Satyapal2017,Weston2017,Barrows2023,Dougherty2024}. 

Numerical simulations predict that a large fraction of the obscuration towards luminous AGN and quasars is due to intense on-going dust-obscured star formation, which can be enhanced in gas-rich galaxy mergers \citep[e.g.,][]{Hopkins2008,Blecha2018}. Observationally, it has been shown that the host galaxy can contribute to a substantial fraction of the obscuration towards the AGN \cite[e.g.,][]{Buchner2017,Circosta2019,Gilli2022,Alonso2024}. Focusing on mid-IR selected quasars, \cite{Chen2015} and \cite{Andonie2022,Andonie2024} showed that the fraction of obscured quasars increases with SFR such that the majority of quasars are obscured at the highest SFRs ($>300$~$M_{\odot}$~yr$^{-1}$). Using high-resolution ALMA observations, \cite{Andonie2024} found that the star formation at the highest SFRs is so compact ($\approx$~0.5--3~kpc radius) that dust within the host galaxy can obscure the quasar even up to Compton-thick column densities ($N_{\rm H}>1.5\times10^{24}$~cm$^{-2}$). There is also evidence that the dust-covering factor in the nuclear regions is larger in the most heavily obscured systems, potentially due to chaotic gas inflows and/or radiation-pressure driven dusty outflows \citep[e.g.,][]{Ramos2011,Yamada2021,Yamada2024}, and some obscured quasars may be completely buried so that even their AGN-powered narrow-emission lines are weak or undetected \citep[e.g.,][]{Greenwell2022,Greenwell2024_OQQ}. At the extreme end of the obscured quasar population are Hot Dust Obscured Galaxies (hot DOGs; see Table~\ref{Table:extreme_populations}), which are amongst the most luminous AGN known and are heavily obscured with large dust covering fractors \citep[e.g.,][]{Wu2012,Stern2014,Assef2015,Fan2016_CF}. They appear to be growing close to their maximum rates but are not exclusively merger driven \citep[][]{Fan2016,Wu2018,Diaz_Santos2021}.

Red quasars, systems where the accretion-disk emission is seen in the optical waveband but is at least partially obscured by dust (up-to $A_{\rm V}\approx$~5 mags), are often the lynch-pin in quasar-evolutionary scenarios. They are expected to represent a key transition population caught between the fully obscured and unobscured quasar phases where the quasar is still actively removing the gas and dust along the line of sight through quasar-driven outflows (i.e.,\ winds and jets). Red quasars were first identified by selecting radio-detected quasars several decades ago \citep{Webster1995} and appear to represent a substantial fraction ($>$~10--20\%) of the overall quasar population \citep[e.g.,][]{Banerji2012,Glikman2012}. The colurs of red quasars makes them difficult to distinguish from other ``contaminant" populations in the optical waveband, limiting the reliable selection of red quasars from most large optical surveys such as the SDSS to systems with modest amounts of dust obscuration \citep[$A_{\rm V}<1$~mag;][]{Richards2003,Fawcett2022}, with the exception of Extremely Red Quasars (ERQs), selected using optical--mid-IR colours but identified in modest numbers in the SDSS survey \citep[e.g.,][]{Ross2015,Hamann2017}; see Table~\ref{Table:extreme_populations}. Consequently, many red-quasar samples are constructed using specific selection criteria sensitive to larger levels of dust obscuration, often requiring extensive follow-up observational campaigns even to characterise relatively small samples and often lacking consistently selected control samples. 

The majority of red quasar studies over the last decade have shown that they have the properties expected for a transition population \citep[prevalance for galaxy mergers, accreting at high Eddington ratios, and driving powerful outflows; e.g.,][]{Urrutia2012,Glikman2012,Glikman2015,Banerji2015,DiPompeo2018,Calistro2021,Stacey2022,Kim2024}. ERQs appear to represent the extreme end of the red quasar distribution with the largest luminosities, highest-velocity outflows, and absorbing column densities \cite[e.g.,][]{Zakamska2016,Perrotta2019,Ma2024,Vayner2024}. However, depending on the selection of the ``control" sample, normal blue quasars can also be identified with similar properties to those of red quasars \citep[e.g.,][]{Zakamska2019,Villar2020}. 

Recent research exploiting the excellent source statistics of the SDSS quasar survey and defining careful control samples (but confirmed by other quasar-selection approaches) made the surprising discovery that red quasars emit excess but comparatively weak radio emission over that found for blue quasars
\citep[e.g.,][]{Klindt2019,Fawcett2020,Fawcett2022,Calistro2021,Rosario2021,Glikman2022,Yue2024}. The radio emission is predominantly produced on scales smaller than the host galaxy ($<$~2--10~kpc) with no significant differences between red and blue quasars with either extended radio emission or radio-loud emission. As mentioned in \S\ref{sec:QSOaccretion}, this excess radio emission is connected to the dust extinction towards the quasar, suggesting it may be caused by shocks due to interactions between low power jets/quasar-driven winds with the gas/dust in the circumnuclear/ISM environment \citep[e.g.,][]{Hwang2018,Calistro2021,Calistro2023,Fawcett2023,Petley2024}. This therefore appears to be AGN feedback ``in action" where gas/dust is being removed along the line of sight, as expected for a ``transition" population. There are surprisingly few other differences in the overall quasar SEDs other than tentative excess mid-IR emission \citep[potentially ``dust-entrained" outflows; ][]{Calistro2021} suggesting this ``blow-out" phase is brief.



The large-scale environment can also help determine the relationship between different quasar populations (see \S\ref{sec4:clustering} for an overview). Based on the two-point correlation function, \cite{Petter2022} found no significant clustering differences for SDSS-selected quasars at $0.8<z<2.2$ as a function of optical colour, with the reddest and bluest quasar populations found to have characteristic dark-matter halo masses of $M_{\rm halo}\approx3\times10^{12}$~$h^{-1}$~$M_{\odot}$; this result was confirmed by CMB lensing. Interpreted within an evolutionary sequence, these results indeed suggest any timescale differences between red and blue quasar phases must be relatively short. By contrast, as already highlighted in \S\ref{sec4:clustering}, most two-point clustering analyses of obscured and unobscured quasars find that obscured quasars reside in more massive dark-matter halos  \citep[$M_{\rm halo}\approx10^{13}$~$h^{-1}$~$M_{\odot}$;][]{Donoso2014,DiPompeo2016,Powell2018,Petter2023}, a result again backed up by CMB lensing \citep[e.g.,][]{DiPompeo2016,Petter2023}. These results cannot be  explained by the standard orientation model and argue for a more complex evolutionary connection between obscured and unobscured quasars where the obscuration often occurs on larger scales than the dusty ``torus" (see \S\ref{sec4:clustering}), in broad agreement with recent results \citep[][]{Andonie2022,Andonie2024}. However, large-scale optical spectroscopic confirmation is required to improve on the current photometric redshifts for the majority of the obscured quasars and to rule out a significant population of non-quasars ``contaminating" the clustering signal (see Footnote~\ref{foot:MOSobscured}).

\subsection{The high-redshift frontier of quasars}\label{sec:QSOhighz}
\label{sec:highzqso}

High-redshift ($z>4$) quasars trace the early growth phase of SMBHs and provide key constraints on the formation of BH ``seeds". However, their identification is hindered by two main factors: their rarity, with a number density of $\rm Gpc^{-3}$, and at the very highest redshifts, the significant absorption of their radiation in the observed optical waveband, by material in the foreground intergalactic medium \citep[][for a review]{Fan2023}. Identifying these objects has thus typically required sensitive surveys covering large volumes with wide-field imaging, such as the SDSS \citep{2001AJ....122.2833F,2003AJ....125.1649F}, the Canada-France-Hawaii Telescope Legacy Survey (CFHTLS) \citep{2007AJ....134.2435W,Willott2010}, and the Panoramic Survey Telescope and Rapid Response System 1 \citep[Pan-STARRS 1;][]{2016ApJS..227...11B}.

\begin{figure*}
	\centering 
 	\includegraphics[width=0.8\textwidth, angle=0]{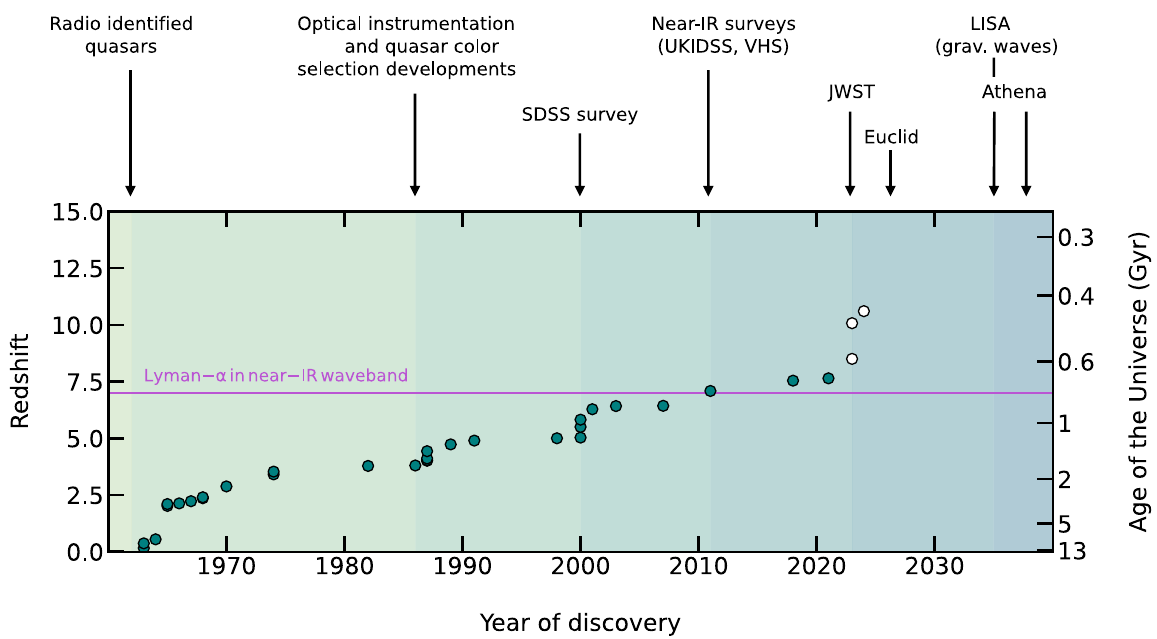}
	\caption{Record-breaking quasar redshifts as a function of their year of discovery, starting from the first ``distant" quasar at $z=$~0.158 \citep{Schmidt1963} to the highest-redshift non-{\it JWST}-identified quasar at $z=$~7.642 \citep{2021ApJ...907L...1W} and the highest-redshift candidate AGN identified by {\it JWST} at $z=$~10.603 \citep{2024Natur.627...59M}; the record-breaking redshifts for the {\it JWST} systems are indicated with an open symbol \citep{2023ApJ...957L...7K,Goulding2023,2024Natur.627...59M,2024NatAs...8..126B}. Key developments that have allowed for the identification of higher-redshift quasars are indicated on the top axis and the redshift where Ly~$\alpha$ enters the near-IR waveband is plotted as a horizontal magenta line. Quasars at $z\geqslant 5$ began to be uncovered about 20 years ago and are now being characterized by the latest telescopes: {\it JWST}, {\it Euclid}, and soon {\it Roman}. Looking ahead, the gravitational wave observatory {\it LISA} will enable the discovery of merging BHs at any redshift, complementing other new telescopes like {\it Athena} and {\it AXIS} that will detect the AGN X-ray emission.} 
	\label{fig_highz_QSO_discovery}%
\end{figure*}

Fig.~\ref{fig_highz_QSO_discovery} charts the history of record-breaking quasar redshifts from the first identified quasar \citep{Schmidt1963} to the present day \citep[cf.][for record-breaking galaxy redshifts]{Stark_2025}. The $z\approx$~5 barrier was broken in 1999 \citep[][]{Fan1999} and the first $z>7$ quasar was identified in 2011 \citep[][]{2011Natur.474..616M}, requiring near-IR photometry; the highest-redshift quasar lies at $z=7.642$ \citep{2021ApJ...907L...1W}. {\it JWST} has pushed the identification of potential AGN to $z\approx$~10 with sensitive near--IR spectroscopy \citep{Goulding2023,2024NatAs...8..126B,2024Natur.627...59M,Napolitano2024}. See Fig.~\ref{fig_highz_QSO_spectra} for spectra of the three $z\approx$~10 systems, all of which have relatively narrow emission lines (FWHM$<$1000~km~s$^{-1}$). The highest-redshift {\it JWST} AGN with clear broad Balmer emission lines (FWHM$>$1000~km~s$^{-1}$) lie at $z=$~8.500 \citep[][]{2023ApJ...957L...7K}, $z=$~8.632 \citep{Tripodi2024}, $z=8.679$ \citep[][]{2023arXiv230308918L}, and $z=9.288$ \citep{Taylor_2025_z9p3_LRD}.

\subsubsection{Increasing the demographics of high-redshift AGN}
\label{sec:highz_demographics}

Fig.~\ref{fig_QSO_parameter_space} provides a comprehensive census of the bolometric luminosities and SMBH masses of high-redshift quasars. To date $>4000$ and $>200$ quasars have been identified at $z>4$ and $z>6$ \citep[][]{Perger2017,Fan2023}, respectively.\footnote{We note that the apparent paucity of quasars at $z\approx$~5.3--5.7 is due to the similarity of their observed optical colours with those of late-type stars, making their selection and identification particularly challenging \citep{Yang2017}.} Substantially fewer high-redshift quasars have SMBH masses due to the need for near-IR spectroscopy to obtain reliable virial SMBH mass tracer emission lines (e.g.\ Mg~II; see \S\ref{sec:BHmass}). The majority of the $z>4$ quasars are extremely luminous ($L_{\rm bol}\geqslant 10^{46}\,\rm erg/s$) and host massive black holes ($M_{\rm BH}\approx10^9$~$M_{\odot}$). We refer to these systems as {\bf bright quasars}: they represent the tip of the active SMBH population, although in terms of their multi-wavelength and spectral properties they are similar to those of luminous quasars at lower redshifts; see \S\ref{sec:QSOaccretion}. A major push over the last decade has been to identify and characterise fainter high-redshift quasars to gain a less biased view on the early assembly of more typical SMBHs, their environments, and co-evolution with their hosts.

At the time of writing, about a hundred faint high-redshift (non-{\it JWST}) quasars with bolometric luminosities ranging from $L_{\rm bol}=10^{45}$ to $10^{46}\,\rm erg/s$ 
have been identified at $z=$~6--7 in dedicated deep quasar surveys, such as CFHQS \citep[Canada-France High-redshift Quasar Survey;][]{2007AJ....134.2435W,Willott2010} and SHELLQs 
\citep[Subaru High-redshift Exploration of Low-Luminosity Quasars;][]{2016ApJ...828...26M,2018ApJS..237....5M}. 
We refer to these systems as {\bf faint quasars}: the identification of these faint quasars has enabled the characterization of the quasar luminosity function at $z=5$ \citep[e.g.,][]{McGreer2018,2020ApJ...904...89N} and $z=6$ \citep{Matsuoka2018}, down to rest-frame UV $M_{1450}\sim -22.3$. These faint high-redshift quasars are powered by SMBHs with a wide range of masses ($M_{\rm BH}\sim 10^{7.5}-10^{9}\, \rm M_{\odot}$) and Eddington ratios 
\citep[$f_{\rm Edd}=0.1-1$;][]{Onoue2019,Takahashi2024}; in broad terms they appear to be a mix of the lower Eddington-ratio tail of the more luminous $z\approx$~6 quasar population and lower mass, higher-Eddington ratio quasars. Comparisons to the extrapolation from observed X-ray luminosity functions, primarily constrained up to $z\sim 4$ \citep[see e.g.,][for new constraints at $z>4$]{2024A&A...685A..97P}, indicate that optical--near-IR photometric selection approaches may overlook a large fraction of the quasar population at $z\geqslant 6$ due to obscuration \citep{2018MNRAS.473.2378V}.

{\it JWST} has now pushed the identification of high-redshift AGN populations down to bolometric luminosities of $\sim 10^{43}-10^{44}\, \rm erg/s$. We refer to these systems as {\bf JWST AGN}: their spectral and multi-wavelength properties differ from those of comparable-redshift quasars, making them less-reliable candidate AGN; however, the presence of an AGN appears the most probable solution for the majority. Within less than 2 years, already $>$~100 UV-faint {\it JWST} AGN at $z\geqslant 4-8.5$ have been identified with weak broad ($\approx$~1000--4000~km~s$^{-1}$) Balmer emission lines \citep[e.g.,][]{2023arXiv230905714G,2023arXiv230200012K,2023arXiv230801230M,Ubler2023b,2023arXiv230605448M,Taylor_2024} and a similar number of narrow-line systems \citep[][]{scholtz_jades_2023,Mazzolari2024_NLAGN,Treiber2024}, due to these systems being common even in small ($\ll$~deg$^{2}$) {\it JWST} fields; see Fig.~\ref{fig_QSO_parameter_space}.
Below we summarise the basic properties of the key detected {\it JWST} AGN populations.

\vspace{0.1cm}
\noindent {\bf Broad-line JWST AGN:} $\approx$~5--10\% of the systems spectroscopically identified in high-redshift galaxy searches have broad Balmer emission lines \citep[][]{2023arXiv230311946H,2023arXiv230801230M,Treiber2024}. Spectroscopically these systems are not like normal quasars: they are metal poor and lack strong high-ionisation emission lines (such as C~IV$\lambda$1549), properties more similar to those of AGN identified in local metal-weak dwarf galaxies \citep[][]{Trefoloni2024}; see also \S\ref{sec4:dwarfs}. Assuming the broad emission lines are due to virialised gas around an accreting SMBH (see \S\ref{sec:BHmass}), they have masses of $M_{\rm BH}\approx10^6$--$10^8$~$M_{\odot}$ and Eddington ratios of $\approx$~0.1--1; we note the absence of broadened forbidden lines makes a sole outflow origin for the broad lines unlikely. A narrow, typically blueshifted, absorption feature is seen in the Balmer lines for $\approx$~10--20\% of the systems, interpreted as a high column density of gas and potentially related to an outflow and/or the outer region of the BLR \citep{2024arXiv240907805I,Juodzbalis2024,Kocevski2024_LRD_sample,Lin2024_JWST,2023arXiv230605448M, Ji_2025}; note, $<0.1$\% of low-redshift broad-line AGN show this feature \citep[][]{Lin2024_JWST}.

Very few systems are X-ray detected, either individually or from deep stacking analyses, indicating they are systematically X-ray weak by at least 1 order of magnitude \citep[][]{Kocevski2024_LRD_sample,Lambrides2024,2024arXiv240500504M}; none are radio detected but this is consistent with them being radio quiet \citep[][]{Mazzolari2024_radio}. The origin of the X-ray weakness is unclear: it may be due to heavy Compton-thick absorption (although given the high rest-frame energies probed at $z>4$, $>$~2.5--40~keV, the absorption would need to be extreme) or the systems may be intrinsically X-ray weak. The most-promising (and explored) option for the intrinsically X-ray weak scenario is super-Eddington accretion which is expected to be faint due to an intrinsically ``soft" X-ray SED (like NLS1s) and absorption from the slim accretion disc \citep{2024ApJ...976L..24M,2024arXiv241203653I,Pacucci2024,King2025}. The lack of high-ionisation emission lines is consistent with a super-Eddington scenario \citep{Lambrides2024} and would make these {\it JWST} systems similar to X-ray weak WLQs; see \S\ref{sec:QSOaccretion}. An implication of the super-Eddington accretion scenario is the luminosity and/or the SMBH masses are overestimated by a factor of at least a few \citep[e.g.,][]{2024arXiv240500504M,2024ApJ...976L..24M}. About 10--30\% of the broad-line systems are red and have compact near-IR morphologies, with the fraction increasing strongly with redshift \citep[][]{Hainline2024,Inayoshi_2025,Ma_2025_LRDs}: they are referred to as ``little red dots".
 
\vspace{0.1cm}
\noindent {\bf Little Red Dots (LRDs):} significant attention has been focused on understanding the nature of LRDs; see Table~\ref{Table:extreme_populations}. LRDs are selected to be compact in NIRCam imaging ($<$~100--200~pc) and typically have steep red rest-frame optical colours with blue rest-frame UV colours, giving them an unusual ``V" shaped UV--optical SED \citep{2023arXiv230607320L}.\footnote{LRDs were first identified in high-redshift galaxy searches and, prior to spectroscopic confirmation, were interpreted as a massive ($\approx10^{10}$--$10^{11}$~$\rm M_\odot$) and surprisingly ubiquitous ``quiescent" galaxy population at $z\approx$~6 \citep{Labbe2023Nature}, the so-called ``universe breaker" galaxies}.\label{foot:universe_breaker} Depending on the exact selection, $\approx$~20--80\% host broad Balmer emission lines and have rest-frame UV--optical spectral, X-ray, and radio properties similar to the {\it JWST} broad-line systems \citep[][]{2023arXiv230905714G,2024arXiv240610341A,2024arXiv240419010A,Hainline2024,Kocevski2024_LRD_sample,2023arXiv230605448M,Perger2024,Yue_2024_LRDs}. If interpreted as AGN then they appear to be dust obscured (up-to $A_{\rm V}\approx$~5~mag) and the red optical colour is predominantly due to the reddened AGN while the blue UV colour is produced by scattered AGN emission or a young stellar population \citep[][]{Barro2023,2023arXiv230607320L,2023arXiv230905714G,2023arXiv230200012K,Killi2024}; $\approx$~30\% of LRDs have extended rest-frame UV emission, as expected from host-galaxy emission, which is often asymmetric potentially due to galaxy interactions \citep[][]{Kocevski2024_LRD_sample,Rinaldi_2024}. However, LRDs are weak at rest-frame near-IR wavelengths, suggesting a lack of obscuring dust on nuclear scales and therefore inconsistent with a standard obscured AGN scenario \cite[][]{2024arXiv240610341A,Li_2025_LRDs,Perez_Gonzalez2024,2024ApJ...968...34W,Casey2024,Li_2025_LRDs}. Few LRDs are variable in the rest-frame UV--optical casting some doubt on an AGN hypothesis \cite[][]{Kokubo2024,Tee2024,Zhang2024_LRD}. However, spectral monitoring of a $z\approx$~7 multiply lensed LRD over a uniquely long timescale (observed-frame 22 years; rest-frame 2.7 years) identified broad Balmer emission line variations consistent with those expected for a damped random walk model of an accretion disk around a $\approx4\times10^{7}$~$\rm M_{\odot}$ SMBH \citep{Furak_2025}. 

Many LRDs have a spectral break around the wavelength of the Balmer break \citep[][]{Labbe2024_balmerbreak,Ma2024,Wang2024_Balmerbreak}. Analysing a large sample of LRD {\it JWST} spectra, \citet[][]{Setton2024} found the inflexion in the ``V" shaped SED corresponds to the Balmer break in $\approx$~50\% of sources; see Fig.~\ref{fig_highz_QSO_spectra} for the break in the composite LRD spectrum. They argued this break is unlikely to occur for dust-reddened AGN with a range of dust extinctions and host-galaxy strengths and, adopting ``Occam's razor", proposed the properties of LRDs are from a single origin, potentially due to late-type stars. However, a growing number of studies have argued that an AGN buried in an extremely dense cocoon of gas (what is sometimes referred to as a ``black-hole star" scenario) would also produce such feature and could simultaneously explain the Balmer absorption lines seen in the spectra of some {\it JWST} AGN \citep[e.g.,][]{2024arXiv240907805I,DeGraaff_2025,Ji_2025,Naidu_2025}. A potential consequence of the large gas density is scattering of the broad emission lines, producing an extended velocity component and artificially increasing SMBH masses by $\approx$~1--2 orders of magnitude, when calculated using the virial SMBH estimator \citep{Rusakov_2025}, although this scenario has been disputed for at least some {\it JWST} AGN \citep{Juodvzbalis_2025}. 
Overall, despite their unusual properties, it therefore appears likely that the majority of LRDs host AGN, potentially representing an early SMBH growth phase which would help explain their strong redshift dependence \citep[][]{Inayoshi_2025} and overmassive SMBHs compared to their hosts; however, as for any broad-band photometric selection, it is quite likely there is some heterogeneity in the LRD population \citep[see, for example][ for some alternative non-SMBH and non-AGN scenarios]{Setton2024,Zwick_2025}.

\vspace{0.1cm}
\noindent {\bf Narrow-line JWST AGN:} an abundant narrow-line AGN population is expected at high redshift with similar properties to the {\it JWST}-identified broad-line systems but lacking the broad Balmer emission lines. Their metal-poor nature means that typical AGN emission-line diagnostics \citep[][]{Baldwin1977,VO1987} are not reliable: indeed, the broad-line systems reside in the same region as low-mass star-forming galaxies. On the basis of new emission-line diagnostics developed to distinguish between high and low ionising continua in high-redshift low-mass galaxies, $\approx$~20\% of {\it JWST}-identified galaxies at $z=$~2--9 are found to have spectral properties suggesting the presence of AGN activity \citep[][]{scholtz_jades_2023,Mazzolari_2024_OIII,Mazzolari2024_NLAGN,Treiber2024}. Their estimated bolometric luminosities ($\approx10^{42}$--$10^{45}$~erg~s$^{-1}$) overlap with those of the broad-line systems but extend down to very low-luminosity systems. Few are X-ray detected, suggesting they are as X-ray weak as the broad-line systems \citep[][]{2024arXiv240500504M,Mazzolari2024_NLAGN}. The AGN signatures are clear in the NIRSpec spectra of some sources but they are weak overall (e.g.,\ see Fig.~1 of \citet{Mazzolari2024_NLAGN}) and further confirmation of AGN in the weaker sources would be valuable.

\vspace{0.1cm}
\noindent {\bf Very high-redshift JWST AGN:} to date, three $z>10$ systems have been proposed as very-high redshift AGN: UHZ-1 at $z=$~10.073 \citep[][]{Goulding2023,Brogan2023}, GHZ9 at $z=$~10.145
\citep[][]{Kovacs2024,Napolitano2024}, and GNz-11 at $z=$~10.603
\citep[][]{2024Natur.627...59M}; see Fig.~\ref{fig_highz_QSO_spectra}. The optical spectrum of UHZ-1 is consistent with co-eval star-forming galaxies with no evidence for an AGN but it is weakly X-ray detected at 2--7~keV, suggesting an optically obscured AGN. GHZ9 has high-ionisation lines, including C~IV, consistent with an AGN and it is X-ray detected at 0.5--3~keV: the derived X-ray luminosity is also consistent with the 440~nm continuum luminosity, giving strong evidence for an AGN. GNz-11 is not X-ray detected but has several high-excitation and high-critical density lines (notably Nitrogen), including modestly broadened emission lines (FWHM~$\approx$~450~km~s$^{-1}$), giving support for the presence of an AGN. Recent {\it JWST} MIRI spectroscopy of GNz-11 detected only narrow H$\alpha$ emission without a clear broad component, casting some doubt on the broad-line AGN classification of GNz-11 \citep[][]{Alvarez2024}.

\vspace{0.1cm}
The rapid progress in identifying and studying the {\it JWST} AGN has come down to the ultra-sensitive near-IR--mid-IR spectroscopic and high-resolution imaging capabilities of the {\it JWST} facility. More work is required to understand why they differ from normal co-eval quasars and to rule out non-AGN solutions. However, at the time of writing, the presence of an AGN seems plausible for the majority. Although such systems 
were theoretically expected at high redshift from BH seed formation and growth models, their number density came as a surprise.
Specifically, the number density of the {\it JWST} AGN is about $10^{-5}\, \rm cMpc^{-3}$ in the UV magnitude range $\rm M_{UV}=-21,-17$ at $z\sim 5$; the exact value depends on the magnitude range, selection, and volume covered by the observations \citep{2023arXiv230605448M,2023arXiv230905714G}. The first constraints on the bolometric luminosity function indicate that the {\it JWST} broad-line AGN are as numerous, or even more numerous, than the AGN predicted by large-scale cosmological simulations when considering all simulated AGN are visible and none obscured \citep{2023arXiv230605448M,2024arXiv240505319H}. 

\begin{figure}[t]
	\centering 
    \includegraphics[scale=0.55]{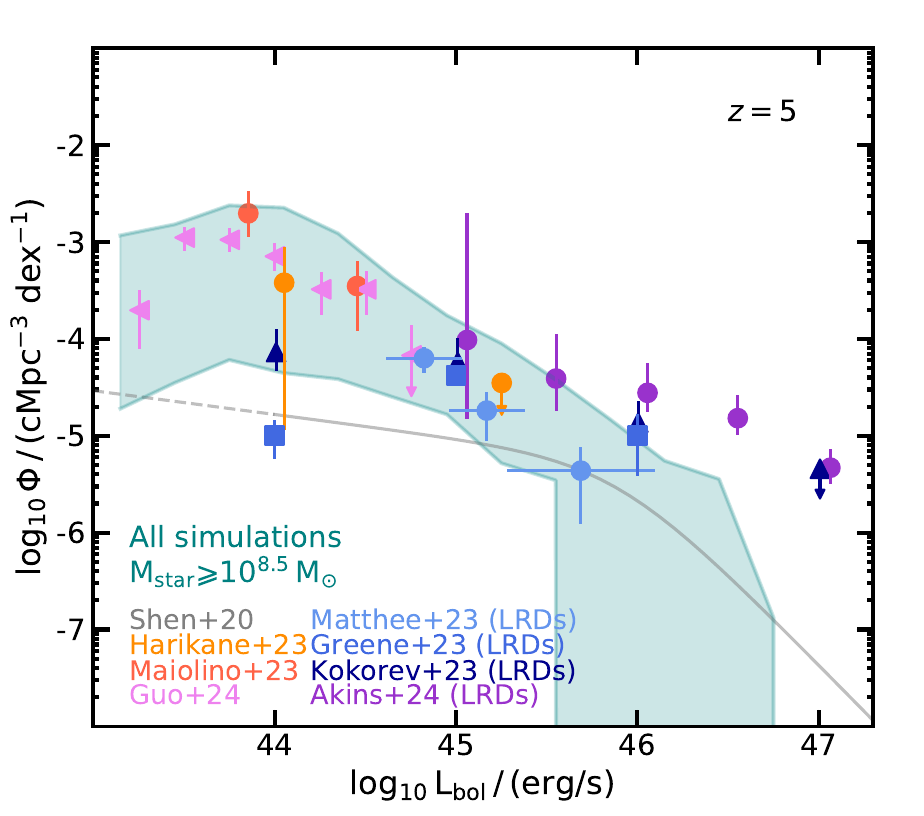}
	\caption{AGN bolometric luminosity function at $z\sim5$. The shaded green region shows the combination of the luminosity functions, including Poisson uncertainties, from the {\sc Horizon-AGN, Illustris (Tng~50, Tng~100, Tng~300), Eagle, Simba}, and {\sc Astrid} large-scale cosmological simulations. Only AGN in galaxies with a total stellar mass of $M_{\star}\geqslant 10^{8.5}\, \rm M_{\odot}$ are considered. Symbols represent observational constraints from {\it JWST} for a sample of AGN candidates at $z\sim 5$ with a redshift range of $z\approx$~4--6.5  \citep{2023arXiv230605448M,2023arXiv230905714G,2023arXiv230801230M,2023arXiv230311946H,2024arXiv240610341A,2024arXiv240919205G}. The grey curve is the pre-{\it JWST} luminosity function at $z\sim5$ \citep{2020MNRAS.495.3252S}. The predicted luminosity function is lower than the observational constraints in some simulations, which in some cases only includes broad-line AGN (or a subset), highlighting the tension between theoretical models and {\it JWST} observations. {\it Source:} adapted from Fig.~4 of \cite{2024arXiv240505319H}.}
	\label{fig:AGN_highz_LF}
\end{figure}

We illustrate this in Fig.~\ref{fig:AGN_highz_LF} with the compilation of bolometric luminosity functions derived from different observational samples (explicitly selecting LRDs or broader AGN candidate samples) and the range probe by the AGN population of eight large-scale cosmological simulations \citep[{\sc Horizon-AGN, Illustris, Eagle, TNGs, Simba, Astrid};][]{2014MNRAS.444.1453D,2014MNRAS.445..175G,Schaye:2015,2018MNRAS.473.4077P,Nelson:19,Dave:2019,2022MNRAS.513..670N}. 
Including both broad-line AGN from various {\it JWST} surveys and narrow-line AGN identified in \citet{scholtz_jades_2023} and the best fit of \cite{2024ApJ...974...84G} yields the same conclusions when compared to semi-analytical models. The comparison between theoretical models (i.e. cosmological simulations and semi-analytical models) and observational constraints indicates that we either need to refine our theoretical models of BH physics to facilitate SMBH growth
or to understand why some AGN candidates may not actually be AGN or why their derived properties (SMBH mass and AGN luminosity) may be overestimated, for example.
Theoretically, this includes processes such as BH formation, gas accretion, super-Eddington regime, SMBH dynamics and coalescence. Additionally, sustaining such efficient SMBH growth is likely closely linked to galaxy physics, such as feedback processes, morphology of the host, the gas reservoir and the rapid replenishment of cold gas.

\subsubsection{Galactic and large-scale environments of high-redshift quasars and {\it JWST} AGN}
\label{sec:environment}
Over the last decade since AH12, following the discoveries of high-redshift quasars (Fig.~\ref{fig_highz_QSO_discovery}), the community has begun investigating their environments to understand their fuelling mechanisms and how they could have assembled so quickly. 
Significant progress has already been made in several key aspects, thanks to advances from new instruments such as ALMA and VLT-MUSE, as well as developments in theoretical studies. This includes the creation of dedicated semi-analytical models following the merger trees of the progenitors of quasars \citep{2014MNRAS.444.2442V} and cosmological simulations of quasar environments \citep{Costa:14a}, as described in \S\ref{sec:highz_progress}  
and in the \citet{2017PASA...34...31V} review, for example.

Quasars are thought to inhabit highly biased regions of the Universe, where SMBH growth is fuelled by an increased number of mergers and the convergence of filaments supplying fresh gas. In the $\Lambda$CDM cosmology, this hypothesis would be supported by an observed increase in galaxy counts within the quasar's field of view. 
However, the observations of Lyman break galaxies (LBGs, probing a redshift uncertainty of $\Delta z\sim1$) and more recently searches for lower-mass Lyman-alpha emitters (LAEs, $\Delta z\sim 0.1$) revealed a diversity of environments from overdense regions \citep[e.g.,][]{2005ApJ...622L...1S,2005ApJ...626..657W,2006ApJ...640..574Z,2009ApJ...695..809K,2013MNRAS.432.2869H,2014A&A...568A...1M,2014AJ....148...73M,2022ApJ...927..141M} to environments consistent with blank fields
\citep[e.g.,][]{2013ApJ...773..178B,2014MNRAS.442.3454S,2017ApJ...834...83M}. Atypical fields, such as the \citet{2010ApJ...721.1680U} discovery of an overdense ring-like stucture at 3~Mpc from the quasar but no overdensity closer to the quasar, or the low density of LAEs within 3~Mpc (projected distance) found 
in the large-scale and LBG-overdense structure studied in \citet{2018ApJ...856..109O}, could suggest negative feedback on the formation and regulation of galaxies at small physical scales (with potentially a larger effect on low-mass galaxies). Deep-and-wide observations are needed, such as from Subaru/Suprime-Cam or  LBT/LBC, as used in \citet[][with additional observation from MUSE for LAEs]{2020A&A...642L...1M} for the spectroscopic confirmation of the overdensity presented with imaging in \citet{2017A&A...606A..23B}.
[C~II]$\lambda$158~$\mu$m observations on smaller scales with ALMA revealed that a significant fraction of quasars with $M_{\rm BH}\geqslant 3\times 10^{8}\, \rm M_{\odot}$ ($\approx$~16\% of a sample of 25 quasars at $z>5.9$) have a close companion within 100 kpc \citep[see also][]{2017ApJ...848...78F,2017ApJ...836....8T}. ALMA allows for both the discovery of dusty companion galaxies that cannot be selected at any other wavelength, but also to determine accurate [C~II] redshifts.

More recently, large observational campaigns have been designed to provide statistical constraints on the quasar environment by observing, for example, 25 quasars fields with {\it JWST}-NIRCam WFSS \citep{2023ApJ...951L...4W}. While the long-standing absence of a consensus on the environments of high-redshift quasars was in part attributed to reduced observational sensitivity missing galaxies and small fields of view, the first {\it JWST} results demonstrate that the diversity of environments persist even with new deeper observations, including an overdense filamentary structure of 10 [O~III]$\lambda$5007\AA\ emitters at $z=6.61$ \citep{2023ApJ...951L...4W}, dense regions with close (potentially) merging galaxies with {\it JWST} NIRSpec IFU \citep{2023A&A...678A.191M,2024arXiv241011035M}, and fields consistent with average galaxy density \citep{2024arXiv240307986E}. Whether high-redshift quasars reside in a broad range of dark matter haloes and whether their strong feedback suppresses galaxy formation in their surroundings is still to be determined \citep[][for a theoretical perspective from cosmological simulations]{Costa:14a,2024MNRAS.531..930C,2019MNRAS.489.1206H}.\\

At low redshift ($z\leqslant 1$), quasar host galaxies exhibit diverse morphological properties \citep{2020PASJ...72...83I,2015MNRAS.454.4103B,2014MNRAS.440..476F,1997ApJ...479..642B}. However, the brightness of quasars in the optical/near-IR has made it difficult to determine the host galaxies of luminous high-redshift quasars. Taking advantage of the relatively faint millimetre-wavelength emission from the quasars, ALMA has placed constraints on the dynamical mass of the quasars hosts as well as their morphology. Recent [C~II] observations resolving the hosts of $z\geqslant6$ quasars with $\leqslant 1.2\, \rm kpc$ \citep{2021ApJ...911..141N}, revealed that about a third of high-redshift quasars have disturbed [C~II] emission profiles characteristic of galaxy mergers, a third reside in extended disk galaxies (i.e.,\ with smooth velocity gradients consistent with emission arising from rotating disks), and the last third show no clear velocity gradients and are consistent with dispersion-dominated galaxies with turbulent gas (i.e.,\ compact bulge-like galaxies). The diversity of high-redshift quasar morphologies, similar to those at low redshifts suggests a diverse range of mechanisms for SMBH fuelling, not just galaxy mergers.
Very high spatial resolution ALMA imaging down to 200--400~pc, similar to that achieved for nearby galaxies, has been achieved for a few high-redshift quasar hosts and, indeed, reveals complex velocity structure and nearby companions hinting towards gas funnelled to the galaxy centre by interactions, non-regular structures in which gas flow could be driven by bar motions, and surprisingly compact emission with regular disk emission and a lot of gas within the central 100~pc but with no clear signature of the presence of a massive SMBH on the gas dynamics \citep{2022ApJ...927...21W,2020ApJ...904..130V,2019ApJ...882...10N}.
In the near future, more statistical studies are needed to understand the relative role of the galactic morphologies and kinematics, whether they are intrinsically distinct or instead part of an evolutionary path of the quasars, and whether they are linked to distinct channels of SMBH fuelling.
The new exquisite resolution achieved in observations allow comparison with recent cosmological hydrodynamical zoom-in simulations of quasars hosts following 
their gas and stellar kinematics
with a resolution of 50 pc \citep{2022MNRAS.510.5760L}. The simulations show that the host galaxies experience several mergers and a rapid evolution of the gas and stellar distributions, which can go from a perturbed disk, to a compact system, and later settle down into a more extended system in less than 200 Myr.\\

Over the years through the growth of observational samples of quasars with SMBH mass estimates, 
it has become possible to investigate the masses of the SMBHs powering the high-redshift quasars as a function of the dynamical masses of their hosts. Results have shown that bright quasars tend to be located above the $M_{\rm BH}-M_{\star}$ mass scaling relation derived in the local Universe \citep[e.g.,][assuming $M_{\star}=M_{\rm dyn}$]{Kormendy:13}, while faint quasars 
are consistent with, and sometimes lie below, the local relation \citep{2019PASJ...71..111I}.
The stellar component of the hosts, however, remained a missing piece until the advent of {\it JWST} which prevented a direct $M_{\rm BH}-M_{\star}$ comparison to local systems.
Decomposition of the point-source quasar from its host galaxy requires a precise PSF, a method that was successfully employed with HST but limited to quasars at $z\leqslant 2$. The exciting first detections of the stellar light from two quasar hosts with {\it JWST} were presented in \citet{2023Natur.621...51D}. 
This work was quickly followed by \citet{2023ApJ...953..180S,2024ApJ...966..176Y,2024ApJ...964...90S,2024arXiv241011035M}, mostly for bright quasars. Although still with a limited number of systems and large error bars on both $M_{\rm BH}$ and $M_{\star}$, quasars reside mostly in galaxies within the range $M_{\star}\sim 10^{10}-10^{11.5}\, \rm M_{\odot}$, as shown in Fig.~\ref{fig:mbh_mstar_diagram}.
Initial results are consistent with the ALMA constraints and suggest that faint quasars tend to align with the $M_{\rm BH}-M_{\star}$ relation in the local Universe, while bright quasars have higher $M_{\rm BH}/M_{\star}$ ratios, although not yet significantly different from the most massive SMBHs found in local large elliptical galaxies; see Fig.~\ref{fig:mbh_mstar_diagram}.
These potential overmassive quasars reignite the debate on whether super-Eddington accretion can explain these systems. Although new high-resolution simlations of quasars suggest that prolonged super-Eddington phases can occur in gas-rich, high-redshift environments \citep{2024A&A...686A.256L}, even in the challenging scenario of relatively high spin and associated strong jet feedback, further investigation is required. This is because we are still unable to resolve the physical processes at very small scales in these simulations (e.g., accretion at the Bondi radius, early expansion of feedback-shocked gas).\\

\noindent 

In terms of their galaxy properties, {\it JWST} AGN are found in a wide range of hosts with stellar mass estimates from $\sim 10^{8}$ to $10^{11}\, \rm M_{\odot}$ (the majority $<10^{10}\, \rm M_{\odot}$), for both broad-line and narrow-line systems \citep[e.g.][]{2023arXiv230801230M,2023arXiv230311946H,scholtz_jades_2023,Kocevski2024_LRD_sample}. However, it is important to caveat that contamination from the AGN leads to uncertainties in the stellar-mass estimate: the impact is less severe for the narrow-line candidates, which is dominated by the galaxy but the AGN can ``contaminate" the Balmer, impacting SFR estimates. Indeed, uncertainties on both SMBH and stellar mass estimates can be as much as an order of magnitude \citep[][for discussions on using standard SMBH mass methods calibrated against local relations, selection biases, and $M_{\rm BH}-M_{\star}$ measurement uncertainties]{2024arXiv240300074L,2024MNRAS.531..550K,2024A&A...689A.128L}. For example, SMBH mass measurements assume virialised gas, and could be over-estimated if the SMBH is in a super-Eddington regime, undergoing a TDE phase (see \S\ref{sec:TDE}), or a significant fraction of the broadened emission is scattered \citep{Rusakov_2025}. With these caveats in mind, a number of {\it JWST} AGN are found to have $M_{\rm BH}/M_{\star}$ ratios similar to galaxies in the local Universe, although a non-negligible fraction have larger ratios making their SMBHs {\it overmassive} compared to their hosts, similar to that found for bright quasars but at lower masses; see Fig.~\ref{fig:mbh_mstar_diagram}. However, the large uncertainties on the derived quantities preclude strong conclusions on whether the {\it JWST} AGN host overmassive SMBHs and, consequently, their origins (e.g.,\ massive BH seeds, BH seeds forming prior to the host galaxies, efficient growth, atypical host-galaxy evolution). The uncertain origin of the blue and red components of the SED in LRD systems further complicate the estimation of their $M_{\rm BH}/M_{\star}$ mass ratios: if the continuum emission is primarily due to an AGN with absorption by dense neutral gas clumps on the shorter-wavelength side of the Balmer break then estimates of $M_{\star}$ based on a stellar origin for the continuum would be overestimated \citep{2024arXiv240907805I}.
Finally, some of the outliers in the $M_{\rm BH}-M_{\star}$ diagram (for relatively low stellar masses with $\leqslant 10^{9}\, \rm M_{\odot}$) are found {\it not} to be outliers in $M_{\rm BH}-\sigma$ or $M_{\rm BH}-M_{\rm dyn}$ diagrams with respect to what would be expected from observations in the local Universe \citep{2023arXiv230801230M}. 

The majority of the AGN candidates detected by {\it JWST} were identified serendipitously in extensive spectroscopic surveys, in both average fields and quasar fields (with candidates in the foreground and background of the quasars). Despite the modest source statistics and cosmological volumes probed by the {\it JWST} surveys, it already seems likely that the {\it JWST} AGN reside in lower-mass haloes than the high-redshift quasars. Performing angular and projected cross-correlation analyses of 27 {\it JWST} AGN at $5<z<6$ with 679 photometrically selected galaxies, \cite{Arita2025} estimated a typical dark-matter halo mass for the {\it JWST} AGN of $M_{\rm halo}\approx3\times10^{11}$~$h^{-1}$~$M_{\odot}$, an order of magnitude lower than that of faint and bright high-redshift quasars \citep[][]{Arita2023}; see also \S\ref{sec4:environment} \& \ref{sec:QSOphases}. Two-point cross correlation analyses of LRDs with co-eval H$\alpha$ emitting galaxies over $z\approx$~4--6 suggest that LRDs reside in similarly massive dark-matter halos of $M_{\rm halo}\approx$~(1--2)~$\times10^{11}$~$h^{-1}$~$M_{\odot}$ \citep{Lin_2025}; complementary constraints are also placed by \cite{Pizzati2024} who, on the basis of the source density of LRDs at $z\approx$~4--6 and the relative number of coeval dark-matter haloes, argued that LRDs must reside in haloes of $M_{\rm halo}<5\times10^{11}$~$h^{-1}$~$M_{\odot}$ to avoid a halo occupancy of $>100$\%. These results indicate that the {\it JWST} AGN reside in dark-matter haloes an order of magnitude less massive than quasars (see \S\ref{sec:QSOphases}), suggesting that they represent either an entirely different population or a much-earlier SMBH growth phase; however, we note that at least one $z\approx$~7 LRD is found in a much denser large-scale environment \citep{Schindler_2024_LSS}, suggesting some diversity in LRD environments.

On smaller spatial scales, there is evidence for a number of {\it JWST} dual AGN candidates at $z=4-6$ \citep{2023arXiv230801230M} and $z=3-3.5$ \citep[][]{2023arXiv231003067P,2023A&A...679A..89P}; see \S\ref{sec4:mergers} for dual AGN in regular quasars and AGN. This highlights the potential increased role of SMBH mergers in the growth of SMBHs at high redshift, which is theoretically expected to be much less significant than the dominant gas accretion channel, and the importance of galaxy interactions in triggering phases of enhanced gas accretion.

\subsubsection{Origins of the seeds of massive black holes: models and observational constraints}
\label{sec:seeds}
\begin{figure}
	\centering 
	\includegraphics[width=0.48\textwidth, angle=0]{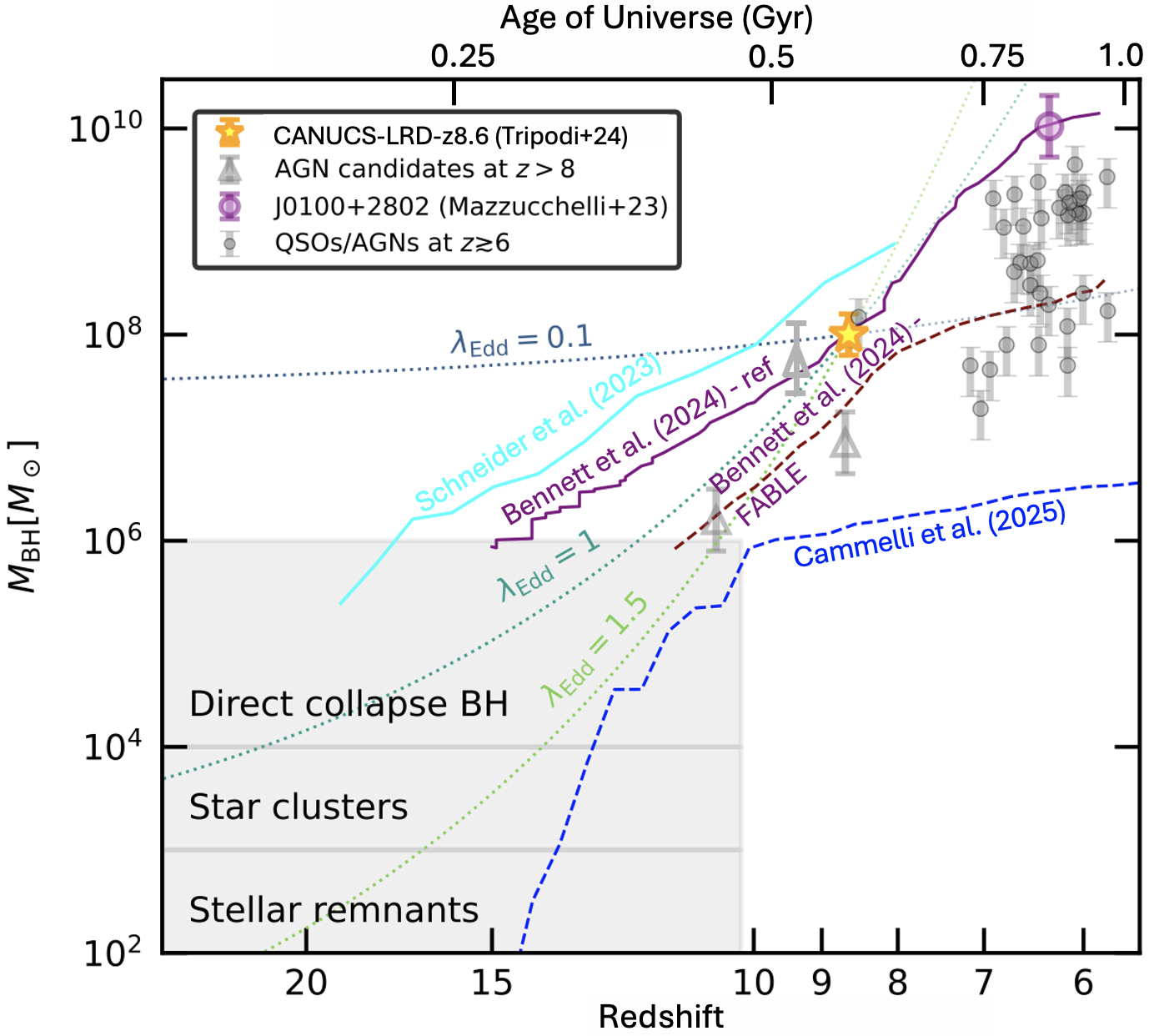}
	\caption{Black-hole mass versus redshift for several $z>6$ quasars (grey circles) and $z>8$ {\it JWST} AGN (grey triangles), focused on the $z=$~8.632 broad-line system from \citet[][]{Tripodi2024} (yellow star) and the extreme $z=$~6.327 quasar from \citet{Mazzucchelli2023} (magenta circle) showing that such massive BHs could be related through Eddington-limited accretion from a direct collapse BH seed. Other potential sub Eddington and super Eddington scenarios are also shown from different seed masses as well as early SMBH growth simulations \citep[][]{Bennett:24} and semi-analytical models \citep[][]{2023arXiv230512504S,Cammelli2025}. {\it Source:} adapted from Fig.~4 of \cite{Tripodi2024} and updated for recent $z>8$ {\it JWST} AGN.} 
	\label{fig:BHgrowth_seeds}%
\end{figure}

Prior to {\it JWST}, observational facilities seeking evidence of the first BHs (i.e., the black hole seeds), could only probe the very tip of the BH distribution at $z\geqslant 5$ with the observations of extremely luminous and rare quasars with a number density of $\sim 1\, \rm Gpc^{-3}$, and powered by BHs of $M_{\rm BH}\geqslant 10^{8}\, \rm M_{\odot}$. The existence of quasars at $z\geqslant 6$  more than a decade ago and the quasars discovered at $z=7.5-7.6$ a few years ago suggest that their progenitors, referred to as BH seeds, 
must have formed with masses ranging from a few hundred solar masses (``light seeds") up-to a million solar masses 
(``heavy seeds") in the very early Universe. Some of the new {\it JWST} discoveries, such as the broad-line AGN at $z>$~8.5 with $M_{\rm BH}\sim 10^{8}\, \rm M_{\odot}$ \citep{2023ApJ...957L...7K,Tripodi2024,Taylor_2025_z9p3_LRD}, push even further the question of BH formation; see Fig.~\ref{fig:BHgrowth_seeds}.

Contemporary theories on BH seed formation can be categorized into two main groups \citep[][for recent reviews]{2020ARA&A..58...27I,2021NatRP...3..732V}. First, seeds may have originated before even the formation of the first galaxies, arising from the collapse of high peak density fluctuations or cosmic string loops \citep[e.g.,][]{2020ARNPS..70..355C}. These ``cosmological'' pathways predict seed formation as early as inflation and exhibit a broad range of initial masses ranging from 1 gram to several $10^{5}\, \rm M_{\odot}$, contingent on the precise time of formation. 
Alternatively, more widely studied pathways involve the formation of BH seeds alongside the first galaxies, possibly within either primeval or more evolved galaxies. The formation occurs through the gravitational collapse of a single stellar object or through the hierarchical merger of stars or stellar black holes. Each avenue for seed formation, including those described further below, carries its own set of advantages and drawbacks to explain the massive SMBHs sitting in the heart of almost all galaxies today. Importantly, these pathways are not theoretically mutually exclusive: the origins of the observed population of SMBHs may be diverse.

In the first theoretical endeavours, BHs were postulated to have formed from the remnants of the gravitational collapse of the first-generation stars, known as Pop~III \citep{2001ApJ...551L..27M,Klessen2023_popIII}. However, only the most massive Pop~III stars would leave behind a seed with the capability to overcome the numerous challenges in evolving into a central SMBH. Furthermore, pair-instability supernovae occurring in massive Pop~III stars with $140-260 \, \rm M_{\odot}$ might not leave any remnants, leaving only stars with a mass of $\geqslant 260, \rm M_{\odot}$ to form ``light seeds'' of a few $\sim 100\, \rm M_{\odot}$ at best. Understanding the potential contribution of this scenario to the overall BH population is further hindered by uncertainties regarding the initial mass function of Pop~III stars, the shallow gravitational potential well of the host mini-halos, which could impede gas retention and, consequently, prevent efficient accretion onto the black holes. Additionally, challenges include the extended sinking time of low-mass BHs toward the galactic centre and the possible recoil of BHs from the central regions. The formation of more massive ``heavy seeds" ($\sim 10^{4-5}\, \rm M_{\odot}$) would ease some of these important hurdles and make the existence of high-redshift quasars $<1$~Gyr since the Big Bang ($z>6$) less complex to elucidate \citep[e.g.,][]{2003ApJ...596...34B,2006MNRAS.370..289B,2006MNRAS.371.1813L}. Under specific conditions, gas cooling and fragmentation can be delayed in primordial atomic cooling halos and instead collapse monolithically into one single protostar. With a sufficiently large accretion rate, a protostar can grow into a Super Massive Star (SMS) and collapse into a ``heavy seed". Conditions to avoid gas fragmentation have been increasingly studied over the last decade: intense photo-dissociating (Lyman-Werner) radiation from a neighboring galaxy \citep[suppressing H$_{2}$ cooling, e.g., ][]{2014MNRAS.445.1056V}, dynamical heating from rapid halo mergers \citep[increasing the heating rate,][]{2019Natur.566...85W}, or large residual baryonic streaming motions from recombination (preventing gas infall and contraction into dark matter haloes), turbulence, and even mergers of evolved galaxies \citep{2019RPPh...82a6901M,2024ApJ...961...76M}.
Another important pathway is the formation of seeds with $\sim 10^{2-3}\, \rm M_{\odot}$ through hierarchical mergers of massive stars in compact and dense stellar clusters \citep[e.g.,][]{2009ApJ...694..302D,2017MNRAS.472.1677S,Rantala_2025}. {\it JWST} probes the existence of such clusters as early as $z\sim 10$ \citep{2024Natur.632..513A,2023ApJ...945...53V,2024arXiv240208696M}. Relatively massive BH seeds formed in stellar clusters would benefit from favourable dynamics and environment for rapid growth. 

Although most of the theoretical pathways for BH seed formation have been around for quite some time\footnote{The paper of Martin Rees (1978) remains a solid and enlightening foundation of most contemporary theories \citep{1978Obs....98..210R}.}, the last decade has been crucial to assess their detailed physical aspects (e.g.,\ fragmentation of gas clouds or their collapse into protostars, accretion onto protostars and whether feedback processes inhibit growth) through cosmological simulations and semi-analytical models \citep[e.g.,][]{2015MNRAS.450.4350I,2016ApJ...832..134C,2016MNRAS.461..111R,2019Natur.566...85W,2022Natur.607...48L}. 

What have we learned from recent observations with {\it JWST}?
So far, {\it JWST} has provided three key insights. First, it has pushed the redshift record for the earliest observed SMBHs in the Universe to $z\approx$~8.5--10.6; see Figs.~\ref{fig_QSO_parameter_space} \& \ref{fig_highz_QSO_discovery}. The SMBH masses of these systems are highly uncertain but broadly appear to lie in the range of $\approx10^7$--$10^8$~$M_{\odot}$, measured from the broad emission lines for the $z>$~8.5 broad-line systems (see \S\ref{sec:BHmass}) and estimated from the AGN luminosity and assuming an Eddington ratio for the $z\approx$~10 narrow-line systems. However, these SMBH masses are broadly consistent with that required to explain the previously known high-redshift quasars \citep{2023arXiv230512504S}. In other words, in terms of their growth history, these systems at $z \sim 8.5-10$ could be the progenitors of the high-redshift quasars \citep[see discussion in][]{2023MNRAS.521..241V}; see Fig.~\ref{fig:BHgrowth_seeds}. The occupation fraction — the number of BHs in bins of galaxy properties such as stellar mass or velocity dispersion — can theoretically provide insights on the origin of BHs \citep{greene_intermediateMass_2020}. However, more theoretical work is required to generate precise predictions for each potential BH formation channel, which can then be compared to observational data. Preliminary results from {\it JWST} suggest a potentially high number density of AGN (see \S\ref{sec:highz_demographics}), and as these systems and the host galaxies in which they reside become better understood, these observations will enable more accurate constraints on the AGN occupation fraction.
One of the most striking discoveries by {\it JWST} is the potentially high $M_{\rm BH}/M_{\star}$ mass ratios in a significant population of AGN across a wide redshift range ($z \geqslant 4$). If confirmed, these observations could suggest the presence of heavy seeds \citep{2024ApJ...960L...1N} and/or faster super Eddington or Eddington limited growth of SMBHs relative to the stellar populations of their host galaxies. However, the recent identification of a {\it JWST} broad-line AGN with $M_{\rm BH}\approx10^7$~$M_{\odot}$ in a $z=7.04$ metal-poor galaxy ($<1$\% solar metallicity) argues against substantial super Eddington, or even Eddington limited accretion, for at least one system as the required gas inflow would drive significant star formation and hence metal enrichment \citep[][]{Maiolino_metal_poor_2025}. If a substantial fraction of the {\it JWST} AGN population are also found to reside in metal-poor galaxies then this would point towards more exotic BH formation scenarios such as primordial BHs to explain these systems \citep[][]{Maiolino_metal_poor_2025}; see \S\ref{sec:highz_demographics} for more details. Therefore, at the time of writing, efficient growth of light BHs with phases of super-Eddington accretion, Eddington-limited growth of heavy seeds, and primordial origins of BHs, all remain viable options to explain the {\it JWST} AGN and it is quite likely that no one scenario explains the whole population  \citep{2023arXiv230512504S,Bennett:24,2024A&A...690A.182D,2024ApJ...960L...1N}.

The seeds of BHs still fall below the sensitivity of {\it JWST} for which only AGN powered by $M_{\rm BH}\geqslant 10^{6}\, \rm M_{\odot}$ are detectable and for which a SMBH mass estimate can be derived. However, by the mid-2030s the future {\it LISA} gravitational wave space antenna \citep{2023LRR....26....2A} will uncover all merging BHs in the mass range $\sim 10^{4}-10^{7}\, \rm M_{\odot}$ at any redshift (larger range at low redshift), while ground-based new-generation interferometers such as the Einstein Telescope \citep{Einstein2023} will complement the discovery space with BHs of $10-10^{4}\, \rm M_{\odot}$ (i.e.,\ seed-mass BHs). These interferometers will open the door to new exciting discoveries, but will remain ``silent" to non-merging BHs and will need to be completed by novel telescopes sensitive to electromagnetic emission from BH mass accretion. Next-generation X-ray telescopes such as {\it NewAthena} \citep{cruise_newathena_2024} and {\it AXIS} \citep{2023SPIE12678E..1ER}, expected to be operational on a similar timescale as {\it LISA} and the Einstein Telescope, will hunt for faint high-redshift AGN powered by low-mass BHs, potentially as low as $\sim 10^{4-5}\, \rm M_{\odot}$, and will help confirm the nature of the high-redshift AGN candidates being uncovered by {\it JWST}.\\


\subsubsection{Implication of recent discoveries on reionization}

The epoch of reionization refers to the period during which the high-redshift IGM is converted from a fully neutral state to a fully ionized state. Prior to {\it JWST}, the prevailing wisdom was that reionization was primarily driven by star formation from the first dwarf galaxies, with AGN activity playing only a minor role. There are uncertainties about when reionization ended, although through the analysis of 70 quasar sightlines at $z>5.5$, it appears to be around $z=5.3$ \citep{2022MNRAS.514...55B}, and even greater uncertainty about when it exactly began \citep[i.e.,\ the redshift of the first Pop~III stars, the first ionizing objects;][]{Klessen2023_popIII}. However, although numerous, the AGN candidates discovered by {\it JWST} are globally too faint to contribute significantly to reionization, with an upper limit of $30\%$ to the ionizing budget \citep{2024arXiv240111242D,Jiang_2025_reionisation}. \\

\subsection{Summary of the key drivers of progress}
\label{sec:highz_progress}

Many of the great advances in our understanding of the rapid and early growth of SMBHs over the last decade have been facilitated by significant increases in data and improvements in techniques, allied to 
theoretical model developments which have provided both the physical insight and clear observational predictions. The large order-of-magnitude increase in the number of identified quasars plus greater multi-wavelength datasets (from X-ray--radio wavelengths) has allowed for systematic comparisons of observed properties as a function of key physical parameter (e.g.,\ AGN luminosity; SMBH mass; Eddington ratio) with sufficient source statistics to control for potential biases. It is now clear that the observed properties of quasars are tightly connected to the accretion-disk emission, which changes with SMBH mass and Eddington ratio (see Fig.~\ref{fig:Accretion-flows}), and the emission ``felt" by the gas in the vicinity of the SMBH is modified depending upon opacity and orientation. This is unsurprising of course, given the accretion disk is the origin of the power of quasars, but connecting the observed properties of quasars to the underlying physics has required decades of observational and theoretical research.\footnote{It is perhaps sobering to note that if the gas in our galaxy did not absorb far-UV emission then these key discoveries may have been made decades earlier!} The greater source statistics have also allowed for a better characterisation of different quasar populations, critical for determining the role of environment (host and large scale) on the properties of quasars and whether quasars evolve through discrete phases, particularly when allied to careful population studies utilising control samples. These studies have mostly focused on the unobscured quasar population but, over the next decade, the operation of ambitious large-scale spectroscopic programmes targetting X-ray, infrared, and radio AGN will address this imbalance, providing obscured quasar samples in the hundred thousands--millions (see Footnote~\ref{foot:MOSobscured}).

Facilities have been a key driver of scientific progress for high-redshift AGN and quasars. Observations from {\it JWST} have expanded the redshift--luminosity discovery plane for AGN (and candidate AGN), allowing for rapid advances in our understanding of the early growth of SMBHs. The NIRSpec instrument has been crucial in driving these advances since the wide wavelength coverage, high spectral resolution, and high sensitivity allows for not only the identification of these high-redshift sources but also the determination of their physical properties. Larger field of view optical--near-IR imaging (e.g.,\ Subaru Hyper Supreme Cam; PanSTARRs; DESI-legacy survey) and multi-object spectrographs, with increased spectroscopic multi-plexing (e.g.,\ SDSS-III, IV; DESI), have allowed for more efficient observing programmes and greater data volumes, providing unprecedented source statistics across the luminosity--redshift plane. New multi-wavelength facilities over the last decade (e.g.,\ LOFAR in the radio; ALMA in the submm/mm; VLT-MUSE in the optical) have provided the broader multi-wavelength context to understand the physical properties of quasars and high-redshift AGN and their broader host galaxy and large-scale environments. An associated advance to these facility developments has also been greater community involvement by developing large programmes that exploit a large fraction of the available observing time but deliver data products to allow the wider astronomical community to exploit the data, driving efficient scientific progress.

Rapid advancements in computational power over the last decade have also enabled the execution of large simulations with 100~comoving Mpc box sizes, such as {\sc Eagle, Horizon-AGN, and Illustris}. This has allowed for the evolution of thousands of galaxies and AGN to be captured across cosmic time (\S\ref{sec4:gal-properties} \& \S\ref{sec4:varying_growth}). A parallel advance has been an increase in the resolution for computationally expensive zoom-in simulations of high-redshift quasars and their environments \citep{Costa:14a,2024A&A...686A.256L,Angles-Alcazar2021}, critical to accurately trace the physics on smaller spatial and temporal scales; see also \S\ref{sec:simulations}.

%% file: sec6.tex
The previous three sections have investigated the conditions for, and processes of, fuelling SMBH growth.  As described, during episodes of SMBH growth (i.e., when the SMBH is identified as an AGN), a huge amount of energy, momentum, and mass can be released in the form of radiation, accretion disk winds, and relativistic jets of charged particles; see Figs.~\ref{fig:agn_model}, \ref{fig:AGN-schema}, \& \ref{fig:Accretion-flows}. In this section we explore the current understanding of how AGN energy, momentum, and mass can produce a ``feedback" effect on future SMBH growth and the properties of the host galaxies.

The co-existence of AGN and high-velocity gas in galactic nuclei has long been known \citep{Crenshaw:03}.
Evidence has sporadically pointed to large-scale outflows\footnote{Here, we follow \cite{Harrison2024} and define an outflow as ISM/CGM material that has been swept up/entrained by some driving mechanism due to the AGN (such as an accretion disk wind, jet, or radiation pressure).} \citep{Tremonti:07, Morganti:07}.
However, it is only in the last $\approx 15 \, \rm years$ that observations have shown that galaxy-wide impact driven by AGN activity, over scales $\sim 100 \, \rm pc \-- 10 \, \rm kpc$, is \emph{widespread}. In quasars and local Seyferts, this impact takes the form of massive outflows, traced via P-Cygni-type profiles, broad emission lines and high-velocity absorption components \citep{Fischer:10, Feruglio:10, Sturm:11, Rupke:11, Nesvadba:11, Aalto:12, Combes:13, Cicone:14}. 

For a SMBH lying on the local $M_{\rm BH} \-- M_\star$ scaling relation, the cumulative energy released via accretion over the black hole's lifetime exceeds the galactic binding energy by a factor $\sim 100$ \citep{Fabian:12, King:15}. Indeed, even before the discovery of tight scaling relations linking SMBH masses and galaxy properties, energy release via AGN activity (commonly referred to as ``AGN feedback)" had already been identified as a likely process to offset radiative cooling losses and expel gas from local and high-redshift, massive galaxies \citep{Tabor:93, Haehnelt:98, Silk:98}. 

Two central questions underpin the ongoing debate surrounding AGN's role in galaxy formation:
\begin{enumerate}
    \item What fraction of the available energy is effectively transferred to the surrounding gas?
    \item Which galactic component—the ISM or the CGM is most impacted?
\end{enumerate}
This section reviews how our answers to these two central questions have evolved over the past decade. During this time, our understanding of AGN impact on galaxy evolution has been reshaped by parallel advances across several fields. These include a deeper (analytical) understanding of how winds and jets interact with galactic gas, the development of more sophisticated `sub-grid' models for SMBH accretion and AGN feedback, higher-resolution simulations performed using new, more accurate (radiation-)hydrodynamic solvers. Additionally, the explosion of AGN-driven outflow detections—fuelled by the golden age of Integral Field Unit (IFU) spectroscopy and the maturity of ALMA—has further transformed our understanding. 

\begin{figure*}
    \centering
    \includegraphics[width=\textwidth]{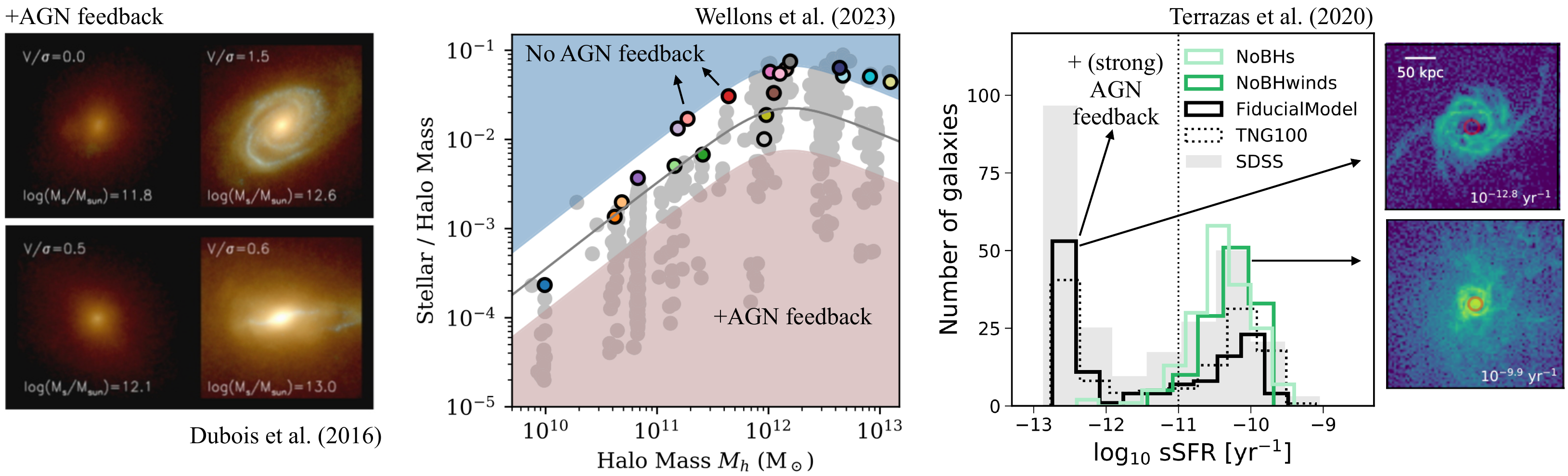}
    \caption{Despite the use of different hydrodynamic solvers, different models for star formation, stellar feedback, interstellar medium, virtually all cosmological models of galaxy evolution must invoke feedback from AGN in order to produce a realistic population of massive galaxies. According to cosmological hydrodynamic simulations, AGN feedback regulates the gas content of massive haloes and their central galaxies, modulating morphological transformations (left-hand panel), the stellar-to-halo mass ratio (central panel) and quiescent fraction of massive galaxies (right-hand panel), among a long catalogue of other galaxy properties (see text). On the left-hand panel (adapted from Fig.~2 of \citealt{Dubois:2016}), the annotations give the halo mass (at the bottom) and the ratio of rotation to dispersion velocities for the stellar component. The central panel shows results from \cite{Wellons:23} (adapted from their Fig.~2), who compare simulations that do not include AGN feedback (coloured points), with simulations including various AGN feedback prescriptions (grey points) and with the semi-empirical relation (solid curve). The right-most plot shows the specific SFR (sSFR) for different runs the IllustrisTNG simulations (empty histograms; adapted from Fig.~1 of \citealt{Terrazas:2020}) and observations (filled histogram). The fiducial simulation (black histogram) includes two modes of AGN feedback, which are required to produce the quiescent (low sSFR) galaxy population. The two smaller panels show example star-forming and quenched galaxies from the simulations (adapted from Fig.~3 of \citealt{Terrazas:2020}).}
    \label{fig:sec6-effectsOfFeedback}
\end{figure*}

Since feedback from AGN was first invoked on theoretical grounds \citep{Tabor:93, Haehnelt:98, Silk:98}, we start this section by reviewing the state-of-the-art theoretical treatment of AGN feedback in (large-scale) cosmological models of galaxy formation (\S\ref{sec:AgnFeedbackEffects}). In \S\ref{sec:AgnFeedbackMechanisms}, we `zoom-in' on what has changed in our understanding of the small-scale processes driving AGN feedback. In \S\ref{sec:AGNFeedbackImpact}, we assess what this progress has taught us about the impact of AGN on galaxy evolution, while in \S\ref{sec:AGNFeedbackObservations} we summarise the new observational insights into AGN feedback, reflecting on how these constrain the role of AGN in galaxy evolution.

\subsection{AGN feedback in galaxy formation models}
\label{sec:AgnFeedbackEffects}

Every successful cosmological model of galaxy evolution now incorporates feedback from AGN. Despite different hydrodynamic solvers in numerical simulations, distinct models for star formation, the ISM and supernova feedback, there is broad consensus that AGN feedback is required to produce a realistic galaxy population in massive haloes with $M_{\rm vir} \gtrsim 10^{12} \, \rm M_\odot$ \citep{Vogelsberger:2014, Schaye:2015, Dubois:2016, Weinberger:2017, Dave:2019, Wellons:23, Dolag:2025}.

The hot, gaseous atmospheres that exist within massive haloes have long cooling times, naturally suppressing gas inflow towards central galaxies.
But this suppression is insufficient without an additional source of energy such as AGN \citep{Sijacki:07, vanDeVoort:2011, Beckmann:2017, Feldmann:2023}. Processes unrelated to AGN have been invoked as sources of feedback, including Type Ia supernovae, asymptotic giant branch (AGB) stellar winds, magnetic fields, (non-AGN) cosmic rays, thermal conduction, gravitational heating and morphological quenching \citep{Martig:09, Dekel:09}. Some key processes (e.g. thermal conduction) still do not feature in state-of-the-art cosmological boxes. While they all plausibly contribute to suppressing star formation in massive haloes, AGN feedback is thought to be dominant \citep{Su:19}.

\subsubsection{Galaxy formation without AGN feedback}

Models fail to capture key aspects of galaxy evolution when AGN feedback is omitted; see Fig.~\ref{fig:sec6-effectsOfFeedback}. Massive galaxies would convert too many baryons into stars. They would be gas-rich, actively star-forming, and would fail to `quench' \citep{Weinberger:18, Terrazas:2020, Wellons:23}. As shown in the left-hand panel in Fig.~\ref{fig:sec6-effectsOfFeedback}, they would exhibit blue optical colours, while AGN feedback ensures that massive galaxies become gas-poor, passive and red \citep[][]{DiMatteo:05, Springel:05,Somerville:08, Gabor:11}. By suppressing cooling in massive galaxies, AGN feedback sets the high end of the stellar mass- and luminosity functions \citep{Scannapieco:04, Bower:06,Vogelsberger:14}. Feedback from AGN may also mediate the morphological transition from discs to ellipticals \citep{Dubois:2016, Sparre:17}, likely by preventing discs from re-forming after major mergers. The morphology of the stellar component would also likely be affected, as the lack of AGN feedback facilitates the formation of compact stellar cores that do not as easily form in the presence of strong AGN feedback \citep{Peirani:17, Choi2018, vanderVlugt:19, Cochrane2024}. 
Differences arise also in the chemical enrichment properties of galaxy populations (e.g. the $\alpha$-enhancement of stellar populations), which can be better reconciled with observational constraints if AGN feedback is present \citep{Taylor:15, Segers:16}.
Discrepancies are evident also at the scale of the CGM. The absence of strong feedback in massive galaxies results in denser gaseous atmospheres \citep{Davies:2020, Voit:24}, producing more X-ray emission than is observed \citep{Choi:15}. At high redshift, the absence of AGN feedback would prevent bright quasars from being unobscured in the UV \citep{Ni:20, Vito:22, Bennett:24}, preventing the formation of extended Ly$\alpha$ nebulae \citep{Costa:22}. 

\subsubsection{Modelling AGN feedback}
In cosmological volumes of $\gtrsim 50^3 \, h^{-3} \, \mathrm{Mpc}^3$, spatial resolution has improved over the past decade from $\sim 1 \, \mathrm{kpc}$ to $\sim 100 \, \mathrm{pc}$ \citep[e.g.,][]{Nelson:19}. Despite these advances, this resolution still falls short of resolving SMBH accretion and feedback processes, which occur on scales of $\lesssim 10^{-2} \, \mathrm{pc}$ (see Fig.~\ref{fig:agn_model}). In state-of-the-art cosmological simulations, the various effects of AGN feedback described in \S\ref{sec:AgnFeedbackEffects} are the outcome of `sub-grid' models. 

The purpose of AGN feedback sub-grid models is to parametrize the effects of unresolved processes, such as winds, jets, and AGN radiation, on the scales that simulations can resolve. Since they operate at scales of $\sim 100 \, \mathrm{pc}$, the sub-grid models used in large cosmological simulations cannot be straightforwardly connected to smaller-scale processes like jets and accretion disk winds, which are launched on sub-pc scales (see Fig.~\ref{fig:AGN-schema}). As a result, these models rely on assumptions about how such interactions affect gas on much larger scales. Moreover, sub-grid models are often partly designed with the need to overcome problems of numerical origin in view \citep[e.g.][]{Booth:09}.

A common approach to simulating AGN feedback is to continuously inject thermal energy around sink particles representing SMBHs  \citep{DiMatteo:05, Springel:05} and is standard for modeling feedback in quasar phases, i.e., ``quasar feedback'', in simulations like {\sc Magneticum} \citep{Steinborn:15, Dolag:2025}, {\sc IllustrisTNG} \citep{Weinberger:2017}, and {\sc NewHorizon} \citep{Dubois:21}. Yet, continuous thermal energy injection can suffer from numerical cooling losses, dispersing energy into dense gas where it dissipates before a significant effect occurs \citep{Booth:09, Weinberger:18}.
To address this, \citet{Booth:09} proposed injecting thermal energy in bursts, raising the gas temperature above a threshold ($T \gtrsim 10^8 \, \mathrm{K}$) to ensure longer cooling times. This model, adopted in simulations such as cosmo-OWLS \citep{LeBrun:14} and {\sc Eagle} \citep{Schaye:2015}, leads to feedback occurring in intense, periodic bursts.

Other models abandon thermal energy injection altogether and, instead, inject energy in kinetic form to represent quasar feedback. This approach can produce stronger feedback, provided the energy is not immediately thermalized \citep{Costa:2020}. Injecting momentum can cause acceleration, while offsetting cooling losses \citep[e.g.,][]{Choi:12}. Notably, \citet{Choi:15} and \citet{Farcy:2025} show that kinetic quasar feedback can yield a similarly realistic $M_{\rm BH} \-- M_{\star}$ relation as thermal feedback, while exerting a much stronger influence on halo gas and the AGN host galaxy.

When continuous thermal injection is used to model quasar feedback, simulations indeed often rely on enhanced feedback during low accretion rate phases. A clear example is the {\sc IllustrisTNG} simulation, where AGN feedback at high accretion rates is modeled with continuous thermal feedback, and, at low accretion rates, is modeled through (i) the injection of kinetic energy and (ii) a pulsating rather than continuous injection. This approach is key to enabling the quenching of massive galaxies in the simulation \citep{Weinberger:18, Terrazas:2020}, as shown in the right-most panel of Fig.~\ref{fig:sec6-effectsOfFeedback}. \citet{Weinberger:2017} emphasize this model remains agnostic about underlying microphysics, although wide-angle winds and jets are highlighted as possible physical motivations for their approach.

Large-scale simulations are often performed with multiple such sub-grid models \citep[e.g.][]{Rennehan:2024, Husko:2024}. {\sc NewHorizon} \citep{Dubois:21} and {\sc Simba} \citep{Dave:2019}, for instance, include both a `wind mode', proceeding via bipolar injection of momentum (in {\sc Simba}) or thermal energy (in {\sc NewHorizon}), and a `jet mode' producing higher-velocity, kinetically-dominated, collimated outflows. {\sc Simba} further includes an `X-ray mode whereby a combination of thermal and kinetic energy is deposited into gas to mimic X-ray heating and radiation pressure, i.e. without explicit radiative transfer. Understanding to what extent the flows produced by such models at the current resolution scale of cosmological simulations ($\gtrsim 100 \, \rm pc$) can result from the impact of jets, radiation and winds with the ISM on smaller scales is an important future milestone in the field.

\subsection{Physical mechanisms}
\label{sec:AgnFeedbackMechanisms}
From a theoretical perspective, the last decade has witnessed growing consensus that AGN feedback has to proceed via `energy-driven' large-scale outflows \citep{King:15}. \emph{Importantly, this conclusion appears to hold regardless of whether these outflows are powered by jets or winds.} 
This insight was brought about by advances in analytic theoretical modeling \citep{King:11, Zubovas:12, Faucher-Giguere:12, Thompson:15, Hartwig:18}, higher-resolution simulations probing the impact of jets and winds on multiphase media \citep{Wagner:12, Wagner:13, Mukherjee:16, Ward2024, Sivasankaran:2025}, but also by new `cross-talk' between analytic models and numerical simulations for winds \citep{Nayakshin:10, Costa:14} and jets \citep{Bourne:17, Talbot:21, Husko:2022}.

\begin{figure}
    \centering
    \includegraphics[width=\linewidth, trim={0 2.3cm 0 0.1cm},clip]{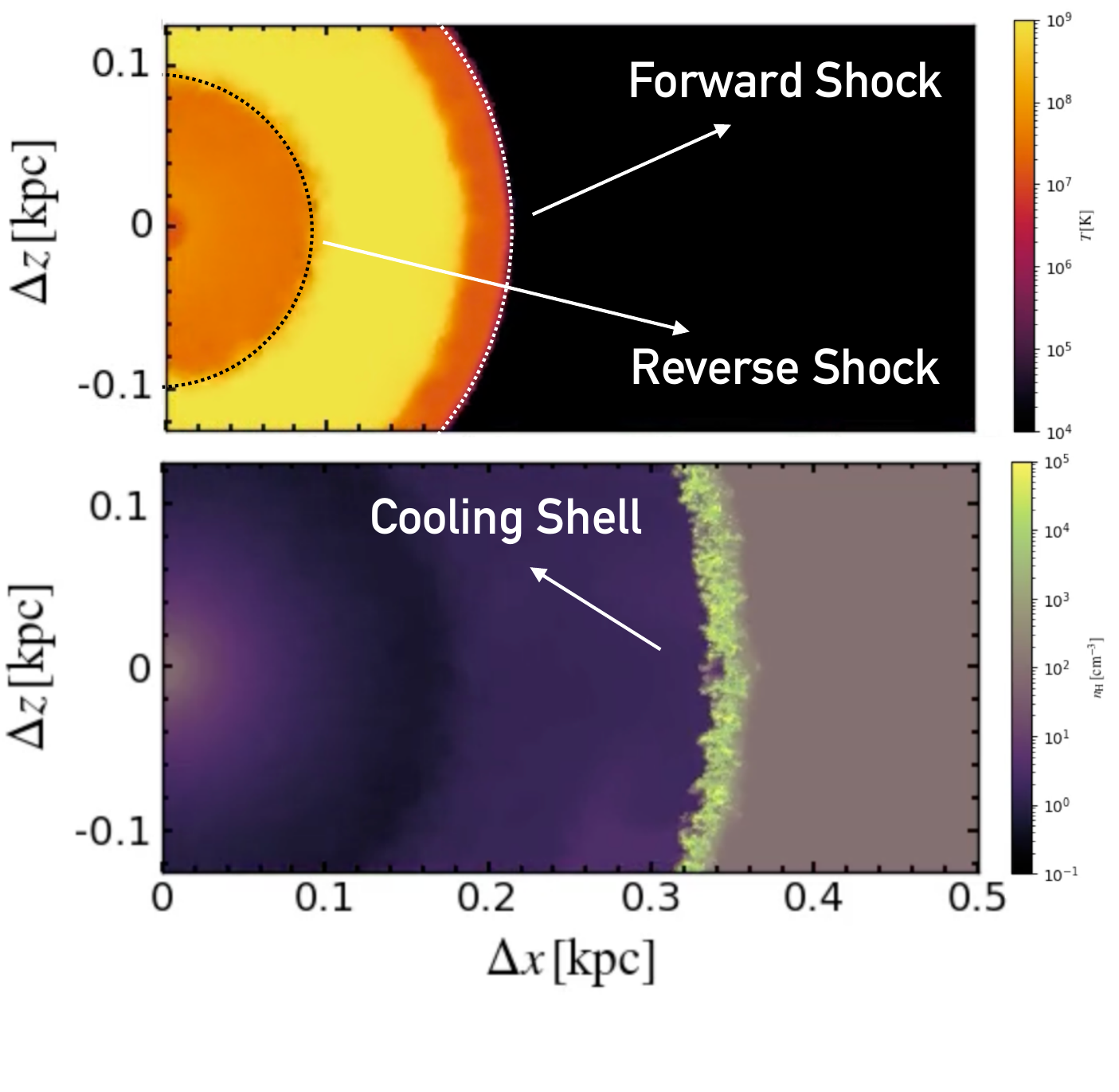}
    \caption{Energy-driven outflow powered by a spherical wind with velocity $v = 10,000 \, \rm km \, s^{-1}$ propagating through an ambient medium of constant density. The top panel shows gas temperature and the bottom panel shows gas density (at a later time). The backward facing reverse shock (black circle) converts the bulk of the wind's kinetic energy to thermal energy, producing a hot bubble that expands in an energy-conserving fashion. The denser swept-up ambient gas shell bounded at large radii by the forward shock (white circle) cools radiatively at late times (bottom panel). {\it Source:} produced from simulation results of} \citet{Costa:2020}.
    \label{fig:sec6-energyDrivenSchematic}
\end{figure}

\subsubsection{Energy- vs. momentum-driven outflows}
A hot bubble forms when wind or jet fluid passes through a strong shock that converts the bulk of its kinetic energy into thermal energy (see Fig.~\ref{fig:sec6-energyDrivenSchematic}). For winds and jets with velocities $\gtrsim 0.1c$, the temperature of the post-shock fluid can exceed $\sim 10^{10} \, \rm K$. If cooling losses are negligible in the shocked wind/jet fluid, its energy can, in principle, be fully transferred to ambient gas. The resulting outflow is termed `energy-driven' or `energy-conserving'. In an `energy-driven' outflow, gas expands, doing `PdV' work on intervening ambient gas, sweeping it up and driving it outwards at speeds $\gtrsim 1000 \, \rm km \, s^{-1}$ \citep{King:11, Zubovas:12, Faucher-Giguere:12, Costa:14, Richings:18}. 

\begin{figure*}[!t]
    \centering
    \includegraphics[width=0.7\textwidth]{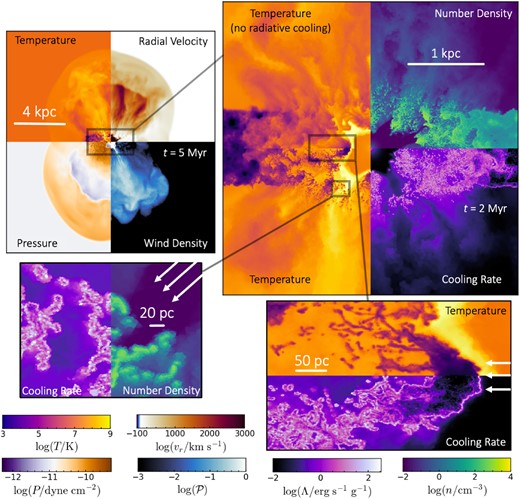}
    \caption{In an energy-driven outflow, ambient gas is driven out via adiabatic expansion of a hot, energy-conserving bubble. These outflows impact galaxies on different spatial and temporal scales. Large-scale bubbles propagate through halo material, clearing out the low-density CGM. Within the ISM, these outflows trigger shock waves that compress the ISM. Shear flows between the energy-driven bubble and ISM clouds destroy these clouds, although, under certain conditions, cold material can reform within the outflow. {\it Source:} reproduced from Fig.~2 of \citet{Ward2024}.}
    \label{fig:sec6-energyDrivenOutflow}
\end{figure*}

Much of the debate in the preceding decade has centred around whether these hot bubbles indeed retain their thermal energy or whether they cool \emph{catastrophically}, to produce `momentum-driven' or `momentum-conserving' outflows. In a `momentum-driven' outflow, the large-scale outflow is no longer accelerated via adiabatic expansion of an inner shocked bubble, relying instead on (conserved) residual momentum from the wind \citep{King:03}. {\bf}
Partially-radiative outflows also exist. Here, the shocked wind fluid cools more slowly, on a timescale comparable to the outflow time \citep{Faucher-Giguere:12}.

Whether outflows are energy- or momentum-driven determines the strength of the outflow, its kinematics and phase structure.
Energy conservation between a small-scale wind, with a kinetic wind power $L_{\rm w}$, and the swept-up ISM, expelled at velocity $v_{\rm out}$ and momentum flux $\dot{P}_{\rm w}$, gives
\begin{equation}
    \dot{P}_{\rm out} \, \sim \, L_{\rm w} / v_{\rm out} \, .
\end{equation}
Momentum conservation alone gives
\begin{equation}
    \dot{P}_{\rm out} \, \sim \, L_{\rm w} / v_{\rm w} \, .
\end{equation}
If there is significant entrainment of ambient gas, the outflowing mass must slow down, giving $v_{\rm out} \ll v_{\rm w}$ and thus $L_{\rm w} / v_{\rm w} \ll L_{\rm w} / v_{\rm out}$. Therefore \emph{energy-driven outflows exert stronger forces on ambient gas than momentum-driven outflows} \citep{King:05}. Indeed, observational efforts have centred on identifying outflows with enhanced momentum fluxes (see \S\ref{sec:AGNFeedbackObservations}).
The energy coupling efficiencies are also different. While energy-driven winds inject energy at a rate $\sim L_{\rm w}$ into ambient gas, the energy coupling efficiency is lower by a factor $v_{\rm out} / v_{\rm w} \, = \, 10^{-1} \left(v_{\rm out} / 10^3 \, \rm km \, s^{-1}\right) \left(v_{\rm w} / 10^4 \, \rm km \, s^{-1}\right)^{-1}$.

It is based on momentum-driven solutions that \citet{King:03} reproduces both the normalisation and the slope of the observed $M_{\rm BH} \-- \sigma$ relation. The model predicts
\begin{equation}
    M_{\rm BH} \, = \, 3.9 \times 10^8 \left( \frac{f_{\rm gas}}{0.17} \right) \left( \frac{\sigma}{200 \, \rm km \, s^{-1}} \right)^4 \, ,
    \label{eq:MsigmaTheoretical}
\end{equation}
in striking agreement with e.g., \citet{Kormendy:13}, who find the empirical relationship to be
\begin{equation}
    M_{\rm BH} \, = \, 3 \times 10^8 \left( \frac{\sigma}{200 \, \rm km \, s^{-1}} \right)^{4.4} \, .
\end{equation}

The momentum-driven outflow picture has since been challenged. \citet{Faucher-Giguere:12} argue that two temperature effects expected in the shocked wind fluid prevent the catastrophic cooling required for a momentum-driven solution. Using hydrodynamic simulations, \citet{Costa:2020} show with hydrodynamic simulations that fast winds often thermalise outside their cooling radius, which causes them not to `skip' a momentum-driven phase. 

\citet{Nayakshin:14} argues that even purely energy-driven outflows may produce a $M_{\rm BH} \-- \sigma$ relation as in Eqn.~\ref{eq:MsigmaTheoretical}. In a medium comprising gas clouds and low density channels, energy-driven bubbles take paths of least resistance, pushing the cold gas via ram pressure (see Fig.~\ref{fig:sec6-energyDrivenOutflow}). In a more complex setting such as a porous medium \citep{Wagner:13, Ward2024} or the cosmological environment of a bright quasar \citep{Costa:14}, energy-driven outflows couple their energy less efficiently than in homogeneous media, where their impact is maximal \citep{Costa:14, Ward2024}.
Crucially, \emph{a momentum flux $\sim L/c$ appears to be insufficient on scales $\gtrsim 100 \, \rm pc$}. This conclusion has been drawn repeatedly in the last $\sim 15 \, \rm years$. For instance, \citet{Debuhr:11, Debuhr:12} require momentum fluxes $\sim 10 L/c$ to effectively impact the galactic ISM and reproduce a realistic $M_{\rm BH} \-- \sigma$ relation in simulations of major mergers. In their cosmological simulations, \citet{Costa:14, Costa:18} also find that momentum fluxes $\sim 10 L/c$ are required to power strong outflows in quasar-host galaxies.

When pieced together, the theoretical developments of the last decade point to AGN feedback via energy-driven bubbles. Whether mediated via (inclined) jets or winds, a low-density, hot $\gtrsim 10^9 \, \rm K$, energy-conserving fluid is likely to be shaping the evolution of massive galaxies. What the last decade has also taught us is that this outflow phase is almost impossible to detect directly \citep{Nims:15}. Observations of AGN outflows typically probe much denser and colder gas phases. The recent development in our understanding of how this dense phase comes about is reviewed next.

\subsubsection{Formation of multi-phase structure}
Observations of AGN-driven outflows (see \S\ref{sec:AGNFeedbackObservations}) indicate efficient entrainment of warm ($\sim 10^4 \, \rm \, K$) and cold, molecular ($T \sim 100 \, \rm K$) gas. Gas passing a strong shock with velocity $\gtrsim 200 \, \rm km \, s^{-1}$ shock heats the gas to a temperature $\gtrsim 10^6 \, \rm K$. Collisions of ambient gas with outflows at velocities $> 1000 \, \rm km \, s^{-1}$ naturally produces hot gas.
Central to research in the last decade has been the origin of the cool gas component. 
Pre-existing ambient cold gas clouds should be rapidly destroyed, before they are fully entrained in the outflow \citep{Klein:94, Marinacci:10, Scannapieco:15, Zhang:17}. 
The likely small ($\lesssim 10 \, \rm pc$) cloud sizes places stringent resolution demands that cannot be met in cosmological boxes, requiring smaller-scale idealised simulations. 
In the context of AGN-driven outflows, two main origin scenarios for cold gas outflows have emerged from these studies:

\begin{description}
    \item[Formation of cold gas in shock fronts:] The propagation of a strong forward shock compresses and accelerates ambient gas (see Fig.~\ref{fig:sec6-energyDrivenSchematic}). Enhanced radiative cooling in swept-up ambient gas causes the condensation of a warm, dense gas phase with $T \sim 10^4 \, \rm K$ \citep{Zubovas:14, Costa:14}. Using hydrodynamic simulations performed with an on-the-fly chemical network, \citet{Richings:18} show that cooling proceeds down to temperatures $\sim 100 \, \rm K$, at which point the outflow becomes dominated by molecular gas. In this scenario, newly-formed cold gas is comoving with the fast hot outflow and is thus less affected by destructive shear. But there are several difficulties: (i) dust, needed to catalyse the formation of molecules, may be destroyed by the strong shock and thermal sputtering \citep{Ferrara:16, Barnes:20} and (ii) this mechanism requires the energy-driven bubble to be effectively trapped by the ambient medium \citep{Nims:15, Ward2024} and does not operate efficiently if hot gas vents out \citep{Costa:15}.
    \item [Cloud destruction, mixing and re-cooling:] Fig.~\ref{fig:sec6-energyDrivenOutflow} shows a recent example of a simulated AGN-driven outflow, produced by the collision between a fast ($v_{\rm w} \sim 10^4 \, \rm km \, s^{-1}$) wind and a clumpy disc \citep[from][]{Ward2024}. There is coexistence  over six orders of magnitude range in gas temperatures and densities. Here, the coldest, densest gas does not form in shocks, but within mixing layers comprising cloud debris and outflow material that cool radiatively on an outflow timescale \citep{Marinacci:10, Armillotta:16, Gronke:18}. While it produces fast-moving cold gas, the hot AGN-driven bubble retains most of its energy, remaining energy-conserving \citep{Ward2024}.
\end{description}

\subsection{The impact of AGN on galaxy evolution}
\label{sec:AGNFeedbackImpact}

AH12 distinguish two AGN feedback modes: (i) a `superwind mode' (often termed `quasar-' or `wind' mode) which operates in luminous AGN and expels cold gas from galaxies, and (ii) a `radio mode' which is mediated via jets and maintains halo gas hot. More physical `sub-grid' models for AGN winds and jets, and influx IFU observations conducted over the last $\sim 10 \, \rm years$ blur this neat two-mode division. 

\begin{figure}
    \centering
    \includegraphics[width=0.9\linewidth]{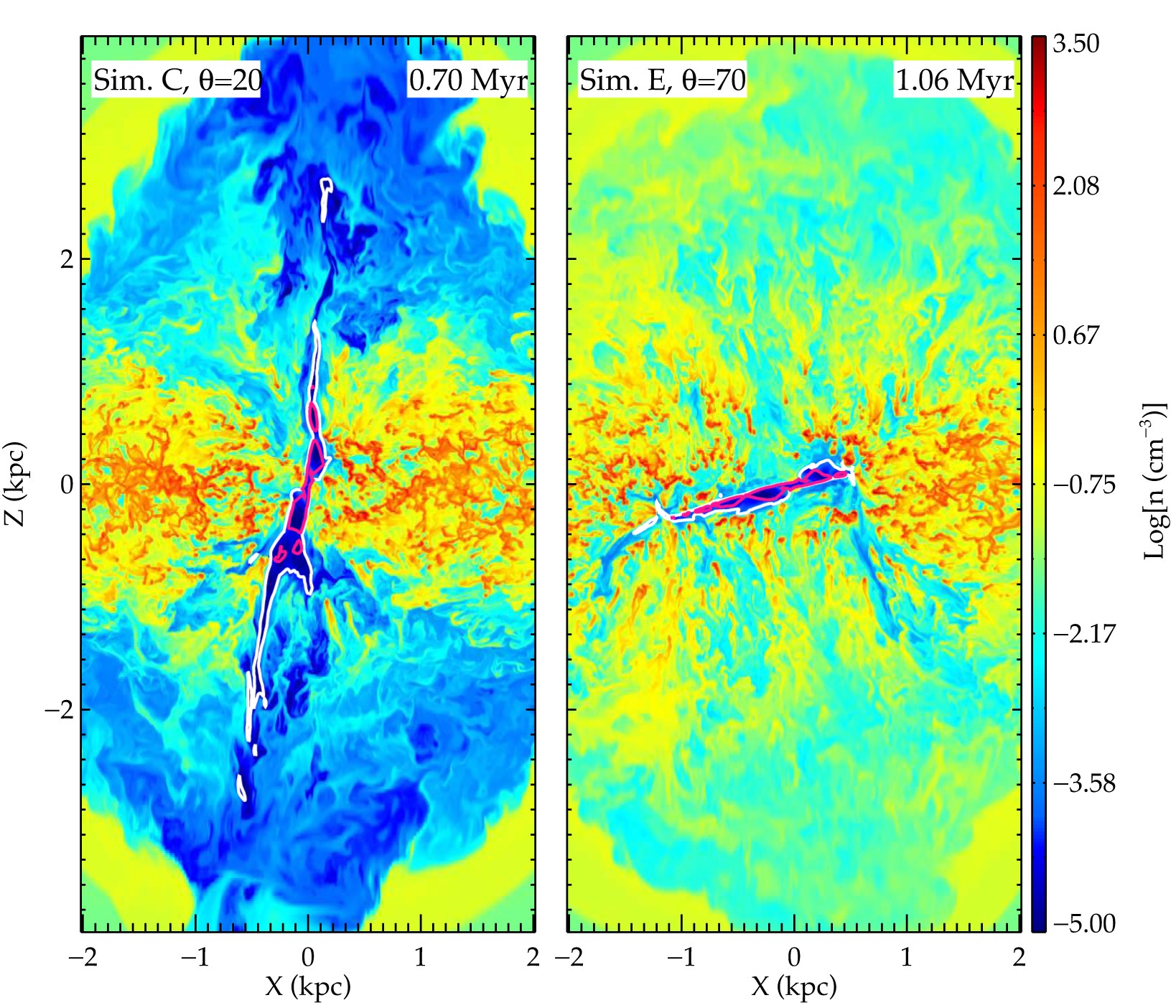}
    \caption{Simulation of relativistic low-power jet interacting with inhomogeneous disc galaxy. Such jets, like quasar winds, inflate energy-driven bubbles that carve out cavities in galactic nuclei and clear out halo material. Even in configurations where jets propagate along the disc plane, the outflow becomes bipolar on larger scales. {\it Source:} reproduced from Fig.~16 of \citet{Mukherjee:18}.}
    \label{fig:sec6-JetOutflow}
\end{figure}

\subsubsection{Feedback channels}

Whether driven by jets or winds, AGN-driven outflows produce multi-faceted effects operating on different timescales:

\begin{figure*}
    \centering
    \includegraphics[width=0.95\linewidth]{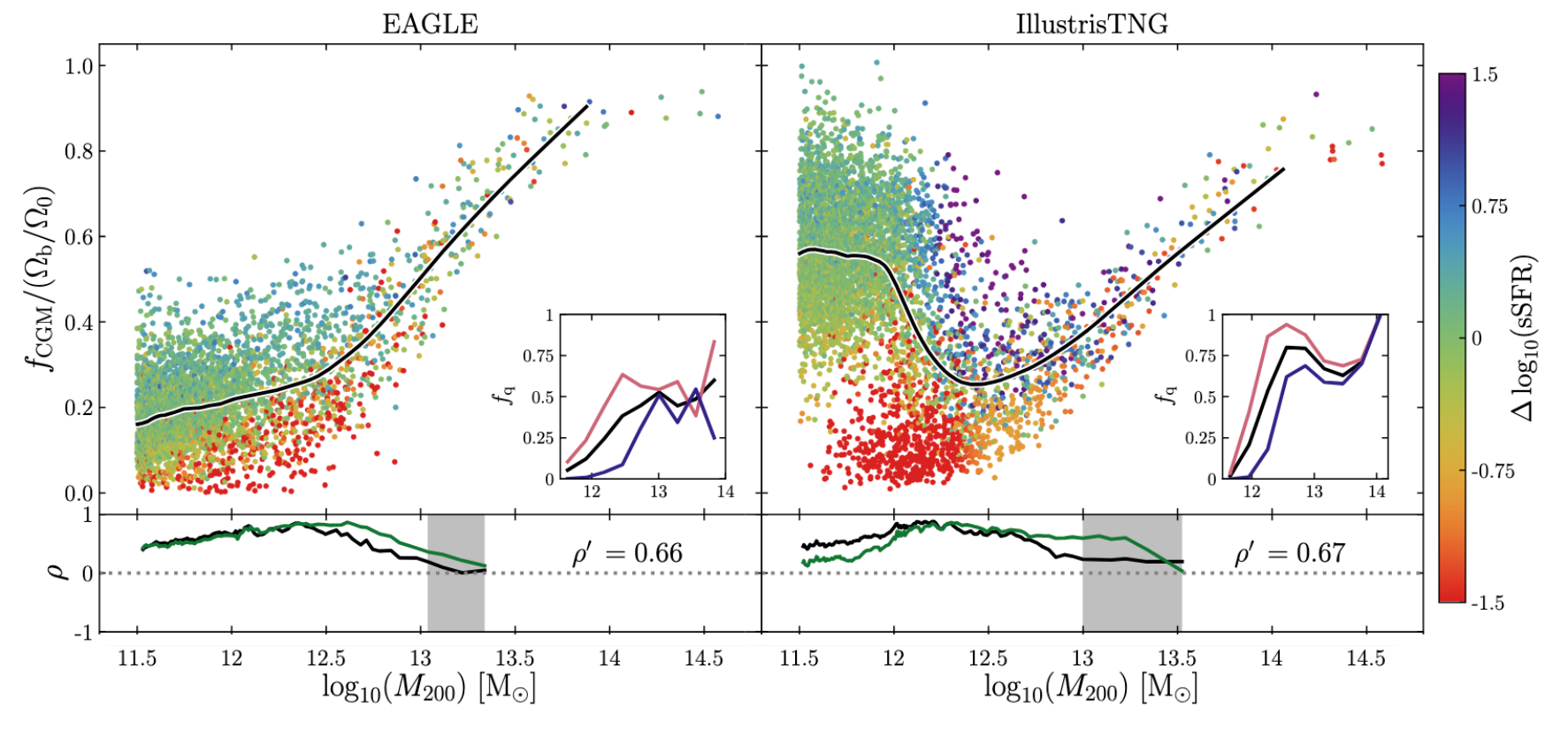}
    \caption{In cosmological simulations, massive galaxies quench when AGN-driven outflows clear out the CGM. Shown here is the fraction of total halo mass in the CGM as a function of halo mass in the state-of-the-art simulations {\sc Eagle} (left-hand panel) and {\sc IllustrisTNG} (right-hand panel). The colour gives specific SFRs. This drops when the CGM mass fraction drops as well as when the SMBH mass increases (not shown here). {\it Source:} adapted from Fig.~2 of \citet{Davies:2020}.}
    \label{fig:sec6-CGMexpulsion}
\end{figure*}

\begin{description}
    \item[Central gas cavities:] At the smallest scales resolved in hydrodynamic simulations, outflows clear out material, carving low-density cavities in the galactic nucleus \citep{Gabor:2014, Costa:14, Costa:2020, Torrey:2020, Raouf:23, Mercedes-Feliz:23}. The size of these cavities, which is of order $\sim 100 \, \rm pc$, is sensitive to AGN wind strength and the AGN outburst duration. The size of central cavities is also very sensitive on numerical resolution \citep{Bourne:15, Curtis:2016} and exactly how the AGN feedback is modelled.
    \item[Ejection of the ISM:] Outflows take paths of least resistance \citep{Costa:14, Bieri:2017, Nelson:19}. Even if energy is communicated in the form of a collimated jet at small scales, outflows on larger scales are shaped by the properties of the ambient medium through which they propagate \citep{Wagner:12,Tanner:2022}. This means it is possible for jets propagating along the plane of a galactic disc to produce bipolar bubbles (Fig.~\ref{fig:sec6-JetOutflow}) that escape along the disc poles \citep[e.g.][]{Mukherjee:16, Mukherjee:18, Talbot:22}. The typical outcome of simulations following the propagation of jets and winds in systems comprising galactic discs is a weak ejective feedback effect \citep{Gabor:2014, Costa:2020, Torrey:2020}. In configurations where gas has a high covering factor around the AGN (e.g.,\ the aftermath of a merger), outflows should clear out larger ISM masses \citep{Faucher-Giguere:12}, although it remains unclear whether feedback via ejection of the ISM will become dominant even in this case \citep[e.g.][]{Costa:18, Mercedes-Feliz:23}. 
    \item[Clearing out the halo:] Once they break out of their host galaxy, AGN-driven outflows push into tenuous, galactic halo gas (Fig.~\ref{fig:sec6-CGMexpulsion}). Here the effect of outflows is to reduce the density of the CGM and/or raise its temperature \citep{Angles-Alcazar:2017, Costa:18}. In simulations, the expulsion of the CGM coincides with quenching \citep{Davies:2020, Costa:2020, Oppenheimer:2020, Terrazas:2020, Voit:24}. After the CGM is cleared out, accretion from the halo stalls and star formation in the central galaxy drops due to reduced availability of gas fuel on a timescale of $\sim 100 \, \rm Myr \-- 1 \, \rm Gyr$ \citep{Oppenheimer:2020},
\end{description}

\subsubsection{`Radio' vs.\ `Quasar' and  `Ejective' vs.\ `Preventive' feedback}

Mechanisms such as quasar-powered winds produce both `ejective' and `preventive' feedback\footnote{The term `ejective' feedback refers to the removal of material from the ISM, whilst the term `preventative' feedback refers to regulating the rate at which gas can accrete onto the host galaxy from the wider halo, to ultimately form part of future ISM.} \citep{Costa:2020, Zinger:2020}. Winds powered at small scales produce energy-conserving bubbles that expand both into the ISM and the CGM \citep{Davies:2020}. 
As they propagate into the more isotropic and tenuous halo, these bubbles isotropise and push out significant amounts of halo gas. The density decrease caused by halo gas ejection leads to longer cooling time-scales, which suppress long-term gas accretion onto the central galaxy, producing strong preventive feedback \citep[e.g.][]{Davies:2020}. 
The tendency for energy-conserving bubbles to escape through low density channels means that it is even possible that quasar-mode feedback is predominantly preventive. This question remains open to debate. Answering it will require improvements both in the modelling of the ISM and AGN feedback processes in numerical simulations.

At low Eddington ratios, feedback was pictured to be communicated via radio jets. Their high degree of collimation was assumed to imply that jets produce no effect on the interstellar medium. Instead, the reasoning went, they interact primarily with halo gas, where they inflate over-pressurised cavities that expand, do work and heat halo material. The effect was thought to be mainly `preventive'.
While the impact of jets on suppressing cooling in hot haloes is established \citep[e.g.][]{Weinberger:17b}, jets, when highly inclined with respect to the disc, can have a similar effect as quasar winds \citep{Wagner:12, Mukherjee:18, Tanner:2022, Talbot:22}, powering bubbles that clear out gas from the nucleus, but expand to take paths of least resistance. This is especially true for low velocity, low power jets, which are more easily confined, and deposit their energy within the host galaxy ISM, where-as higher power jets deposit most of their energy in the surrounding halo. As \citet{Wagner:16} put it, \emph{the interaction of young jets in a radio galaxy with the ISM leads to quasar-mode, energy-driven, mechanical, negative or positive feedback.}. In summary, quasars winds and jets can provide both preventative ('maintenance') and ejective feedback, and it is not trivial to separate observed AGN populations into contributing to distinct feedback modes \citep[see discussion in][]{Harrison2024}.

\subsection{Observational work on AGN feedback}
\label{sec:AGNFeedbackObservations}

As recently reviewed by \cite{Harrison2024}, there has been considerable effort over the last three decades to constrain the mechanisms of AGN feedback, measure AGN-driven outflow properties, and search for evidence of feedback `in action'. This requires a multi-faceted set of observations, covering detailed multi-wavelength observations of individual targets (e.g., to understand the physics of how AGN drive multi-phase outflows) through to population studies (e.g., to look for global evidence of AGN feedback). As detailed below, this work has lead to considerable progress over the last decade, in particular, in understanding the challenges involved in interpreting observations of AGN-driven outflows, and finding `smoking gun' evidence for global AGN feedback. 

\subsubsection{Searching for the impact of AGN feedback}
Despite many earlier attempts to directly search for the impact of feedback on galaxies hosting luminous AGN \citep[see review in][]{Harrison:17}, the last decade of work has resulted in the consensus, that there is no strong reason to assume the impact of AGN feedback on a star-forming galaxy's global properties is observable, while the SMBH is still active. This is, at least partly, because AGN feedback also suppresses BH growth and the AGN will turn itself off, potentially before impacting upon the host galaxy. In the CGM expulsion scenario, it also takes $>100 \, \rm Myr$ for the effects of AGN feedback to become evident. Consequently, there will be no correlation between the state of the AGN at that time with the episode that brought about quenching. The gas reservoir will indeed become depleted, long after a bright AGN has switched off. Indeed, state-of-the-art cosmological simulations like {\sc IllustrisTNG} or {\sc Eagle} typically predict that the brightest AGN reside in gas-rich star-forming galaxies, while gas-poor, quenched galaxies have faint AGN \citep{Scholtz:18,Piotrowska:2022, Ward:2022}. Therefore, the fact that gaseous reservoirs are not typically suppressed around the brightest AGN is not in contradiction with theoretical predictions. According to these models, it is the SMBH mass, which is proportional to integrated injected feedback energy, that predicts whether a galaxy is quenched \citep{Terrazas:2016}. Indeed, the signatures of AGN feedback on a population level are likely better found in the galaxy population as a whole, irrespective of a currently active SMBH. The latest results from {\it JWST} of very early ($z\sim$3--4), massive quenched galaxies, are an interesting population to test against different models for AGN feedback in relation to the timescale of growing the SMBHs and shutting down star formation \citep[e.g.,][]{DEugenio:2024,Nanayakkara:2024,Scholtz_2024_ALMA}.  

On the other hand, localised and transient feedback effects on host galaxy properties can be seen in specific galaxies hosting in-situ luminous AGN. For example, a local region of depleted molecular gas, or a change in the distributions of young stellar populations in the vicinity of an outflow or jet is sometimes observed \citep{Harrison2024}. Nonetheless, it is likely that multiple AGN episodes would ultimately be responsible for {\em global} star formation quenching. Furthermore, a global role of regulative AGN feedback in the most massive galaxies, is observationally evidenced by combining radio observations (as a tracer of AGN) and X-ray observations of the intracluster, or intragroup medium. Observations over the last decade have confirmed earlier work (and expanded to higher redshift), that the required amount of heating is available from AGN to offset cooling for massive early-type galaxies \citep{McDonald2019,Hardcastle2019}.   

\subsubsection{Observed multi-phase outflows}

\begin{table*}[]
    \centering
    \begin{tabular}{lccll}
        \textbf{Gas phase} & \textbf{Temp} & \textbf{Density} & \textbf{Typical tracers} & \textbf{Scales} \\
         & [K] & [particles cm$^-3$] &   &  \\[2pt] \hline \\[-8pt]
        Hot ionised & 10$^5$ - 10$^8$  & 10$^6$ - 10$^8$ & X-ray absorption lines & $<$1 pc \\
        Warm ionised & 10$^3$ - 10$^5$ & 10$^2$ - 10$^5$ & emission lines in UV, optical, NIR & $<$1 pc to $\sim$1 Mpc \\
        Neutral atomic & 10$^2$ - 10$^3$ & 10$^1$ - 10$^2$ & H$\,$\textsc{i} absorption, Na$\,$\textsc{I}$\,$\textsc{D} absorption, [C$\,$\textsc{ii}] emission lines & $\sim$1 pc to $\sim$1 Mpc \\
        Molecular & 10$^1$ - 10$^3$ & $>$10$^3$ & OH, CO, H$_2$ lines & $\sim$1 pc to $\sim$1 Mpc \\ \hline
    \end{tabular}
    \caption{Gas phases in which outflows are observed, and their associated tracers and properties.}
    \label{tab:ism_phases}
\end{table*}

\begin{figure}[t]
    \centering
    \includegraphics[width=0.45\textwidth]{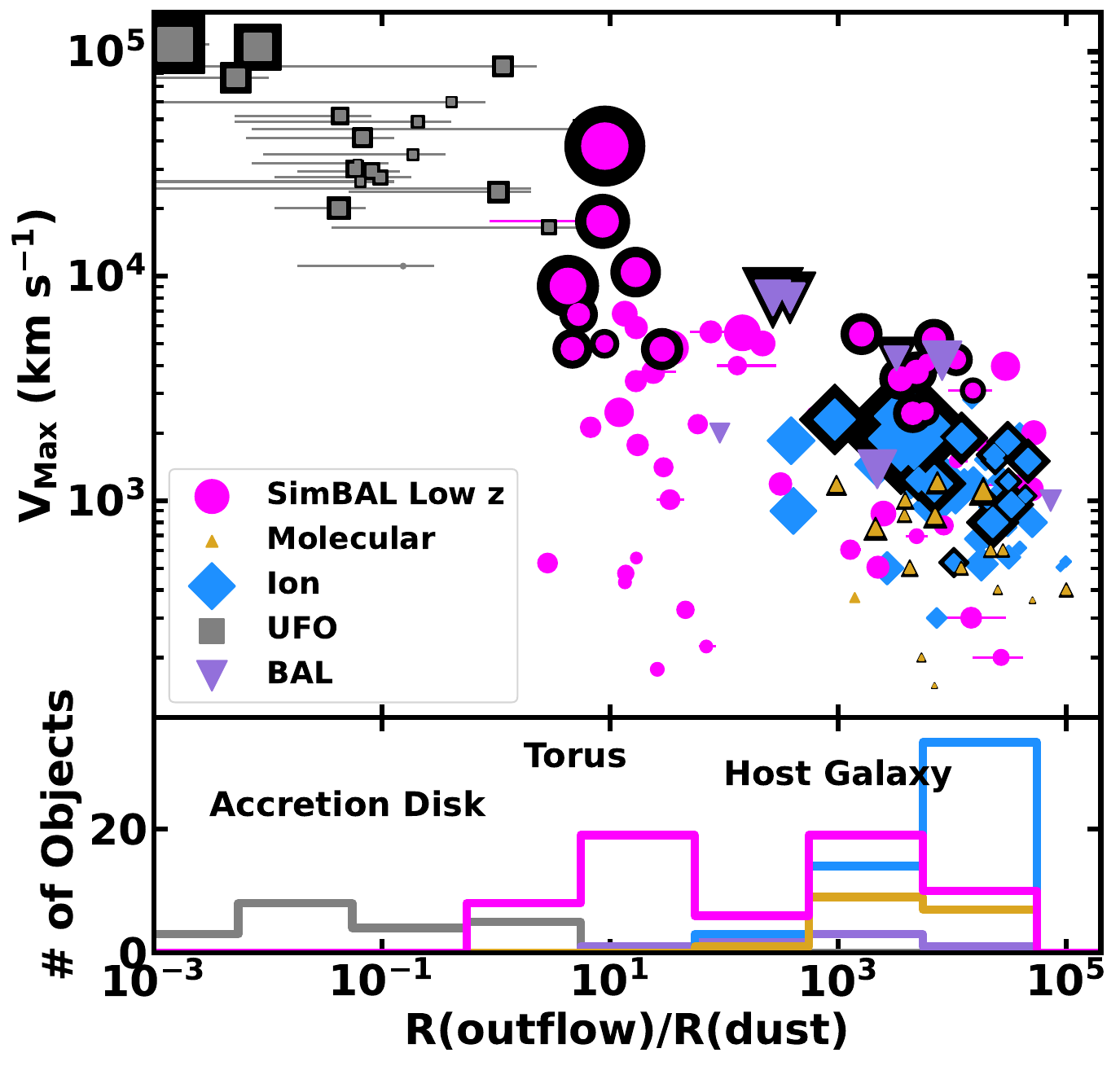}
    \caption{Top: Outflow velocity as a function of normalized radius \citep[dust sublimation radius $R_{\textrm{sub}} = 0.2L_{46}^{1/2}\,$pc;][]{laor_draine1993}. Grey, purple, blue, and yellow markers show objects from \cite{Fiore2017}, specifically, ultrafast outflows observed in the X-ray band, high-ionization BALQSO outflows from the literature, ionized outflows detected in emission lines, and outflows detected via molecular transitions, respectively. Magenta circles show results from analysis of low redshift FeLoBAL quasars \cite{choi2022} using SimBAL \cite{Leighly2018}. The size of the point scales with the bolometric luminosity, which spans $43.1<$~log($L_{\rm bol}$/erg~s$^{-1}$)~$<48.0$. The black circles show objects with inferred $L_{\rm bol}/L_{\rm KE} > 0.5$\% that have the potential for significant host-galaxy impact through AGN feedback \citep{2010MNRAS.401....7H}. Bottom: Number of outflows per decade of $R/R_{\textrm{sub}}$. Low-redshift FeLoBAL outflows span the largest range, and appear to be concentrated near the torus and at larger distances from the central engine. Kindly produced by K.~Leighly.}
    \label{fig_obsoutflows}
\end{figure}

Theoretical models naturally predict a multi-phase gas outflow structure (see Table~\ref{tab:ism_phases} for the different gas phases). However, it is difficult to know if the different {\it observed} outflows are all related to the same underlying cause. Larger statistical studies by necessity often focus on a single survey / waveband, but there have been attempts to build a more holistic picture of multi-phase outflows in AGN. For example, \cite{Fiore2017} investigated outflow properties for molecular, ionised, broad absorption line (BAL; see Footnote~\ref{foot:BALQSO}), and X-ray absorption line outflows \citep[Ultra-Fast Outflows; UFOs; see][]{Tombesi2010} for 94 AGN. However, the same objects are not covered by the different outflow phases. Fig.~\ref{fig_obsoutflows} shows the maximum velocity of the different types of outflows in this sample (also including new BALQSO measurements), as a function of the dust sublimation radius with point size scaled by bolometric luminosity (K.~Leighly, priv.~comm.). There are clear correlations between the outflow and AGN parameters, but the relationships may differ between the various types of outflows. 

Until recently, nearly all outflow studies focused solely on a single tracer, providing an incomplete picture of the multi-phase content of the outflows \citep{Cicone2018}. Even among single tracer studies, there may be more than one causal factor driving the outflows. For example, using SimBAL, \cite{choi2022} demonstrated that BALQSOs split into two sub-populations with distinct outflow properties when considering emission line properties \citep[i.e., using Eigenvector 1 as a proxy for Eddington ratio;][]{Boroson1992}. These single-tracer studies are still the only viable way to effectively undertake large statistical studies of outflows \citep[e.g.,][]{Mullaney2013,ForsterSchreiber2014,Avery2021}. This is because the cross section of observations of different phases of outflows drastically reduces the number of sources with information available at the required wavebands. 

It is critical to have spatially-matched information across multiple different phases of outflows, to understand how they are related, and gain a more complete understanding of their energy and mass content. This is expensive in terms of telescope time and person power, since the analyses require expertise across a wide range of techniques. However, the last few years has seen some progress in this area, typically by studying individual sources with detailed multi-wavelength information. These have helped pinpoint many important aspects of the properties and drivers of multi-phase outflows. For example, ISM modifications that depend on radio jets \citep[e.g.][]{Girdhar2022}, or where there is spatial correlation between tracers of different phases of outflows \citep{Lansbury2018,Venturi2023,Speranza2024}. Advances have recently been made towards building up larger samples of carefully selected samples with multiple tracers of outflow phases \citep[e.g.,][]{Fluetsch2021,Yamada2021,Riffel2023,Speranza2024}, but it is yet unclear how much these studies can be extrapolated to the general population. 

Understanding how efficiently an AGN can couple their energy output to the host galaxy ISM is crucial for assessing the potential impact on the host galaxy, and testing different feedback models. However, a lot of work over the last few years has highlighted that with the limitations of typical observations, it is extremely difficult to measure outflow properties, such as energetics, and momentum rates reliably \citep[e.g.,][]{Rose:2018,RDavies:2020,Lutz2020,Harrison2024}. When simulations are to mimic observational experiments, many further challenges become apparent, including the potential to miss most of the energy content (without a tracer of the hottest phases) and to mis-interpret energy-driven outflows as momentum-driven outflows \citep{Ward2024}. Nonetheless, the future is promising, as cutting-edge simulations, can now be used more directly with observations to help determine what can, and can not, be inferred from specific observations that will have incomplete information (e.g., limited gas phases, limited spatial information etc.). 

Understanding the physical drivers of multi-phase outflows (e.g., jets; radiation pressure; accretion-disk winds) is also challenging observationally, but improvements in observations have resulted in significant progress. Recently, the advent of more sensitive radio surveys like the LOFAR Two-metre Sky Survey \citep[LoTSS;][]{lotss2019,lotss2022} has allowed studies to expand at least in terms of including radio information \citep{rankine2021,petley2022,Kukreti2024}. For a typical synchrotron source, LoTSS is almost an order of magnitude deeper than the Faint Images of the Radio Sky at Twenty cm \citep[FIRST;][]{Becker1995}, and delves into the canonically `radio-quiet' population (see \S~\ref{sec4:radio-selection}. 
While this does not yet provide spatially matched information, LoTSS is currently undergoing post processing to provide sub-arcsecond resolution imaging of sources with $S_{144\textrm{MHz}}>10\,$mJy \citep{2022A&A...658A...1M}. Images where radio emission is spatially resolved can help provide information on the interaction between jets and multi-phase outflows \citep[e.g.,][]{Tadhunter2014,Girdhar2022,Venturi2023}. Advances in simulations are also helping provide direct observational predictions for different outflow driving mechanisms \citep[e.g.,][]{Meenakshi2024,Ward2024} 

The last decade has seen a shift towards more multi-wavelength studies which look at different outflow gas tracers in the bigger picture context, but still for relatively small samples. Large statistical studies are limited to one or two outflow tracers at a time. To really make progress on how outflows in different phases are related to each other and their origin, we need larger samples with spatially resolved information on multi-phase outflows, and more direct comparisons with simulations to try to unpick their underlying cause(s).



\subsection{Summary of the key drivers of progress}

From a theoretical point of view, the last decade has seen a significant transition from a simplified approach and understanding of AGN feedback, to more sophisticated models focused on understanding the  underlying physical mechanisms, advances driven by improvements in facilities and techniques. Early models and simulations implemented simplistic AGN feedback prescriptions as a required means to present for the formation of too many massive galaxies. However, advances in both understanding and simulation development has seen more focus on testing various physically-motivated AGN feedback prescriptions (e.g., radiative driving on dust, accretion disk winds, low power jets). Furthermore, the field is moving on from assuming two simplistic, and isolated,  modes of feedback (i.e., an ``ejective'' mode associated quasars and a ``preventative'' mode associated with radio lets from low accretion systems). It is now clear that mutliple feedback channels are possible, from a variety of different mechanisms, across different accretion rates. Furthermore, sophisticated modelling of AGN-driven outflows now suggest that energy-driven outflows are more relevant than momentum-driven outflows, with both accretion-disk winds and (low power, inclined) jets able to produce similar effects. 

The predicted dominant hot phase of outflows remains a challenge to test directly with observations (it is too tenuous); yet the last few years are seeing more attempts to use simulations to directly predict observational tracers of the sub-dominant colder phases (e.g., by incorporating additional models to enable predictions of emission lines from warm ionised and cold molecular gas phases). The general ``closing of the gap'' between observations and simulations is proving to be crucial for both understanding the limitations of observations, and testing specific AGN feedback models.

From an observational point of view, the explosion of multi-wavelength data, through a greater collection of data particularly IFU spectroscopy with new facilities (Gemini; VLT-MUSE; VLT-VIMOS), has driven progress. A key facet of these advances, is moving to a more complete census of typical (more common) AGN populations, such as those which are `radio quiet', facilitated by a new generation of radio surveys such as those undertaken by LOFAR. Another key facet, is the ability to obtain a detailed (sensitive and spatially-resolved) multi-phase picture of the ISM. Whilst still in the early days, we are already having a much better census of how AGN impact on the multi-phase ISM across a diversity of AGN samples (including extreme quasars, those hosting low power jets, and those with low power winds). Nonetheless, work is still required to understand how these measurements can definitely inform the long-term impact on the ISM in these host galaxies, as we overcome challenges with systematic uncertainties for moving from measured to derived properties. 

The last decade has also seen the community move on from assuming there is a simple ``smoking gun'' piece of evidence for suppressed star formation directly associated with luminous AGN, a conceptual advance. A greater understanding of the likely timescales involved has seen more work searching for indirect evidence of AGN feedback, such as studying the population of non-active quenched galaxies. Using such population-level observations of galaxy populations, to test specific predictions (which are not `tuned') from different AGN feedback models will be an important are of progress over the coming years.